%% file: thesis.tex
\documentclass[final,12pt]{macthesis} 

\usepackage[final,dvips]{graphicx} 	
\usepackage{longtable}			
\usepackage{bm}			
\usepackage{subfigure}
\usepackage{amsmath}
\usepackage{pgf}		

\usepackage{multirow}
\usepackage{array}
\usepackage{enumerate}
\usepackage{enumitem}
\usepackage{mdwlist}
\usepackage{tabu}
\usepackage{bold-extra}						
\usepackage[pagestyles,raggedright]{titlesec}
\usepackage{indentfirst}
\usepackage{afterpage}
\usepackage{epstopdf}

\usepackage[framed,numbered,autolinebreaks,useliterate]{mcode} 		

\usepackage{hyperref}
\hypersetup{
    linktocpage,
    colorlinks={true},
    citecolor=blue,
    linkcolor=blue,
}

\newpagestyle{main}[\small\scshape]{%
  \headrule
    \sethead[\thepage][][\chaptertitlename\enspace\thechapter.\quad\chaptertitle]{\thesection.\quad\sectiontitle}{}{\thepage}
    \sethead[\thepage][][]{\chaptertitlename\enspace\thechapter.\,\,\,\chaptertitle}{}{\thepage}
}
\renewcommand{\bibname}{References}  

\begin{document}

\bibliographystyle{ieeetr}

\newcommand{\C}[1]{#1}
\newcommand*{\Scale}[2][4]{\scalebox{#1}{$#2$}}      
\newcommand{\vv}[1]{\ensuremath{\mathbf{#1}}}        
\newcommand{\mx}[1]{\ensuremath{\mathbf{#1}}}        
\newcommand{\sinc}{{\rm sinc}}                       
\newcommand{\myeq}[1]{equation~(\ref{#1})}
\newcommand*\mycaption[3]{\caption[#1#3]{#1#2}}     

\newcommand*{\V}[1]{\vec{#1}}
\newcommand{\Eq}[2]{\begin{equation} \label{#1} #2 \end{equation}}
\newcommand{\Eqaa}[3]{\begin{eqnarray} \label{#1} #2 \nonumber\\ #3 \end{eqnarray}}
\newcommand{\Eqaaa}[4]{\begin{eqnarray} \label{#1} #2 \nonumber\\ #3 \nonumber\\ #4 \end{eqnarray}}
\newcommand{\Fig}[4]{\begin{figure}[!htb] \centering \includegraphics[width=#2\textwidth]{#3} \caption{#4} \label{#1} \end{figure}}

\include{thesis_frontmatter}

\include{Chap_Introduction}

\include{Chap_TFBG_0_Introduction}
\include{Chap_TFBG_1_Derive}
\include{Chap_TFBG_2_FD_num}

\include{Chap_TFBG_4_FD_conclussion}
\include{Chap_TFBG_5_Orthogonality}

\include{Chap_TFBG_CMT_1_Derive}

\include{Chap_TFBG_CMT_2_Grating}
\include{Chap_TFBG_CMT_3_Spectra}
\include{Chap_TFBG_CMT_4_Polarization}

\include{Chap_Exp_Opt_Setup_Polarization}

\include{Chap_Material_Metals}
\include{Chap_Material_Mixtures}

\include{Chap_Material_DDA}

\include{Chap_Material_Conclusion}

\include{Chap_NP_Lit}
\include{Chap_NP_Scat_Analit}
\include{Chap_NP_Results}

\include{Chap_Chem_Deposition}

\include{Chap_Chem_Deposition_Methods}

\include{Chap_Chem_Results}

\include{Chap_Summary}

\appendix{MatLab Code. The full vectorial complex mode solver.}
\include{Apendix_Code_MatLab}

\appendix{Mathematica Code for Mie scattering}
\include{Append_Code_Mie}

\newpagestyle{back}[\small\scshape]{%
  \headrule
    \sethead[\thepage][][]{\chaptertitle}{}{\thepage}
}  
\pagestyle{back}


\end{document}

%% file: thesis_frontmatter.tex
\begin{frontmatter}

\bigtitle{The sensitivity enhancement of tilted fibre Bragg grating sensors with polarization dependent resonant nano-scale coatings.}
\pagesinfront{xxxxx}
\pagesinbody{xxxxx}
\author{Aliaksandr Bialiayeu}
\degreeyear{2014}
\degree{Doctor of Philosophy}
\prevdegrees{}
\field{Electronics}

\maketitle

\begin{abstract}
\input{Abstract}
\end{abstract}

\begin{acknowledgements}
\input{Acknowledgments}
\end{acknowledgements}

\tableofcontents
\listoffigures
\listoftables

\begin{nomenclature}
\input{nomenclature}
\end{nomenclature}

\end{frontmatter}

%% file: Abstract.tex
\addcontentsline{toc}{chapter}{Abstract}
Fibre Bragg grating sensors have emerged as a simple, inexpensive, accurate, sensitive and reliable platform, a viable alternative to the traditional bulkier optical sensor platforms.
In this work we present an extensive theoretical analysis of the tilted fibre Bragg grating sensor (TFBG) with a particular focus on its polarization-dependent properties.

We have developed a highly efficient computer model capable of providing the full characterization of the TFBG device in less then $3$~minutes for a given state of incident light polarization. As a result, the polarization-dependent spectral response, the field distribution at the sensor surface as well as the fine structure of particular resonances have become accessible for theoretical analysis. 
As a part of this computer model we have developed a blazingly fast full-vector complex mode solver, capable of handling cylindrical waveguides of an arbitrary complex refractive index profile.

Along with the theoretical study we have investigated optical properties of the TFBG sensor with application to polarisation-resolved sensing. 
We proposed a new method of the TFBG data analysis based on tracking the grating transmission spectra along its principle axes, which were extracted from the Jones matrix.

In this work we also propose a new method of enhancing the TFBG sensor refractometric sensitivity limits, based on resonant coupling between the TFBG structure resonances and the local resonances of nanoparticles deposited on the sensor surface. The $3.5$-fold increase in the TFBG sensor sensitivity was observed experimentally.

%% file: Acknowledgments.tex
\addcontentsline{toc}{chapter}{Acknowledgments}
I would like to express my sincere gratitude to Dr. Jacques Albert for his support, patience, help, encouragement and guidance.  
He has shown me how to approach my work as a scientist and has reminded me of the high standards and quality of research.
Working with Dr. Jacques Albert had a profound impact on me as a scientist and as a person.
I wish to extend my most sincere thanks Dr. Anatoli Ianoul, I always knew I could count on his help and support.  

During my Ph.D program I was fortunate to build some true friendships. In particular, I am very grateful to Ksenia Yadav, Yanina Shevchenko, Nur Ahamad and Alec Millar who shared my everyday life. I would also like to acknowledge my colleagues. In particular, Alexander Andreyuk, Albane Laronche, Zahirul Alam and Nina Mamaeva. I am especially grateful to Lingyun Xiong for his willingness to help.

I am sincerely thankful to my parents who encouraged me throughout the years, and I deeply 
appreciate their support. 

Finally, I acknowledge the generous financial assistance from Carleton University and Institut national d’optique (INO). 
My past six years at Carleton University have been a particularly enjoyable part of my life.

%% file: nomenclature.tex

\begin{longtable}[l]{p{50pt} p{300pt}}
\textbf{Symbol}	& \textbf{Description} \\

$\vec{E}$, $\vec{H}$ & Electric and Magnetic fields\\
$\epsilon$, $\mu$ & Dielectric and magnetic permittivity\\
$c$ & Velocity of light in vacuum\\
$\omega$ & Angular Frequency\\
$\beta$ & Propagation constant\\
$n$ & Refractive index\\
$N_{eff}$ & Effective refractive index\\
$k$ & Wave propagating constant\\
$\nabla \cdot$ & The Divergence operator\\
$\nabla \times$ & The Curl operator\\
$\partial_t$  & Partial derivative with respect to the variable $t$\\
$\Delta \epsilon$ & Dielectric perturbation\\
$\Delta \beta$ & Phase mismatch\\
$\delta_{mn}$ & Dirac delta\\
$C_{mn}$ & Matrix coefficients of the coupling strength\\
$m$ & Mode azimuthal symmetry number\\
$\lambda$ & Wavelength\\
$L$ & Length of the fibre\\
$I$ & Intensity\\
$\rho_j$ & Eigenvalues\\
$\sigma_j$, $\V{u_j}$ & Principal values and corresponding principal vectors\\
$J_m$ & Bessel functions of the first kind\\
$J_l(r)$ & Spherical Bessel functions,\\ 
$N_l(r)$ & Spherical Neumann functions,\\ 
$H_l^{(1)}(r)$ & Spherical Hankel functions of the first kind\\
$H_l^{(2)}(r)$ & Spherical Hankel functions of the second kind\\

\end{longtable}

\subsection*{Acronyms}
\begin{longtable}[l]{p{50pt} p{200pt}}
AFM & Atomic Force Microscope\\
APTMS & ((3-aminopropyl)trimethoxysilane\\ 
CVD & Chemical Vapor Deposition\\
EM & Electromagnetic field \\
FBG & Fiber Bragg Grating\\
IR & Infrared Radiation \\
NIR & Near Infrared Radiation \\
LPG & Long Period Grating\\
RIU & Refractive Index Unit\\
ODE & Ordinary Differential Equation\\ 
OVA & Optical Vector Analyzer\\ 
SEM & Scanning Electron Microscope\\
SMF-28 & Single mode fiber (Corning SMF-28)\\
SNR & signal to noise ratio \\
SPR & Surface Plasmon Resonance\\
LSPR & Local Surface Plasmon Resonance\\ 
TFBG & Tilted Fiber Bragg Grating\\
TE & Transverse Electric\\
TM & Transverse Magnetic\\
UV & Ultraviolet Radiation\\

\end{longtable}

%% file: Chap_Introduction.tex
\chapter{Introduction}
In the presented work we study the tilted fibre Bragg grating (TFBG) sensor and present new methods of improving its sensitivity.
The TFBG sensor has proven to be a simple, inexpensive, accurate, sensitive, and reliable platform~\cite{Albert:2012}, with a variety of applications, including chemical and biological sensing~\cite{Shevchenko:2007, Homola:2008}. 

The TFBG sensor is based on a standard telecommunication fibre with a tilted grating inscribed inside the core. 
The telecommunication fibre is an optical cylindrical waveguide consisting of a high refractive index  core surrounded by a cladding layer with a lower refractive index. 
The light propagating in the fibre core is confined by the total internal reflection phenomenon.

In our research the 1-cm-long TFBGs were inscribed in the hydrogen-loaded core of a standard telecom single mode optical fibre (Corning SMF-28) using the phase mask technique and intense ultraviolet light (pulsed KrF excimer laser at $193~nm$ or $248~nm$).
The TFBG sensor coated with small spherical particles is shown schematically in Figure~\ref{TFBG_introduction}.
\Fig{TFBG_introduction}{0.6}{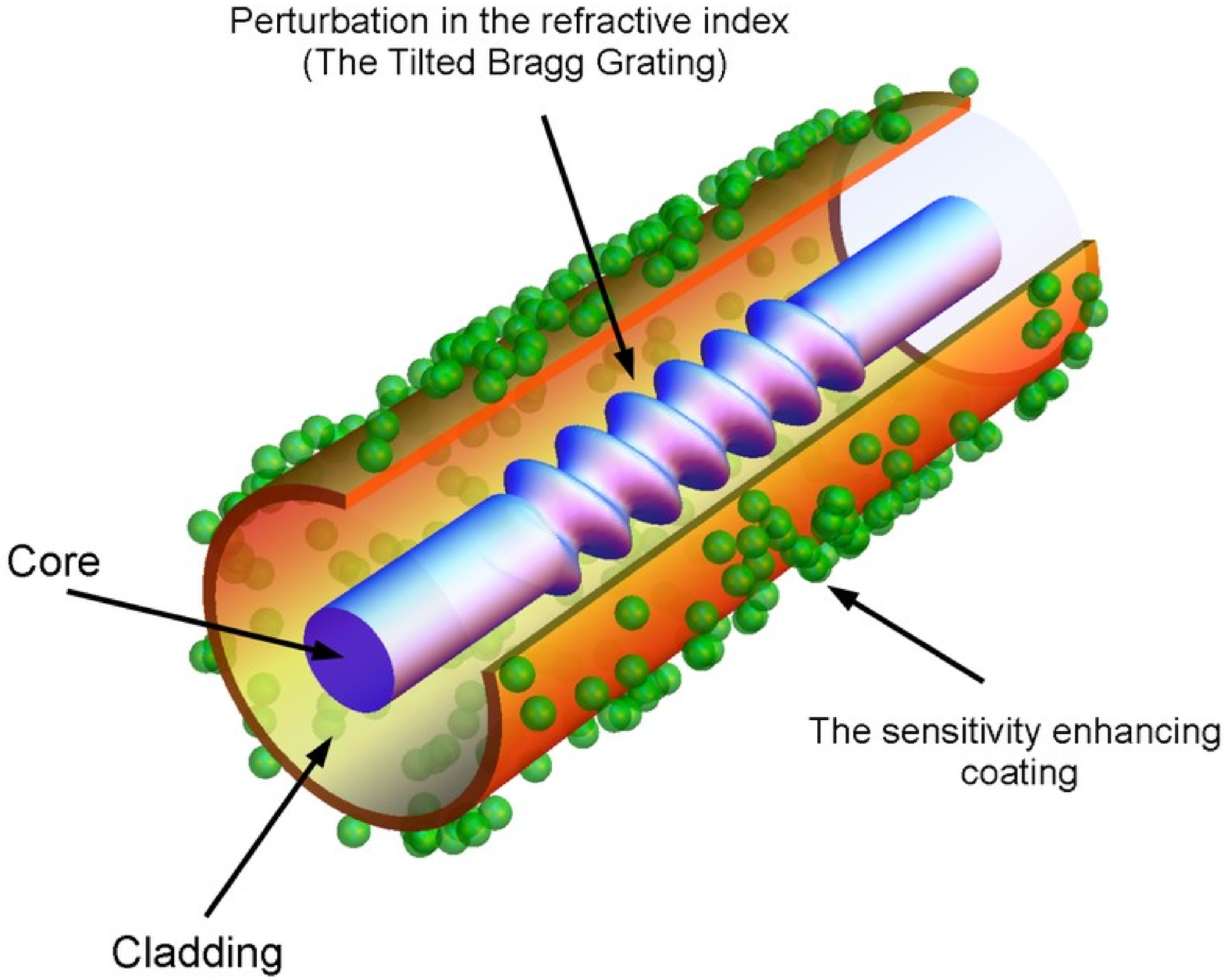}
{Schematic representation of the TFBG sensor coated with sensitivity enhancing layer of nanoparticles.}

The inscribed grating planes are slightly tilted relative to the fibre cross-section, which allows to couple the forward-propagating light from the fibre's core to the backward-propagating cladding modes~\cite{Lee_00, Laffont:2001}.
More than a thousand propagating cladding modes are usually excited.
Each cladding mode can be viewed as a propagating electromagnetic wave, with a unique wave vector and field distribution, thus interacting differently with the outside medium.

The coupling between the core and the cladding modes, mediated by the grating, produces discrete narrow attenuation bands in the transmitted spectrum (the \C{``resonances''}) and since the cladding modes are guided by the fibre-surrounding medium interface, these resonances are extremely sensitive to the refractive index of the surrounding medium~\cite{Chan:2007}.
The energy coupled to a particular cladding mode as well as the corresponding propagation constants can be precisely measured by acquiring the transmission spectrum of the grating. The spectrum is shown in Figure~\ref{TFBG_introduction_spectr}, consisting of a set of narrow resonances sensitive to the refractive index change of environment.
\Fig{TFBG_introduction_spectr}{0.8}{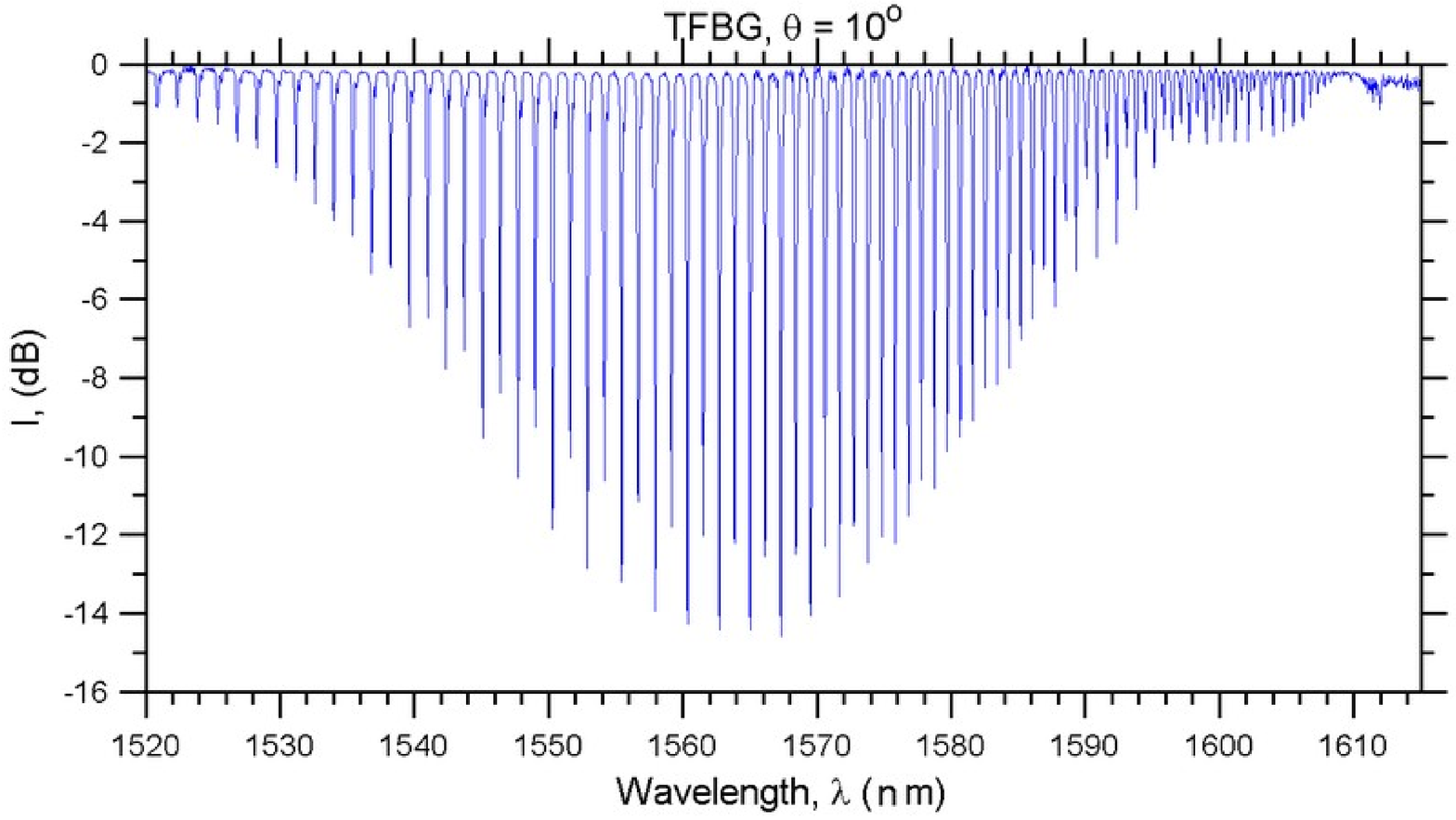}
{A typical spectrum of a $10$ degree TFBG sensor.}

The TFBG sensor has a unique set of advantages, including the possibility to isolate temperature effects and develop compact and robust refractometers with a wide operating range~\cite{Chan:2007, Guo:2009}. 
In addition, the simultaneous probing of the external medium at various wavelengths with modes having unique polarization properties and incidence angles is possible.

The sensitivity of the TFBG sensor can be further improved if modes of the sensor can be coupled to an external resonant system, such as a nano-scale coating, with properties sensitive to the surrounding medium. 
The cladding modes excited by \C{the TFBG structure} have a non-zero evanescent field at the cladding boundary, therefore the light can tunnel outside the fibre into a coating layer. 
If the phase matching condition is met the resonant coupling between the cladding modes and resonances of the coating layer can occur.  
When some modes (among the large number of cladding modes accessible) are phase matched to the surface plasmon resonance (SPR) of the outer surface of the metal coating, the light energy incident from these cladding modes is efficiently coupled to the surface plasmon-polariton oscillations~\cite{Shevchenko:2007}. 
The energy coupled to the SPR causes a drastic change in the TFBG transmission spectrum at wavelengths corresponding to the resonant modes.
The principle of operation of the TFBG SPR sensor is analogous to the well-known Kretschmann-Raether setup~\cite{Matsubara:1988, Raether:1967}, except that the cladding modes use wavelength and angle scanning to probe for the SPR phase matching condition (because each cladding mode couples at a different wavelength and \C{``strikes''} the cladding boundary at a different incident angle).

Finally, a further increase of the selectivity of the modes coupling to SPR is provided by controlling the light polarization.
This discovery led the way to a strong increase in the accuracy with which we can follow SPR shifts associated with small refractive index changes of the outer medium surrounding the fibre by using the polarization-dependent loss (PDL) of the TFBG transmission~\cite{Caucheteur:11}. In particular, it was found that narrowband resonances (100 pm spectral bandwidth (BW)) with refractometric sensitivities (S) of ${350~nm/RIU}$ (refractive index units) showed up in the PDL spectrum.
These resonances have some of the highest figures of merit (${S/BW = 3500~RIU^{-1}}$) for SPR sensors of any kind~\cite{Roh:2011}. 

Several resonances in the transmission spectrum are shown in Figure~\ref{TFBG_introduction_deposition}. 
The resonances were continuously observed during the process of nano-scale metal coating deposition.
The complicated structure of the resonances, with strong dependence on the film thickness, is proportional to the time of the deposition is clearly seen. 
\Fig{TFBG_introduction_deposition}{0.9}{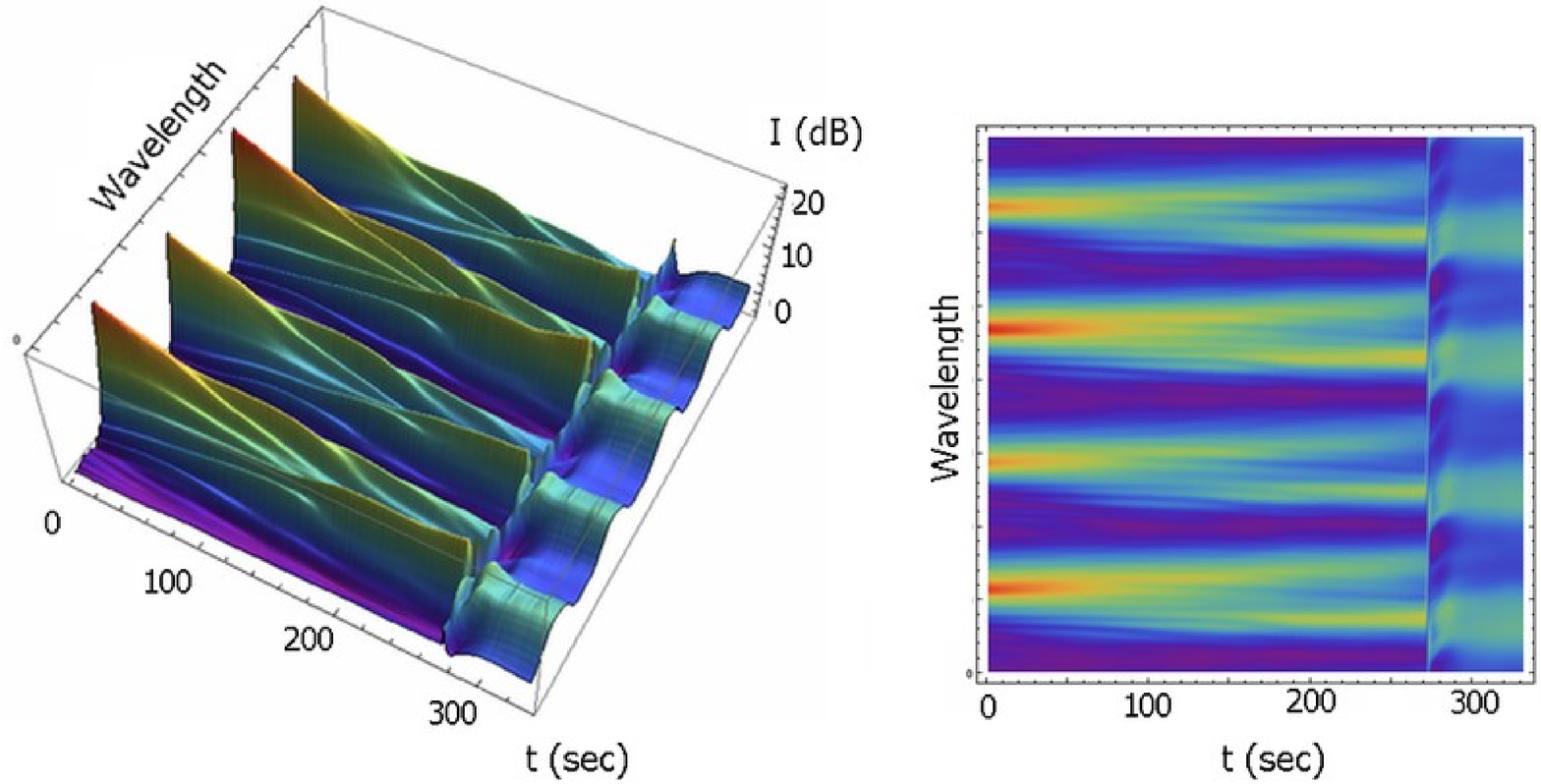}
{Evolution of the TFBG spectral response during the silver \C{nanoparticle} deposition followed by the 
continues film formation.}

Recently it was discovered that the sensitivity of TFBG sensors can be improved by coating its surface with randomly oriented silver nanowires. 
Such a coating would allow for a large number of cladding modes to interact with the deposited nanowires, exciting localized surface plasmon resonances (LSPR) when the appropriate phase matching condition is met. 
The sensitivity of the TFBG resonances to the external refractive index change was increased by a factor of~$3.5$ even though the surface coverage of the nanowires was less than $14\%$.

In the presented work we investigate methods of improving the TFBG sensor sensitivity by coating its surface with various types of nano-scale films, including metal films and nanoparticle coatings. 
We also discuss the polarization properties of the sensor and present the developed highly sensitive measurement technique.   

However, the main objective of our work was to simulate the behaviour of the TFBG sensor. 
The sensor is an interesting object with more than a thousand interacting modes and a complex polarization-dependent response, as can be seen from 	ures~\ref{TFBG_introduction_spectr} and~\ref{TFBG_introduction_deposition}. 
The simplicity of the spectral response hides many non-trivial physical effects that we might wish to decode. 
Our goal here was to understand the TFBG sensor behaviour, interpret the polarization-dependent spectral response and create a computer model which would allow us to predict the outcome of experiments.
\C{Moreover}, the process of improving the sensor sensitivity required the rather tedious experimental work of scanning the parameter space by \C{probing particles} with different geometrical shapes, such as spheres, cubes, cages and wires, varying the nanoparticle's size and the deposition densities. 
A theoretical guidance which would allow the prediction of an optimal set of parameters, \C{such as the film material, thickness}, composition and morphology is highly desirable. 

\chapter*{The thesis Organization}
This work is organized in the following manner. 
Chapter~1 provides an introduction to the problem and motivation.

In Chapter~2 we explain how we developed a full-vector complex mode solver which we applied to circularly symmetric optical waveguides in order to compute scalar and vectorial modes. 
The exact and approximate solutions were compared and the orthogonality of the modes was verified. 
\C{An interesting analogy} between TE and TM modes in slab waveguides and modes in cylindrical structures was established. 
The analysis of a \C{weakly-guided} approximation and the exact solution allowed us to build a connection between the scalar and vectorial equations. The split in eigenvalues was discussed and analyzed thoroughly. To our knowledge, this particular analysis has not been reported previously.

Although a number of commercial mode solver software packages is available, the software interfaces usually impose certain limitations on the process automation. For example, to simulate a TFBG spectral response more than a thousand interacting modes should be considered. These modes should be computed, often an various frequencies, stored and be available for further processing.
If the optimal parameters are searched, a number of separate computations must be done sequentially. In addition, in order to study a particular property of the sensor only the modes with this particular property are required, thus these modes can be targeted discriminately without computation of the remaining modes. Thus, by targeting only specific modes the simulation speed can be increased dramatically.

Chapter~3 is based upon solutions obtained in Chapter~2. The modes obtained for the non perturbed case were used as basis functions to solve the problem of tilted Bragg grating structure. The polarization-dependent effects were investigated. The results of the simulations were compared with the experimental measurements.
\C{The emphasis} was set on polarization dependency of the coupling coefficients, which were computed for all possible angles of incident core mode polarization, to our knowledge this result has not been previously reported.

Chapter~4 provides a complete experimental characterization of the TFBG structure. The experimental measurement techniques with application to polarization-based sensing were developed.

In Chapter~5 we start our discussion on the possible enhancement of TFBG sensor sensitivity by coating its surface with nano-scale films. 
We review optical properties of various materials, metals in particular.

Chapter~6 provides a detailed overview of nanoparticle optical properties, as well as several simulation methods, in particular the Mie theory. 

In Chapter~7 we present our ideas on enhancing the sensitivity of TFBG sensors, and conduct a search for parameters that would provide the optimal nanoparticle-based coating.

Chapter~8 presents experimental results from various nano-scale film coatings and their associated sensitivity enhancements.
 
Lastly, Chapter~9 is a summary of the work findings.


%% file: Chap_TFBG_0_Introduction.tex
\chapter{A full-vector complex mode solver for circularly symmetric optical waveguides}
\chaptermark{A Full-vector complex mode solver}
\label{Mode_Solver}

\section{Introduction}

In this chapter we present a simple yet efficient and exact approach for fast and accurate modeling of waveguides with cylindrical or elliptical symmetry, with arbitrary real and imaginary refractive index profiles.

The analytical solution for the simplest case of a cylindrical dielectric waveguide, as well as modes classification and field plots were presented by Snitzer in 1961~\cite{snitz:61}. 
Shortly after, direct numerical integration techniques were presented, in a noticeable paper published by Dil and Blok~\cite{dil:1973} on a numerical solution of four coupled Maxwell's equation in radial parabolic dielectric waveguides.

It is well known that in the case of waveguide with a small refractive index contrast the weakly guided approximation can be used, allowing the system of Maxwell's to be reduced to the Helmholtz equation, which can more easily be solved~\cite{yam:97, Yen-Chung:1997}. Here we consider the general problem of full vectorial mode solutions of waveguides with high refractive index contrasts and a non-zero imaginary part of the refractive index,~\textit{i.e.} we consider an arbitrary complex permittivity profiles.

Several methods have been proposed earlier. 
Usually the waveguide profile is subdivided into a number of piecewise homogeneous concentric layers, and the system of Maxwell's equations are solved in each layer. 
The solutions are next connected with the help of boundary conditions for the tangential electric and magnetic fields at the layer boundaries.

In cylindrical coordinates the solutions for each layer can be represented in terms of Bessel and modified Bessel functions, and then connected through a 4x4 transfer matrix, which incorporates the boundary conditions. This is the so-called transfer matrix method proposed by Yeh and Lindgren~\cite{Yeh:1977}.
The modes propagation constant are obtained by finding the roots of a polynomial, over Bessel functions, with a degree proportional to the number of layers. 
This method becomes increasingly prohibitive with regards to computational resources as well as to numerical stability with the increase in number of layers. 
The transfer matrix method becomes even more complicated if the permittivity posses an imaginary part or the modes are leaky. In such a case the roots have to be searched on the complex plane.

Although its limitations, the transfer matrix method can be successfully applied to waveguide structures consisting only of a few uniform layers~\cite{Tsao:89, Uwe:2009, Monerie:1982} or to structures consisting of infinitely many alternating uniform layers, the so-called Bragg fibres~\cite{Yeh:78}. The computational complexity of the matrix method can be reduced by replacing Bessel functions with their asymptotic expressions, or if the periodic cylindrical layers are approximated as planar Bragg stacks~\cite{Xu:00}. 

Another possibility is to use the pseudospectral method, also called the spectral collocation method, where a solution is searched in terms of sine functions~\cite{Marcuse:1992}, Chebyshev-Lagrange~\cite{PoJuiChiang:2008, Song:10}, Laguerre-Gauss~\cite{Huang:2003, Shangping:04}, Hermite-Gauss~\cite{Weisshaar:1995} interpolating polynomials, or in terms of some other suitable basis functions.
The pseudospectral method can be implemented either for the entire domain or separate basis functions can be chosen for each uniform layer. In each case the basis functions are chosen with regards to boundary conditions. If the problem is solved for the entire domain the approximation of mode fields might require hundreds or even thousands of functions, and the convergence is generally a problem~\cite{Shangping2:04}. 
The multidomain method is in a way similar to the transfer matrix method. The dielectric interface conditions should be fulfilled, thus the functions between different layers are again connected through the transfer matrix at the boundaries~\cite{Song:10}.
In both cases the system of algebraic equations results in numerical eigenvalue problem, with eigenvectors consisting of the expansion coefficients for a particular mode.  
A simpler but more resource hungry method was proposed in~\cite{Chiang:02, Horikis:06} where instead of the global functions for each homogeneous region a finite different method was applied, and  then the transverse electric or magnetic fields were matched at the radial index discontinuities. The comparison between the transfer matrix method and the pseudospectral method was conducted in~\cite{Shangping2:04, Song:10} and definitely favours the pseudospectral method.

Yet another possibility is to use a standard finite difference (FDM) or finite element (FEM) method for the entire 2D waveguide profile, with the Yee's cell~\cite{Yee:1966} adapted for the cylindrical symmetry~\cite{Yeh:1975, Saitoh:2002, Zhang:2005}.
A number of commercial simulation software based on the finite difference and finite element methods are available.
This approach does not require understand of engineering or physical aspects of the problem. A given problem is viewed as a \C{``blackbox''} accepting the input data~\textit{e.g.} geometrical shape, an operational wavelength,~\textit{etc.} and returning the required numerical results. 
However, the use of commercial software imposes certain limitations, for example it might be challenging to automate the process of obtaining a series of solutions for a varying parameter. 
It should also be noted that the FDM and FEM method are the slowest and the most resource hungry, especially if used for 2D or 3D problems. 

Here we present a simple yet efficient and fast numerical method. First the system of Maxwell's equations was reduced to only 2 coupled ordinary differential equations for the electric field. The variation of the dielectric permittivity was incorporated in the equation such that equations become suitable for numerical integration through the entire domain, in contrary to the piecewise homogeneous methods described above. Next the equations were transformed into a system of algebraic equations with the help of a finite difference method. Finally the problem was reduced to finding eigenvalues and eigenvectors of a five-diagonal matrix. The eigenvalue problem was effectively solved with the standard iterative method commonly applied to large sparse linear systems. 
The proposed method requires less grid points if the refractive index discontinuity is approximated with a smooth function at an interval of about $1~nm$ at the position of discontinuity.

%% file: Chap_TFBG_1_Derive.tex
\section{The solutions for a cylindrical waveguide}
We start with Maxwell's equations written in the MKS system of units, for a source \C{free region}~\cite{Jackson:1998}:
\begin{subequations}
\begin{align}
\label{eq_f_Mxs1a}
        \nabla \times \V H(\V r,t) &= \frac{\epsilon_o \epsilon(\V r)}{c} \partial_t \V E(\V r,t),\\
\label{eq_f_Mxs1b}
        \nabla \times \V E(\V r,t) &= - \frac{\mu_o}{c} \partial_t \V H(\V r,t),
\end{align}
\end{subequations}
\begin{subequations}
\begin{align}
\label{eq_f_Mxs2a}
        \nabla \cdot \V H(\V r,t) &= 0,\\
\label{eq_f_Mxs2b}
        \nabla \cdot \left( \epsilon(\V r) \V E(\V r,t)\right) &= 0.
\end{align}
\end{subequations}

Assuming that fields $\V E(\V r,t)$ and $\V H(\V r,t)$ are varying harmonically with time, \textit{i.e.} proportional to the oscillating term $e^{i \omega t}$, we can take the time derivative and next apply the curl operator to both equations (\ref{eq_f_Mxs1a}) and (\ref{eq_f_Mxs1b}). 
Next plugging ${\nabla \times \V H}$ term into second equation~(\ref{eq_f_Mxs1b}) and ${\nabla \times \V E}$ term into first equation~(\ref{eq_f_Mxs1a}) we split Maxwell's equations into two subsystems, separating ${\V E}$ and ${\V H}$ fields:
\Eqaa{eq_f_Mxs3}
	{\nabla \times \nabla \times \V E(\V r) &=& \epsilon(\V r) k_o^2 \V E(\V r),}
	{\nabla \cdot \left( \epsilon(\V r) \V E(\V r)\right) &=& 0,}
and
\Eqaa{eq_f_Mxs4}
	{\nabla \times \frac{1}{\epsilon(\V r)} \nabla \times \V H(\V r) &=& k_o^2 \V H(\V r),}
	{\nabla \cdot \V H(\V r) &=& 0,}
where ${k_o = \frac{\omega}{c}} = \frac{2 \pi}{\lambda}$, and ${\omega = \frac{1}{\sqrt{\epsilon_o \mu_o}}}$.

Each system of equations~(\ref{eq_f_Mxs3}) and~(\ref{eq_f_Mxs4}) contains the full description of electromagnetic field. We can either use system~(\ref{eq_f_Mxs3}) to find $\V E$ field and next express field $\V H$ in terms of ${\nabla \times \V E}$ by using~(\ref{eq_f_Mxs1b}), or alternatively, the field $\V H$ can be obtained from equation~(\ref{eq_f_Mxs4}) and field $\V E$ can be found in terms of ${\nabla \times \V H}$ with help of~(\ref{eq_f_Mxs1a}).

Here we prefer to use system~(\ref{eq_f_Mxs3}) over~(\ref{eq_f_Mxs4}) to avoid complications caused by the ${\nabla \times \frac{1}{\epsilon(\V r)} \nabla \times \V H(\V r)}$ term.

Now considering the vector identity~\cite{Arfken:1985}:
\Eq{}
{\nabla \times \nabla \times \V E = \nabla (\nabla  \V E) - \nabla^2 \V E,}
and expansion of the Gauss equation~(\ref{eq_f_Mxs2b}):
\Eq{}
{\nabla \left( \epsilon \V E \right) = \epsilon \nabla \V E + \V E \cdot \nabla \epsilon = 0,}
we obtain the following identity:
\Eq{}
{\nabla \times \nabla \times \V E = - \nabla \left(\V E \cdot \frac{\nabla \epsilon}{\epsilon}\right) - \nabla^2 \V E,}
Thus system~(\ref{eq_f_Mxs3}) can be written as:
\Eq{eq_f_Mxs5}
{\nabla^2 \V E + \epsilon k_o^2 \V E = -  \nabla \left(\V E \cdot \frac{\nabla \epsilon}{\epsilon}\right)}
The equation~(\ref{eq_f_Mxs5}) is all that is needed to describe electromagnetic waves in inhomogeneous nonmagnetic media. Once the electric field $\V E$ is found the corresponding magnetic field can be expressed in terms of ${\nabla \times \V E}$ as follows from~(\ref{eq_f_Mxs1b}) equation.

\subsection{Weakly guided approximation}
Let us assume that the rate at which function ${\epsilon(\V r)}$ is changing \C{with the change} of $\V r$ coordinate is insignificant, \textit{i.e.} ${\nabla \epsilon(\V r)}$ is smaller than $\epsilon(\V r)$. Then the right hand side of equation~(\ref{eq_f_Mxs5}) can be neglected~\cite{Snyder:80}:
\Eq{eq_f_Mxs6}
{\nabla^2 \V E + \epsilon(\V r) k_o^2 \V E = \V 0}
This is the so-called weakly guided approximation. The equation~(\ref{eq_f_Mxs6}) provides full description of electromagnetic waves in the case of small ${\nabla \epsilon(\V r)}$.

Let us write equation~(\ref{eq_f_Mxs6}) in cylindrical coordinates: 

\Eq{eq_f_Mxs6_cyl}{
\begin{pmatrix}
	\nabla^2 - \frac{1}{\rho^2} + \epsilon k_o^2 & - \frac{2}{\rho^2}\partial_\phi & 0\\
	\frac{2}{\rho^2}\partial_\phi & \nabla^2 - \frac{1}{\rho^2} + \epsilon k_o^2 & 0\\
	0 & 0 & \nabla^2 + \epsilon k_o^2
\end{pmatrix}
\begin{pmatrix}
E_\rho\\
E_\phi\\
E_z
\end{pmatrix} = \V 0.
}
Here $\nabla^2 \psi$ is the Laplacian for a scalar function $\psi$ in cylindrical coordinates:
\Eq{}
{\nabla^2 \psi = \left(\partial_\rho^2 + \frac{1}{\rho} \partial_\rho + \frac{1}{\rho^2} \partial_\phi^2 + \partial_z^2\right)\psi,}
and $\nabla^2 \V A$ is the vector Laplacian in cylindrical coordinates:
\Eq{eq_vecLap}
{\nabla^2 \V A = \left(\frac{1}{\rho} \partial_\rho(\rho \partial_\rho) + \frac{1}{\rho^2} \partial_\phi^2 + \partial_z^2\right) \left( A_\rho \hat \rho + A_\phi \hat \phi + A_z \hat z \right),}
The equation~(\ref{eq_f_Mxs6_cyl}) can be obtained \C{from}~(\ref{eq_vecLap}) by taking into account that unit vectors themselves are functions of coordinates, and noting that in cylindrical coordinates there are two nonzero derivatives of the unit vectors: ${\partial_\phi \hat \rho = \phi}$ and ${\partial_\phi \hat \phi = - \rho}$, the remaining derivatives are zero.

We note that equation~(\ref{eq_f_Mxs6_cyl}) decouples into two independent equations,\\ 
the vectorial equation:
\Eq{eq_cyl_vec}{
\begin{pmatrix}
	\nabla^2 - \frac{1}{\rho^2} + \epsilon k_o^2 & - \frac{2}{\rho^2}\partial_\phi\\
	\frac{2}{\rho^2}\partial_\phi & \nabla^2 - \frac{1}{\rho^2} + \epsilon k_o^2\\
\end{pmatrix}
\begin{pmatrix}
E_\rho\\
E_\phi\\
\end{pmatrix} = \V 0,
}
and the scalar equation: 
\Eq{eq_cyl_scal}
{ \nabla^2 E_z + \epsilon k_o^2 E_z = 0.}

Here we are interested in circularly symmetric optical waveguides, assuming that the waveguide structure is uniform along the $z$ coordinate, and searching for a periodic solution in $\phi$,  \textit{i.e.} ${u(\rho,\phi,z) \sim u(\rho) e^{j\beta z} e^{j m \phi}}$, we can can replace $\partial_z$ with $j \beta$ and $\partial_\phi$ with $j m$. Thus the equations~(\ref{eq_cyl_vec}) and~(\ref{eq_cyl_scal}) can be written in form of the vector eigenvalue problem:
\Eq{eq_cyl_vec_eig}{
\begin{pmatrix}
d_\rho^2 + \frac{1}{\rho} d_\rho - \frac{m^2+1}{\rho^2} + \epsilon k_o^2 & - \frac{j2m}{\rho^2}\\
\frac{j2m}{\rho^2} & d_\rho^2 + \frac{1}{\rho} d_\rho - \frac{m^2+1}{\rho^2} + \epsilon k_o^2
\end{pmatrix}
\begin{pmatrix}
E_\rho\\
E_\phi\\
\end{pmatrix}
= \beta^2 \begin{pmatrix}
E_\rho\\
E_\phi\\
\end{pmatrix},
}
and the scalar eigenvalue problem:
\Eq{eq_cyl_scal_eig}
{\left(d_\rho^2 + \frac{1}{\rho} d_\rho - \frac{m^2}{\rho^2} + \epsilon k_o^2\right)E_z = \beta^2 E_z.}
Here $\lambda = \beta^2$ are unknown eigenvalues and $\beta$ is the propagation constant.

The equation~(\ref{eq_cyl_scal_eig}) is well known and is often considered for the weakly guided approximation case while the equation~(\ref{eq_cyl_vec_eig}) is rarely used, see for example~\cite{Black:2010, Su:1986,Lu:08}. 

Once any of these two equations are solved, the remaining components of the electric field $\V E$ can be found from Gauss's law~(\ref{eq_f_Mxs2b}), and next the magnetic field~$\V H$ can expressed in terms of electric field with use of equation~(\ref{eq_f_Mxs2b}).

\subsection{The exact solution for cylindrical waveguides}
The exact solution can be obtained if term ${\nabla \epsilon}$ is no longer neglected. The electromagnetic field in such a case is described by equation~(\ref{eq_f_Mxs5}):
\Eq{eq_f_Mxs7}
{\nabla^2 \V E + \epsilon k_o^2 \V E = -  \nabla \left(\V E \cdot \frac{\nabla \epsilon}{\epsilon}\right).}
Let us assume that dielectric permittivity varies only along the the radial direction ${\epsilon = \epsilon(\rho)}$, then
\small
\Eqaaa{}
{\V E \cdot \frac{\nabla \epsilon}{\epsilon} &=& E_\rho \frac{\partial_\rho \epsilon(\rho)}{\epsilon(\rho)} + E_\phi \frac{1}{\rho} \frac{\partial_\phi \epsilon(\rho)}{\epsilon(\rho)} + E_z \frac{\partial_z \epsilon(\rho)}{\epsilon(\rho)} =}  
{&=& E_\rho \frac{\partial_\rho \epsilon(\rho)}{\epsilon (\rho)} =} 
{&=& E_\rho \cdot (\ln \epsilon (\rho))',}
and
\Eq{}{
\nabla  \left(\V E \cdot \frac{\nabla \epsilon}{\epsilon}\right) = \nabla \left( E_\rho (\ln \epsilon )'\right) = \begin{pmatrix}
(\ln \epsilon )''E_\rho + (\ln \epsilon )'\partial_\rho E_\rho\\
(\ln \epsilon )' \frac{1}{\rho} \partial_\phi E_\rho\\
(\ln \epsilon )' \partial_z E_\rho
\end{pmatrix}.
}
\normalsize
Now equation~(\ref{eq_f_Mxs7}) can be rewritten in cylindrical coordinates:
\small
\Eq{}{
\begin{pmatrix}
	\nabla^2 - \frac{1}{\rho^2} + \epsilon k_o^2 + (\ln \epsilon )'' + (\ln \epsilon )'\partial_\rho& - \frac{2}{\rho^2}\partial_\phi & 0\\
	\frac{2}{\rho^2}\partial_\phi + (\ln \epsilon )' \frac{1}{\rho} \partial_\phi & \nabla^2 - \frac{1}{\rho^2} + \epsilon k_o^2 & 0\\
	(\ln \epsilon )' \partial_z & 0 & \nabla^2 + \epsilon k_o^2
\end{pmatrix}
\begin{pmatrix}
E_\rho\\
E_\phi\\
E_z
\end{pmatrix} = \V 0.
}
\normalsize
The first two equations can again be separated, as they do not depend on the $E_z$ component.

Replacing $\partial_z$ with $j \beta$ and $\partial_\phi$ with $j m$ as previously, we obtain a slightly different from of equation~(\ref{eq_cyl_vec_eig}). The vectorial eigenvalue problem~(\ref{eq_cyl_vec_eigexact}), now depends on the rate of change of dielectric permittivity ${\epsilon'(\rho)}$:

\Eq{eq_cyl_vec_eigexact}
{\Scale[0.87]{
\begin{pmatrix}
d_\rho^2 + \frac{1}{\rho} d_\rho + (\ln \epsilon )' d_\rho - \frac{m^2+1}{\rho^2} + \epsilon k_o^2 + (\ln \epsilon )'' & - \frac{j2m}{\rho^2}\\
\frac{j2m}{\rho^2} + \frac{jm}{\rho}(\ln \epsilon )' & d_\rho^2 + \frac{1}{\rho} d_\rho - \frac{m^2+1}{\rho^2} + \epsilon k_o^2
\end{pmatrix}\begin{pmatrix}
E_\rho\\
E_\phi\\
\end{pmatrix}
= \beta^2 \begin{pmatrix}
E_\rho\\
E_\phi\\
\end{pmatrix}
}
}
Here we note again that terms ${\epsilon'(\rho)}$ were previously ignored in the weakly guided approximation~(\ref{eq_cyl_vec_eig}). 
In the following section we compare the results following from the weakly guided approximation~(\ref{eq_cyl_vec_eig}) and from the exact formulation~(\ref{eq_cyl_vec_eigexact}).

Finally, once the fields~$E_\rho$ and~$E_\phi$ are known from equation~(\ref{eq_cyl_vec_eigexact}), the remaining~$E_z$ component can be found from the Gauss's equation~(\ref{eq_f_Mxs2b}) written in cylindrical coordinates:
\Eq{}
{E_z = - \frac{1}{j\beta} \left(\frac{1}{\epsilon \rho} \frac{d}{d \rho} (\rho \epsilon E_\rho) + \frac{jm}{\rho} E_\phi\right),}
and the magnetic field $\V H$ can now be expressed in terms of electric field ${\V E = (E_\rho, E_\phi, E_z)}$ with help of equation~(\ref{eq_f_Mxs2b}):
\Eq{}
{\V H = j \frac{c}{\omega \mu_o} \nabla \times \V E.}

\clearpage
\subsection{TE and TM modes in slab waveguides}

Here we show that our approach is valid, by deriving the well known equation for slab waveguides.
Again, let us start with equation~(\ref{eq_f_Mxs5}):
\small
\Eq{eq_f_Mxs10}
{\nabla^2 \V E + \epsilon k_o^2 \V E = -  \nabla \left(\V E \cdot \frac{\nabla \epsilon}{\epsilon}\right).}
Assuming that the slab waveguide is uniform along the~$y$ and $z$ axis, the dielectric permittivity can only be function of the~$x$ coordinate~${\epsilon = \epsilon(x)}$, thus equation~(\ref{eq_f_Mxs10}) can be written in Cartesian coordinate system as follows:
\Eq{eq_f_Mxs11_cyl}{
\begin{pmatrix}
	\nabla^2 + \epsilon k_o^2 & 0 & 0\\
	0 & \nabla^2 + \epsilon k_o^2 & 0\\
	0 & 0 & \nabla^2 + \epsilon k_o^2
\end{pmatrix}
\begin{pmatrix}
E_x\\
E_y\\
E_z
\end{pmatrix}
= - \begin{pmatrix}
(\ln \epsilon )''E_x + (\ln \epsilon )'\partial_x E_x\\
(\ln \epsilon )' \partial_y E_x\\
(\ln \epsilon )' \partial_z E_x
\end{pmatrix}.
}
\normalsize
Considering the waveguide symmetries, uniformity along the~$y$ and $z$ axis, and assuming that the wave is propagating along the~$z$ axis, we can can replace $\partial_z$~with~$j \beta$ and $\partial_y$~with~$0$.
The scalar Laplacian~$\nabla^2 \psi$ can now be written in the following form:
\Eqaa{}
{\nabla^2 \psi &=& \left(\partial_x^2 + \partial_y^2 + \partial_z^2\right)\psi =}
{&=& \left(d_x^2 - \beta^2\right)\psi,}
and equation~(\ref{eq_f_Mxs11_cyl}) can be rewritten in the the form of vectorial eigenvalue problem:
\small
\Eq{eq_f_Mxs11_cyl2}{
\begin{pmatrix}
	d_x^2 + (\ln \epsilon )'d_x + \epsilon k_o^2 + (\ln \epsilon )''& 0 & 0\\
	0 & d_x^2 + \epsilon k_o^2 & 0\\
	(\ln \epsilon )' \beta & 0 & d_x^2 + \epsilon k_o^2
\end{pmatrix}
\begin{pmatrix}
E_x\\
E_y\\
E_z
\end{pmatrix}
= \beta^2 \begin{pmatrix}
E_x\\
E_y\\
E_z
\end{pmatrix}.
}
\normalsize
The first 2 equations are not coupled and can be written separately:
\Eq{eq_f_TM}
{\left(d_x^2 + (\ln \epsilon )'d_x + (\ln \epsilon)'' + \epsilon k_o^2\right)E_x = \beta^2 E_x}
\Eq{eq_f_TE}
{\left(d_x^2 + \epsilon k_o^2\right)E_y = \beta^2 E_y}
The equation~(\ref{eq_f_TE}) is well known equation for the TE modes in slab waveguide~\cite{kats:2006, Chin-Lin:2007}.

The equation~(\ref{eq_f_TM}) can be recognized as equation for TM modes if we rewrite it in a slightly different form: 
\Eq{eq_f_TM2}
{\left( \epsilon d_x \frac{1}{\epsilon} d_x + \epsilon k_o^2 \right)(\epsilon E_x) = \beta^2 (\epsilon E_x),}
where we have considered that
\Eq{}
{d_x \frac{1}{\epsilon} d_x (\epsilon u) = \left(\frac{1}{\epsilon}(\epsilon u)'\right)' = \left(u' + \frac{\epsilon'}{\epsilon} u \right)' = u'' + (\ln \epsilon)' u' + (\ln \epsilon)'' u}

We note that in the case of slab waveguide the system of equations~(\ref{eq_f_Mxs11_cyl2}) for the transverse field components~$E_x$ and~$E_y$ decouples into separate equations, independently describing TE and TM modes, unlike in the case of cylindrical waveguide where the system of equations~(\ref{eq_cyl_vec_eigexact}) is coupled.

We also observe that in the case of weakly guided approximation the equations for TM and TE mode~(\ref{eq_f_TM2},\ref{eq_f_TE}) become identical, as 
\Eq{}
{\epsilon d_x \frac{1}{\epsilon} d_x u(x)\sim d_x^2 u(x),}
thus the degeneracy occurs,~\textit{i.e.}~the two different set of eigenfunctions ${\epsilon E_x(x)}$ and ${E_y(x})$ corresponding to the same eigenvalues.


%% file: Chap_TFBG_2_FD_num.tex
\section{The numerical method}

In this section we describe the numerical procedure we implemented to obtain a vectorial mode solution and propagating constants of a dielectric waveguide with circular symmetry. 
The waveguide can be of an arbitrary refractive index profile,~\textit{i.e.} it can consist of any number of radially stratified layers made of various materials including absorbing material such as metals with the dominant imaginary part in the refractive index. 
In the following sections these modes, in other words eigenfunctions, are used as a basis for a more general problem of circularly symmetric dielectric waveguide with a small perturbation along the $z$ axis.

\subsection{The scalar modes}
Let us start our analysis with the weakly guided approximation and the scalar Helmholtz equation~(\ref{eq_cyl_scal_eig}):
\Eq{the_scal_eig_eq}
{\left(d_\rho^2 + \frac{1}{\rho} d_\rho - \frac{m^2}{\rho^2} + \epsilon k_o^2\right)E_z(\rho) = \beta^2 E_z(\rho).}
Rewriting the above equation in a slightly different notation we get: 
\Eq{eq_eig1}
{\left[\frac{1}{\rho} d_\rho (\rho d_\rho) + U^m(\rho) \right]R_k^m(\rho) = (\beta_k^m)^2 R_k^m(\rho),}
where
\Eq{eq_pot_barier}
{U^m(\rho) = k_o^2 n_o^2(\rho) - \frac{m^2}{\rho^2}.}
The eigenfunctions are denoted as $R_k^m(\rho)$ and eigenvalues as $\beta_k^m$, known as propagation constants in waveguide theory.
It should be noted that for each index $m$ a set of different solutions will be obtained, keeping this in mind, let us omit the index $m$ from the notation for simplicity.

First, let us build a matrix representation of equation~(\ref{eq_eig1}). This can be achieved by introducing an uniform grid $(\rho_1, \rho_2, ..., \rho_N)$ and representing the unknown function $R(\rho)$ as a vector with components $R_j = R(\rho_j)$. 
Next, applying the finite difference method we can rewrite equation~(\ref{the_scal_eig_eq}) in the following form:  
\Eqaa{eig_operat}
{&\hat L& R(\rho) = \beta^2 R(\rho),}
{&\hat L& = d_\rho^2 + \frac{1}{\rho} d_\rho - \frac{m^2}{\rho^2} + \epsilon k_o^2,}
with the central difference approximation:
\Eq{eq_f_FD}
{ a_j R_{j-1} + b_j R_j + c_j R_{j+1} = \beta^2 R_j, }
where
\Eqaaa{eq_f_FD_cof}
{ a_j &=& \frac{1}{h^2} - \frac{1}{\rho_j} \frac{1}{2h},}
{ b_j &=& -\frac{2}{h^2} - \frac{m^2}{\rho_j^2} + k_o^2 n_j^2,}
{ c_j &=& \frac{1}{h^2} + \frac{1}{\rho_j} \frac{1}{2h}.}
Here $n_j = n(\rho_j)$ is the refractive index profile of a given structure. The refractive index can be a complex number to take into account absorbing materials.

Thus equation~(\ref{eig_operat}) can be written in the matrix form:
\Eq{}{\lbrack L \rbrack \V R = \beta^2 \V R,}
where $\lbrack L \rbrack$ is the sparse 3-diagonal matrix with $b_j$ on the main diagonal, $a_j$ on the lower sub-diagonal and $c_j$ on the upper sub-diagonal.

Considering that equation~(\ref{eq_f_FD}) is the second order differential equation, the two proper boundary conditions should be imposed:

\begin{enumerate}
\item The condition at infinity $ \rho \to \infty $.

This condition can be imposed by introducing an extra point $R_{N+1}$, outside the computational domain.
Here we are interested in guided modes, thus we can assume that there is no field at infinity and simply impose the zero boundary condition. 
By choosing a sufficiently large computational window we ensure that the field value is negligible at the computational boundary. 
\Eq{}{R_{N+1} = 0.}
Plugging this condition into~(\ref{eq_f_FD}) we have
\Eq{}{ a_N R_{N-1} + b_N R_N + c_N \cdot 0 = \beta^2 R_N. }
We note that this condition is already imposed by the initial equation~(\ref{eq_f_FD}), thus there is no need to modify this equation. 

\item The condition at $ \rho = 0 $.
\begin{enumerate}
\item Assuming that $m = 1, \pm 2, \pm 3, ... $ we should require
\Eq{eq_FD_b2}{R(\rho = 0) = 0.}
By introducing an imaginary point $R_0$ outside the computational domain on the left, we can write:
\Eq{eq_f_B2}{ a_1 R_0 + b_1 R_1 + c_1 R_2 = \beta^2 R_1, }
considering~(\ref{eq_FD_b2}) we get:
\Eq{}{ a_1 \cdot 0 + b_1 R_1 + c_1 R_2 = \beta^2 R_1,}
Again, we see that this condition is already imposed be the default form of the equation~(\ref{eq_f_FD}). 

\item Assuming that $m = 0$ we get a maximum in the field distribution at the center of the waveguide, and thus should require a zero derivative of the field: 
\Eq{eq_f_B1}{ \frac{d R(\rho)}{d \rho}|_{\rho = \rho_1} = 0,}
using the central difference the above relation can be written as: 
\Eq{}{R_0 = R_2 = 0.}
Plugging the above condition into~(\ref{eq_f_B2}) we get:
 \Eq{}{ b_1 R_1 +( c_1 + a_1) R_2 = \beta^2 R_1, }
Therefore the matrix $\lbrack L \rbrack$ has to be modified:
 \Eq{eq_f_B3}{L_{12} = L_{12} + a_1 }
This is the only modification of~(\ref{eq_f_FD}) which is required.
    
\end{enumerate}
\end{enumerate}

Finally considering modification~(\ref{eq_f_B3}) the problem~(\ref{eq_f_FD}) can be stated in the matrix form: 
\Eq{mat_eig}
{\lbrack \hat{L} \rbrack \V R = \beta^2 \V R. }
Here we note that the zero boundary condition can be imposed in both cases if the problem is solved on the internal ${\rho \in [-R,R]}$ instead of ${\rho \in [0,R]}$.

The problem~(\ref{mat_eig}) is a well known numerical eigenvalue problem, which can be solved with variety of numerical techniques. 
Because we are dealing with a 3-diagonal sparse matrix, very efficient iterative methods for large sparse linear systems can be implemented. We used MATLAB software, which allowed us to consider up to one million elements. 
Even for a very large matrix of $10^6 \times 10^6$ elements, the required eigenvalues and corresponding eigenvectors, in the range of interest, were found in less than $1$ minute.
The matrix does not actually includes all of the $10^{12}$ numbers, but rather $3 \cdot 10^6$ numbers written in terms of three separate vectors corresponding to the three matrix diagonals. 
The computational algorithm is based on the iterative routine converging from a \C{random guess~\cite{Golub:1996}} towards the exact eigenvectors and corresponding eigenvalues in the given range.
If, instead of a sparse matrix, a dense matrix is used, the computation might take a prohibitive amount of time for the particular problem presented in this work (for example, it would take a couple of days to compute eigenvalues of a $5 000 \times 5 000$ matrix in Matlab with the default computational routine).

After the numerical eigenvalue problem is solved, a set of eigenvalues $\beta_k$ and corresponding eigenvectors $\V R_k$ \C{are} accessible.
It is convenient to use a potential barrier analogy with quantum mechanics. 
The potential energy function is defined by equation~(\ref{eq_pot_barier}). 
Basically, the \C{square of refractive} index profile $n^2(\rho)$ plays a role similar to the role of a potential energy barrier in quantum mechanics. 
The eigenvalue can be though of as \C{``energy states''}, and should be expressed in the same units as the potential barrier, or refractive index.
\C{The eigenfunctions} $\V R_k(\rho)$ represent the electric field profile, that is $E_z$ component if the scalar equation~\ref{the_scal_eig_eq} is considered, or $E_\rho$ and $E_\phi$ components for the vectorial case, as will be discussed later.

Here we introduce the so-called effective refractive indices  \textit{i.e.} $N_{eff,k} = k_0 \beta_k$, and plot the normalized eigenfunctions $\V R_k(\rho)$ centering them at the corresponding $N_{eff,k}$ with analogy to quantum mechanics.
The result is shown in Figure~\ref{Modes_1}.

\Fig{Modes_1}{1}
{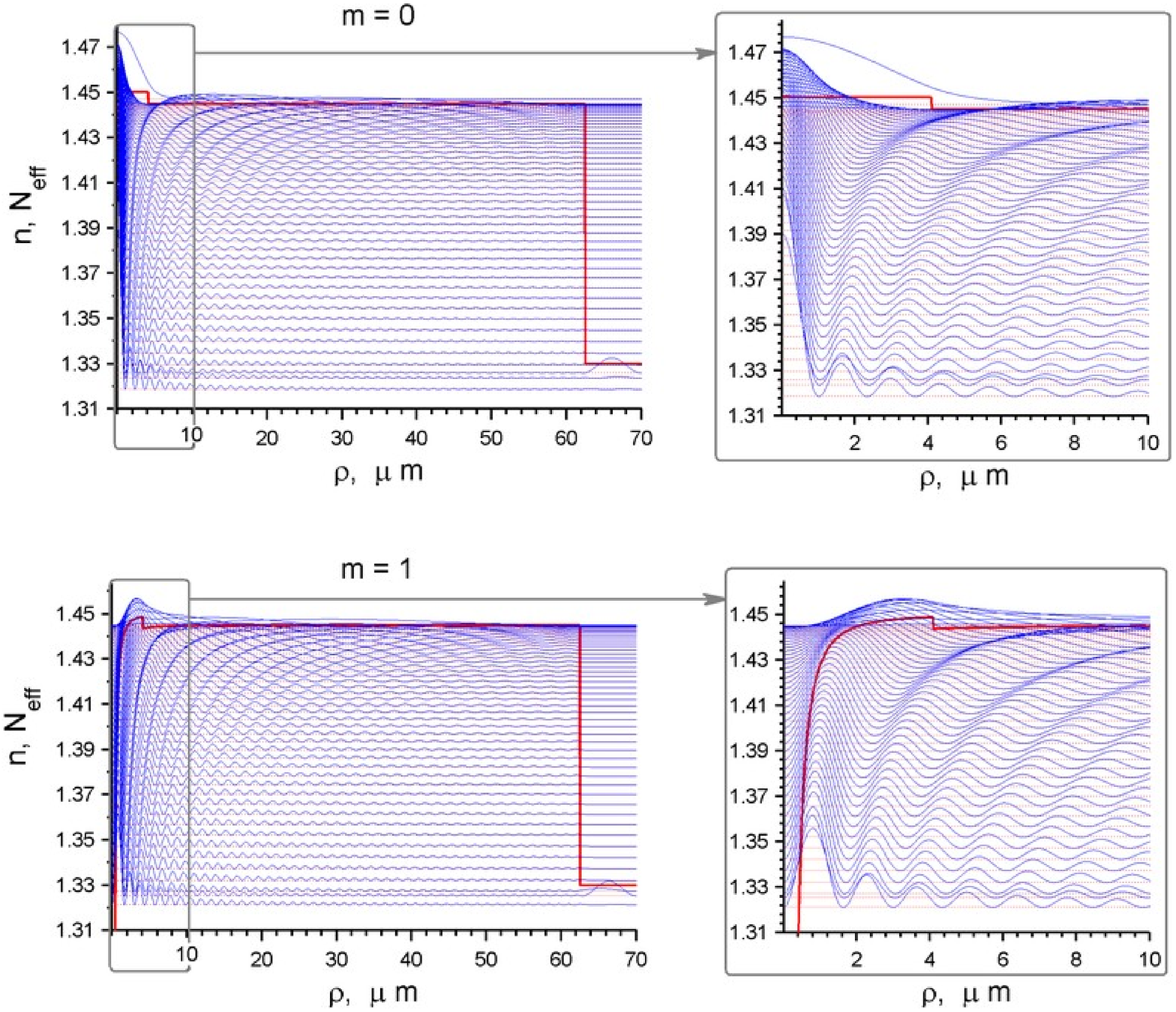}{The refractive index profile of SMF-28 fibre immersed in water. The eigenvalues and eigenfunctions are plotted for $m=0$ and $m=1$ }

As can be seen from Figure~\ref{Modes_1} that in the case when $N_{eff,k}$ becomes smaller than the refractive index of the surrounding medium, the field energy is leaking outside the fibre boundaries, and hence the energy is no longer confined inside the fibre as shown in Figure~\ref{Modes_diag}. Such modes are called the leaky modes.
The proposed method should be modified for the proper treatment of leaky modes. The zero boundary condition
should be replaced with a numerical absorbing boundary condition, ensuring that the incident travelling waves are absorbed and no energy is reflected back into the waveguide, and hence the formation of standing waves is prohibited. Such absorbing boundary layer is called the perfectly matched layer (PML). An efficient numerical technique for implementing the absorbing boundary condition in the FDTD method was developed by Mur in 1981~\cite{Mur:1981}.

\Fig{Modes_diag}{0.7}
{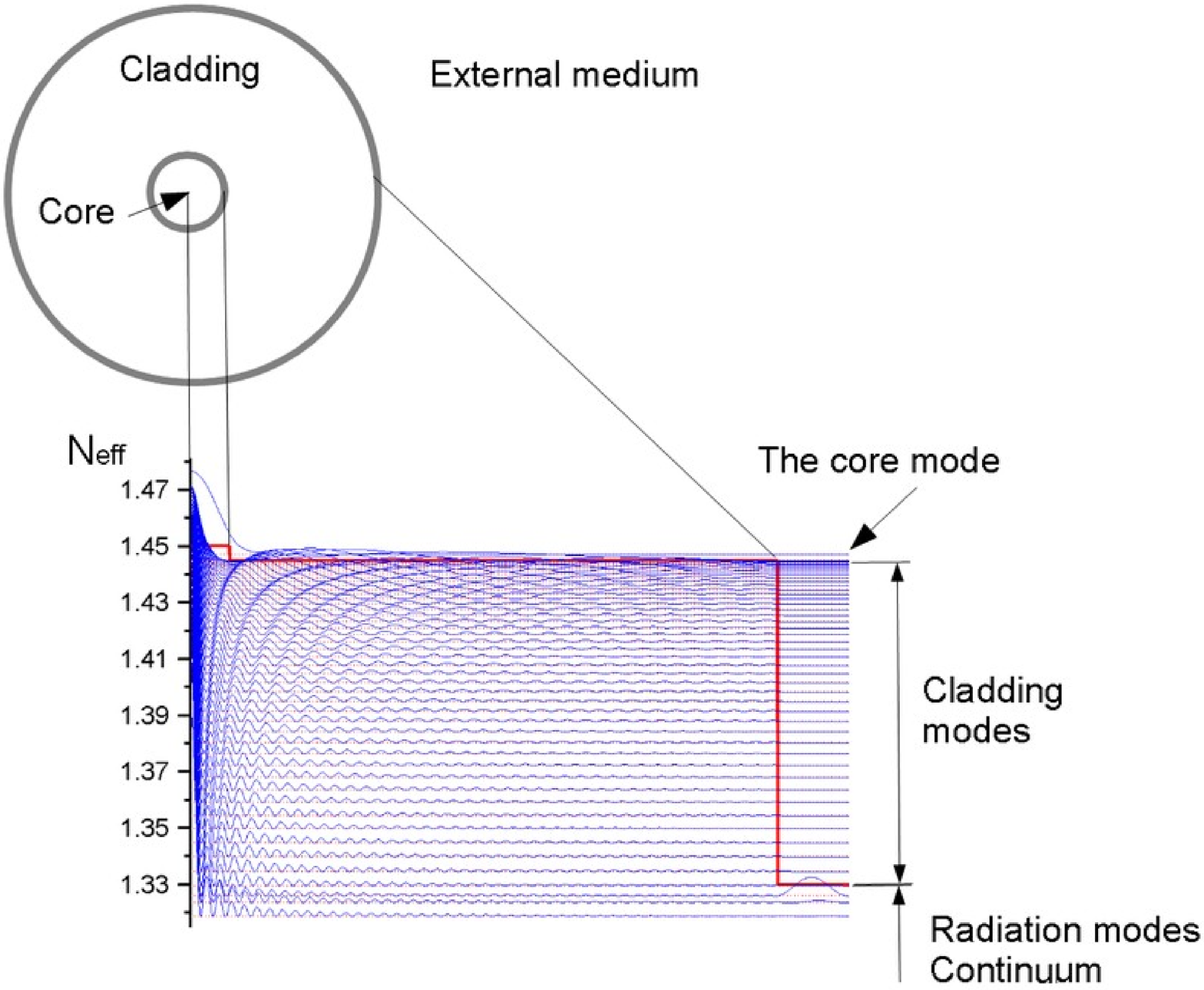}{ The fibre cross-section and corresponding refractive index profile with modes bounded inside the fibre.}

We can note the further similarities with quantum mechanics. The number $m$ in the equation~(\ref{eq_eig1}) plays a role similar to the role of orbital quantum number in quantum mechanics. 
The larger the angular momentum of a particle, the closer its wavefunction is to the periphery, which is also consistent with classical mechanics. 
Unless the critical value is exceeded, the particle stays confined by central potential, although some fraction of its energy leaks outside the potential barrier boundaries. 
The described effect can be observed in Figure~\ref{Modes_2}. 
The potential function $U^m(\rho)$ is defined by the equation~(\ref{eq_pot_barier}) and is shown in red color. 
We can note in Figure~\ref{Modes_2} how significant is the role of angular momentum potential ${U^m(\rho)= \frac{m^2}{\rho^2}}$. 

\Fig{Modes_2}{0.5}
{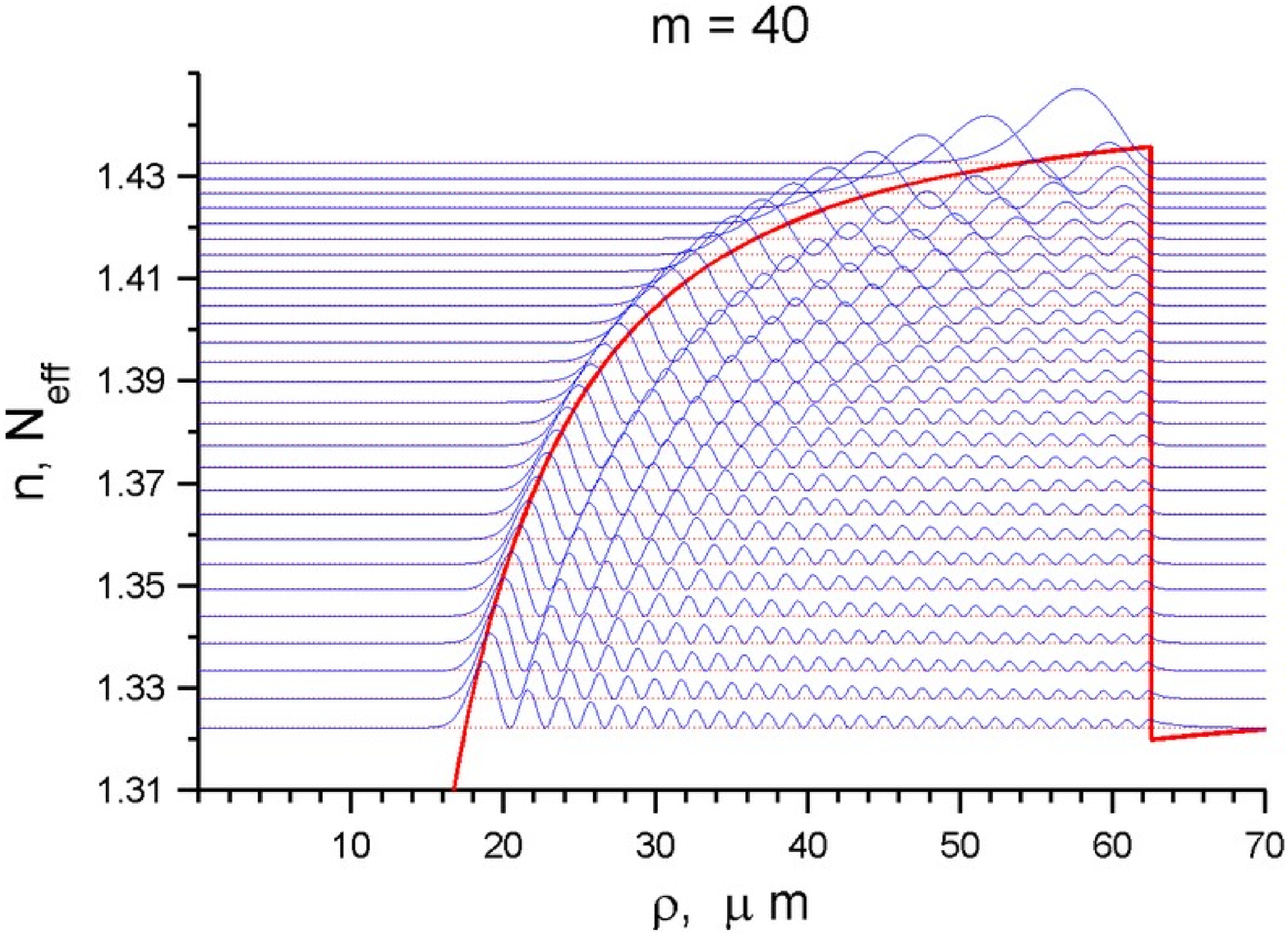}{The radial eigenfunctions $R_k^m(\rho)$ (shown in blue color) and the potential well function $U^m(\rho)$ (shown in red color) for $m=40$.}

Finally, the complete basis functions $\psi_k^m(\rho,\phi)$ can be constructed by considering the angular dependence as well:
\Eq{}
{\psi_k^m(\rho,\phi) = R_k^m(\rho)e^{j m\phi}.}
A particular solution can be constructed by choosing an orbital number $m$, thus defining the potential function $U_k^m(\rho)$, and picking a particular radial eigenfunction $R_k^m(\rho)$ for the chosen mode family $m$. 
An example of such particular solution is shown in Figure~\ref{Modes_3D}.
\Fig{Modes_3D}{0.9}
{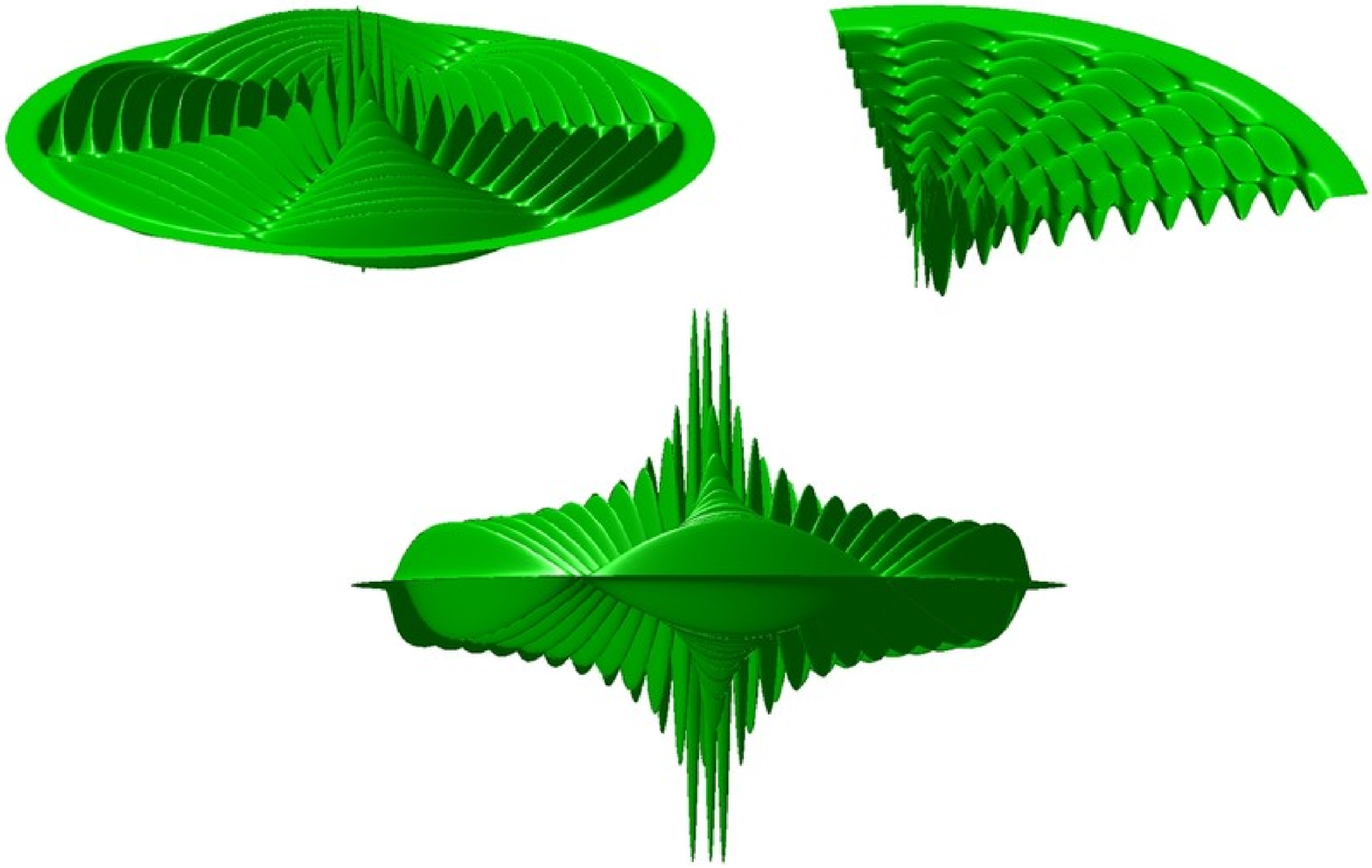}{One of the basis function (mode) $\psi_{10}^3 (\rho, \phi)$ , at $n=10$ (${N_{eff} = 1.4181}$) and ${m=3}$.}

\subsection{The vectorial modes}
Let us first consider the vectorial equation~(\ref{eq_cyl_vec_eig}) for the weakly guided approximation:
\Eq{FD_matrix1}
{
\begin{pmatrix}
d_\rho^2 + \frac{1}{\rho} d_\rho - \frac{m^2+1}{\rho^2} + \epsilon k_o^2 & - \frac{j2m}{\rho^2}\\
\frac{j2m}{\rho^2} & d_\rho^2 + \frac{1}{\rho} d_\rho - \frac{m^2+1}{\rho^2} + \epsilon k_o^2
\end{pmatrix}
\begin{pmatrix}
E_\rho\\
E_\phi\\
\end{pmatrix}
= \beta^2 \begin{pmatrix}
E_\rho\\
E_\phi\\
\end{pmatrix}.
}
Again, we can implement the finite difference method to replace the system of two second order coupled ordinary differential equations~(\ref{FD_matrix1}) with a system of algebraic equations:  
\Eq{}
{[\hat M] \V E_T = \beta^2 \V E_T.}
Before constructing the finite difference implementation we note that matrix in~(\ref{FD_matrix1}) contains complex numbers. Although the complex numbers does not impose any limitation on the numerical method, the computational speed can be increased if the matrix is real. The memory requirement is also reduced by a factor of two for the real numbers.
Multiplying the first equation in~(\ref{FD_matrix1}) by imaginary unit $j$ and pulling it under the $E_\rho$ component we get:
\Eq{FD_matrix2}
{
\begin{pmatrix}
d_\rho^2 + \frac{1}{\rho} d_\rho - \frac{m^2+1}{\rho^2} + \epsilon k_o^2 & \frac{2m}{\rho^2}\\
\frac{2m}{\rho^2} & d_\rho^2 + \frac{1}{\rho} d_\rho - \frac{m^2+1}{\rho^2} + \epsilon k_o^2
\end{pmatrix}
\begin{pmatrix}
j E_\rho\\
E_\phi\\
\end{pmatrix}
= \beta^2 \begin{pmatrix}
j E_\rho\\
E_\phi\\
\end{pmatrix}.
}
The operator matrix now is converted to a pure real form, unless the material permittivity $\epsilon$ is complex. The same equation was previously obtained elsewhere~\cite{Lu:08, Black:2010}.

The finite difference matrix $[\hat M]$ can be constructed in the following way:
\Eq{}
{
[M] \V E_T
= \beta^2 \V E_T,
}
where
\Eqaaa{FD_diag}
{[\hat M] &=& \begin{pmatrix}
\hat L & \V a \\
\V a & \hat L
\end{pmatrix}, }
{a_i &=& \frac{2m}{\rho_i^2},}
{\hat L &=& [D]^2 + \frac{1}{\rho} [D] - \frac{m^2+1}{\rho^2} + \epsilon k_o^2,}
here ${[D]^2}$ and ${[D]}$ are the second order and the first order finite difference matrices, respectively, assembled with the help of the central difference approximation in a fashion similar to~(\ref{eq_f_FD}) and~(\ref{eq_f_FD_cof}).
We note that operator $\hat L$ becomes identical to the scalar operator~(\ref{eig_operat}) if ${m^2+1}$ is replaced by $m^2$.

The structure of matrix $[\hat M]$ is shown in Figure~\ref{sparse}, where the number of nodes was limited to ${N = 10}$ for the convenience of representation, however for a practical application at least several thousands of elements should be considered. 

In can clearly be seen that the matrix is sparse with only five main diagonals. 
The right most and left most diagonals represent term ${\frac{2m}{\rho^2}}$ which is responsible for the coupling  between $E_\rho$ and $E_\phi$ field components. In the case when ${m = 0}$ the matrix becomes a three diagonal, with uncoupled $E_\rho$ and $E_\phi$ components.

\Fig{sparse}{0.5}
{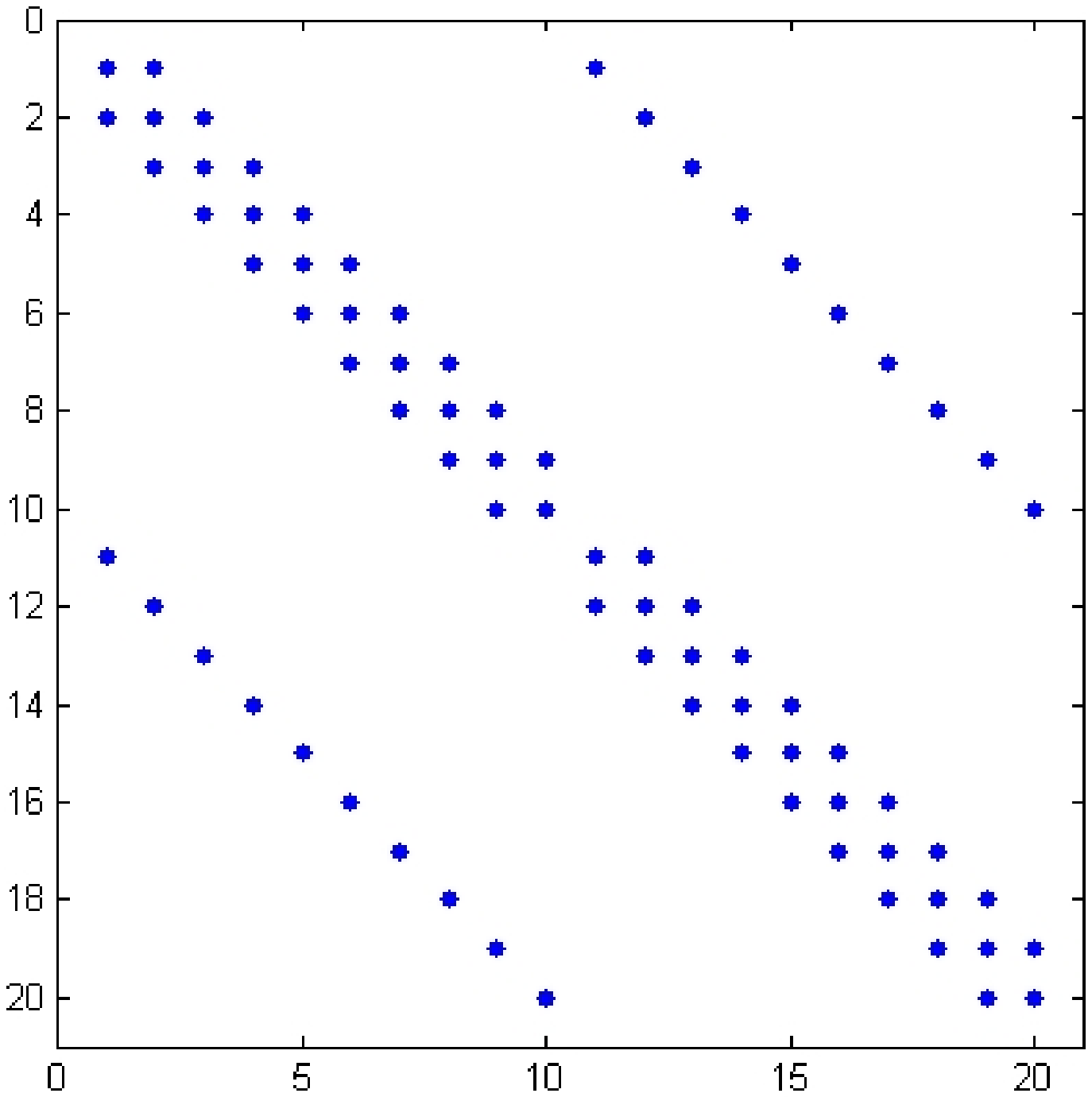}{The structure of the sparse matrix $[\hat M]$ for ${N = 10}$ and ${m \ne 0}$.}

In the similar way we can approach the exact equation~(\ref{eq_cyl_vec_eigexact}): 

\Eq{}
{\Scale[0.87]{
\begin{pmatrix}
d_\rho^2 + \frac{1}{\rho} d_\rho + (\ln \epsilon )' d_\rho - \frac{m^2+1}{\rho^2} + \epsilon k_o^2 + (\ln \epsilon )'' & \frac{2m}{\rho^2}\\
\frac{2m}{\rho^2} + \frac{m}{\rho}(\ln \epsilon )' & d_\rho^2 + \frac{1}{\rho} d_\rho - \frac{m^2+1}{\rho^2} + \epsilon k_o^2
\end{pmatrix}\begin{pmatrix}
j E_\rho\\
E_\phi\\
\end{pmatrix}
= \beta^2 \begin{pmatrix}
j E_\rho\\
E_\phi\\
\end{pmatrix},
}
}
here we pulled the imaginary unit $j$ out of the matrix.

The system of algebraic equation takes the following form: 
\Eq{}
{
\begin{pmatrix}
\hat L_2 & \V a \\
\V b & \hat L
\end{pmatrix} \V E_T
= \beta^2 \V E_T,
}
where $\V a$ and $\hat L$ are defined previously in~(\ref{FD_diag}), and 
\Eqaa{}
{b_i &=& a_i + \frac{m}{\rho}(\ln \epsilon )',}
{\hat L_2 &=& \hat L + (\ln \epsilon )' [D]  + (\ln \epsilon )''.}
The matrix for the exact case has the same structure as shown in Figure~\ref{sparse}.

Alternatively, the exact system of equations~(\ref{eq_cyl_vec_eigexact}) can be rewritten in the Sturm-–Liouville form: 
\Eq{}
{\Scale[0.95]{
\begin{bmatrix}
\epsilon \rho \frac{d}{d \rho} \left(\frac{1}{\rho \epsilon}\frac{d}{d \rho} \right) + \left(- \frac{m^2}{\rho^2} + \epsilon k_o^2\right) & \epsilon \frac{2m}{\rho^2} \\
\frac{1}{\epsilon}\left(\frac{2m}{\rho^2} + \frac{m}{\rho}(\ln \epsilon )'\right) & \rho \frac{d}{d \rho} \left(\frac{1}{\rho}\frac{d}{d \rho}\right) + \left(- \frac{m^2}{\rho^2} + \epsilon k_o^2\right)
\end{bmatrix}
\begin{pmatrix}
j \epsilon \rho E_\rho\\
\rho E_\phi\\
\end{pmatrix}
= \beta^2 \begin{pmatrix}
j \epsilon \rho E_\rho\\
\rho E_\phi\\
\end{pmatrix},
}
}
where we assumed 
\Eqaa{fd_subst}
{\left(\frac{1}{r}(r u)'\right)' &=& u'' + \frac{1}{r} u' - \frac{1}{r^2} u,}
{\left(\frac{1}{r \epsilon}(r \epsilon u)'\right)' &=& u'' + \frac{1}{r} u' + (\ln \epsilon)' u' - \frac{1}{r^2} u + (\ln \epsilon)'' u.}
We discuss this subject in more detail in the following section.
Next, the numerical finite difference method explicitly designed for Sturm-–Liouville problems
can be applied. 
The main idea of this approach is to derive a central difference equation for the ${\frac{d}{d \rho}\left(p(\rho) \frac{d}{d \rho}\right)}$ operator.

In the next section we compare results obtained by \C{numerically} solving the scalar (\ref{eq_cyl_scal}) and vector (\ref{eq_cyl_vec}) equations for the weakly guided approximation with the exact solution obtained from (\ref{eq_cyl_vec_eigexact}).

%% file: Chap_TFBG_4_FD_conclussion.tex
\section{Discussion}
\label{modes_discussion}

In this section we verify our results and discuss properties of the modes.
First let us start by verifying the developed numerical method.
The analytical solution can be easily obtained for slab waveguides, as shown for example in~\cite{kats:2006, Yariv:2007}. 

The solution for TE and TM modes guided in a thin symmetric glass slide immersed in water as well as the dispersion curves were presented in~\cite{kats:2006}.

\Fig{DisperCurvTETM}{0.9}{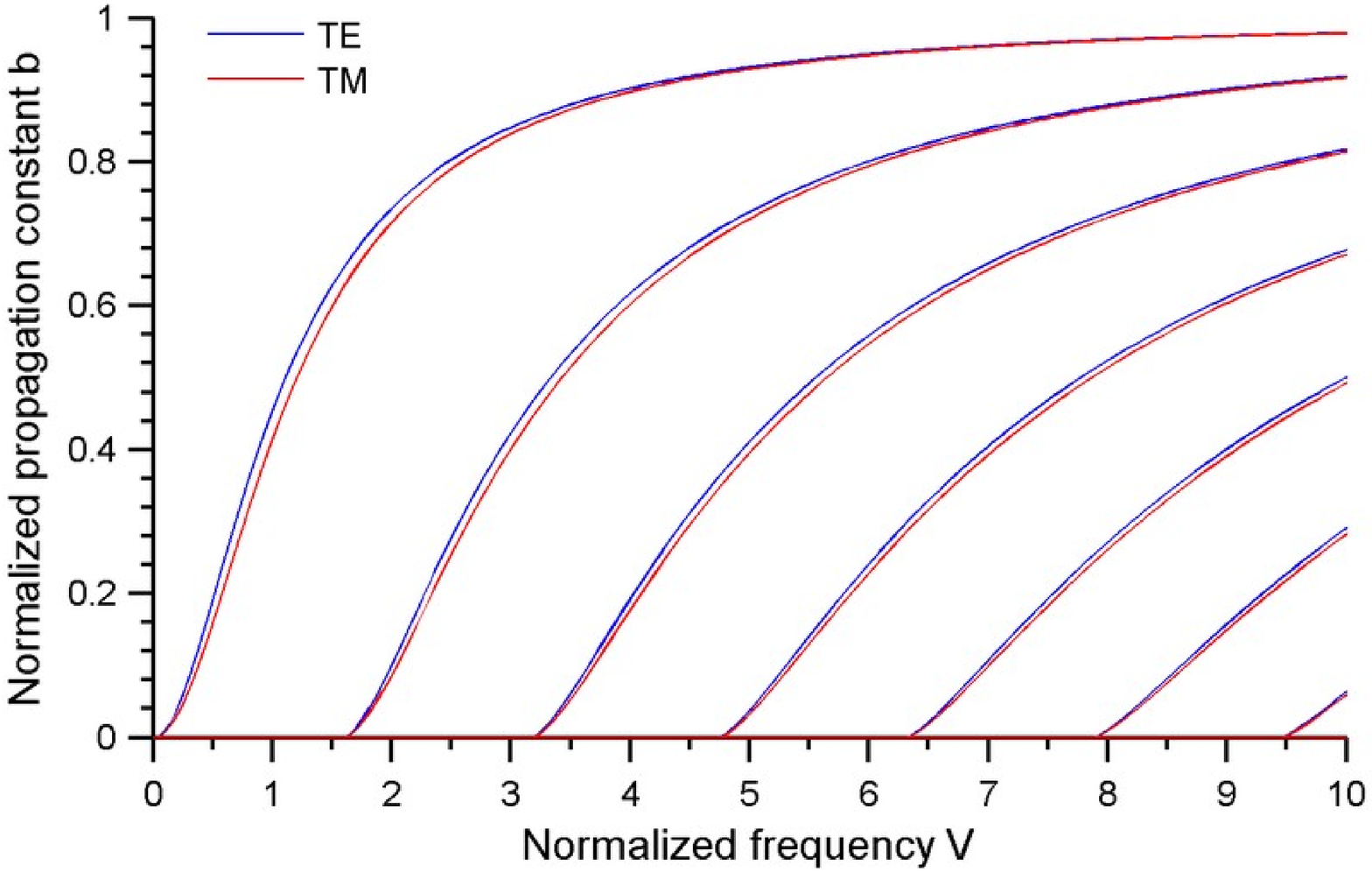}
{The dispersion curves of a symmetric glass slide immersed in water. The waveguide and the medium refractive indices are ${n_{WG} = 1.45}$ and ${n_{Med.} = 1.33}$, respectively.}

Defining the normalized frequency $V$ and normalized propagation constant $b$ as: 
\Eqaa{}
{V &=& R k_o \sqrt{n_1^2 - n_2^2} \sim \frac{R}{\lambda},}
{b &=& \frac{N_{eff}^2 - n_2^2}{n_1^2 - n_2^2} \sim N_{eff}^2,}
and solving equations~(\ref{eq_f_TM}, \ref{eq_f_TE}) numerically using the proposed method we plot the dispersion curves for TE and TM modes, as shown in Figure~\ref{DisperCurvTETM}. The identical graph was obtained in~\cite{kats:2006}.
Here $R$ is the waveguide width (or diameter), $n_1$ and $n_2$ are the refractive indices of the waveguide and the surrounding medium, respectively. In the similar manner we can verify the results for cylindrical waveguides, see for example results obtained elsewhere~\cite{snitz:61, dil:1973, Su:1986, Lu:08, kats:2006, Yariv:2007}.

It was shown in the previous section that the Maxwell equations can be decoupled into the scalar~(\ref{eq_cyl_scal}) and vector~(\ref{eq_cyl_vec}) equations if the weakly guided approximation applies. 
Solving these equations independently we obtain dispersion curves as shown in Figure~\ref{DisperCurvScal} and Figure~\ref{DisperCurvVect}.

The eigenvalues for the scalar and vectorial problems are identical. 
Indeed, this result can be expected as both equations were obtained from the same initial system of Maxwell's equations, and are in a way simply a different representation of the same physical system.

We note that solutions to the vectorial equation are degenerate, except for the modes with the zero angular part,~\textit{i.e.} ${m = 0}$. The equations for the $E_\rho$ and $E_\phi$ field components become uncoupled at ${m = 0}$, thus $E_\rho$ component can take arbitrary values independent on the value of $E_\phi$ component.
In other words the modes belonging to the ${m = 0}$ family can have an arbitrary polarization, with no restriction imposed by the cylindrical symmetry of the problem. 
In such cases modes can be represented either by a superposition of two orthogonal linearly polarized waves or as a superposition of left and right circularly polarized waves. In the literature these modes are known as linearly polarized (LP) modes~\cite{snitz:61, Black:2010}, although the term circularly polarized (CP) modes would also be appropriate~\cite{Black:2010}.

The mode profile at ${m = 0}$ is shown in Figure~\ref{vec_scal_1}. 
Indeed, it can be noted that two modes~${\begin{pmatrix} E_\rho\\ 0\\ \end{pmatrix}}$ and~${\begin{pmatrix} 0\\ E_\phi\\ \end{pmatrix}}$ have an identical propagation constant, thus can be superposed to represent an arbitrary polarization, including the circularly polarization. 
Hence the widely used term the linearly polarized modes are in a way misleading. A more detailed discussion on this subject can be found in~\cite{Black:2010}.

\clearpage
\Fig{DisperCurvScal}{0.9}{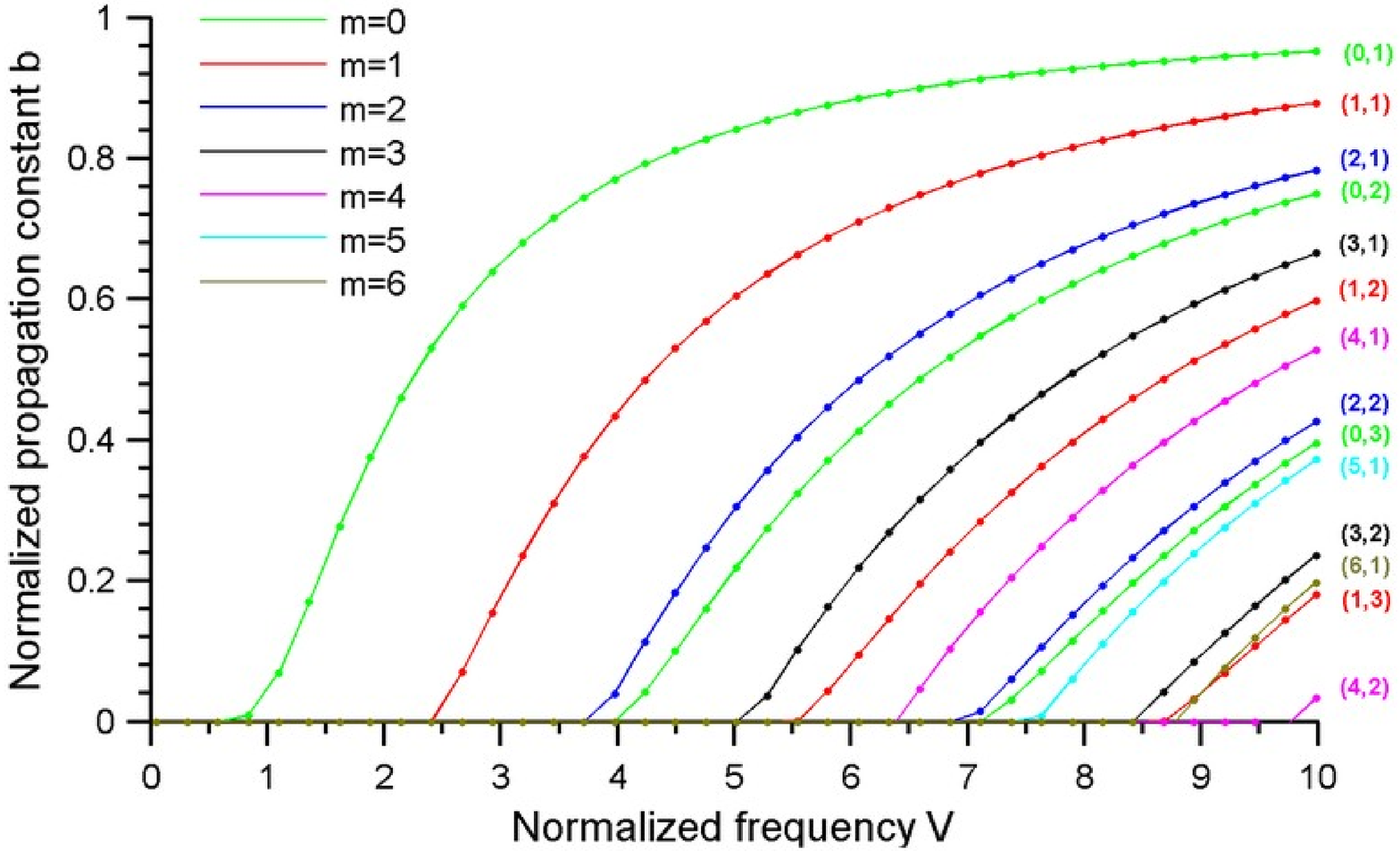}
{The dispersion curves of the scalar~(\ref{eq_cyl_scal}) modes, here ${n_{WG} = 1.45}$ and ${n_{Med.} = 1.33}$. }

\Fig{DisperCurvVect}{0.9}{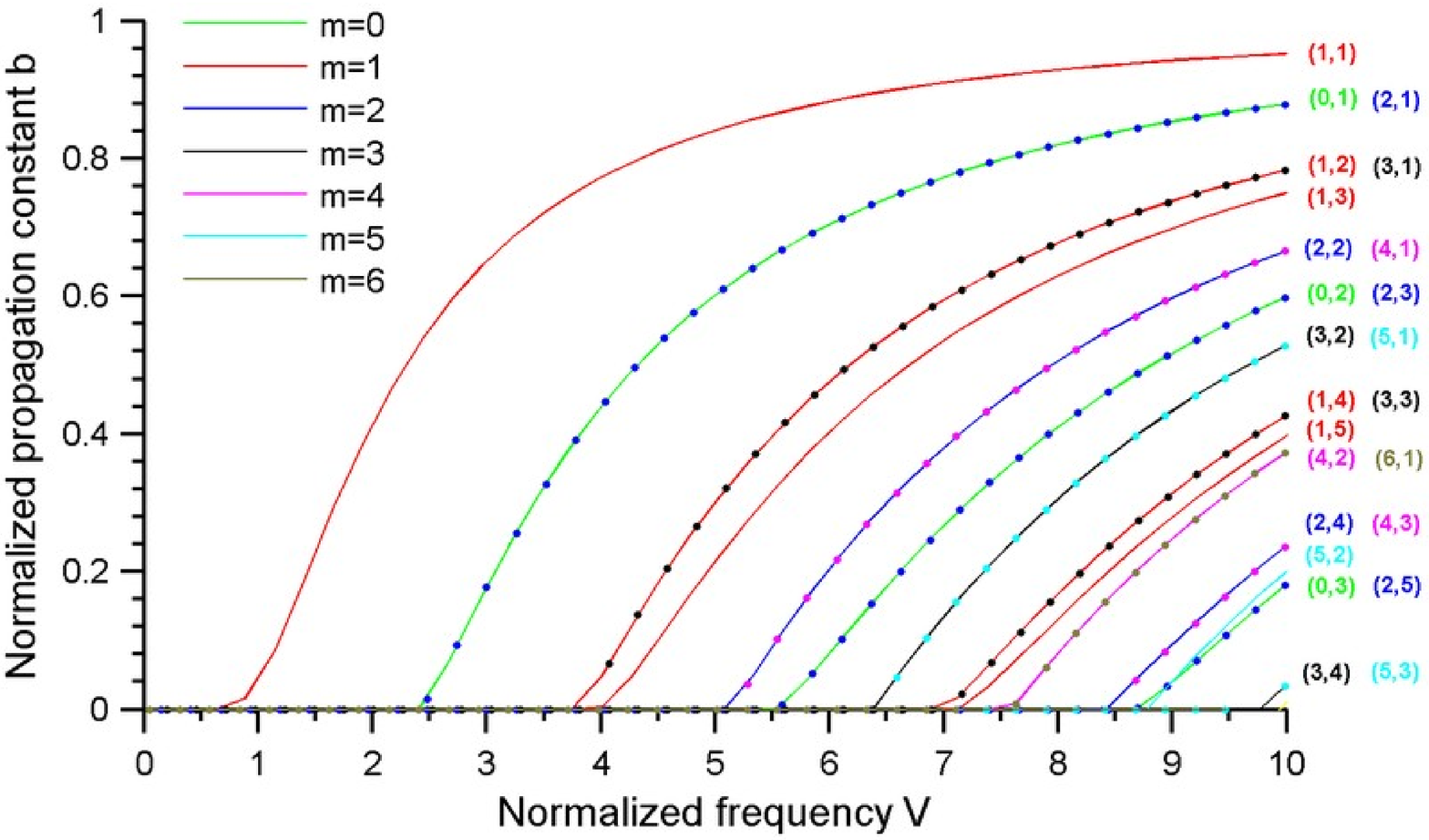}
{The dispersion curves of the vector modes~(\ref{eq_cyl_vec}). The waveguide and medium refractive indices are ${n_{WG} = 1.45}$ and ${n_{Med.} = 1.33}$, respectively.}

\clearpage
\Fig{vec_scal_1}{0.9}{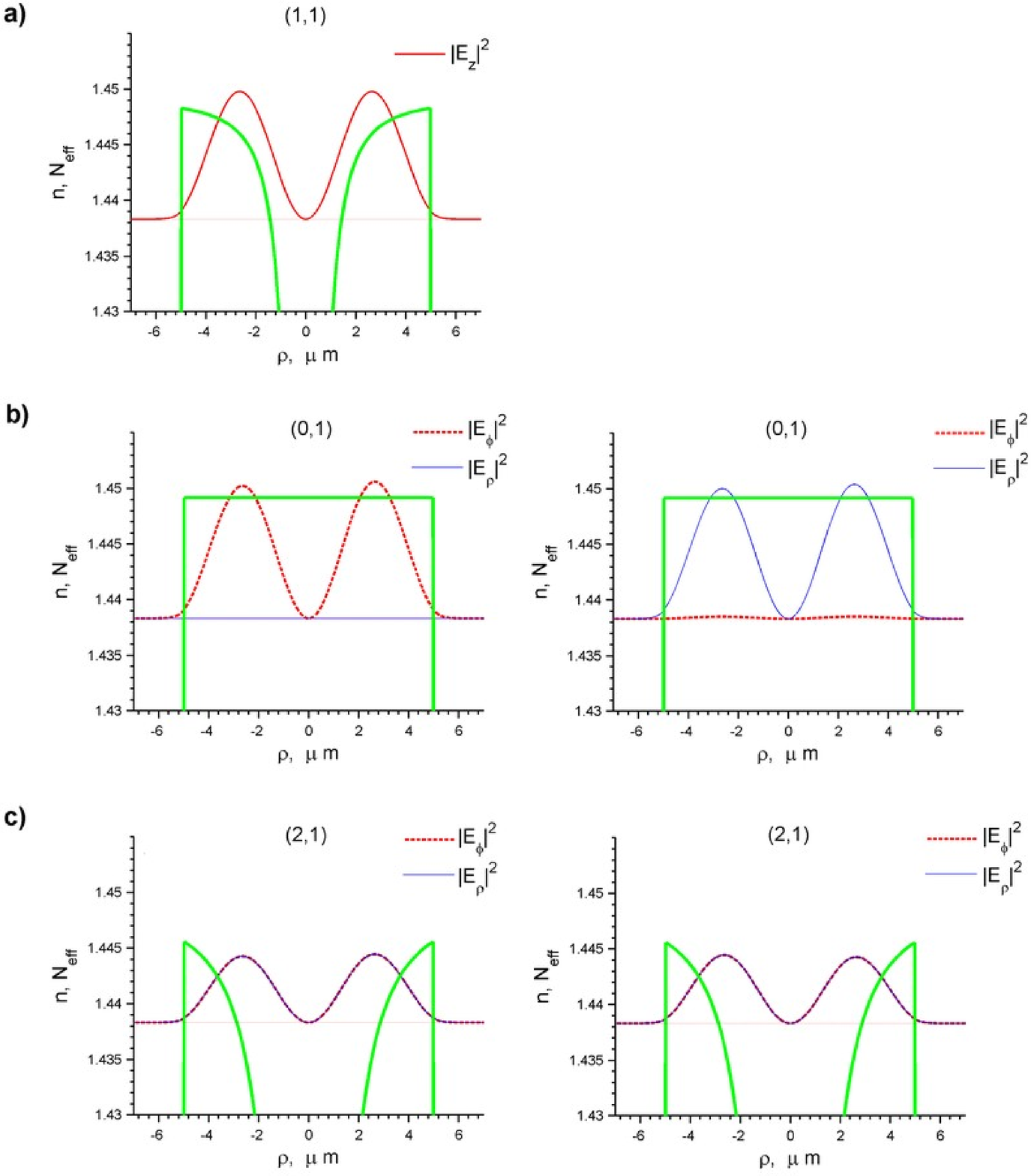}
{The degeneracy of modes: the scalar ${(1,1)}$ mode and two vectorial ${(0,1)}$ and ${(2,1)}$ modes. The waveguide and medium refractive indices are ${n_{WG} = 1.45}$ and ${n_{Med.} = 1.33}$, respectively. The potential barrier is depicted with green color}
\clearpage

In addition we note that the first mode in a cylindrical waveguide, or the core mode in a single mode fibre, is obtained by assuming that ${m = 0}$ in the scalar case or ${m = 1}$ in the vectors case, as can be seen from Figures~\ref{DisperCurvScal} and~\ref{DisperCurvVect}. In the vectorial case the second mode is observed at ${m = 0}$ , with which the so-called LP modes are associated. Therefore the LP modes can propagate only in the cladding of a single mode fibre.

In the case when the weakly guided approximation can be applied, each mode can be obtained by solving either the scalar equation~(\ref{eq_cyl_scal}) or vector equation~(\ref{eq_cyl_vec}). The vectorial modes are mostly degenerate.
It is interesting to compare radial profiles of such identical modes. For example, in Figure~\ref{vec_scal_1} the radial profile of ${(1,1)}$ scalar mode and ${(0,1)}$, ${(2,1)}$ vectorial modes are shown. All three modes have an identical propagation constant.
We note that the LP modes ${(0,1)}$, with the radial and angular components decoupled, coincide with the ${(2,1)}$ mode, with coupled $E_\rho$ and $E_\phi$ components.
It is interesting to note that all the modes have the identical radial profile, even though the scalar mode was obtained by solving the scalar equation at ${m = 0}$ for the $E_z$ component, where the vectorial modes were solved for $E_\phi$ and $E_\phi$ at ${m = 1}$ and ${m = 2}$.
Each mode ${\begin{pmatrix} E_\rho\\ E_\phi\\ \end{pmatrix}}$ is normed to unity, hence the difference in amplitudes between ${(0,1)}$ and ${(2,1)}$ modes. The ${(0,1)}$ mode has only single nonzero component: ${\begin{pmatrix} E_\rho\\ 0\\ \end{pmatrix}}$ or ${\begin{pmatrix} 0\\ E_\phi\\ \end{pmatrix}}$, whereas in the mode ${(2,1)}$ the both components $E_\rho$ and $E_\phi$ are nonzero.

\Fig{vec_scal_2}{0.9}{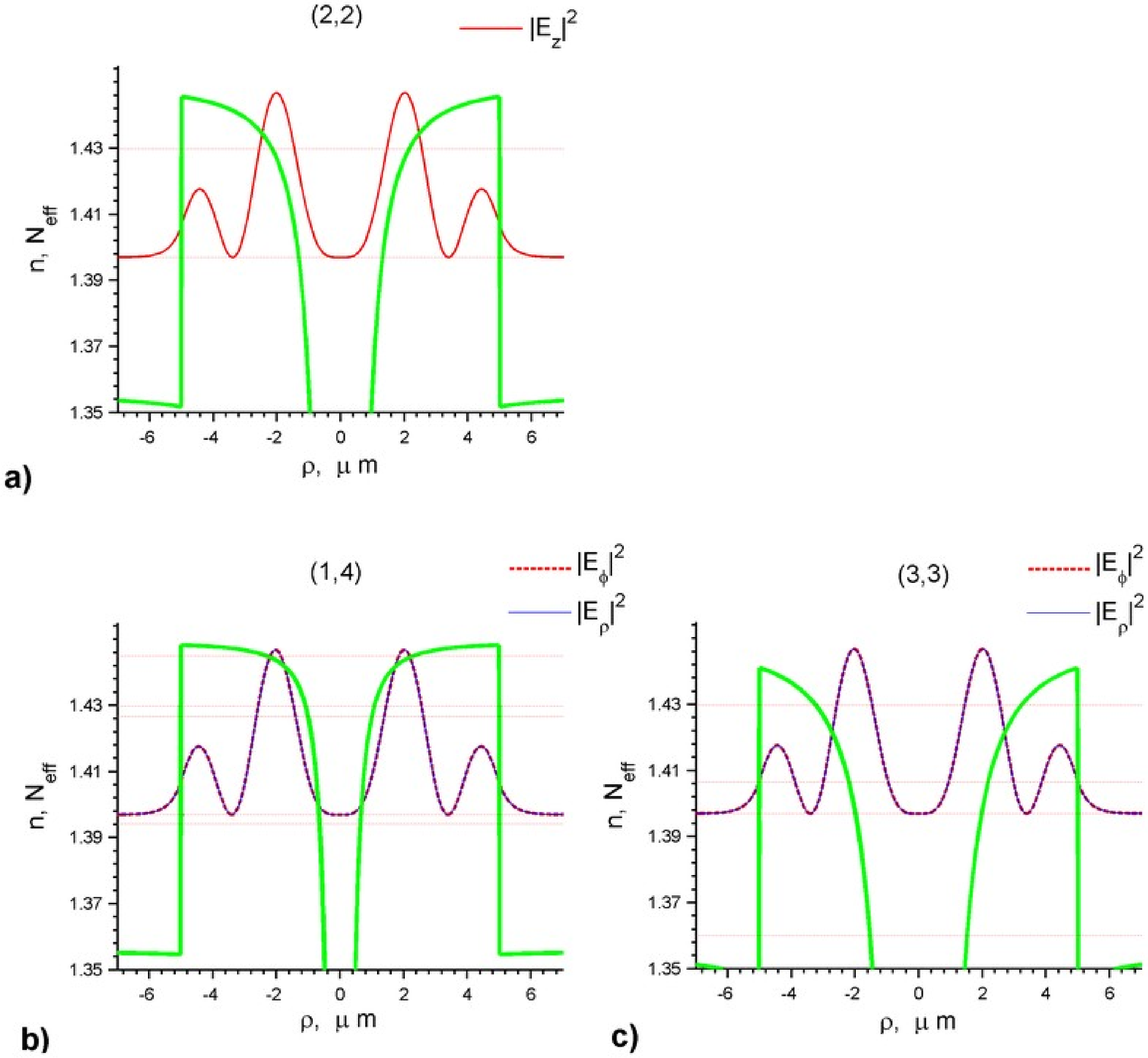}
{The degenerate modes: the scalar ${(2,2)}$ mode and two vectorial ${(1,4)}$ and ${(3,3)}$ modes. 
Here the waveguide with ${n_2 = 1.45}$ is immersed in water ${n_{Med.} = 1.33}$. }

Let us consider the vectorial modes with ${m \ne 0}$. 
The field components $E_\rho$ and $E_\phi$ are always coupled in this instance, thus one can not be chosen independently of another. 
In the case of weakly guided approximation the modes are mostly degenerate. 
For example, the radial profile of ${(2,2)}$ scalar mode and ${(1,4)}$, ${(3,3)}$ vectorial modes are shown in Figure~\ref{vec_scal_2}. Again, we note the similarity in the radial profile of these modes.

If an exact solution is considered the degeneracy is removed. 
The modes with a different $m$ number start to behave differently, diverging apart from each other with the increase in refractive index difference between the waveguide and the medium, as shown in Figure~\ref{Disper_m1m3_Split}.
The modes ${(1,2)}$ and ${(3,1)}$, ${(1,4)}$ and ${(3,3)}$, ${(1,6)}$ and ${(3,5)}$ no longer coincide, as in the case of weakly guided approximation. 
In the same graph the single modes ${(1,1)}$, ${(1,3)}$ and ${(1,5)}$, without a pair mode, are also plotted. 
In general, each mode obtained for a small refractive index difference, splits into two modes, each with a distinct dispersion curve.

\Fig{Disper_m1m3_Split}{0.9}{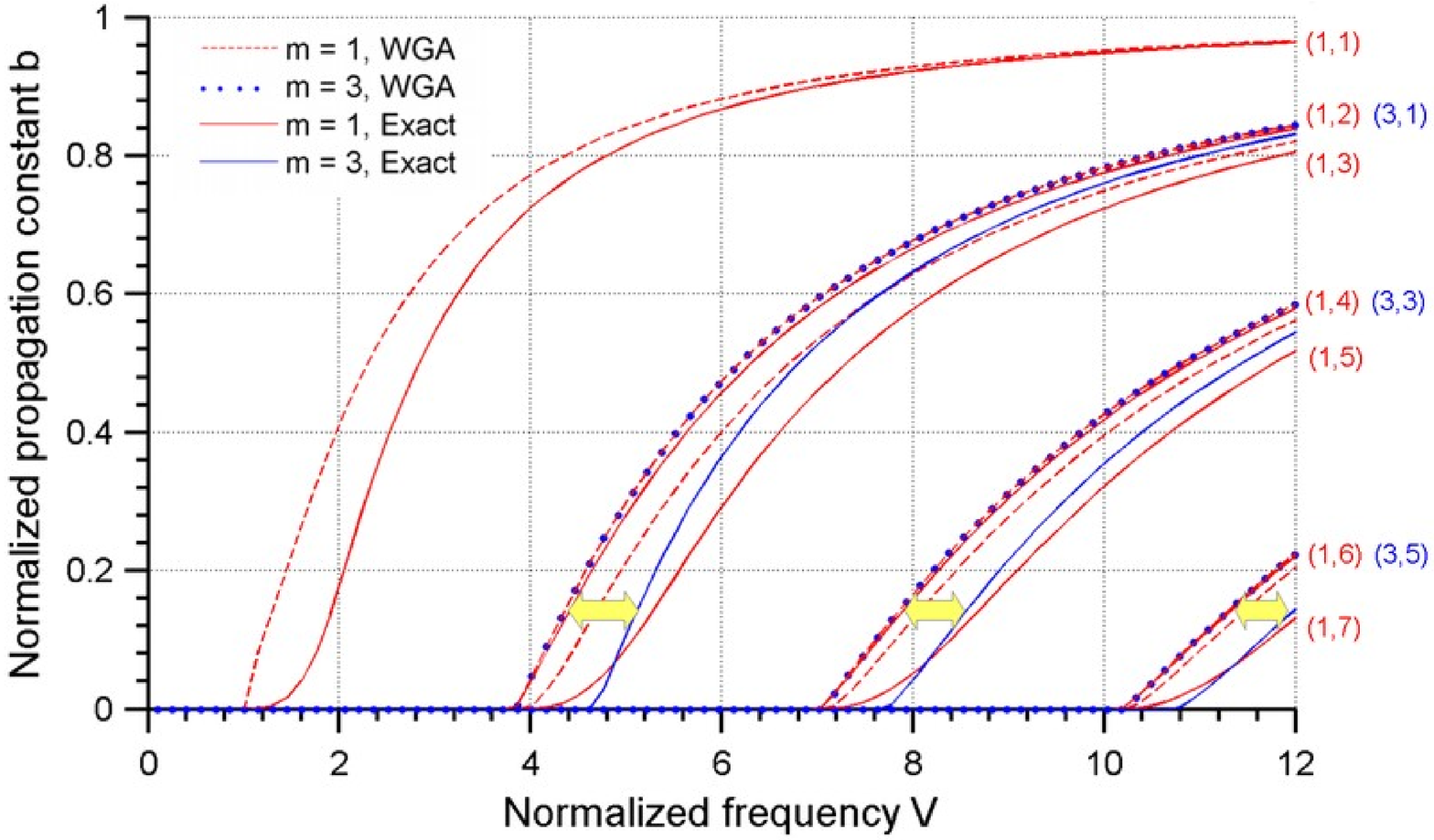}
{The split in dispersion curves between $m = 1$ and $m = 3$ modes. 
\newline The waveguide and the medium refractive indices are ${n_{WG} = 3}$ and ${n_{Med} = 1.33}$, respectively.}

From Figure~\ref{Disper_m1m3_Split} we note that not only the dispersions curves of different mode families $m$ diverge apart from each other, but even the modes belonging to the same family start to behave differently, for example the ${(1,3)}$, ${(1,5)}$ and ${(1,7)}$ modes.
This effect can be studied in more detail by plotting the dispersions curves corresponding to different refractive index differences ${\Delta n = n_{WG} - n_{Med}}$ between the waveguide and the medium, as shown in Figure~\ref{Disper_m2_spread} for the ${m = 2}$ family.

\Fig{Disper_m2_spread}{0.9}{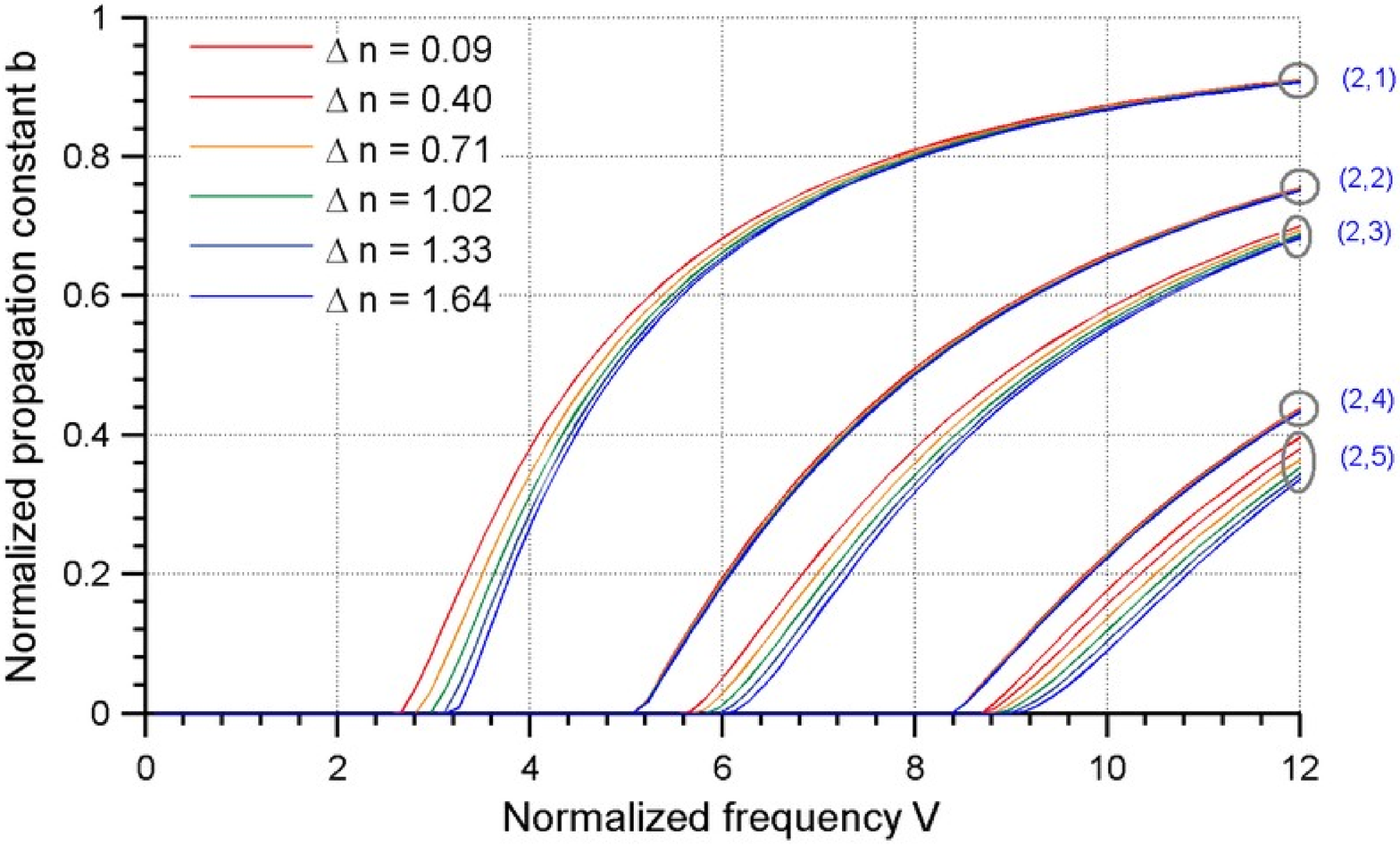}
{The dispersion curves, obtained by solving the exact vectorial equation, for various refractive index ratios ${\Delta n = n_{WG} - n_{Med.}}$ at ${m = 2}$.
The waveguide is immersed in water ${n_{Med.} = 1.33}$.}

We note that different modes inside the same mode family $m$, behave differently. 
For example, the modes ${(2,2)}$ and ${(2,4)}$ have barely changed in location on the dispersion plane, whereas modes ${(2,1)}$, ${(2,4)}$ and ${(2,5)}$ are highly divergent, especially at the cutoff region. We discuss this effect for the ${m = 0}$ case in more detail later.

We conclude our discussion here by pointing that if ${m \ne 0}$, and the refractive index difference between the waveguide and the medium is significant, so that the weakly guided approximation can no longer be applied, the majority of modes splits into two separate modes, thus the degeneracy is removed. 
Once this property is understood we can move on to a more complicated case of ${m = 0}$.

If the vectorial equation is solved for the case of ${m = 0}$ the degeneracy observed for a small refractive index difference is removed as was discussed above. 
But now not only the modes with different $m$ numbers are distinctly seen as separated (for example, as was shown in Figure~\ref{Disper_m1m3_Split} ), but additionally the modes inside the same family ${m = 0}$ are split into two subfamilies. 
Thus a single mode is split \C{into three} distinct modes, as shown in Figure~\ref{Disper_m0_Split}.

\Fig{Disper_m0_Split}{0.9}{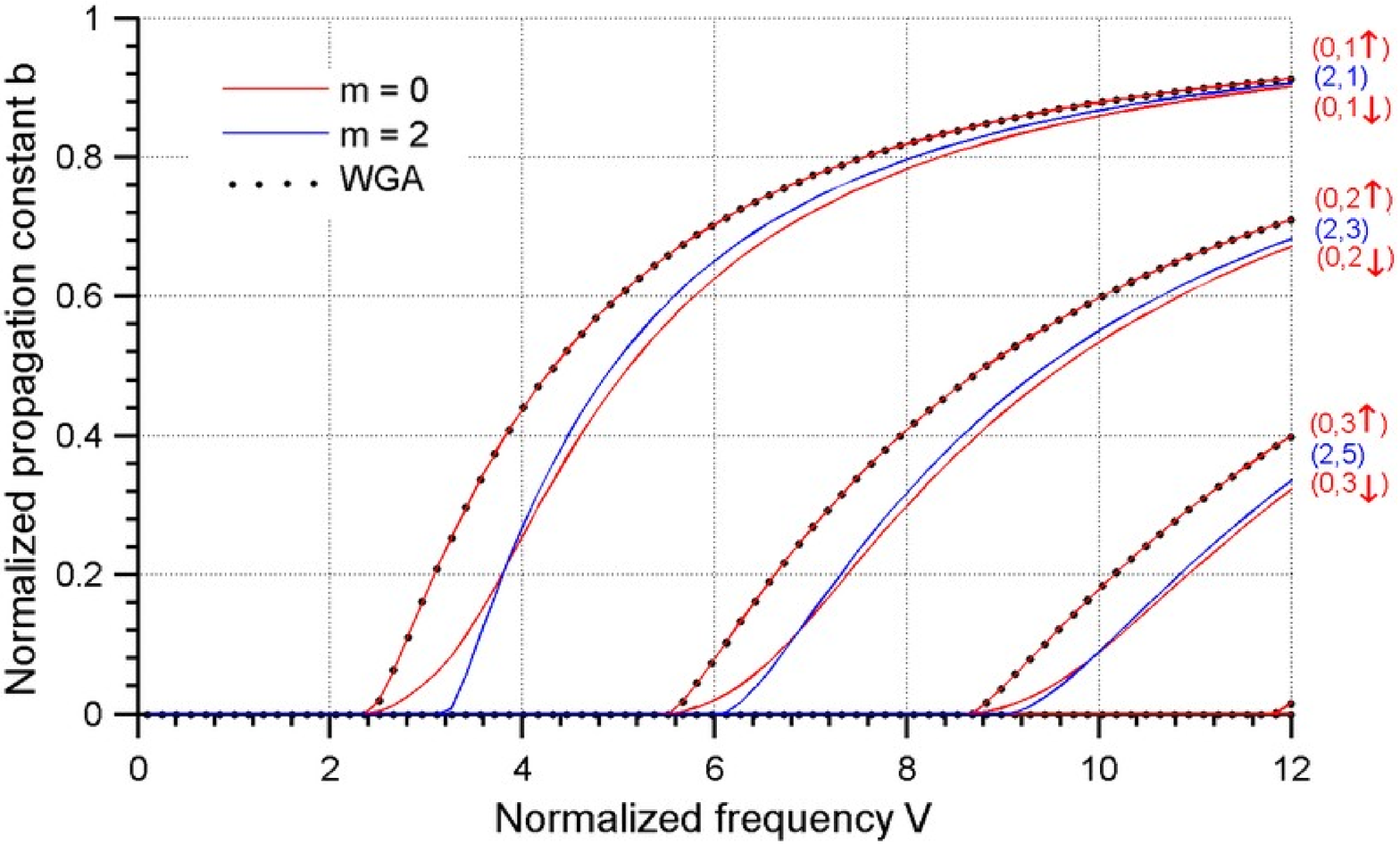}
{The split in dispersion curves between ${m = 0}$ and ${m = 2}$ mode families, and split between the modes inside the same family at ${m = 0}$. The waveguide and the medium refractive indices are ${n_{WG} = 3}$ and ${n_{Med.} = 1.33}$, respectively.}

The split between modes belonging to different families ${m = 0}$ and ${m = 2}$ was discussed above. Now, we focus our attention to the extra split within the same family of modes at ${m = 0}$. 
To understand this split we shall go back to the vectorial equation~(\ref{eq_cyl_vec_eigexact}):
\Eq{}
{\Scale[0.87]{
\begin{pmatrix}
d_\rho^2 + \frac{1}{\rho} d_\rho + (\ln \epsilon )' d_\rho - \frac{m^2+1}{\rho^2} + \epsilon k_o^2 + (\ln \epsilon )'' & - \frac{j2m}{\rho^2}\\
\frac{j2m}{\rho^2} + \frac{jm}{\rho}(\ln \epsilon )' & d_\rho^2 + \frac{1}{\rho} d_\rho - \frac{m^2+1}{\rho^2} + \epsilon k_o^2
\end{pmatrix}\begin{pmatrix}
E_\rho\\
E_\phi\\
\end{pmatrix}
= \beta^2 \begin{pmatrix}
E_\rho\\
E_\phi\\
\end{pmatrix},
}
}

which splits into two uncoupled separated differential equations at ${m = 0}$:
\Eq{split_1}
{\left(d_\rho^2 + \frac{1}{\rho} d_\rho + (\ln \epsilon )' d_\rho + \epsilon k_o^2 + (\ln \epsilon )'' \right)E_\rho = \beta^2 E_\rho,}
and
\Eq{split_2}
{\left(d_\rho^2 + \frac{1}{\rho} d_\rho + \epsilon k_o^2 \right)E_\phi = \beta^2 E_\phi.}
In the weakly guided approximation case the term ${(\ln \epsilon )'}$ can be neglected, thus equation~(\ref{split_1}) becomes identical to equation~(\ref{split_2}), hence the same propagation cases are obtained from both equations, and hence the corresponding modes are degenerate. 

However, if the refractive index difference between the waveguide and the medium is significant, the term ${(\ln \epsilon )'}$ can no longer be neglected, and thus different propagation constants would result from the solution to equations~(\ref{split_1}) and~(\ref{split_2}). The initial degeneracy would be removed. 

The result for two modes at ${m = 0}$ is shown in Figure~\ref{Mode_m0_split} for a significant refractive index difference. 
The modes~${\begin{pmatrix} E_\rho\\ 0\\ \end{pmatrix}}$ and~${\begin{pmatrix} 0\\ E_\phi\\ \end{pmatrix}}$ are obtained by solving equation~(\ref{split_1}) and~(\ref{split_2}), respectively. The split between the corresponding propagation constants is clearly seen. 
Thus the modes no longer coincide and hence can no longer be used to represent an arbitrary polarization, such as linear polarization, therefore the typically used term \C{``LP modes''} does not completely reflect the physical properties of such modes. 

\Fig{Mode_m0_split}{0.7}{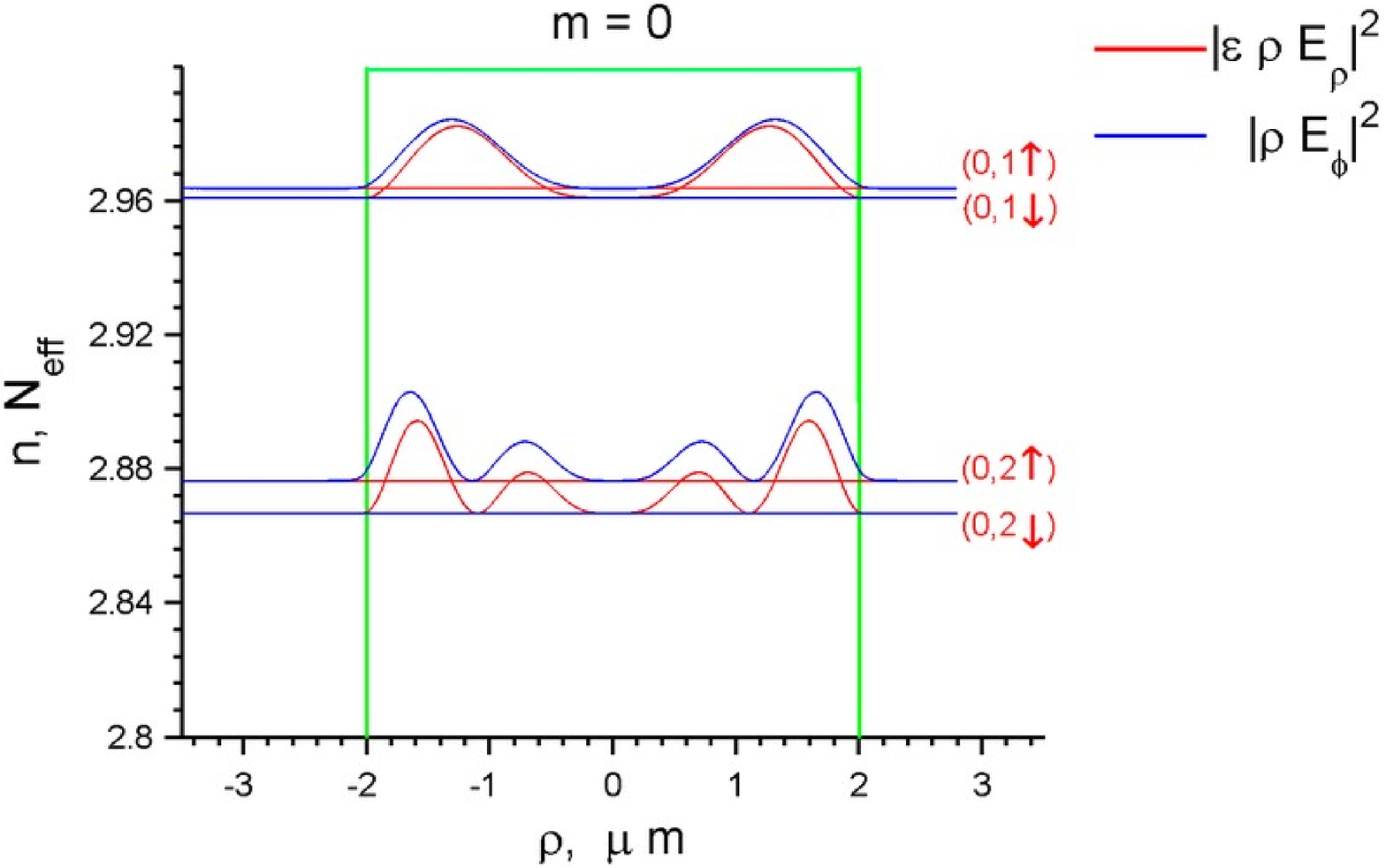}
{The split between the eigenvalues of pure radially polarized $E_\rho$ and pure angular polarized $E_\phi$ modes of the $m = 0$ family. 
The potential barrier is depicted with green color, the core refractive index ${n_2 = 3}$ is surrounded by the cladding with ${n_1 = 1.33}$.} 

The vector $\V E$ of the ${\begin{pmatrix} E_\rho\\ 0\\ \end{pmatrix}}$ mode is orientated radially,~\textit{i.e.} is aligned transversally with respect to the interface along the $\hat \rho$ vector, whereas the vector $\V E$ of the mode~${\begin{pmatrix} 0\\ E_\phi\\ \end{pmatrix}}$ is \C{aligned} tangentially to the interface, along the $\hat \phi$ vector. The remaining $E_z$ component gives a small contribution to both radial and tangential components.

The equations~(\ref{split_1}) and~(\ref{split_2}) can be rewritten in a more elegant Sturm-–Liouville form:
\begin{subequations}
\begin{align}
\label{cyl_TE}
	\left( \rho \frac{d}{d \rho} \left(\frac{1}{\rho}\frac{d}{d \rho}\right) + \epsilon k_o^2 \right) v &= \beta^2 v,\\
\label{cyl_TM}
	\left( \epsilon \rho \frac{d}{d \rho} \left(\frac{1}{\rho \epsilon}\frac{d}{d \rho} \right) + \epsilon k_o^2 \right) u &= \beta^2 u.
\end{align}
\end{subequations}
Here ${v = \rho E_\phi}$ and ${u = j \epsilon \rho E_\rho}$. The equations were derived with the help of~(\ref{fd_subst}). 

We note the similarity between equations~(\ref{cyl_TM}, \ref{cyl_TE}) and the eigenvalue equations for the TE~\ref{eq_f_TE} and TM~\ref{eq_f_TM2} modes in the case of slab waveguides:

\begin{subequations}
\begin{align}
\label{slab_TE}
	\left( \frac{d^2}{d x^2} + \epsilon k_o^2\right)v &= \beta^2 v\\
\label{slab_TM}
	\left( \epsilon \frac{d}{d x} \left( \frac{1}{\epsilon} \frac{d}{d x} \right) + \epsilon k_o^2 \right)u &= \beta^2 u,
\end{align}
\end{subequations}

Here ${v = E_y}$ and ${u = \epsilon E_x}$. Because of this similarity the solutions obtained form~(\ref{cyl_TE}) and~(\ref{cyl_TM}) equations sometimes called TE and TM modes, respectively. 
The role of $E_x$ and $E_y$ components in a slab waveguide is played by $E_\rho$ and $E_\phi$ components in the case of a cylindrical waveguide.

Both equations~(\ref{slab_TM}) for the TM modes in the case of a slab waveguide and the equation~(\ref{cyl_TM}) for the TM--like modes in the cylindrical waveguide case contain terms ${\epsilon \frac{d}{d x} \left( \frac{1}{\epsilon} \frac{d}{d x} \right) }$ and ${\epsilon \rho \frac{d}{d \rho} \left(\frac{1}{\rho \epsilon}\frac{d}{d \rho} \right)}$, respectively, are sensitive to the rate of change of permittivity. 
Therefore, we conclude that the TM and TM--like modes should be particular sensitive to changes in refractive index.

\Fig{Disper_m0_spread}{0.9}{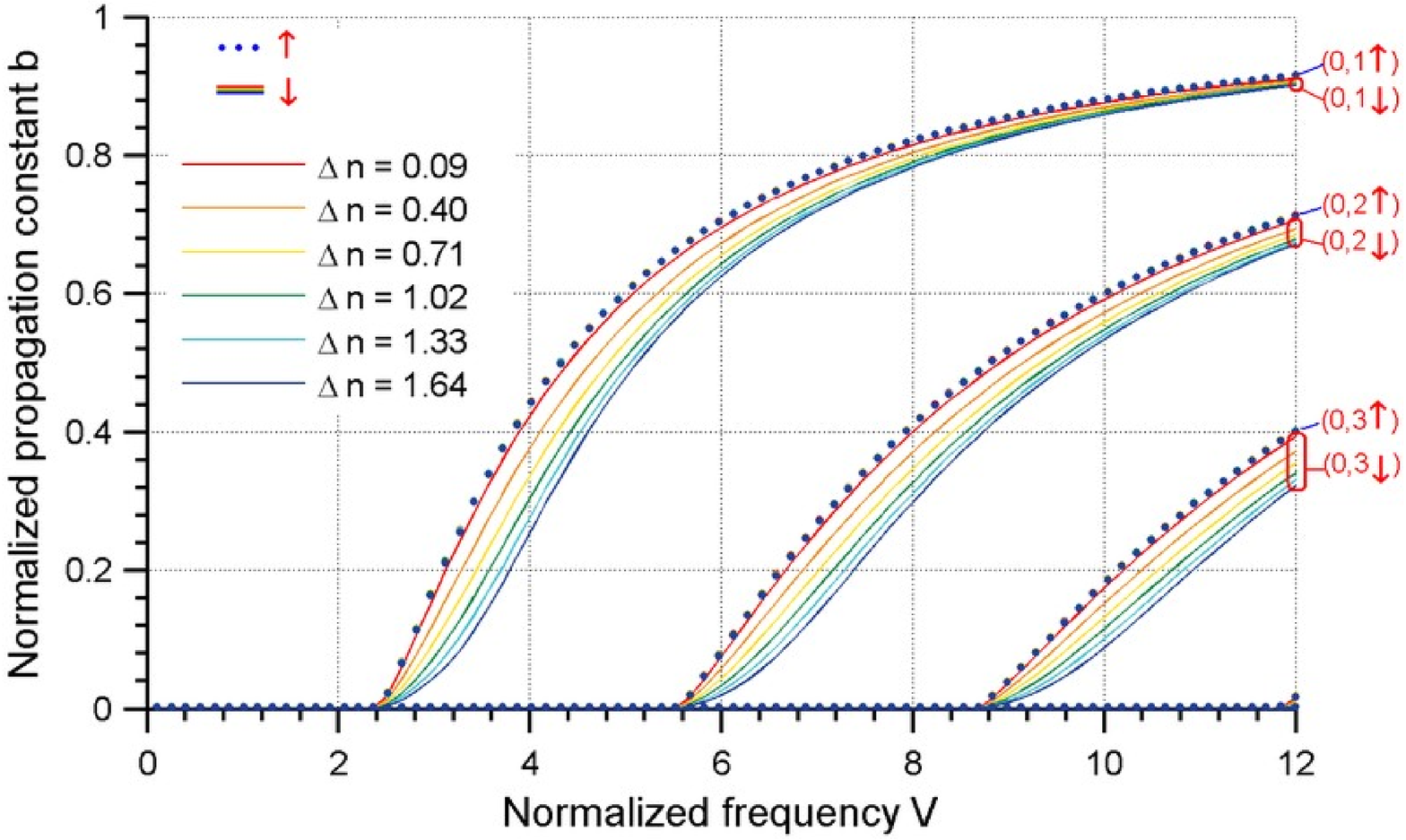}
{The dispersion curves, obtained by solving the exact vectorial equation, for various refractive index ratios ${\Delta n = n_{WG} - n_{Med.}}$ at ${m = 0}$. The waveguide is immersed in water ${n = 1.33}$.}

The dispersion curves for various index contrasts are shown in Figure~\ref{Disper_m0_spread}.
Indeed, it can be clearly seen that the TM--like modes for the $E_\rho$ component are seen shifting significantly with the increase in index contrast, whereas the TE--like modes for the $E_\phi$ component are almost unmovable.
The operator ${\epsilon \rho \frac{d}{d \rho} \left(\frac{1}{\rho \epsilon}\frac{d}{d \rho} \right)}$ is sensitive to changes in the refractive index profile, affecting the $E_\rho$ solution.
On the other hand, the rate of change in the refractive index is not incorporated in the $E_\phi$ solution, hence if the corresponding propagation constant is plotted along the ${b = \frac{N_{eff}^2 - n_2^2}{n_1^2 - n_2^2}}$ axis the corresponding dispersion curve should stay intact.

%% file: Chap_TFBG_5_Orthogonality.tex
\section{The orthogonality of the basis functions}

In the next Chapter we are going to look for a solution to the more general problem of a waveguide with a small perturbation along the $z$ axis.
The modes obtained for the unperturbed case might be used as a basis function in terms of which the general solution can be expressed. 
The calculations can be significantly simplified if the basis functions are orthogonal. 
In this section we derive the orthogonality relation in terms of a weighted product, ensuring the orthogonality of modes. 

To our knowledge the orthogonality relation for the vectorial modes of the form ${\V E = \begin{pmatrix} E_\rho\\ E_\phi\\ \end{pmatrix}}$ in cylindrical waveguides with an arbitrary radial refractive index profile is not available in literature.

\subsection{The orthogonality relation for the scalar modes}
Let us start by checking the orthogonality property of scalar modes.
The equation~(\ref{eq_cyl_scal_eig}) 
\Eq{}
{\left(\frac{d^2}{d \rho^2} + \frac{1}{\rho} \frac{d}{d \rho} - \frac{m^2}{\rho^2} + \epsilon k_o^2\right)E_z = \beta^2 E_z}
can be rewritten in the self-adjoint or Sturm-–Liouville form~\cite{Vladimirov:1971, MorseFeshbach:1953}:
\Eq{SL_base}
{\left[\frac{d}{d \rho} \left(p(\rho)\frac{d}{d \rho}\right) + q(\rho) + \lambda w(\rho)\right]e(\rho) = 0,}
that is
\Eq{SL_base_2}
{\left[ \frac{d}{d \rho} \left(\rho \frac{d}{d \rho}\right) + \left(\epsilon k_o^2 \rho - \frac{m^2}{\rho} \right) - \beta^2 \rho \right]e(\rho) = 0,}
where 
\Eqaaa{}
{p(\rho) &=& \rho}
{q(\rho) &=& \epsilon k_o^2 \rho - \frac{m^2}{\rho}}
{w(\rho) &=& \rho}

It can be proven that the eigenfunctions $e_k(\rho)$ obtained by solving equation~(\ref{SL_base}), written in the Sturm-–Liouville form, are orthogonal with respect to the weighted function $w(\rho)$,~\textit{i.e.}:
\Eq{eq_SL2}
{<e_k|w|e_j> =  \int_a^b w(\rho) e_k(\rho)e_j(\rho) d\rho = \delta_{kj},}
where $e_k(\rho)$ and $\lambda_k$ are the eigenfunctions and eigenvalues of~(\ref{SL_base_2}), respectively.

Here we provide a short proof of the fact that the scalar modes obtained from equation~(\ref{SL_base}) are indeed orthogonal, and later we will build our original prove for the vectorial case upon it.
Detailed discussions of the scalar case can be found in~\cite{Vladimirov:1971} and~\cite{ MorseFeshbach:1953}, although a more general proof for the vectorial case was not provided. 

We start with the eigenvalue problem written in the Sturm-–Liouville form~(\ref{SL_base}):
\Eq{}
{L e_k(\rho) = \lambda_k w(\rho) e_k(\rho).}
where
\Eq{}
{L = \frac{d}{d \rho}\left(p \frac{d}{d \rho}\right) + q.}
Considering
\Eqaa{}
{L u &=& \mu w u,}
{L v &=& \lambda w v,}
We can construct the following expression:
\Eq{}
{v L u - u L v = (\mu - \lambda) w u v.}
Inserting the operator $L$ and using the Lagrange's Identity~\cite{Vladimirov:1971,MorseFeshbach:1953} we obtain
\Eq{}
{d_\rho(p u d_\rho v - p v d_\rho u) = (\mu - \lambda) w u v.}
After integration over the interval ${\rho \in [a,b]}$
\Eq{}
{\int_a^b d_\rho(p u d_\rho v - p v d_\rho u) = \int_a^b (\mu - \lambda) w(\rho) u(\rho) v(\rho) d\rho,}
we come to the Green's Identity~\cite{Vladimirov:1971,MorseFeshbach:1953}:
\Eq{}
{\left[p(\rho) \left(u(\rho) \frac{d v(\rho)}{d \rho} - v(\rho) \frac{d u(\rho)}{d \rho}\right)\right]_a^b = (\mu - \lambda) \int_a^b w(\rho) u(\rho) v(\rho) d\rho.}
The left hand side vanishes if we consider that the field decays outside the waveguide boundary and is approximately zero at the numerical boundaries.
\Eqaa{}
{u(a) &=& v(a) = 0,}
{u(b) &=& v(b) = 0.}
Finally we obtain
\Eq{}
{0 = (\mu - \lambda) \int_a^b w(\rho) u(\rho) v(\rho) d\rho .}
For non-degenerate modes the above relation can be true only if 
\Eq{}
{\int_a^b w(\rho) u(\rho) v(\rho) d\rho = 0  \text{ , for  } \mu \ne \lambda.} 
Thus we have proven the orthogonality relation for the scalar case~(\ref{SL_base_2}):
\Eq{SL_scal_orthog2}
{<E_{z,\mu}|E_{z,\lambda}> = \int_a^b \rho E_{z,\mu}(\rho) E_{z,\lambda}(\rho) d\rho = 0 \text{ , for  } \mu \ne \lambda.}
In the next section, in a similar manner, we prove the orthogonality relation for the vector case.

\subsection{The orthogonality relation for the vectorial modes}
Let us assume that we have a system of equations: 
\Eq{SL_vect}{
\begin{bmatrix}
	\hat L_1 & a\\
	b & \hat L_2
\end{bmatrix}
\begin{pmatrix}
	u_1\\
	u_2
\end{pmatrix} = \lambda
\begin{bmatrix}
	w_{11} & 0\\
	0 & w_{22}
\end{bmatrix}
\begin{pmatrix}
	u_1\\
	u_2
\end{pmatrix},
}
where operators $\hat L_1$ and $\hat L_2$ are given in the Sturm-–Liouville form:
\Eqaa{}
{\hat L_1 = \frac{d}{d \rho}\left(p_1(\rho) \frac{d}{d \rho}\right) + q_1(\rho),}
{\hat L_2 = \frac{d}{d \rho}\left(p_2(\rho) \frac{d}{d \rho}\right) + q_2(\rho).}
Following the derivation in the previous section let us start with two modes:
\Eqaa{}
{\lbrack M \rbrack \V u &=& \lambda \lbrack W \rbrack \V u,}
{\lbrack M \rbrack \V v &=& \mu \lbrack W \rbrack \V v.}

Our goal here is to find the orthogonality relation between the two modes $\V u$ and~$\V v$.

First, let us construct the scalar product:
\Eq{}
{\V v^\dagger [M] \V u = \lambda \V v^\dagger [W] \V u,}
here ${\V v^\dagger = (\V v^T)^*}$. Rewriting the above relation in coordinate form we obtain:
\Eq{}
{(v_1^*, v_2^*)
\begin{bmatrix}
	\hat L_1 & g\\
	h & \hat L_2
\end{bmatrix}
\begin{pmatrix}
	u_1\\
	u_2
\end{pmatrix} = \lambda (v_1^*, v_2^*)
\begin{bmatrix}
	w_{11} & 0\\
	0 & w_{11}
\end{bmatrix}
\begin{pmatrix}
	u_1\\
	u_2
\end{pmatrix},
}
or
\Eq{}
{v_1^* \hat L_1 u_1 + g v_1^* u_2 + h v_2^* u_1 + v_2^* \hat L_2 u_2 = \lambda \V v^\dagger [W] \V u .}
Now the expression
\Eq{}
{\V v^\dagger [M] \V u - \V u^\dagger [M] \V v = \lambda \V v^\dagger [W] \V u - \mu \V u^\dagger [W] \V v}
can be rewritten in the form
\Eqaa{}
{(v_1 u_2 - u_1 v_2)(g - h) + (v_1 \hat L_1 u_1 + v_2 \hat L_2 u_2) - (u_1 \hat L_1 v_1 + u_2 \hat L_2 v_2) = }
{= (\lambda - \mu) \left(w_{11}u_1 v_1 + w_{22}u_2 v_2\right),}
or rearranging the terms we get:
\Eqaa{}
{(v_1 \hat L_1 u_1 - u_1 \hat L_1 v_1) + (v_2 \hat L_2 u_2 - u_2 \hat L_2 v_2) = }
{= (\lambda - \mu) (w_{11} u_1 v_1 + w_{22} u_2 v_2) + (v_1 u_2 - u_1 v_2)(h-g).}
Finally, applying the Lagrange's Identity we obtain:
\Eqaaa{}
{\left[p_1(\rho)\left(v_1 \frac{d u_1}{d \rho} - u_1 \frac{d v_1}{d \rho}\right)\right]_a^b + \left[p_2(\rho)\left(v_2 \frac{d u_2}{d \rho} - u_2 \frac{d v_2}{d \rho}\right)\right]_a^b =} 
{ = (\lambda - \mu) \int_a^ b (w_{11}(\rho) u_1(\rho) v_1(\rho) + w_{22}(\rho) u_2(\rho) v_2(\rho)) d \rho + }
{ + \int_a^b ((v_1(\rho) u_2(\rho) - u_1(\rho) v_2(\rho))(h(\rho) - g(\rho))) d \rho,}
Assuming as previously that at the computational boundaries all the field components vanish, we get the following expression
\Eqaa{SL_vect_not_ort}
{(\lambda - \mu) \int_a^ b (w_{11}(\rho) u_1(\rho) v_1(\rho) + w_{22}(\rho) u_2(\rho) v_2(\rho)) d \rho  = } 
{= \int_a^b ((v_1(\rho) u_2(\rho) - u_1(\rho) v_2(\rho))(h(\rho) - g(\rho))) d \rho.}
For the two different modes,~\textit{i.e.} ${\mu \ne \lambda}$, the expression 
\Eq{SL_vect_done}
{<u|W|v> = \int_a^b(w_{11}(\rho) u_1(\rho) v_1(\rho) + w_{22}(\rho) u_2(\rho) v_2(\rho)) d \rho \ne 0}
is \C{non zero}, thus we can not consider modes~$\V u$ and~$\V v$ to be orthogonal in the general case.
However, as we will see later, for a small azimuthal number~$m$ we can proximately consider modes to be orthogonal:
\Eq{}
{<u|W|v> \sim 0  \text{ , for  } \mu \ne \lambda.}

Now, let us apply the orthogonality expression~(\ref{SL_vect_done}) to a particular case of modes propagating in cylindrical waveguide of an arbitrary profile~(\ref{eq_cyl_vec_eigexact}):
\Eq{SL_vec_cyl}
{\Scale[0.8]{
\begin{bmatrix}
d_\rho^2 + \frac{1}{\rho} d_\rho + (\ln \epsilon )' d_\rho - \frac{m^2+1}{\rho^2} + \epsilon k_o^2 + (\ln \epsilon )'' - \beta^2 & - \frac{j2m}{\rho^2}\\
\frac{j2m}{\rho^2} + \frac{jm}{\rho}(\ln \epsilon )' & d_\rho^2 + \frac{1}{\rho} d_\rho - \frac{m^2+1}{\rho^2} + \epsilon k_o^2 - \beta^2
\end{bmatrix}\begin{pmatrix}
E_\rho\\
E_\phi\\
\end{pmatrix}
= \V 0
}
}

First, let us rewrite equation~(\ref{SL_vec_cyl}) in the Sturm-Liouville form~(\ref{SL_vect}). 
Considering that
\Eqaa{}
{\left(\frac{1}{r}(r u)'\right)' &=& u'' + \frac{1}{r} u' - \frac{1}{r^2} u,}
{\left(\frac{1}{r \epsilon}(r \epsilon u)'\right)' &=& u'' + \frac{1}{r} u' + (\ln \epsilon)' u' - \frac{1}{r^2} u + (\ln \epsilon)'' u}
we can rewrite the matrix~(\ref{}) equation in the following form:
\Eq{}
{\Scale[0.8]{
\begin{bmatrix}
\frac{d}{d \rho} \left(\frac{1}{\rho \epsilon}\frac{d}{d \rho} \rho \epsilon \right) + \frac{1}{\rho \epsilon}\left(- \frac{m^2}{\rho^2} + \epsilon k_o^2 - \beta^2 \right)\rho \epsilon & - \frac{1}{\rho }\frac{j2m}{\rho^2} \rho \\
\frac{1}{\rho \epsilon}\left(\frac{j2m}{\rho^2} + \frac{jm}{\rho}(\ln \epsilon )'\right)\rho \epsilon & \frac{d}{d \rho} \left(\frac{1}{\rho}\frac{d}{d \rho}\rho\right) + \frac{1}{\rho}\left(- \frac{m^2}{\rho^2} + \epsilon k_o^2 - \beta^2 \right)\rho
\end{bmatrix}
\begin{pmatrix}
E_\rho\\
E_\phi\\
\end{pmatrix}
= \V 0
}
}
Now we can pull $\epsilon \rho $ and $\rho $ under the $E_\rho$ and $E_\phi$ field components, respectively: 
\Eq{SL_vec_cyl_done}{
\begin{bmatrix}
	\hat L_1 & g\\
	h & \hat L_2
\end{bmatrix}
\begin{pmatrix}
	\epsilon \rho E_\rho\\
	\rho E_\phi
\end{pmatrix} = \beta^2
\begin{bmatrix}
	\frac{1}{\epsilon \rho} & 0\\
	0 & \frac{1}{\rho}
\end{bmatrix}
\begin{pmatrix}
	\epsilon \rho E_\rho\\
	\rho E_\phi
\end{pmatrix},
}
here
\Eqaa{}
{g(\rho) &=& -\frac{1}{\rho} \frac{2jm}{\rho^2}}
{h(\rho) &=& \frac{1}{\epsilon \rho}\left(\frac{2jm}{\rho^2} + (\ln \epsilon)' \frac{jm}{\rho}\right)}
\Eqaa{SL_vec_cyl_oper}
{\hat L_1 &=& \frac{d}{d \rho} \left(\frac{1}{\rho \epsilon}\frac{d}{d \rho} \right) + \frac{1}{\rho \epsilon} \left(- \frac{m^2}{\rho^2} + \epsilon k_o^2\right
)}
{\hat L_2 &=& \frac{d}{d \rho} \left(\frac{1}{\rho} \frac{d}{d \rho} \right) + \frac{1}{\rho} \left(- \frac{m^2}{\rho^2} + \epsilon k_o^2\right)}

Thus, the eigenvalue equation for cylindrical waveguide~(\ref{SL_vec_cyl}) can be written in the form of equation~(\ref{SL_vect}). 
Considering that
\Eqaa{}
{h(\rho) - g(\rho) &=& \frac{1}{\epsilon \rho}\left(\frac{2jm}{\rho^2} + (\ln \epsilon)' \frac{jm}{\rho}\right) + \frac{1}{\rho} \frac{2jm}{\rho^2} = }
{&=& \frac{jm}{\rho \epsilon}\left(2\frac{\epsilon+1}{\rho^2} + (\ln \epsilon)'\frac{1}{\rho}\right) = 0  \text{ , for  } m = 0,}
a standard orthogonality relation follows form~(\ref{SL_vect_not_ort}):
\Eq{SL_vect_apr_ort}
{<u|W|v> = \int_a^ b (w_{11}(\rho) u_1(\rho) v_1(\rho) + w_{22}(\rho) u_2(\rho) v_2(\rho)) d \rho = 0  \text{ , for  } \mu \ne \lambda.} 
The equation~(\ref{SL_vect_apr_ort}) approximately holds for small azimuthal numbers $m$, as we will see in the next section. The above expression can be rewritten in terms of vectorial modes  ${\V E = \begin{pmatrix} E_\rho\\ E_\phi\\ \end{pmatrix}}$ in cylindrical waveguides as follows:  
\Eqaa{}
{<u|W|v> = \int_a^ b (w_{11}(\rho) u_1(\rho) v_1(\rho) + w_{22}(\rho) u_2(\rho) v_2(\rho)) d \rho =}
{ = \int_a^b \left(\frac{1}{\rho \epsilon} (\rho \epsilon E_{\rho,\mu})^* (\rho \epsilon E_{\rho,\lambda}) + \frac{1}{\rho} (\rho E_{\phi,\mu})^* (\rho E_{\phi,\lambda})\right) d\rho}

Hence, for the same mode family with a relatively small azimuthal number~$m$:
\Eq{SL_vec_othog2}
{<\V E_\mu|\V E_\lambda> \sim \int_a^b \rho \left( \epsilon^* E_{\rho,\mu}^* E_{\rho,\lambda} +  E_{\phi,\mu}^* E_{\phi,\lambda}\right) d\rho = 0  \text{ , for  } \mu \ne \lambda.}
The above equation is used extensively in the following sections.
We note that the vectorial orthogonality relation~(\ref{SL_vec_othog2}) resembles the scalar orthogonality relation~(\ref{SL_scal_orthog2}). The same weighted function $\rho$ is used in both cases, although the extra weighted function ${\epsilon(\rho)}$ is also used to weight the $E_{\rho}$ component.

\subsection{Numerical verification}
Solving the eigenvalue problem~(\ref{eq_cyl_scal_eig}) or~(\ref{eq_cyl_vec_eigexact}) numerically we obtain a set of basis eigenvectors ${\{ \V e_k(\rho) \}}$. We can verify the solution by checking whether the obtained set of  eigenvectors satisfies the orthogonality relations~(\ref{SL_scal_orthog2}) for the scalar case~(\ref{eq_cyl_scal_eig}), or~(\ref{SL_vec_othog2}) for the vector case~(\ref{eq_cyl_vec_eigexact}).

We note that for a different $m$ number we get a different Sturm-Liouville problem, with a different set of eigenvalues ${\{ \beta_k^m \}}$ and eigenvectors ${\{ \V e_k(\rho)^m \}}$. 
For example, for the scalar case~(\ref{eq_cyl_scal_eig}) we have:
\Eq{eq_SL3}
{\left[d_\rho (\rho d_\rho) + \left(\rho k_o^2 n_o^2(\rho) - \frac{m^2}{\rho} \right) - \lambda_k^m \rho \right]R_k^m(\rho) = 0,}
here index $m$ denotes the family of modes obtained for a given angular momentum defined by the $m$ number. 

According to the equation~(\ref{SL_scal_orthog2}) the orthogonality relation is defined as 
\Eq{eq_SL4}
{<R_k^m|R_j^m> =  \int_0^\infty \rho R_k^n(\rho)R_j^m(\rho) d\rho,}
here we considered the proper boundary conditions on ${[a,b] = [0,\infty]}$.

Thus equation~(\ref{eq_SL3}) allows us to construct a complete set of basis functions, orthogonal in the sense of~(\ref{eq_SL4}). As we noted previously, the orthogonality property will substantially simplify the analysis of a more general problem in the following chapter. 

Now let us construct the overlap matrix ${|R_j^m><R_k^n| = C_{jk}^{mn}}$ representing the overlap integral between $j$-th and $k$-th modes, taken from the $m$-th and $n$-th families. 
\Eq{orthogon_scalar}
{|R_j^m><R_k^m| = \int_0^\infty \rho R_k^m(\rho) R_j^m(\rho) d\rho = C_{jk}^{mm} = 0 \indent if \indent j \ne k.}
For convenience, let us norm each basis function to unity:
\Eq{}
{\int_0^\infty \rho |R_k^m(\rho)|^2 d\rho = 1.}
We can view the normalization procedure as a construction of a new basis:
\Eq{}
{R_k^m(\rho) = \frac{R_k^m(\rho)}{\sqrt{\int_0^\infty \rho |R_k^m\rho)|^2 d\rho}}.}

The resulting overlap matrix $C_{jk}^{mn}$ is shown in Figure~\ref{Modes_orthog}.
As can be seen, all the modes belonging to the same family of modes, with the identical $m$ number, are mutually orthogonal,  regardless of the refractive index profile $n(\rho)$. 
However, if the overlap integral is computed between modes of different families, for instance $R_k^0(\rho)$ and $R_j^{10}(\rho)$, the result is \C{non zero} in most of the cases. The overlap matrix is dense. 
Indeed, each family of modes is a solution to a different Sturm-Liouville problem, uniquely defined by a different potential barrier~(\ref{eq_pot_barier}). 

\Fig{Modes_orthog}{1}
{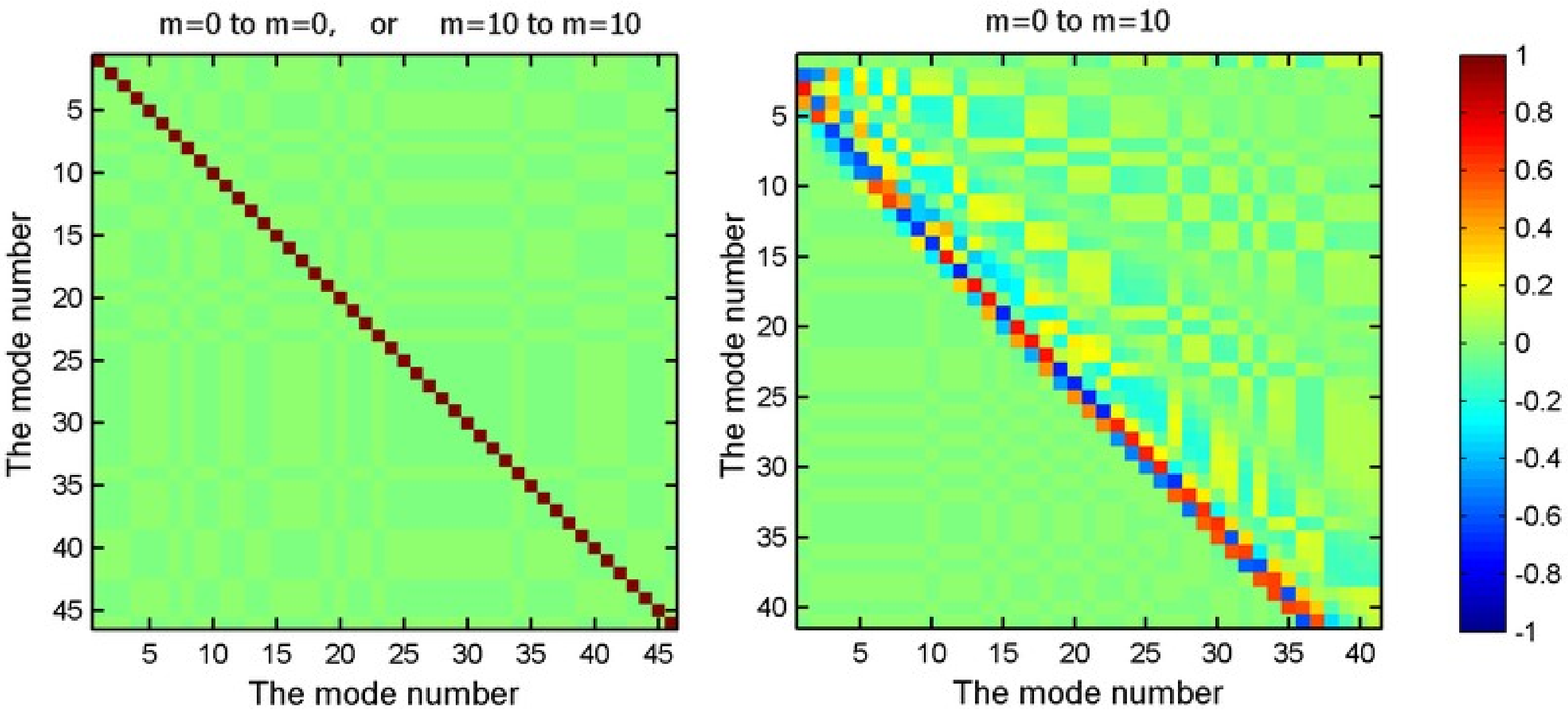}{The overlap matrix $C_{jk}^{mm}$ is computed for SMF-28 fibre immersed in water. The matrix is an identity for modes belonging to the same family of mode ($C_{jk}^{mm} = I$ for ${m=0}$ or ${m=10}$). However, if a different families of modes (${m=0}$ and ${n=10}$) are considered, the overlap matrix $C_{jk}^{mn}$ is dense.}

A similar analysis can be conducted for the vectorial case. The orthogonality relation for a small azimuthal number~$m$ can be written in accordance with~(\ref{SL_vec_othog2}) as follows: 
\Eq{orthogon_scalar}
{C_{jk}^{mm} = |R_j^m><R_k^m| = \int_a^b \rho \left( \epsilon^* R_{\rho, j}^{*m} R_{\rho, k}^m +  R_{\phi, j}^{*m} R_{\phi,k}^m\right) d\rho \sim 0 \indent if \indent j \ne k,}
The radial function can be normed for convenience:
\Eq{}
{R_j^m(\rho) = \frac{R_j^m(\rho)}{\sqrt{\int_a^b \rho \left( \epsilon^* R_{\rho, j}^{*m} R_{\rho, j}^m +  R_{\phi, j}^{*m} R_{\phi,j}^m\right) d\rho}}.}

The overlap matrix $C_{jk}^{mn}$ is shown in Figure~\ref{Modes_orthog_vec}.
\Fig{Modes_orthog_vec}{1}
{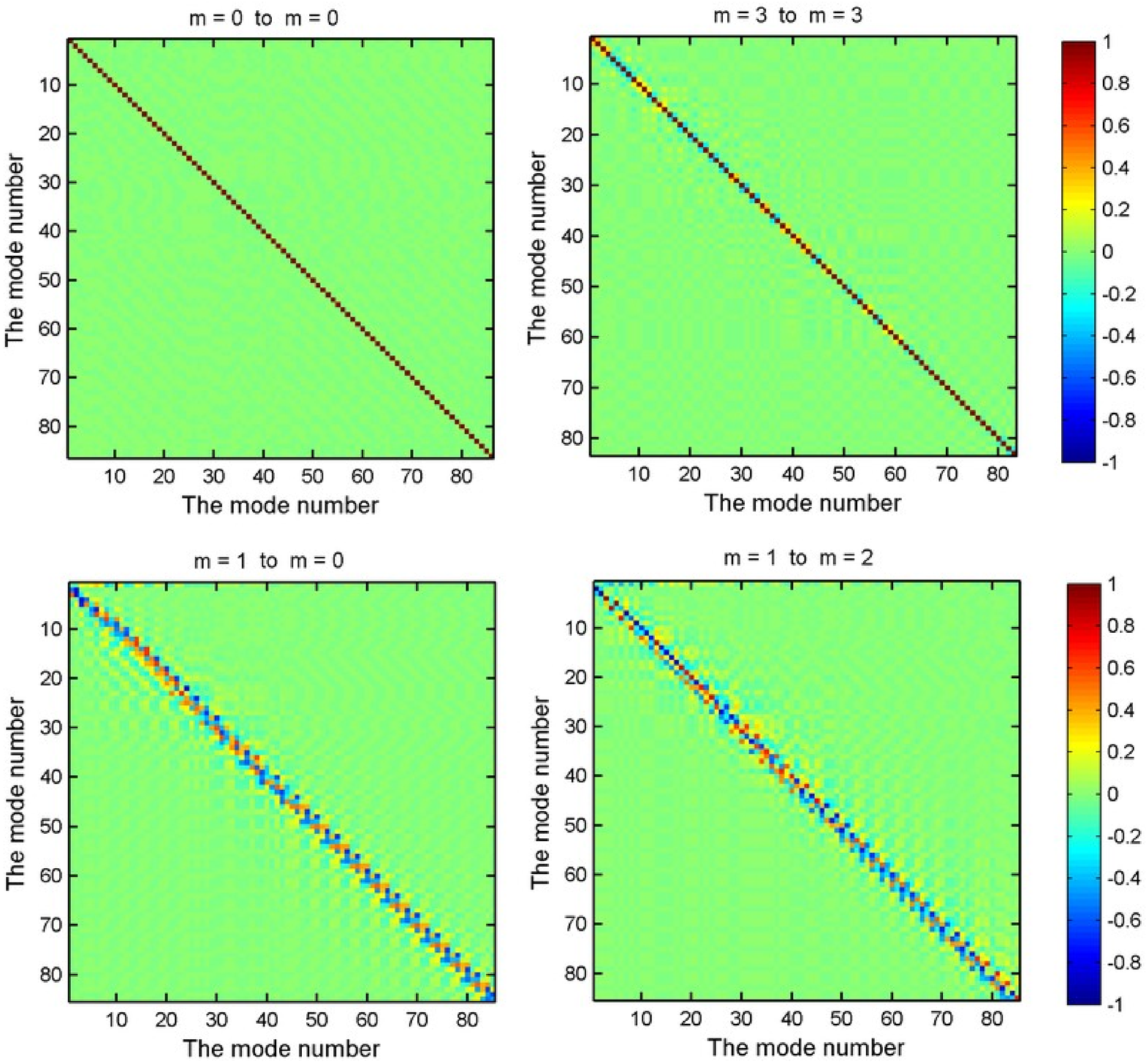}{The overlap matrix for the vectorial case is computed for SMF-28 fibre immersed in water. The overlap matrix $C_{jk}^{mm}$, for the modes belonging to the same family~$m$, is an identity matrix for $m = 0$ or close to the identity for small azimuthal numbers. For different families of modes, the overlap matrix is dense.}

We note that for the vectorial case the number of modes is approximately doubled in comparison with the scalar case due to the mode splitting, as was discussed in Section~\ref{modes_discussion}.

As can be seen from Figure~\ref{Modes_orthog_vec}, the orthogonality relation holds almost exactly for small azimuthal numbers~$m$. In the next section we will see that we need modes only from a few first families with a small~$m$ number. In addition, we will show that due to the phase matching condition, the overlap within a family of modes can be neglected, except for the $m = 1$ family.

We conclude that radial components of the basis functions are not mutually orthogonal for different families of modes,~\textit{i.e.} families with different azimuthal numbers, in both scalar and vectorial cases. However, within a particular family~$m$ the radial components can be considered to be mutually orthogonal.
In the following section we will see that although radial components belonging to different families of modes may not necessarily be mutually orthogonal, the complete basis functions are always orthogonal due to the orthogonality of angular components.

\section{Conclusion}

We conclude this section by stating that the scalar model yields the same result as the vector model if the refractive index contrast is small and hence the weakly guided approximation can be applied. 
However, if the index contrast is high each scalar mode splits, in general, into two separate modes.

In the vectorial case the first mode, with the highest effective refractive index, occurs at~${m=1}$ and is single. The remaining modes come in pairs (for example~${(m=0,m=2)}$,${(m=1,m=3)}$,~\textit{etc.}) and are observed as doublets in the spectrum, which is to say have almost identical dispersion curves in the case of a small refractive index contrast. 
The modes at~${m=0}$ have an additional split due to the independence of radial $E_\rho$ and angular $E_\phi$ components, thus triplets should be observed in the spectrum. 

It is convenient to distinguish modes with the dominant $E_\rho$ and $E_\phi$ components and call them TM and TE--like modes, respectively. 
In the case of ${m = 0}$ we have the full analogy with slab waveguides: the TM--like~(\ref{cyl_TM}) and TE--like~(\ref{cyl_TE}) modes in the cylindrical case are described with similar equations as used for TM~(\ref{eq_f_TM2}) and TE~(\ref{eq_f_TE}) modes in slab waveguides. 
The behaviour of TE and TM--like modes in cylindrical case resembles the behaviour of TE and TM modes in slab waveguides, which is to say TE--like group of modes are relatively unmoved to changes in the refractive index of external medium. Hence, approximately $50\%$ of modes can be obtained correctly with the scalar model even in the case of high refractive index contrast. 

It is interesting to note that the so-called LP modes occur at $m = 0$, where $E_\rho$ and $E_\phi$ components are decoupled. 
Thus, one of the components can be chosen independently of the other. 
Hence, when the refractive index contrast between the core and cladding is small and two modes coincide, the LP (linearly polarized) or CP (circularly polarized ) modes can be contracted by superposing the $E_\rho$ and $E_\phi$ field components. 
However, this procedure is no longer correct if the refractive index contrast is high. 
The modes ${\begin{pmatrix} 0\\ E_\phi\\ \end{pmatrix}}$ and ${\begin{pmatrix} E_\rho\\ 0\\ \end{pmatrix}}$ become separated and can no longer be used as a basis for construction of LP or CP modes.  
The name LP modes can be in a way misleading if the weakly guided approximation is not valid.
In the general case at ${m = 0}$ we have a set modes, coming in pairs, with a slightly different propagation constants and distinct alignment of the vector $\V E$, either radial or tangential. Thus the energy can be coupled separately into ${\begin{pmatrix} 0\\ E_\phi\\ \end{pmatrix}}$ or ${\begin{pmatrix} E_\rho\\ 0\\ \end{pmatrix}}$ modes.

The split between the two LP modes can be viewed as a manifestation of \C{light's} intrinsic degree of freedom. The observed split of degenerate states resembles the Zeeman effect where the electron energy levels are spit due to the electrons intrinsic degree of freedom.


We also note that the presented dispersion curves showed that the modes positioned close to the cutoff region are the most sensitive to the refractive index changes in the surrounding medium. 
The largest split between the TE--like and TM--like modes is also observed at some small distance away from the cutoff region, thus we can particularly target those modes close to the cutoff region to achieve the largest sensor sensitivity.

%% file: Chap_TFBG_CMT_1_Derive.tex
\chapter{Modeling of tilted Bragg grating (TFBG) structures}

The goal of this chapter is to obtain the exact solution to the problem of light propagating through a cylindrical waveguide with a tilted periodic structure inscribed along its longitudinal axis.

\section{Derivation}

The problem geometry is shown in Figure~\ref{TFBG_grating_img_Gold}.
First, let us consider the scalar Helmholtz equation~(\ref{eq_cyl_scal}):

\Eq{eq_Hlmz_scal}{(\nabla^2 + k^2)u(\V r) = f(\V r),}
here $f(\V r)$  is the excitation term. 
Assuming that the excitation is a harmonic function $f(\V r,t) \sim f(\V r)e^{j\omega t}$ , the solution for a linear system should also have a harmonic function $u(\V r,t) \sim u(\V r)e^{j\omega t}$.

\Fig{TFBG_grating_img_Gold}{0.8}{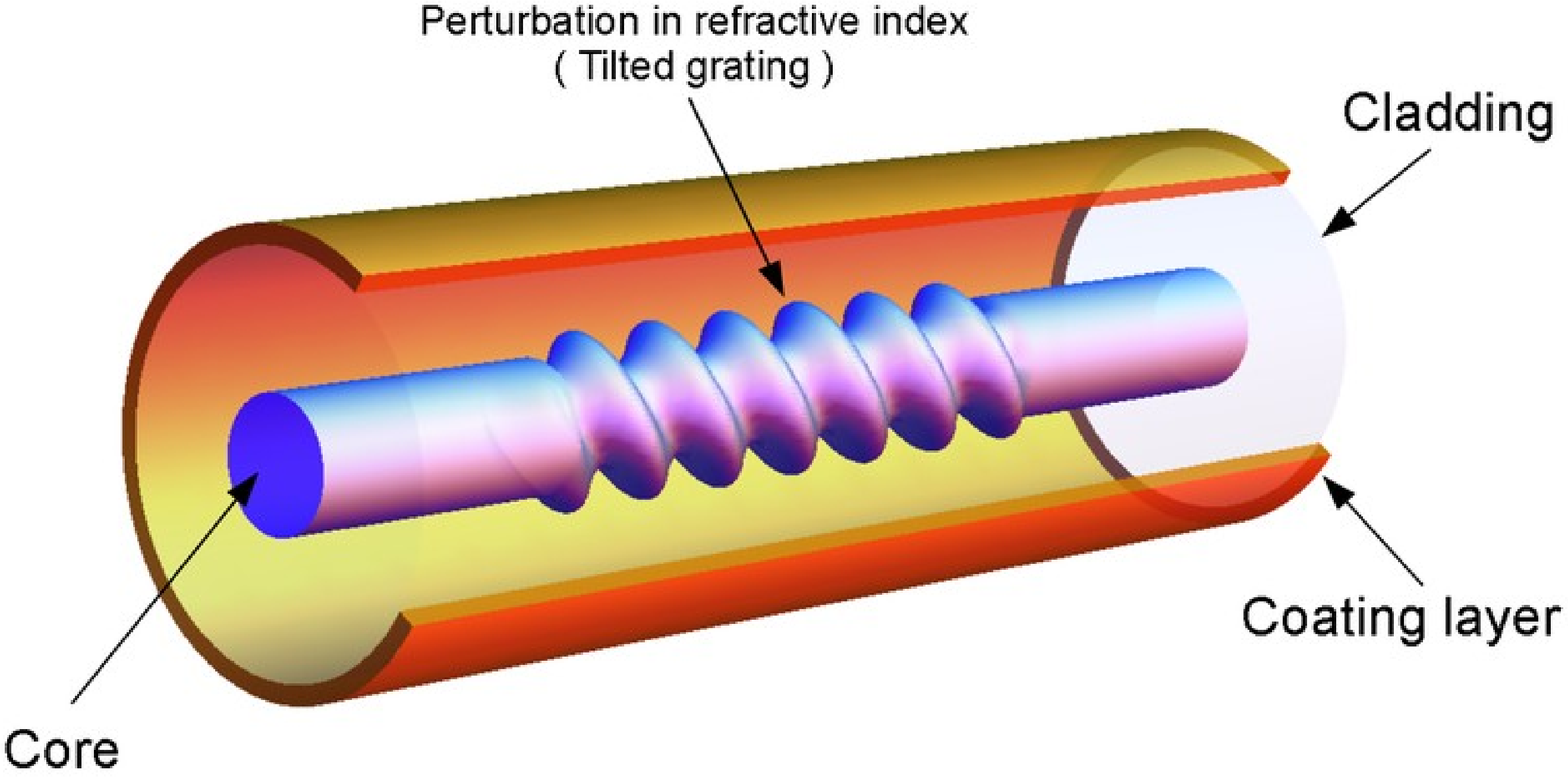}
{The schematic representation of the problem geometry.}

 In a cylindrical coordinate system the equation~(\ref{eq_Hlmz_scal}) can be written in the following form:
\Eq{eq_cyl1}
{\left(\frac{1}{\rho} \partial_\rho (\rho \partial_\rho) + \frac{1}{\rho^2} \partial_\phi^2 + \partial_z^2 + k_o^2 n^2(\rho, \phi, z)\right) u(\rho, \phi, z) = f(\rho, \phi, z),}
or
\Eq{eq_cyl2}
{L u(\rho, \phi, z) = f(\rho, \phi, z),}
We should note that $\rho$, $\phi$ and $z$ coordinates are coupled through the refractive index $n = n(\rho,\phi,z)$ function, dependent on $\rho$, $\phi$ and $z$ coordinates. Thus we have a coupled partial differential equation. 
For a non-tilted grating the coupling would only occur between $z$ and $\rho$ coordinates, but for a tilted grating all \C{three} coordinates ( $z$, $\rho$ and $\phi$ ) are coupled.

If the coupling is relatively small we can use perturbation theory. First we find solutions for the case of unperturbed decoupled equation, and then, using these solutions as a basis we solve the coupled problem in terms of this basis.

It is in our interest to keep the coupled term as small as possible. 
Considering that the perturbation along the $z$ axis is significantly smaller than the perturbation along the radial direction, the following decomposition of the refractive index profile can be written:
\Eq{eq_dn}
{n^2(\rho, \phi, z) \sim n_o^2(\rho) + \Delta(\rho, \phi, z),}
Now the coupling term  $\Delta(\rho, \phi, z)$ is as small as possible.

The next step is to find the basis functions $e_k(r)$ in terms of which we can represent the solution of the coupled problem. 
There are several possible ways the basis function $e_k(r)$ can be chosen, however it is convenient to use the natural basis of the radial part $L_{\rho}$ of the $L$ operator~(\ref{eq_cyl2}).
Such basis functions should satisfy the orthogonality condition~(\ref{Modes_orthog}) with respect to the weighting functions. As well, they should form the complete set of functions, spanning the solution space. 

Before we continue let us split the operator $L$, defined in~(\ref{eq_cyl1}),~(\ref{eq_cyl2}), into the radial and longitudinal parts. 
With this in mind, we search for a solution in the form of the series:
\Eq{eq_sol1}
{u(\rho, \phi, z) = \sum_m c^m(\rho, z) e^{j m\phi}.}
From now on we will use the upper indices to referencing the functions dependent on the angular coordinate.

Ensuring that the function $u(\rho, \phi, z)$ at $\phi=0$ has the same value as at $\phi = 2\pi$ , we conclude that  $m = \pm 1,2,3, ...$ for $\phi \in [0, 2\pi]$.

Plugging the series expansion~(\ref{eq_sol1}) into the Helmholtz equation~(\ref{eq_cyl1}) and considering~(\ref{eq_dn}) we get:
\Eq{eq_sol2}
{\sum_m \left(\frac{1}{\rho} \partial_\rho (\rho \partial_\rho) - \frac{m^2}{\rho^2} + k_o^2 n_o^2(\rho) + \partial_z^2 + k_o^2 \Delta(\rho, \phi, z)\right)c^m (\rho, z)e^{jm\phi} = f(\rho, \phi, z),}
or
\Eq{eq_sol3}
{\sum_m \left(L_\rho^m + \partial_z^2 + k_o^2 \Delta(\rho, \phi, z)\right)c^m (\rho, z)e^{jm\phi} = f(\rho, \phi, z).}
Here the operator acting on $\rho$ is denoted as $L_\rho^m$, and it is the radial part of the $L$ operator:
\Eq{}
{L_\rho^m = \frac{1}{\rho} d_\rho (\rho d_\rho) - \frac{m^2}{\rho^2} + k_o^2 n_o^2(\rho).}
Finally the basis functions can be found by solving the eigenvalue problem:
\Eq{eq_eig_new}
{L_\rho^m e_k^m(\rho) = \lambda_k^m e_k^m(\rho).}
Here $e_k^m(\rho)$ are the eigenfunctions, spanning the solution space,  which we are going to use as the basis function to solve the coupled problem~(\ref{eq_cyl1}). The eigenvalues are defined as $\lambda_k^m$.

Let us assume that the basis functions $e_k^m(\rho)$ were successfully found, and we can continue solving the initial coupled equation~(\ref{eq_cyl1}). 
At this point the coupling appears through $c^m(\rho, z)$ and~${\Delta(\rho, \phi, z)}$ terms in equations (\ref{eq_sol1}) and (\ref{eq_sol2}). 
The coordinates $z$ and $\rho$ can be decoupled by expanding $c^m(\rho, z)$ term into a series over the basis functions $e_k^m(\rho)$, similarly to what we did in~(\ref{eq_sol1}):
\Eq{eq_ser2}
{c_m (\rho, z) = \sum_k C_k^m(z)e_k^m(\rho).}
Upon inserting the series~(\ref{eq_ser2}) into~(\ref{eq_sol3}) we get:
\Eq{eq_ser3}
{\sum_k \sum_m \left(L_\rho^m + \partial_z^2 + k_o^2 \Delta(\rho, \phi, z)\right)C_k^m(z)e_k^m(\rho)e^{jm\phi} = f(\rho, \phi, z).}
The main advantage of our basis now becomes evident. Let us consider the eigenvalue equation~(\ref{eq_eig_new}) and replace ${L_\rho^m e_k^m(\rho)}$ with ${\lambda_k^m e_k^m(\rho)}$ : 
\Eq{eq_ser4}
{\sum_k \sum_m \left(\lambda_k^m + d_z^2 + k_o^2 \Delta(\rho, \phi, z)\right)C_k^m(z)e_k^m(\rho)e^{jm\phi} = f(\rho, \phi, z).}
We are left with the system of coupled ordinary differential equations of only one variable $z$. 

Let us simplify the following derivation by introducing $\psi$ variable:
\Eq{eq_basis}
{\psi_k^m(\rho,\phi) = e_k^m(\rho)e^{jm\phi},}
thus the equation (\ref{eq_ser4}) becomes:
\Eq{eq_ser5}
{\sum_k \sum_m \left(\lambda_k^m + d_z^2 + k_o^2 \Delta(\rho, \phi, z)\right)C_k^m(z) \psi_k^m(\rho,\phi) = f(\rho, \phi, z).}
The upper and lower indices are used to refer to a particular mode family and a particular mode in the given family, respectively.

Now let us consider the orthogonality properties of the $\psi_k^m(\rho,\phi)$ functions:

\Eqaaa{eq_orthog}
{\int_0^{2\pi} \int_0^\infty \rho \psi_i^n {\psi^*}_k^m d\rho d\phi &=& \int_0^{2\pi} \int_0^\infty \rho  (e_i^n(\rho)e^{jn\phi})(e_k^m(\rho)e^{-jm\phi}) d\rho d\phi }
{ &=& \int_0^\infty \rho e_i^n(\rho) e_k^m(\rho)d\rho \int_0^{2\pi} e^{jn\phi}e^{-jm\phi} d\phi }
{ &=& \gamma_{ik}^{nm} \delta_{ik} 2\pi \delta_{mn},}
where the norming factor $\gamma_{ik}^{nm}$ is defined as:
\Eq{eq_norm}
{\gamma_{ik}^{nm} = <e_i^n|e_k^m> = \int_0^\infty \rho e_i^n(\rho) e_k^m(\rho)d\rho.}
Here we have considered the orthogonality relation of harmonic functions and the basis functions~(\ref{eq_SL4}).

The above expression~(\ref{eq_orthog}) is \C{non zero} if and only if $i\ne k$ and $m \ne n$ simultaneously, otherwise it is zero, thus we have proved that $\psi_k^m(\rho,\phi)$ functions are orthogonal. Using this property let us take the scalar product of both sides of equation (\ref{eq_ser5}):
\small
\Eqaaa{}
{\sum_{k,m}(\lambda_k^m + d_z^2)\left[\int \rho {\psi^*}_i^n \psi_k^m d\rho d\phi\right] C_k^m(z) &+& \sum_{k,m}\left[\int \rho {\psi^*}_i^n \psi_k^m k_o^2\Delta d\rho d\phi\right] C_k^m(z) =}
{ &=&  \int \rho {\psi^*}_i^n f(\rho, \phi, z) d\rho d\phi,}
{}
\normalsize
here ${{\psi^*}_i^n(\rho,\phi) = e_i^n(\rho)e^{-jn\phi}}$ is the complex conjugate of $\psi_i^n(\rho,\phi)$ function.

Finally, considering the orthogonality relation we get the system of differential equations: 
\Eq{eq_coup}
{(\lambda_i^n + d_z^2) C_i^n(z)+ \sum_k \sum_m \left[M_{ik}^{nm}(z)\right] C_k^m(z) = F_i^n(z),}
here
\Eqaaa{eq_coup2}
{M_{ik}^{nm}(z) &=& \frac{1}{2\pi} \frac{1}{\gamma_i^n} \int_0^{2\pi} \int_0^\infty \rho k_o^2\Delta(\rho,\phi, z) e_i^n(\rho)e_k^m(\rho) e^{-jn\phi} e^{jm\phi} d\rho d\phi,}
{F_i^n(z) &=& \frac{1}{2\pi} \frac{1}{\gamma_i^n} \int_0^{2\pi} \int_0^\infty \rho f(\rho, \phi, z)e_i^n(\rho)e^{-jn\phi} d\rho d\phi,}
{\gamma_i^n &=& \int_0^\infty \rho e_i^n(\rho) e_i^n(\rho) d\rho.}
Once the system~(\ref{eq_coup}) is solved for $C_i^n(z)$ functions, the final solution can be constructed with help of~(\ref{eq_sol1}) and~(\ref{eq_ser2}):
\Eq{eq_final_sol}
{u(\rho, \phi, z) = \sum_m \sum_k C_k^m(z)e_k^m(\rho) e^{j m\phi}.}

The system of equations~(\ref{eq_coup},\ref{eq_coup2}) allow us to model the process of light propagating through a cylindrical waveguide, with a tilted grating inscribed along its longitudinal axis. 

Let us remark here that equations in system~(\ref{eq_coup}) are enumerated by an unique pair of indices. 
In analogy with quantum mechanics we can refer to the upper index as the orbital quantum number, pointing to a particular mode family with a unique angular momentum; and refer to the lower index as the main quantum number, determining a particular mode in the given family of modes.

In the most general case of vectorial modes~(\ref{eq_cyl_vec_eigexact}) the eigenvalue equation is defined as follows:
\Eq{}
{\Scale[0.87]{
\begin{pmatrix}
d_\rho^2 + \frac{1}{\rho} d_\rho + (\ln \epsilon )' d_\rho - \frac{m^2+1}{\rho^2} + \epsilon k_o^2 + (\ln \epsilon )'' & - \frac{j2m}{\rho^2}\\
\frac{j2m}{\rho^2} + \frac{jm}{\rho}(\ln \epsilon )' & d_\rho^2 + \frac{1}{\rho} d_\rho - \frac{m^2+1}{\rho^2} + \epsilon k_o^2
\end{pmatrix}\begin{pmatrix}
E_\rho\\
E_\phi\\
\end{pmatrix}
= \beta^2 \begin{pmatrix}
E_\rho\\
E_\phi\\
\end{pmatrix}
}
}
or
\Eq{}
{ [L_\rho^m] \V e_k^m(\rho) = \lambda_k^m \V e_k^m(\rho).}
Here ${\epsilon = \epsilon(\rho)}$ is considered to be independent of $z$ and $\phi$ variables. 
The dependence is included as a small perturbation in the ${\Delta(\rho,\phi, z)}$ function on the next step as it is described above. The following derivation is identical except that now we shall consider the vectorial basis functions $\V e_k^m(\rho) = \begin{pmatrix} u_k^m\\ v_k^m\\ \end{pmatrix}$ instead of the scalar basis function ${e_k^m(\rho)}$. 
As a result, the form of equations~(\ref{eq_coup}) is preserved, but the matrix coefficients are computed differently. The scalar product has to be modified in accordance with~(\ref{SL_vec_othog2}): 
\Eq{orthogonal_vector}
{\gamma_{ik}^{nm} = <e_i^n|e_k^m> = \int_0^\infty \left(\epsilon(\rho) u_i^n (\rho) u_i^n (\rho) +  v_i^n (\rho) v_i^n(\rho)\right) d\rho.}
The remaining steps are analogous to the scalar case. 

In this section we showed that the initial problem, defined by the partial differential equation~(\ref{eq_cyl_scal}) or~(\ref{eq_cyl_vec_eigexact}), with coupling along all the three coordinates $z$, $\phi$ and $\rho$, can be successfully reduced to a system of ordinary differential equations, coupled only along the $z$ axis. The coupling was introduced through the matrix elements ${\left[M_{ik}^{nm}(z)\right]}$, varying along the $z$ axis and defined by the grating profile ${\Delta(\rho, \phi, z)}$. 
The basis functions $e_i^n(\rho)$ are defined by the equation~(\ref{eq_SL3}) and depend on the waveguide radial profile. We will determine the matrix elements and basis functions in the following sections.

%% file: Chap_TFBG_CMT_2_Grating.tex
\section{The matrix elements}

In this section we compute matrix elements ${\left[M_{ik}^{nm}(z)\right]}$, defined in equation~(\ref{eq_coup}), for a tilted fibre Bragg grating. Next, energy transfer from the fundamental core mode into a different set of cladding modes can be determined with the help of the coupled-mode theory (CMT). 

The approach based on the coupled-mode theory (CMT) with application to the tilted fibre grating was initially introduced by Erdogan and Sipe~\cite{Erdogan_96}, following by a series of publications presenting numerical results~\cite{Erdogan_97, Lee_00}.
Alternatively, the method based on antenna theory and equivalence theorem~\cite{Schelkunoff_1936}, called the volume current method (VCM), can be implemented as described in~\cite{Jordan_1994, Holmes_1999, Yufeng_2001}, where the scattered field is represented by radiation of an array of elementary current dipole located at the index perturbation. 
Thus, instead of approaching the problem from a pure mathematical point of view, the physical analogy between a perturbation in refractive index and an antenna can be established.
The comparison between CMT and VCM methods was presented in~\cite{Yufeng_2006}.

The tilted fibre Bragg grating is shown schematically in Figure~\ref{TFBG_grating_img_Bare}.
\Fig{TFBG_grating_img_Bare}{0.65}{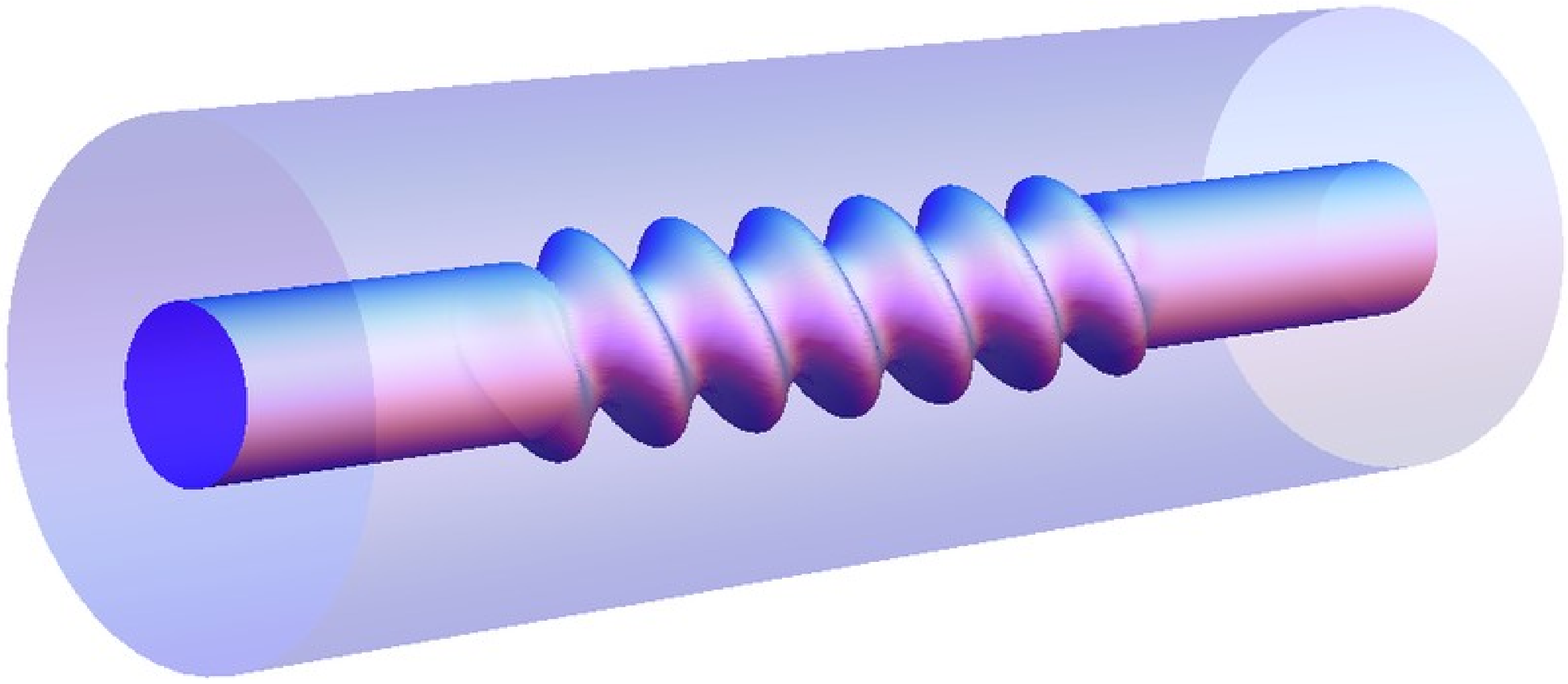}
{Illustration of a tilted fibre grating}

The grating is specially modulated by a sine function written with fringe planes that are blazed with respect to the optical axis.
We will see later that the tilt of the grating is the key element for selective and polarization-dependent light coupling. 

\Fig{TFBG_grating_n}{0.5}{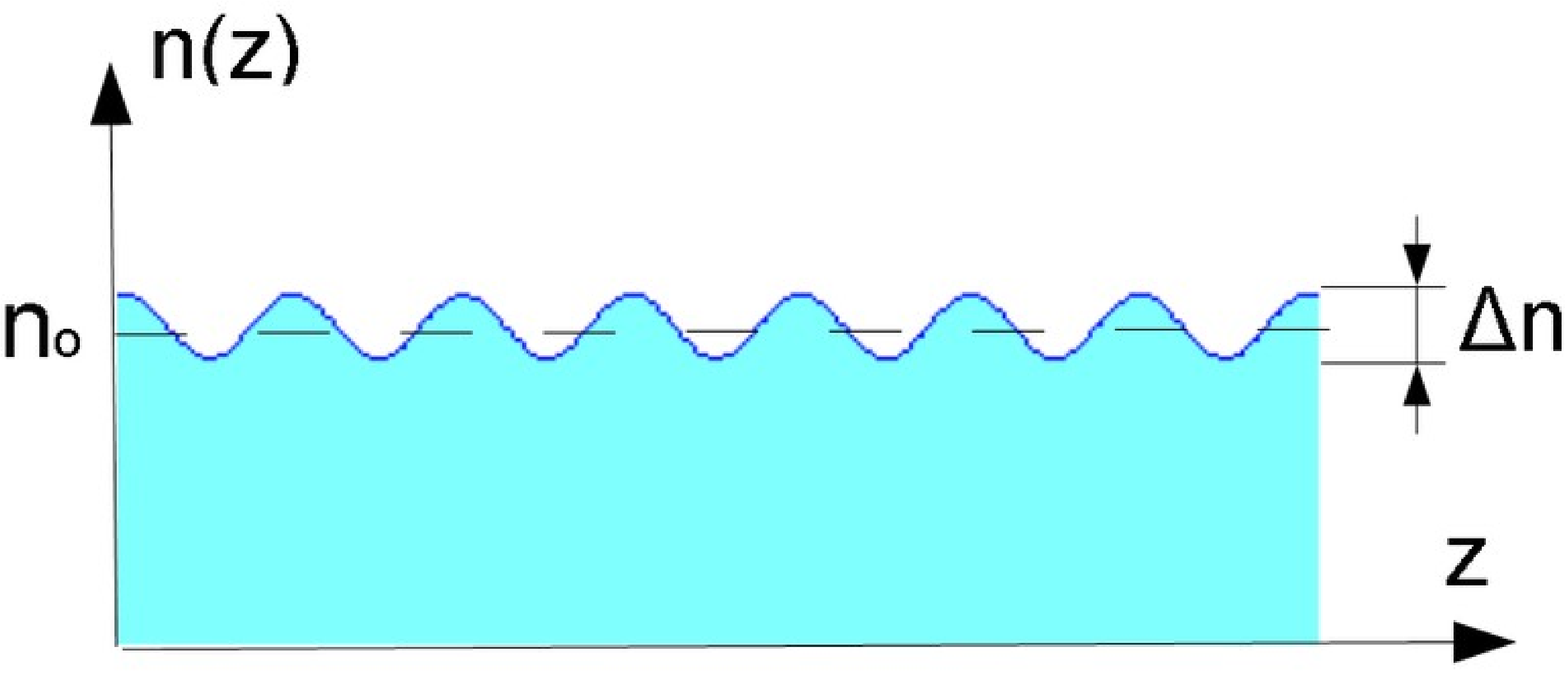}
{Illustration of the refractive index perturbation along the $z$-axis}

Assuming a uniform and periodic grating along the $z$-axis, the refractive index perturbation in the core can be represented in the following form:
\Eq{eq_dn_new} 
{n(\rho,\phi,z) = n_o(\rho) + \Delta n(\rho,\phi,z),}
as shown in Figure~\ref{TFBG_grating_n}.
\Eqaa{}
{n^2 &=& n_o^2 + n_o \Delta n + (\Delta n)^2}
{ &\sim& n_o^2 + n_o \Delta n }
according to~(\ref{eq_dn_new}) 
\Eq{}
{n^2 = n_o^2 + \Delta}
thus
\Eqaa{TFBG_modul}
{\Delta(\rho,\phi,z) &=& n_o(\rho) \Delta n(\rho,\phi,z) =}
{&=& n_o(\rho) \delta(\rho) \cos(K_g z')}

Let us assume that the $z'$-axis is the grating axis, tilted at the angle $\theta_g$ with respect to the fibre axis $z$. The fringe planes of the grating are written parallel to the $y$-axis, as shown in Figure~\ref{TFBG_grating}.

The coordinate system $Oxyz$ can be introduced in such a way that the $z'$-axis transforms into the $z$-axis by the coordinate system rotation above the $y$-axis.

\Fig{TFBG_grating}{1}{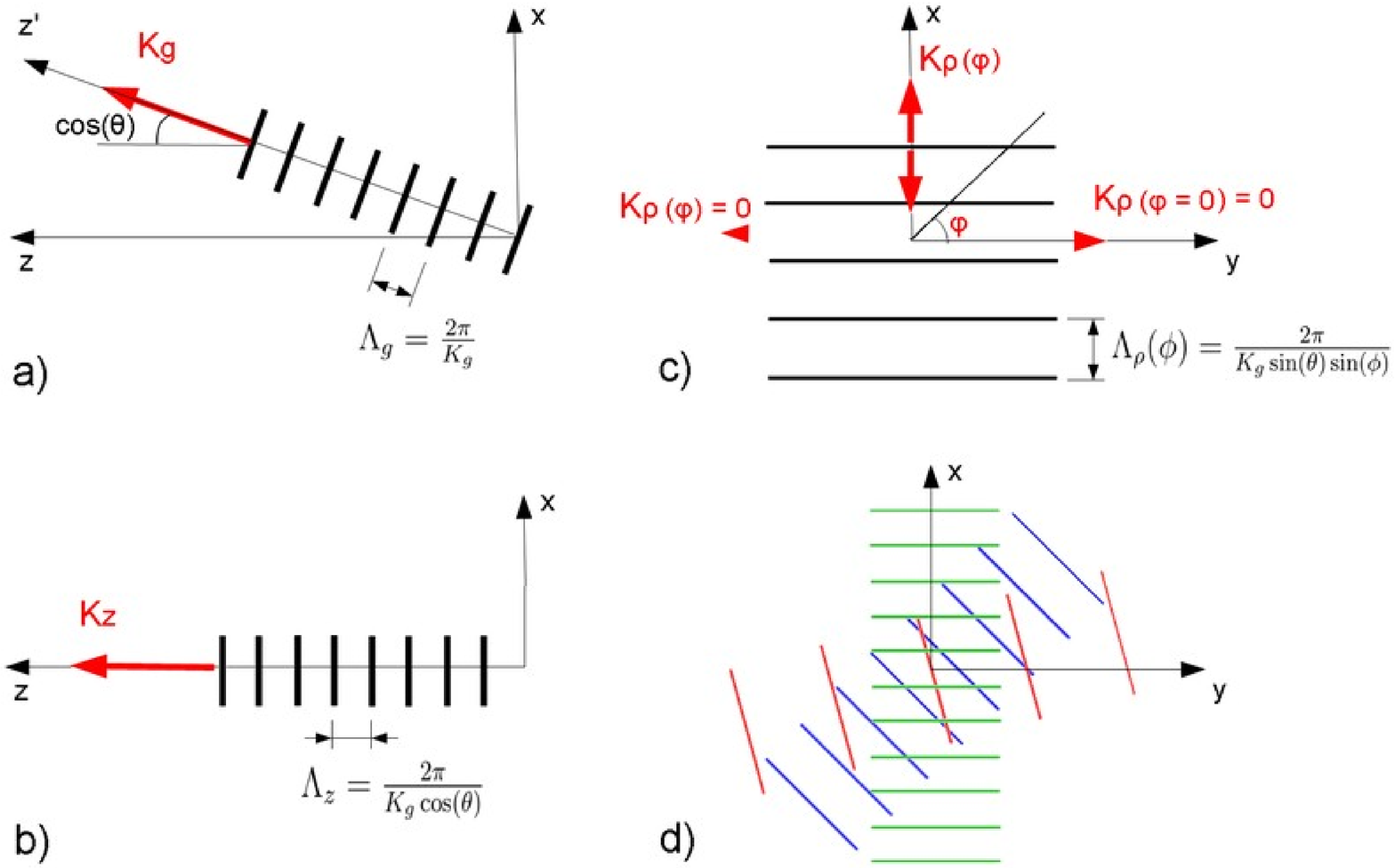}
{a) A schematic representation of a tilted Bragg grating inside the fibre, b) projection onto the $zx$  and c) $xy$ planes, d) superposition of normal gratings taken at various $\phi$ angles.}

Considering the rotation about the $y$-axis
\Eq{TFBG_grating2}
{z' = x \sin(\theta_g) - z \cos(\theta_g)}
the equation~(\ref{TFBG_modul}) can be rewritten in the following form:
\Eqaaa{}
{\Delta(z,\rho,\phi) &=& n_o(\rho) \delta(\rho) \cos \left (K_g x \sin(\theta_g) - K_g z \cos(\theta_g)\right) = }
{ &=& n_o(\rho) \delta(\rho) \cos(- K_z z + K_t (\phi) \rho) =}
{ &=& \frac{n_o \delta}{2} e^{-jK_z z} e^{j K_t ( \phi ) \rho} + c.c.}

Here we considered that ${x = \rho \cos(\phi)}$ as shown in Figure~\ref{TFBG_grating}, and
\Eqaaa{}
{K_g &=& \frac{2\pi}{\Lambda_g},}
{K_z &=& K_g \cos(\theta_g),}
{K_t(\phi) &=& K_x \cos(\phi) = K_g  \sin(\theta_g) \sin(\phi).}

It should be noted that ${K_t(\phi)}$ depends on the angle $\phi$, as shown in Figure~\ref{TFBG_grating}~d), and reaches the maximum value along the~$x$-axis: 
\Eq{}
{K_{t_{max}} = K_t(\phi = \pm \frac{\pi}{2}) = K_g  \sin(\theta_g),} 
and the minimum value along the $y$-axis:
\Eq{}
{K_{t_{min}} = K_t(\phi =0, \pi) = 0,}
hence, the grating period for the transverse grating depends on the $\phi$ angle $\Lambda_\rho(\phi) \in \left[\frac{2\pi}{ K_g \sin(\theta_g)}, \infty \right)$.

Therefore, mathematically such a grating may be described by superposition of infinitely many individual gratings written perpendicular to the fibre axis.

Finally, considering~(\ref{TFBG_grating2}) we can rewrite the matrix elements~(\ref{eq_coup}) in the following form:
\Eq{}{M_{ik}^{nm}(z) = \frac{1}{2\pi} \frac{1}{\gamma_i^n} \int_0^{2\pi} \int_0^\infty k_o^2\Delta(\rho,\phi, z) \V e_i^n(\rho)\V e_k^m(\rho) e^{j(m-n)\phi} d\rho d\phi = \nonumber}
\begin{eqnarray}
&=& \frac{1}{4\pi} \frac{k_o^2}{\gamma_i^n} e^{-jK_z z} \int_0^\infty \V e_i^n(\rho)\V e_k^m(\rho) n_o(\rho) \delta(\rho) \left( \int_0^{2\pi} e^{j K_t ( \phi ) \rho} e^{j(m-n)\phi}d\phi \right) d\rho  + c.c. {}\nonumber \\
&=& \frac{1}{4\pi} \frac{k_o^2}{\gamma_i^n} e^{-jK_z z} \int_0^\infty \V e_i^n(\rho)\V e_k^m(\rho) n_o(\rho) \delta(\rho) \sigma^{mn}(\rho) d\rho  + c.c. {}
\label{coupl_coeff_0}
\end{eqnarray}
Here ${\sigma^{mn}(\rho)}$ is the function dependent only on $\rho$ (a cylindrically symmetric function):
\Eqaaa{coupl_angul}
{\sigma^{mn}(\rho) &=& \int_0^{2\pi} e^{j K_t ( \phi ) \rho} e^{j(m-n)\phi}d\phi = }
{ &=& \int_0^{2\pi} e^{j K_g  \sin(\theta_g) \sin(\phi) \rho} e^{j(m-n)\phi}d\phi = }
{ &=& 2 \pi (-1)^k J_k (\gamma \rho),}
here ${J_k(\gamma \rho)}$ - are Bessel function of first kind, ${k = m-n}$, and ${\gamma = K_g  \sin(\theta_g)}$.

Hence, in the case of a tilted grating the orthogonality between the modes is broken due to the ${n_o(\rho) J_k (\gamma \rho)}$ term, which arises via the perturbation caused by the tilted grating. 

We are mainly interested in the energy transfer from the core modes into the cladding modes, considering the weighted function ${J_k (\gamma \rho)}$ and the fact that the core mode is mainly confined inside the core, and the grating perturbation also exists inside the core, we conclude that the coupling is possible only to a limited number of higher order azimuthal mode families, with azimuthal number $m < 12$, as shown in Figure~\ref{overlap}.

\Fig{overlap}{0.8}{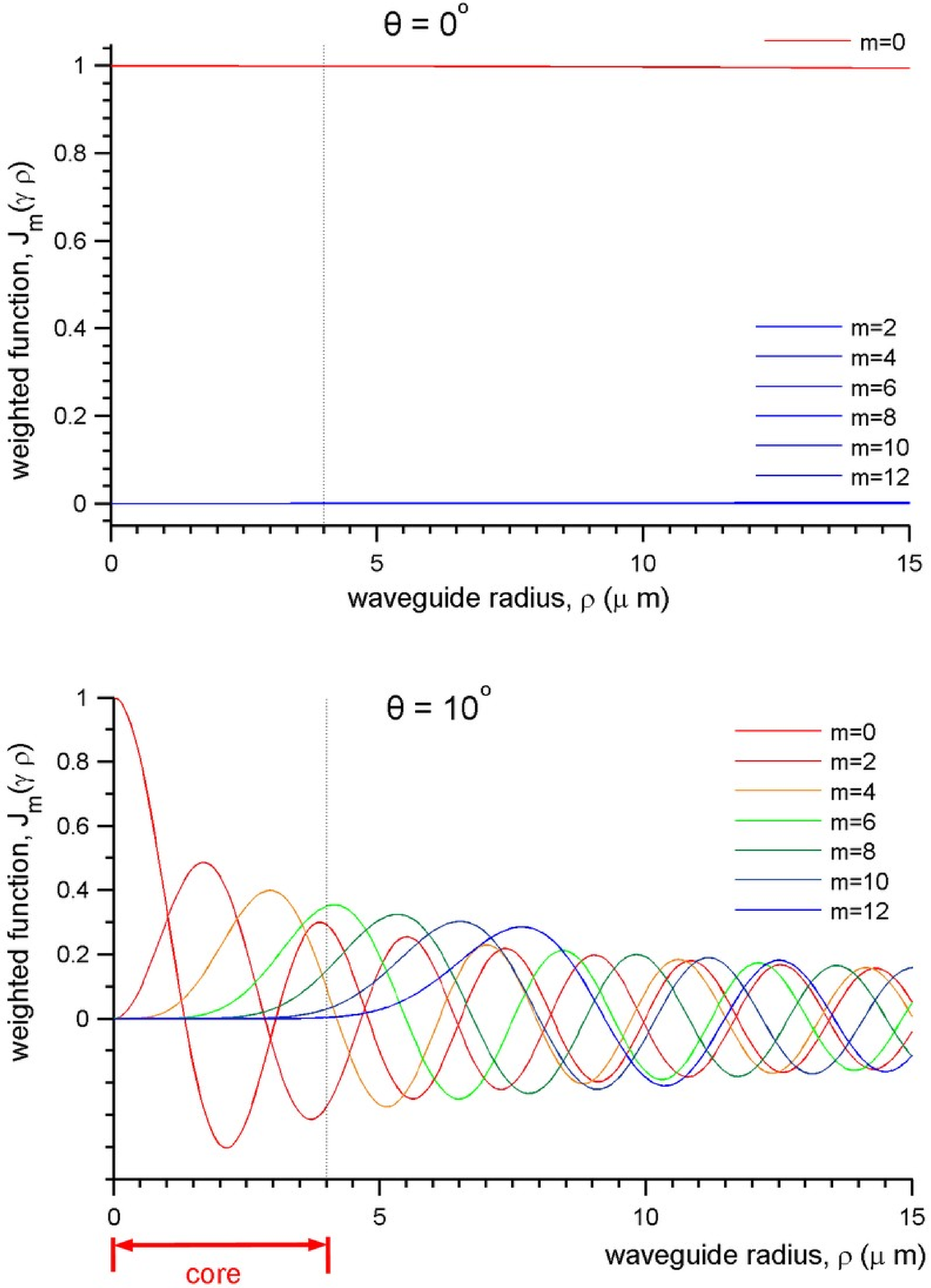}
{The weighted function ${J_m(\gamma \rho)}$ caused by the grating tilt, calculated for ${\theta = 0^o}$ and ${\theta = 10^o}$ grating tilt angles. The grating period is assumed to be ${\Lambda_G = 0.6~\mu m}$, thus for ${\theta = 0^o}$ we get ${\gamma = K_g  \sin(0) = 0}$, and for ${\theta = 10^o}$ we get ${\gamma = 1.81~\mu m^{-1}}$. The number ${m = m_1 - m_2 }$ is the difference between azimuthal order of the first $m_1$ and the second $m_2$ mode families. }

If the grating in non-tilted, the equation~(\ref{coupl_angul}) is reduced to:
\Eq{coupl_angul2}
{\sigma^{mn}(\rho) = \int_0^{2\pi} e^{j(m-n)\phi}d\phi = 2 \pi \delta_{mn},}
thus, the modes belonging to different families are mutually orthogonal, and hence, the energy transfer between such modes is impossible.

We conclude that the expression for coupling coefficients~(\ref{eq_coup}) is reduced to expression describing orthogonality between radial components of modes, except that we have an extra weighted function ${n_o(\rho) \sigma^{mn}(\rho)}$.
The extra weighted function not only breaks orthogonality between modes within a particular family, but also between modes belonging to different families of modes with different azimuthal numbers $m$. 
We note that the radial components of modes belonging to different families were not orthogonal initially, as shown in Figure~\ref{Modes_orthog}, but the complete modes with the angular part included, in the case of a non-tilted grating, were orthogonal due to~(\ref{coupl_angul2}).

According to orthogonality relation~(\ref{SL_vec_othog2}), for the vectorial case the term \\ ${\V e_i^n(\rho)\V e_k^m(\rho)}$ in equation~(\ref{coupl_coeff_0}) should be replaced with:
\Eq{}
{\V e_i^n(\rho)\V e_k^m(\rho) = \epsilon(\rho) \rho u_i^n (\rho) u_k^m (\rho) +  \rho v_i^n (\rho) v_k^m(\rho),}
here $\V e_k^m(\rho) = \begin{pmatrix} u_k^m\\ v_k^m\\ \end{pmatrix}$ are the eigenvectors of non-perturbed problem~(\ref{eq_cyl_vec_eigexact}).

Whereas in the scalar case, in accordance with~(\ref{orthogon_scalar}), we have: 
\Eq{}
{\V e_i^n(\rho)\V e_k^m(\rho) = \rho e_i^n(\rho)e_k^m(\rho),}
with eigenvectors $e_k^m(\rho)$ for the scalar eigenvalue problem~(\ref{eq_cyl_scal_eig}).

Finally, considering the complex conjugate part ${c.c.}$ , and assuming that all the coupling coefficients are known, we can rewrite equation~(\ref{coupl_coeff_0}) in the simple form:  
\Eq{coupl_coeff}
{ M_{ik}^{nm}(z) = \cos(K_z z)[\hat M_{ik}^{nm}],}
here the number ${\hat M_{ik}^{nm}}$ defines overlap (or coupling) between two modes: ${(n,i)}$ and ${(m,k)}$, as it follows from equation~(\ref{coupl_coeff_0}).
All such coupling coefficients are assembled into the matrix ${[\hat M_{ik}^{nm}]}$.

Let us compute coupling coefficients $C_k^m$ between the core and cladding modes (here $m$ is the mode family with azimuthal number $m$, and $k$ is the mode order in the family).
The results are presented in Figures~\ref{Done_2_pi2_Ck_All08},~\ref{Done_2_pi2_Ck_0-3} for $2^o$ degree, Figures~\ref{Done_4_pi3_Ck_All08},~\ref{Done_4_pi3_Ck_0-3} for $4^o$ degree and 
in Figures~\ref{Done_10_pi2_Ck_All},~\ref{Done_10_pi2_Ck_0-4},~\ref{Done_10_pi2_Ck_5-10} for $10^o$ degree gratings. 
We note that it is sufficient to consider only $4$, $5$ and $9$ families of modes for $2^o$, $4^o$ and $10^o$ degree tilted gratings, respectively.

In the following section we apply the coupled mode theory to compute the spectral response of the gratings. The computed coupling coefficients $C_k^m$ are going to used as input parameters.

\Fig{Done_2_pi2_Ck_All08}{0.9}{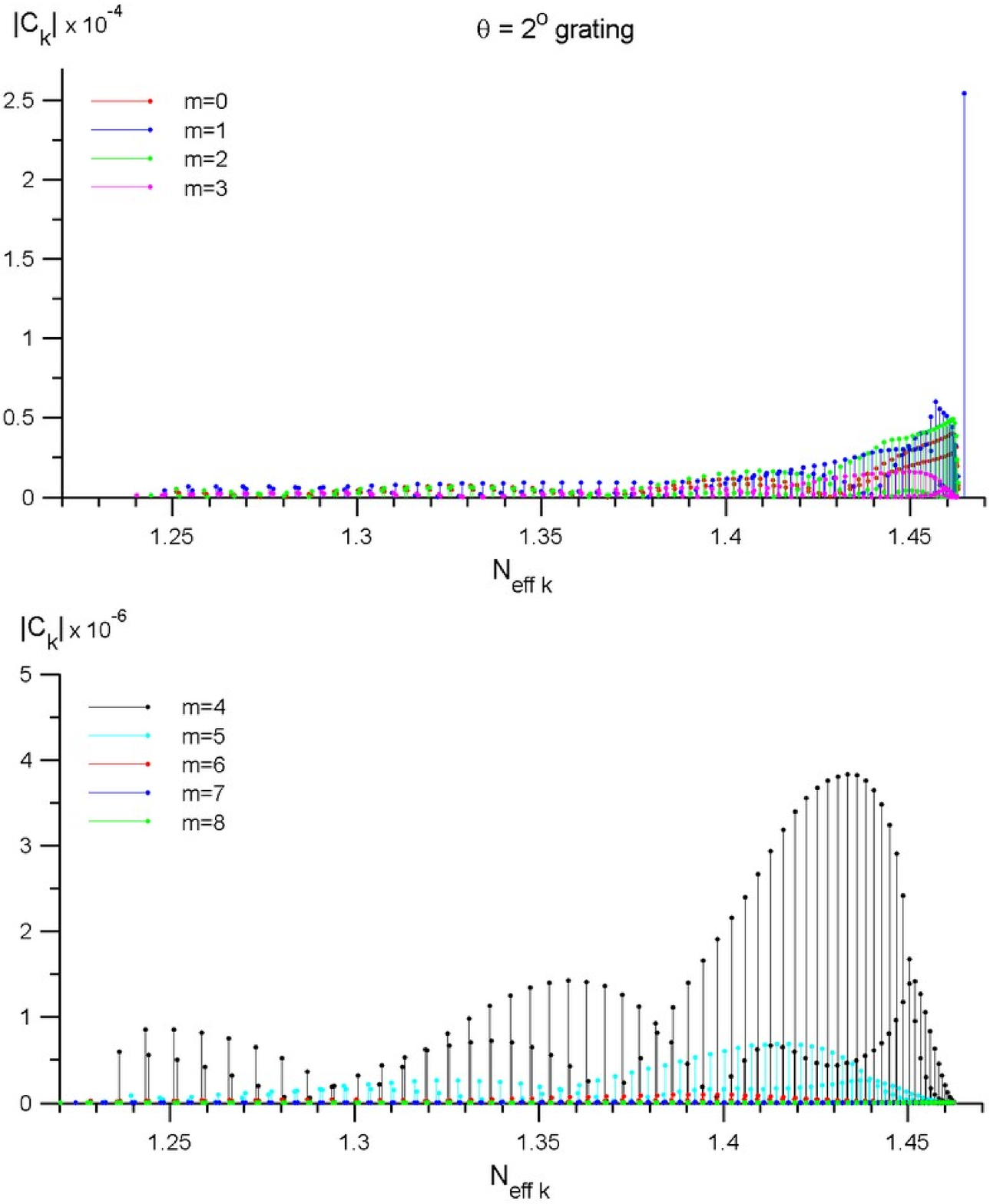}
{The $2^o$ degree grating assisted coupling coefficients between the core and cladding modes, ${\Delta n = 10^{-4}}$.}

\Fig{Done_2_pi2_Ck_0-3}{0.9}{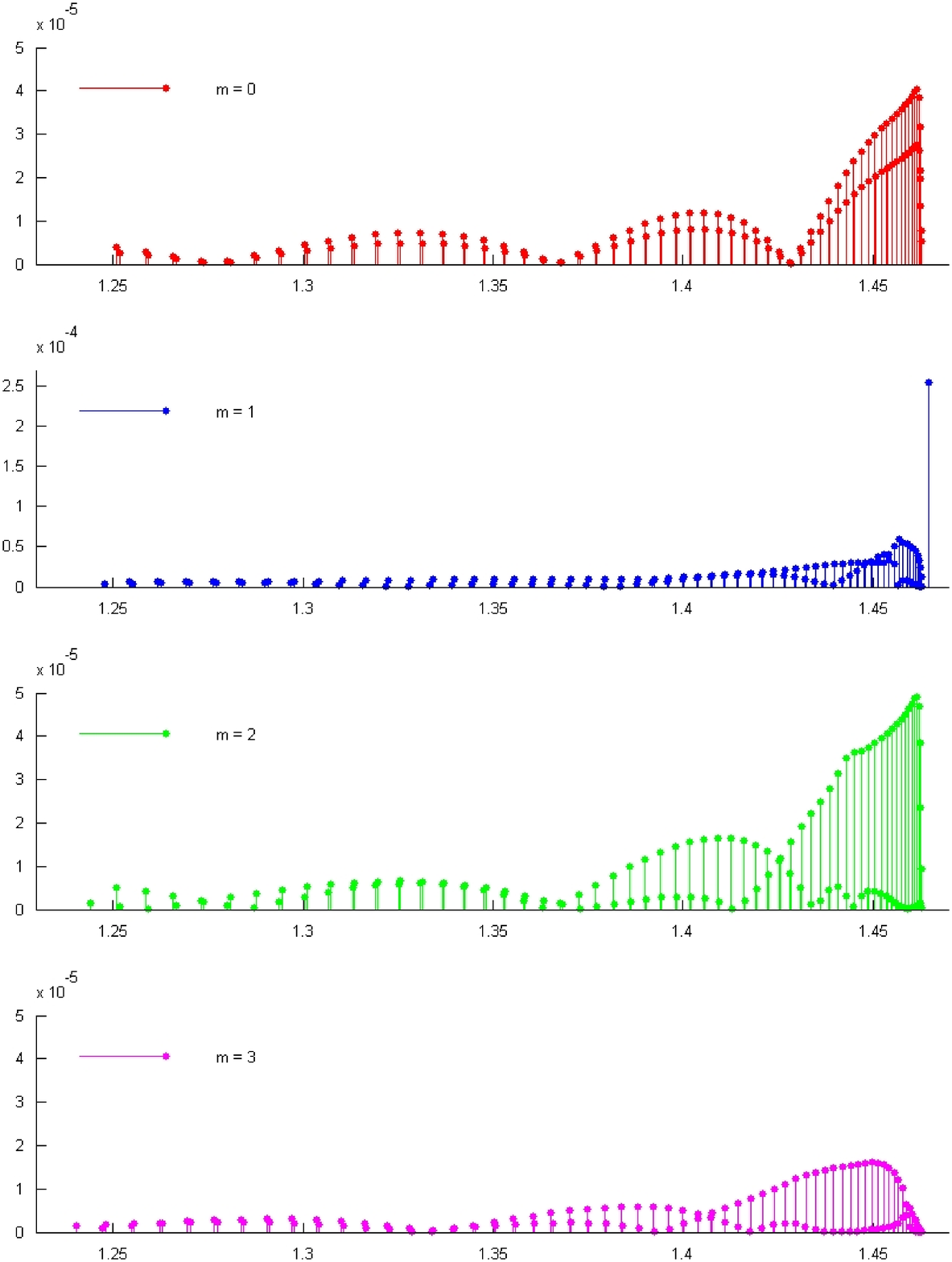}
{The $2^o$ degree grating assisted coupling coefficients between the core and cladding modes, ${\Delta n = 10^{-4}}$.}

\Fig{Done_4_pi3_Ck_All08}{0.9}{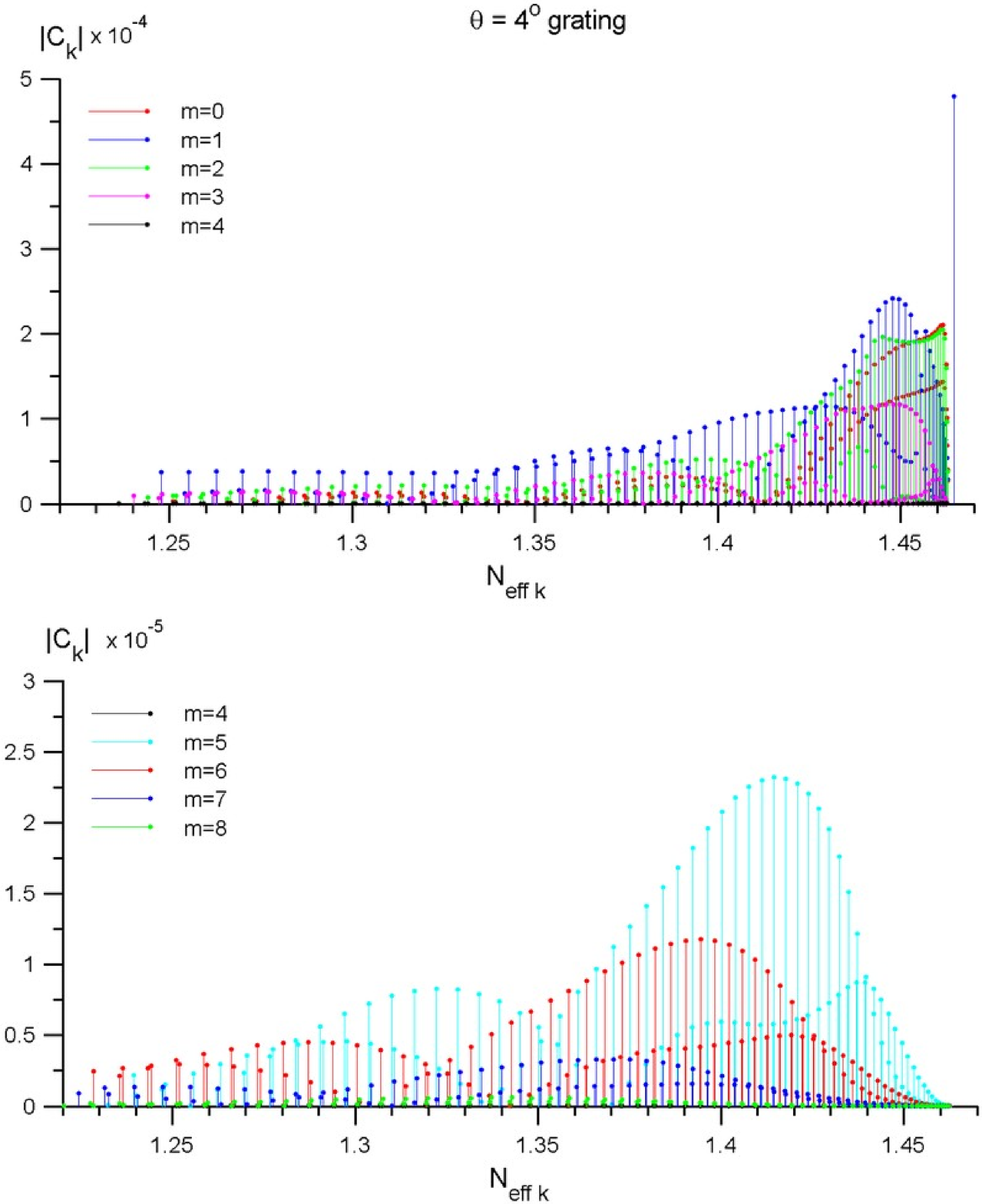}
{The $4^o$ degree grating assisted coupling coefficients between the core and cladding modes, ${\Delta n = 10^{-4}}$.}

\Fig{Done_4_pi3_Ck_0-3}{0.9}{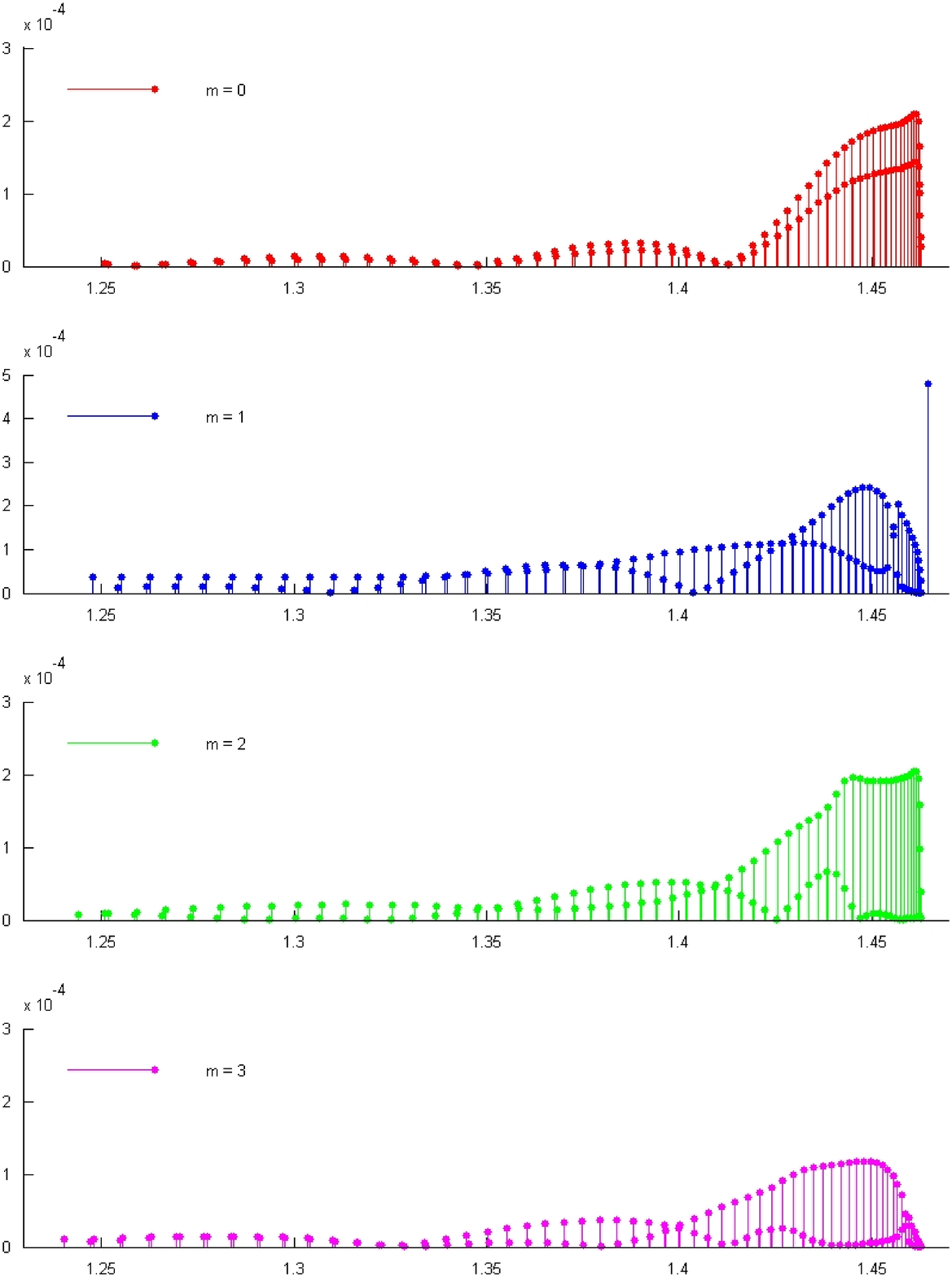}
{The $4^o$ degree grating assisted coupling coefficients between the core and cladding modes, ${\Delta n = 10^{-4}}$.}

\Fig{Done_10_pi2_Ck_All}{0.9}{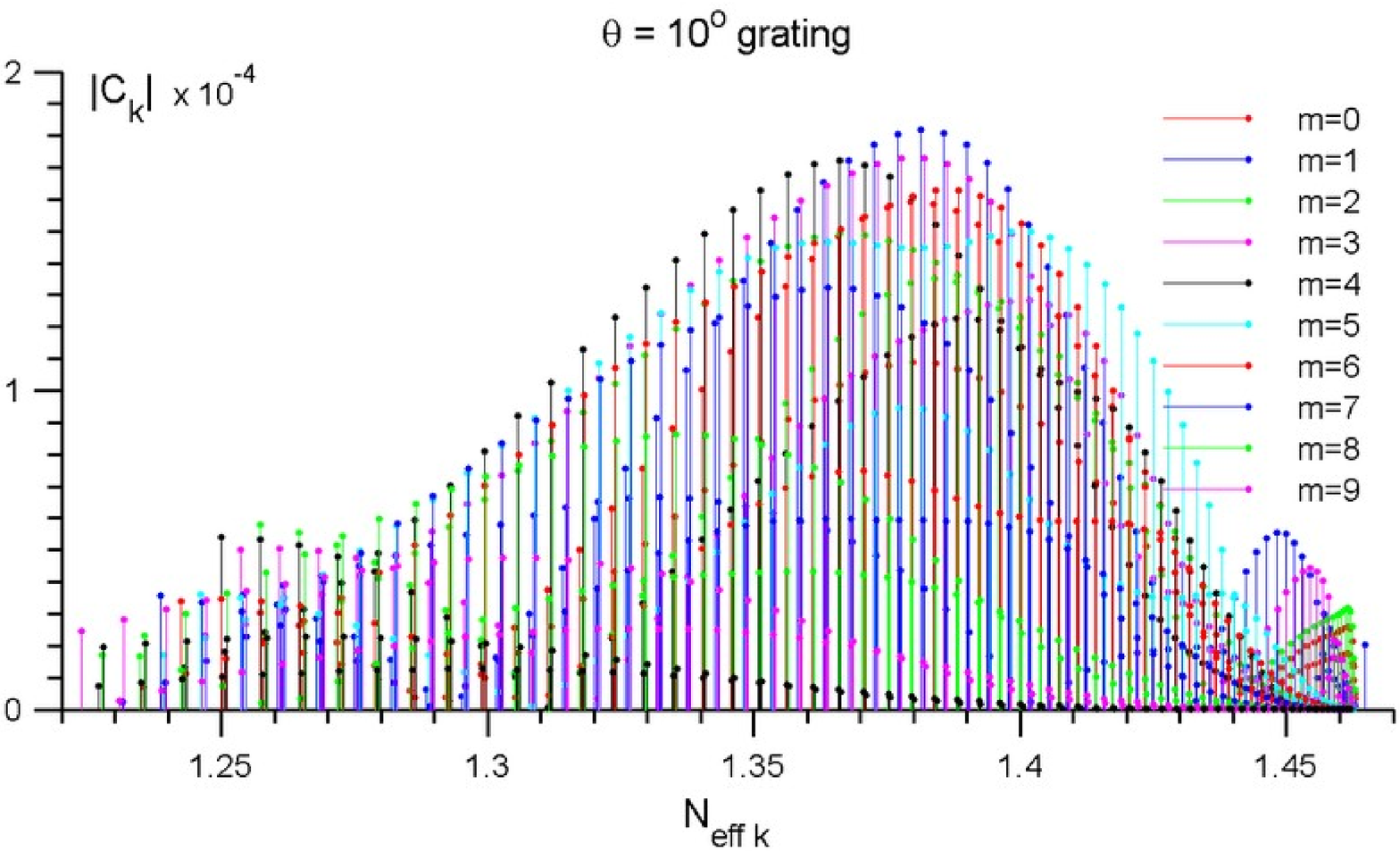}
{The $10^o$ degree grating assisted coupling coefficients between the core and cladding modes, ${\Delta n = 10^{-4}}$.}

\Fig{Done_10_pi2_Ck_0-4}{0.9}{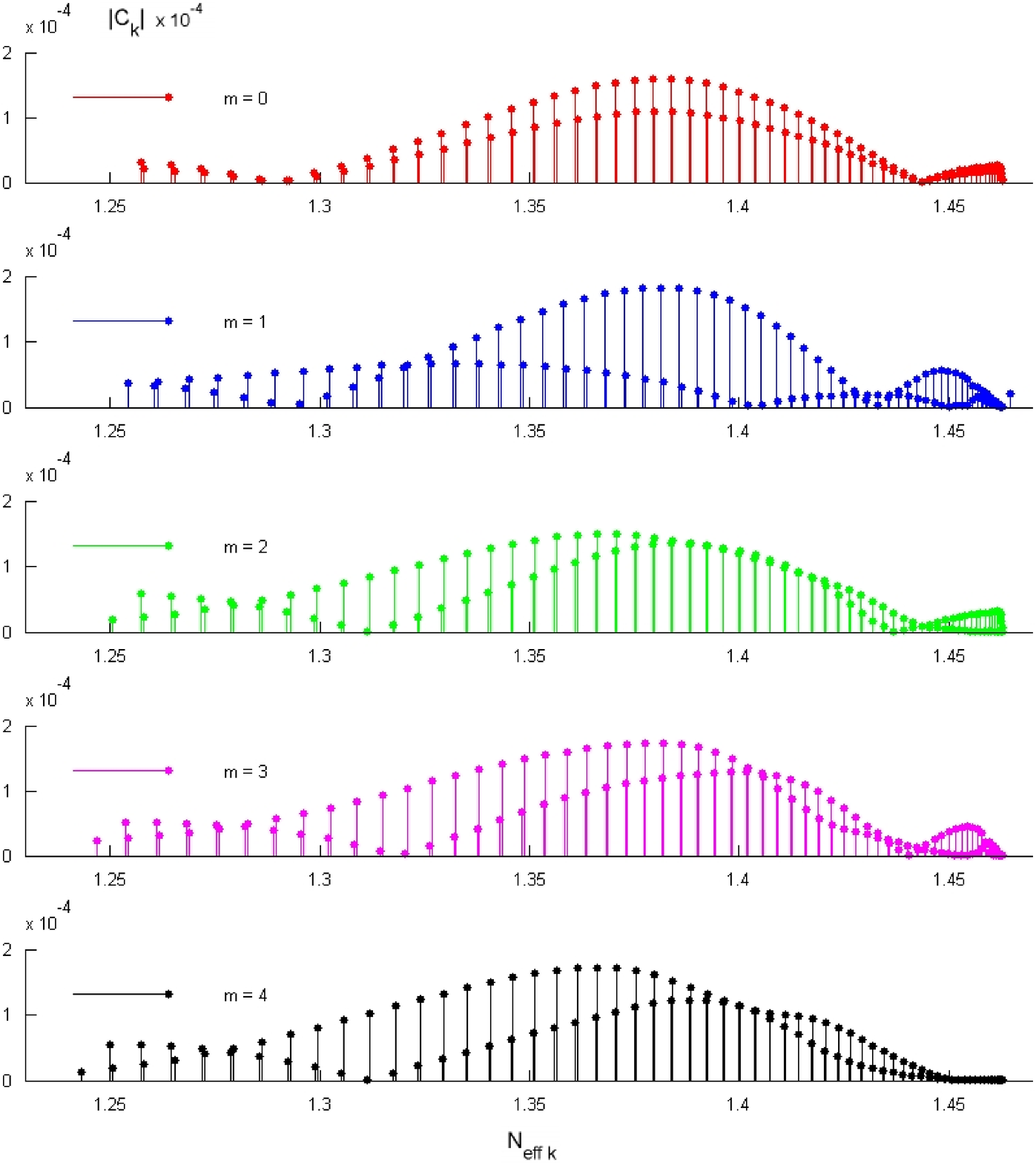}
{The $10^o$ degree grating assisted coupling coefficients between the core and cladding modes, ${\Delta n = 10^{-4}}$.}

\Fig{Done_10_pi2_Ck_5-10}{0.9}{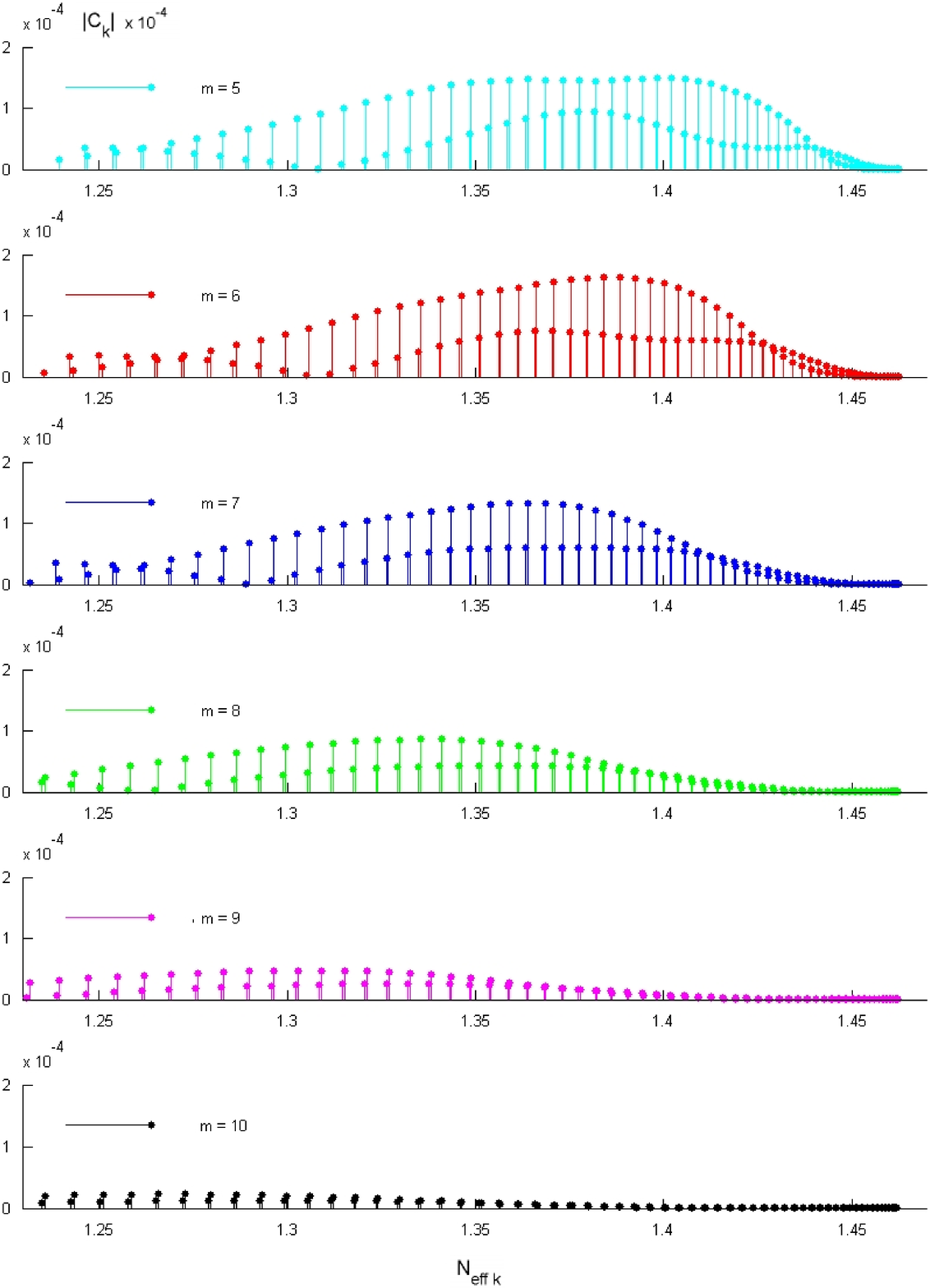}
{The $10^o$ degree grating assisted coupling coefficients between the core and cladding modes, ${\Delta n = 10^{-4}}$.}


%% file: Chap_TFBG_CMT_3_Spectra.tex

\section{Coupled-mode theory,\\ the two modes approximation}


The coupled mode theory was employed in electromagnetics in the early 1950's, and was initially applied to microwave travelling wave devices~\cite{Pierce:54, Miller:54, Schelkunoff:55}. 
In the early 1970's the coupled mode theory had been successfully applied to the modeling of various guided wave optical systems.
The noticeable papers were published by Kogelnik~\cite{Kogelnik:1969, Kogelnik:1972}, Snyder~\cite{Snyder:72}, Yariv~\cite{Yariv:73}, Marcuse~\cite{Marcuse:73} and others. The application to parallel waveguides can be found  in~\cite{Marcuse:71, Taylor:73, Hardy:85}.

In this section we continue our consideration of the cylindrical waveguide with a periodic grating inscribed in its core. 
The grating allows the transfer of energy, initially confined in the fibres core, outside the core into the cladding modes.
It is also known that in our particular case the contra-directional phase matching condition occurs, as we will explain it later.

Let us start with equation~(\ref{eq_coup}). Assuming that there is no external excitation we get:
\Eq{}
{(\lambda_i^n + d_z^2) C_i^n(z)+ \sum_k \sum_m \left[M_{ik}^{nm}(z)\right] C_k^m(z) = 0.}
For convenience we enumerate modes and coefficients with single index, \textit{e.g.} each pair of indices ${(i,n)}$ we will be enumerated with a single index $\alpha$, thus we can rewrite~(\ref{eq_coup}) as
\Eq{eq_coup3}
{(\lambda_\alpha + d_z^2) C_\alpha(z)= - \sum_\gamma \left[M_{\alpha \gamma}(z)\right] C_\gamma(z).}

If the perturbation can be neglected, \textit{i.e.} the right hand side of~(\ref{eq_coup3}) can be canceled ${\left[M_{\alpha \gamma}(z)\right] = 0}$, and hence the solutions would be ${C_\alpha(z) = A_\alpha e^{i \sqrt{\lambda_\alpha} z}}$, where $A_\alpha$ are some constants. 
If the perturbation is switched on, the constants should be replaced with a slowly varying functions $A_\alpha = A_\alpha(z)$ along the $z$ coordinate~\cite{Kogelnik:1969,Yariv:73}.
\Eqaa{eq_coup4}
{C_\alpha(z) &=& A_\alpha(z) e^{i \sqrt{\lambda_\alpha} z} =}
{            &=& A_\alpha(z) e^{i \beta_\alpha z},}
here we denoted $\beta_\alpha : = \sqrt{\lambda_\alpha}$, called the propagation constants.

The final solution to~(\ref{eq_final_sol}) can be written in the following series form:
\Eq{eq_final_sol2}
{u(\rho, \phi, z) = \sum_m \sum_k A_k^m(z)e_k^m(\rho) e^{j m\phi}e^{i \beta_\alpha z}.}

Inserting~(\ref{eq_coup4}) into~(\ref{eq_coup3}), and considering that 
\Eq{}
{d_z^2 C_\alpha(z) = d_z^2 A_\alpha(z)e^{j \beta_\alpha z} = e^{j\beta_\alpha z}[-\lambda_\alpha + j 2 \beta_\alpha d_z + d_z^2]A_\alpha(z),}
we obtain
\Eq{eq_coup5}
{e^{j\beta_\alpha z}[j 2 \beta_\alpha d_z + d_z^2]A_\alpha(z) = - \sum_\gamma \left[M_{\alpha \gamma}(z)\right] A_\gamma(z)e^{j \beta_\gamma z}.}
The matrix coefficients $M_{\alpha\gamma}(z)$ are defined in accordance with~(\ref{coupl_coeff}):
\Eq{}
{[M_{\alpha\gamma}(z)] = [M_{\alpha\gamma}]\cos(K_z z)}

The equation~(\ref{eq_coup5}) can be rewritten in the following form: 
\Eq{eq_coup6}
{[j 2 \beta_\alpha d_z + d_z^2]A_\alpha(z) = - \sum_\gamma \left[M_{\alpha \gamma}\right] A_\gamma(z)  \cos(K_z z) e^{j (\beta_\gamma - \beta_\alpha) z}.}

Now let us consider the oscillating terms:
\Eqaa{eq_phase}
{\cos(K_z z) e^{j (\beta_\gamma - \beta_\alpha) z} &=& \frac{1}{2}\left(e^{j K_z z} + e^{-j K_z z}\right) e^{j (\beta_\gamma - \beta_\alpha)}}
{&=& \frac{1}{2}\left(e^{j (\beta_\gamma - \beta_\alpha + K_z)}+ e^{j (\beta_\gamma - \beta_\alpha - K_z)}\right)}

The matrix terms oscillate periodically and rapidly unless the phase is close to zero, in which case the power coupled between modes accumulates coherently and gives rise to a significant power exchange between the modes.
Thus we can neglect all the matrix terms except a few terms that have a small or zero phase ${\Delta \phi = 0}$ under the exponential functions~\cite{Skorobogatiy:02, Yariv:73, Yariv:2007}.

Considering~(\ref{eq_phase}) and assuming that the energy couples from the core mode with the propagating constant $\beta_\alpha$, to one of the cladding modes with $\beta_\gamma$ propagating constant, there are two possible phase matching conditions:
\Eqaa{eq_phase_match}
{\Delta \phi &=& K_z + \beta_\gamma - \beta_\alpha = 0,}
{\Delta \phi &=& K_z + \beta_\alpha - \beta_\gamma = 0,}
called co-directional and contra-directional coupling, respectively.

From the experiments described in the following sections we know that in our case only the contra-directional coupling occurs, thus we limit ourselves to this particular case. 
Figure~\ref{CMT} shows the phase matching condition, or momentum diagram, of contra-directional coupling.

\Fig{CMT}{0.4}
{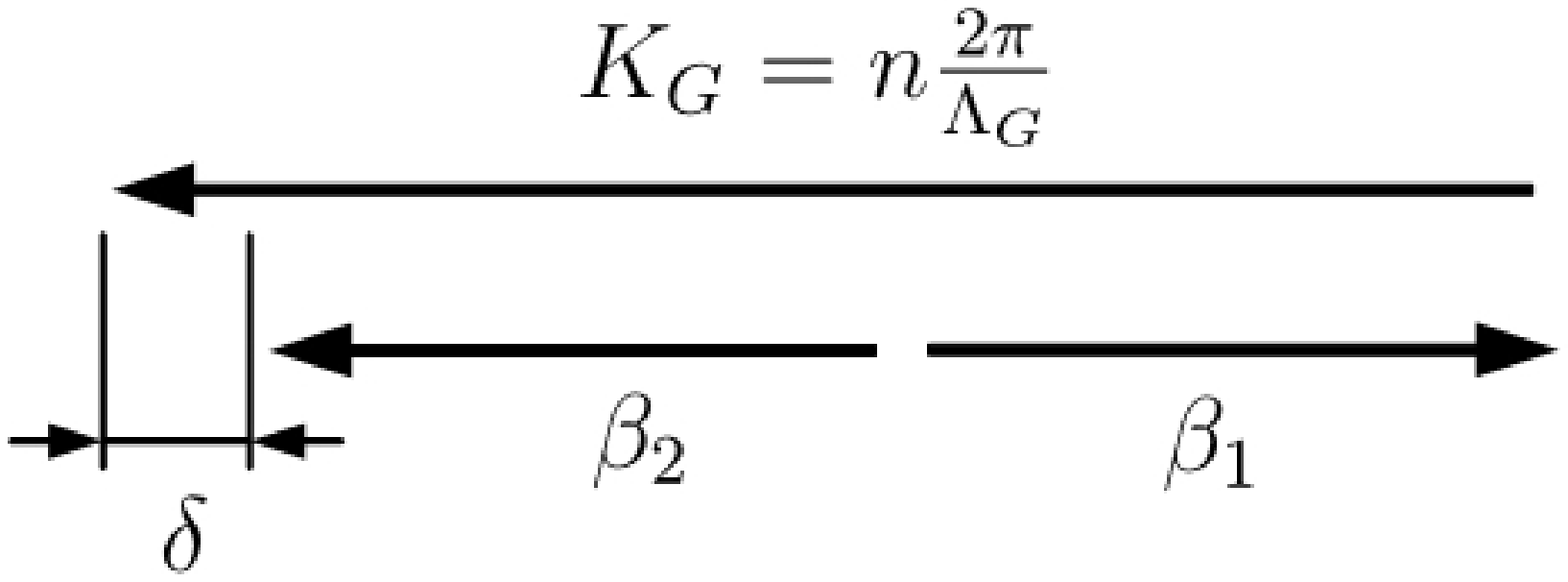}{The momentum diagram of contra-directional coupling (here $\beta_1$ corresponds to the core mode and $\beta_2$ to the cladding mode).}

Now let us make a few extra assumptions. First we assume that instead of considering coupling between all the modes at once, we can consider only two modes, the core mode and one of the cladding \C{modes}. 
This assumption is useful to understand the basic properties of energy transfer from the core mode to a particular cladding mode. Once the energy is transfered to a particular cladding mode it can not be recoupled to an other cladding modes due to the significant phase mismatch. We also assume that the amplitudes $A_\alpha(z)$ in~(\ref{eq_coup4}) are \C{``slowly''} varying, \textit{i.e.} ${d_z^2 A_\alpha(z) \ll \beta_\alpha d_z A_\alpha(z)}$, this is the so-called slowly varying amplitude approximation~\cite{Kogelnik:1969, Yariv:73}.
The resulting equations~(\ref{eq_coup6}) will takes the following form:

\Eqaa{eq_coup7}
{j 2 \beta_1 d_z A_1(z) &=& - \kappa e^{-j \delta z} A_2(z),}
{j 2 \beta_2 d_z A_2(z) &=& - \kappa^* e^{j \delta z} A_1(z).}
Here:\\
$\delta = - K_z + \beta_1 - \beta_2 \to 0$ is the phase mismatch. We should keep in mind that if the propagation constant $\beta_1$ is positive then, due to the contra-directional coupling the $\beta_2$ constant should be taken with the minus sign.\\ 
${\kappa = \left[M_{1 2}\right] = \left[M_{2 1}\right]^* = <\psi_1|\Delta|\psi_2>}$ is the coupling constant, and can be computed in accordance with equation~(\ref{eq_coup}).

We note, by looking at the equation~($\ref{eq_phase}$), that the off-diagonal matrix elements have phases approaching zero in the case when the phase matching condition is satisfied, thus indeed we have condition that allows for coherent power accumulation, which leads to the power redistribution between the modes.

Let us solve equations~(\ref{eq_coup7}) assuming that the phase matching condition is satisfied ${\delta = 0}$, and the proper boundary conditions are given. Considering that the core mode with the amplitude ${A_1(z)}$ to be incident at ${z = 0}$ on the perturbation region ${z \in [0,L]}$ we can set ${A_1(0) = 1}$. The cladding mode ${A_2(z)}$ is \C{``generated''} by the perturbation, hence ${A_2(z)}$ should be set to zero at the end of the perturbation region ${A_2(L) = 0}$. 
The solution of~(\ref{eq_coup7}) (for the case of ${\delta = 0}$) is given by equations~(\ref{eq_coup_sol}) and mode power of the incident and scattered waves are shown in Figure~\ref{TFBG}.
\Eqaa{eq_coup_sol}
{A_1(z)&=& \cosh \left(\frac{\kappa(z-L)}{2 \sqrt{\beta_1 \beta_2}}\right) \cosh^{-1}\left(\frac{\kappa L}{2 \sqrt{\beta_1 \beta_2}}\right),}
{A_2(z)&=& i \sqrt{\frac{\beta_1}{\beta_2}}\sinh \left(\frac{\kappa(z-L)}{2 \sqrt{\beta_1 \beta_2}}\right)\sinh^{-1} \left(\frac{\kappa L}{2 \sqrt{\beta_1 \beta_2}}\right).}

\Fig{TFBG}{0.6}
{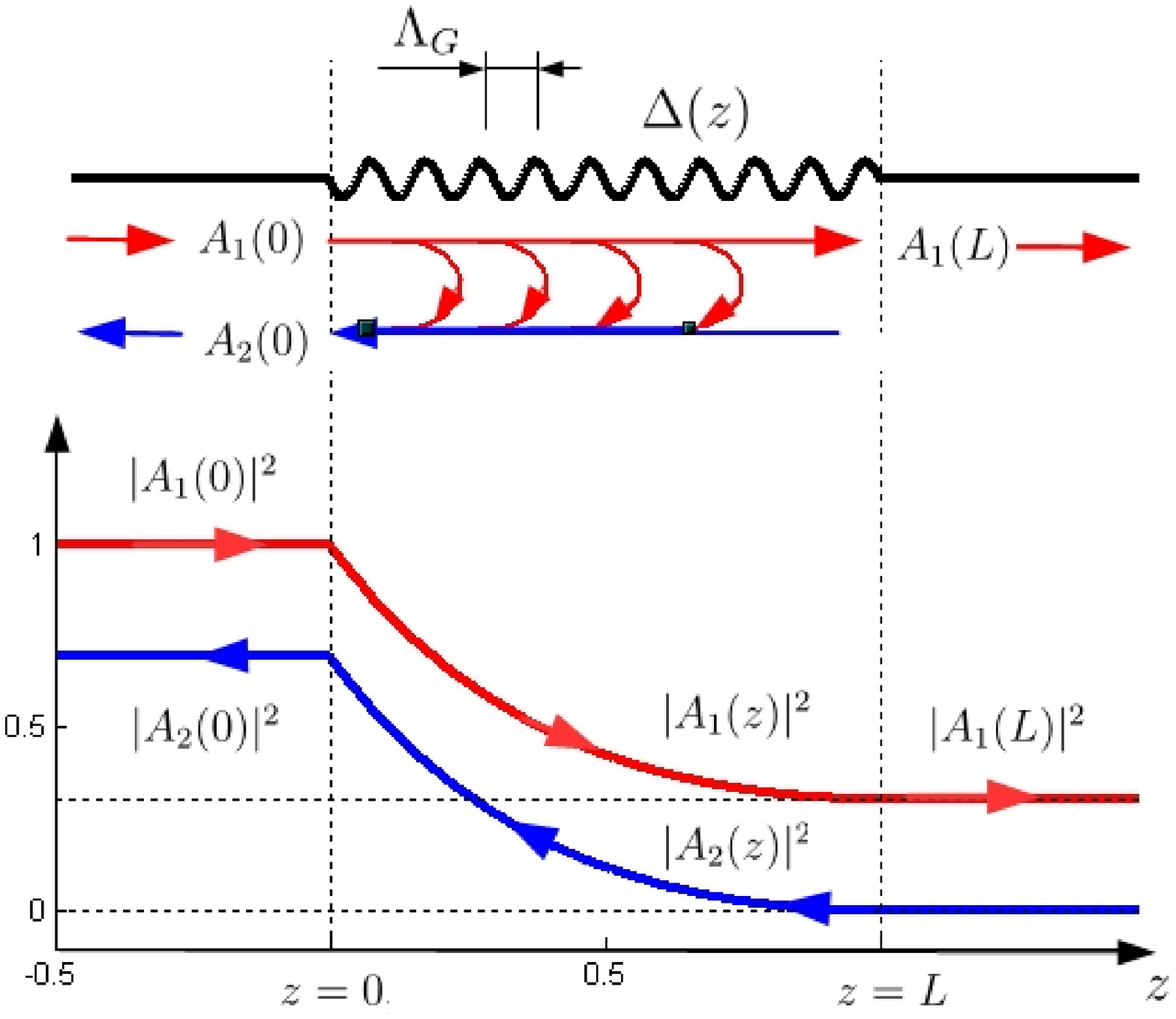}{The transfer of power between the incident core mode ${A_1(z)}$ and back-scatters cladding mode ${A_2(z)}$ in the case of contra-directional coupling. Here ${\frac{k L}{\beta} = 2.4}$ }

Now let us consider the amplitudes $A_1$ and $A_2$ at the perturbation boundary. Setting as previously ${A_1(0) = 1}$ and ${A_2(L) = 0}$ we get for the general case of \C{non zero} phase detuning ( ${\delta \ne 0}$ ) the following solution:

\Eqaa{eq_coup_sol2}
{A_1(L) &=& \frac{\sigma e^{-i L \frac{\delta}{2}}}{\sigma \cosh(L \frac{\sigma}{2})- i \delta \sinh(L \frac{\sigma}{2})},}
{A_2(0) &=& \frac{-\frac{ik}{\beta_2} \sinh(L \frac{\sigma}{2})}{ \sigma \cosh(L \frac{\sigma}{2}) -i \delta \sinh(L \frac{\sigma}{2})}.}
Here 
\Eqaa{}
{\sigma &=& \sqrt{\frac{\kappa^2}{\beta_1 \beta_2}-\delta^2},}
{\delta &=& - K_z + \beta_1 - \beta_2,}
here $L$ is the perturbation length, $\beta_1$ and $\beta_2$ are the propagation constants of the core and cladding modes, respectively, as shown in Figure~\ref{CMT}.  

We have a particular interest in the energy loss due to the coupling from the core to the cladding modes, as this energy loss can be observed experimentally by measuring transmission spectra of the optical system:
\Eqaa{eq_power}
{\eta(x) &=& \frac{|A_1(L,x)|^2}{|A_1(0,x)|^2} =}
{&=& \frac{\gamma^2(x)}{\gamma^2(x) \cosh^2(C \frac{\gamma(x)}{2}) + x^2 \sinh^2(C \frac{\gamma(x)}{2})},}
here $C$ is the coupling parameter defined as ${C := L \sigma}$, ${x := \frac{\delta}{\xi}}$ is the phase detuning parameter, ${\xi := \kappa / \sqrt{\beta_1 \beta_2}}$ and ${\gamma(x) := \frac{\sigma}{\xi} = \sqrt{1 - x^2}}$.

The power loss ${\eta(x)}$ function of the core mode due to the coupling to the cladding mode is shown in Figure~\ref{CMT_varC} for various coupling parameters ${C = L \sigma}$ and the phase mismatch ${x=\frac{\delta}{\xi}}$.

\Fig{CMT_varC}{0.6}
{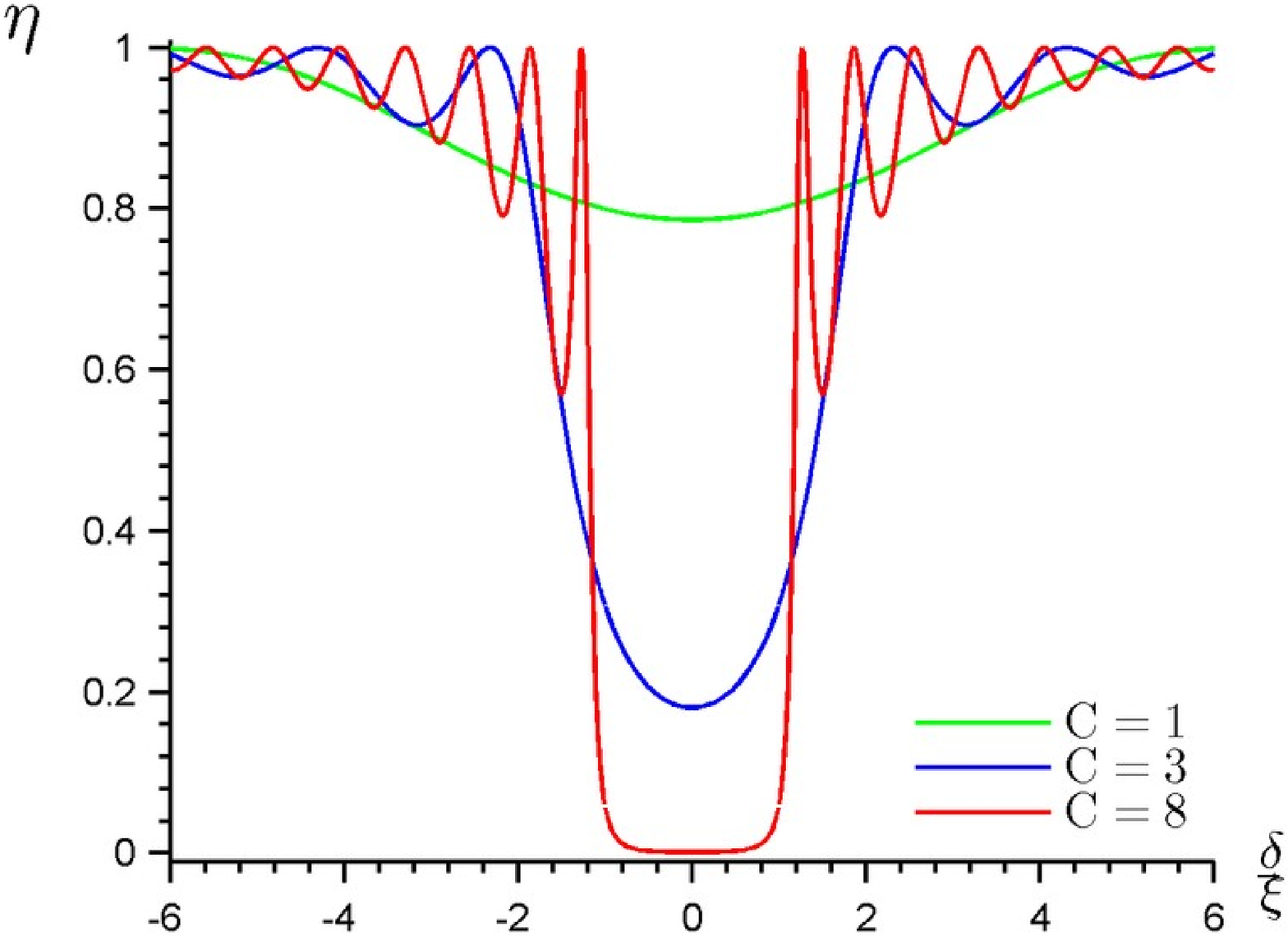}{The power loss of the core mode as a function of the phase mismatch, for various coupling parameters ${C = L \sigma}$.}


In the case of coupling to the lossy mode we note that the peak becomes broader and the base line is \C{shifted}, as shown in Figure~\ref{CMT_Loss2}. 
\Fig{CMT_Loss2}{1}{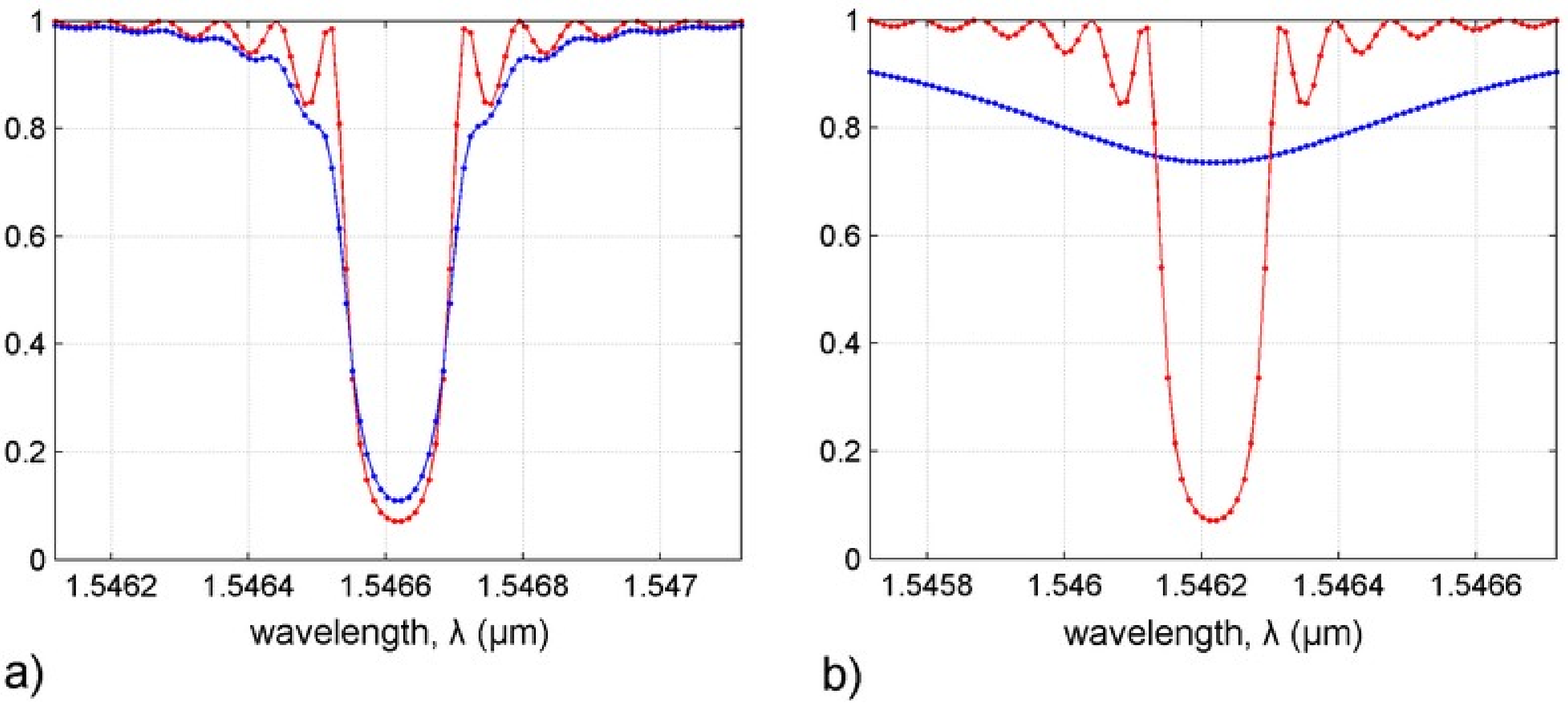}
{The power loss of the core mode as a function of the phase mismatch, for the fibre waveguide coated with a thin film (${h = 100~nm}$) made of lossy material (blue curves) a) ${N_{eff} = 1.3661 - j 6.1165 \times 10^{-4}}$, and b) ${N_{eff} = 1.3669 - j 2.2791 \times 10^{-5}}$. The red curves, used as a reference, corresponds to the non-lossy material ( ${N_{eff} = 1.3661}$ ).}

\clearpage
\section{Coupling between the core mode and many cladding modes in the TFBG}

In this section we apply the coupled mode theory to our problem of interest: the tilted fibre Bragg grating (TFBG) inscribed inside the core of a standard telecommunication fibre SMF-28. 

The phase matching condition in such a case is schematically shown in Figure~\ref{TFBG_phase}, \C{where the core mode} is coupled to a number of cladding modes.

\Fig{TFBG_phase}{0.6}
{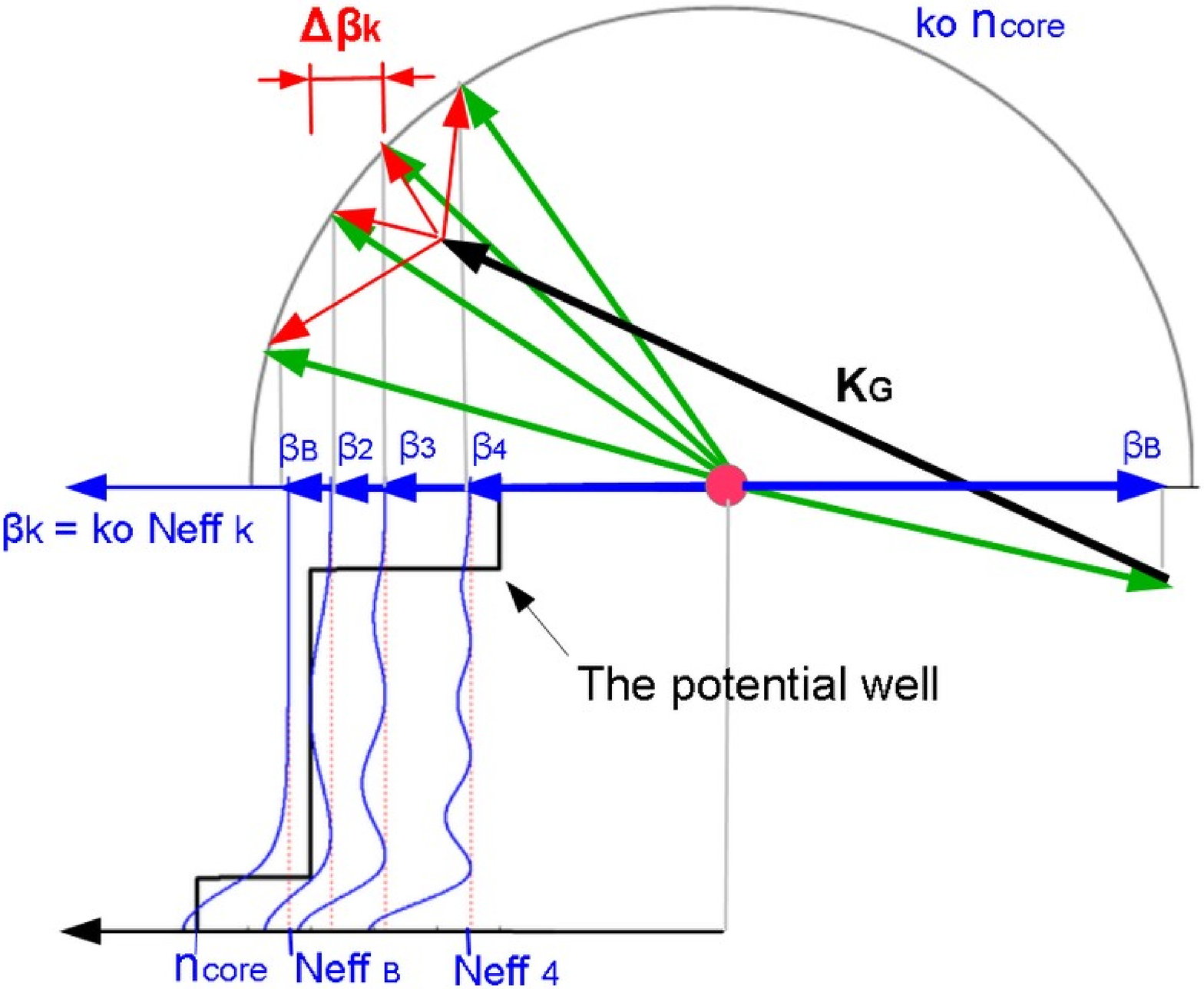}{The phase matching condition in TFBG}

In the case of TFBG the energy confined inside the fibres core is coupled from the core mode into a multiplicity of cladding modes, as shown schematically in Figure~\ref{TFBG_Coupling}. The grating-assisted coupling, for the problem of interest, is only possible for the core and cladding modes propagating in opposite directions. It should also be noted that for the given geometry and operational wavelength the energy coupling between different cladding modes is prohibited due to the significant phase mismatch.

\Fig{TFBG_Coupling}{0.6}
{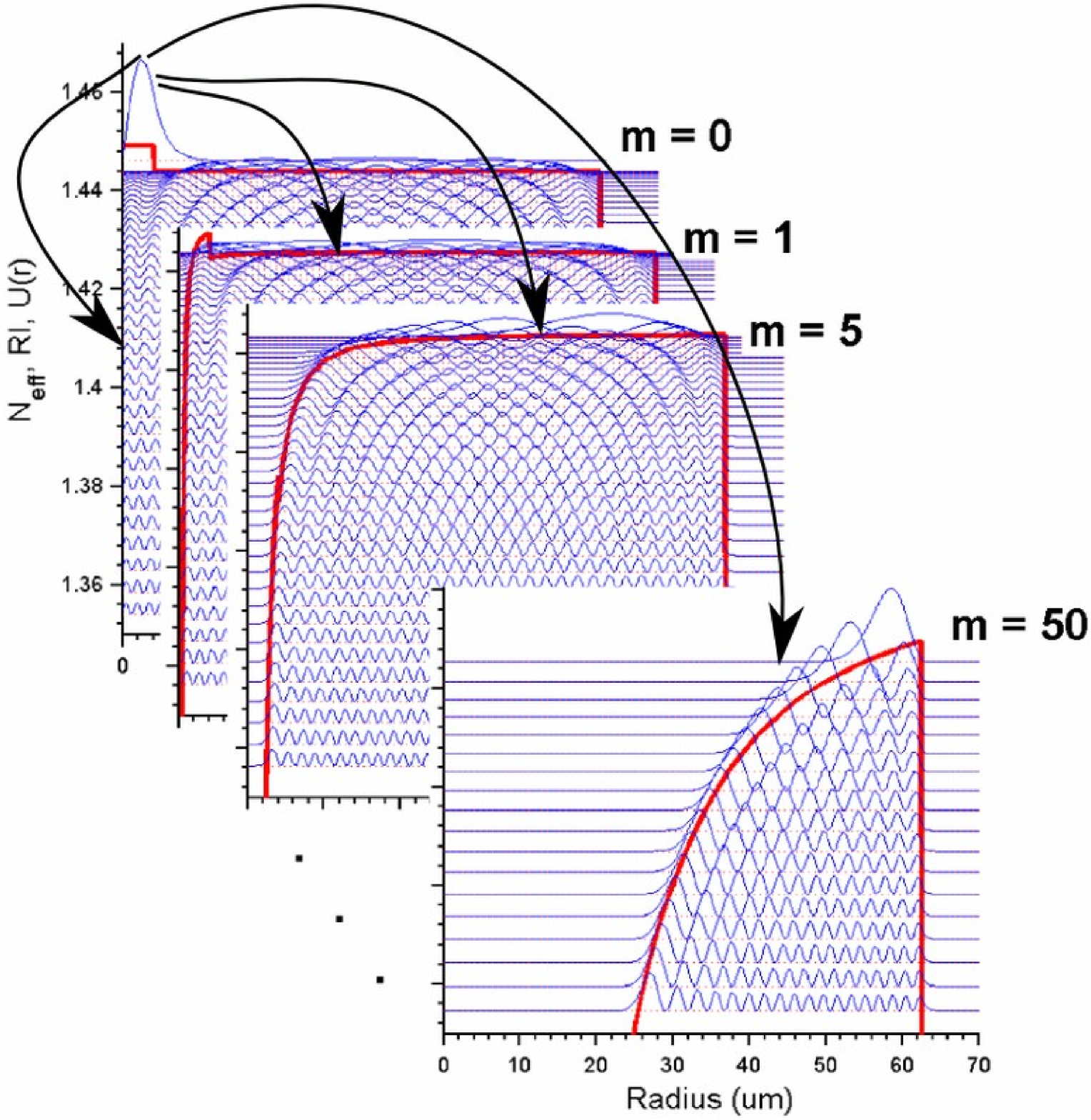}{The schematic ilustration of energy transfer from the core mode into the cladding modes in the TFBG. The various potential barriers correspond to different azimuthal numbers~$m$.}

If a spectrum of TFBG grating was obtained from an experiment, the grating wavenumber $K_G$ can be expressed in terms of the measured value of the Bragg resonance position $\lambda_B$, by assuming that the forward propagating wave in the core is coupled to the backward propagating wave in the core:
\Eq{}
{K_G = 2 \beta_B = 2 N_B \frac{2\pi}{\lambda_B}.}

The phase detuning condition condition~(\ref{eq_phase_match}) takes the following form:
\Eqaa{eq_diphas}
{\Delta k_j(\lambda) &=& K_G - \left(\beta_B(\lambda)+\beta_j(\lambda)\right) =}
{ &=& 2\pi\left(2 \frac{N_B}{\lambda_B} - \frac{N_B + N_j(\lambda)}{\lambda}\right),}
with
\Eqaa{}
{\beta_B(\lambda) &=& N_B k_o = N_B \frac{2\pi}{\lambda},}
{\beta_j(\lambda) &=& N_j(\lambda) k_o = N_j(\lambda) \frac{2\pi}{\lambda}.}
Here $\beta_B$ and $\beta_j$ are propagation wave numbers of the core mode, with the effective refractive index $N_B$, and the $j$-th cladding mode with the corresponding effective refractive index $N_j$, computed at the operational wavelength~$\lambda$.
We have also assumed that $N_\beta(\lambda) \sim constant$ approximately does not depend on the $\lambda$, which is true if the radius of the core is significantly smaller then the radius of the cladding.

The Bragg Condition is obtained by assuming that there is no phase detuning between the modes,~\textit{i.e.} ${\Delta k_j \sim 0}$.
Hence the position of the $j$-th resonance is defined by the following equation:
\Eq{eq_lam_res}
{\lambda_{res_j} = \frac{1}{2}\left(1+ \frac{N_j(\lambda)}{N_B}\right)\lambda_B}

Assuming the operational wavelength $\lambda$ and considering~(\ref{eq_diphas}) we can determine the phase detuning for the $j$-th resonance in the proximity to $\lambda_j$:
\Eqaaa{eq_diphas_lam}
{\delta_j(\lambda_j - \lambda) &=& 2\pi \left(2 \frac{N_B}{\lambda_B} - \frac{N_B + N_j(\lambda)}{\lambda_j}\right) - 2\pi \left(2 \frac{N_B}{\lambda_B} - \frac{N_B + N_j(\lambda)}{\lambda}\right) =}
{ &=& 2\pi(N_B + N_j(\lambda))\left( - \frac{1}{\lambda_j} + \frac{1}{\lambda}\right) \sim}
{ &\sim& 2\pi (N_B + N_j(\lambda))\left(\frac{\lambda - \lambda_j}{\lambda^2}\right).}

\Fig{CMT_many_res}{0.8}
{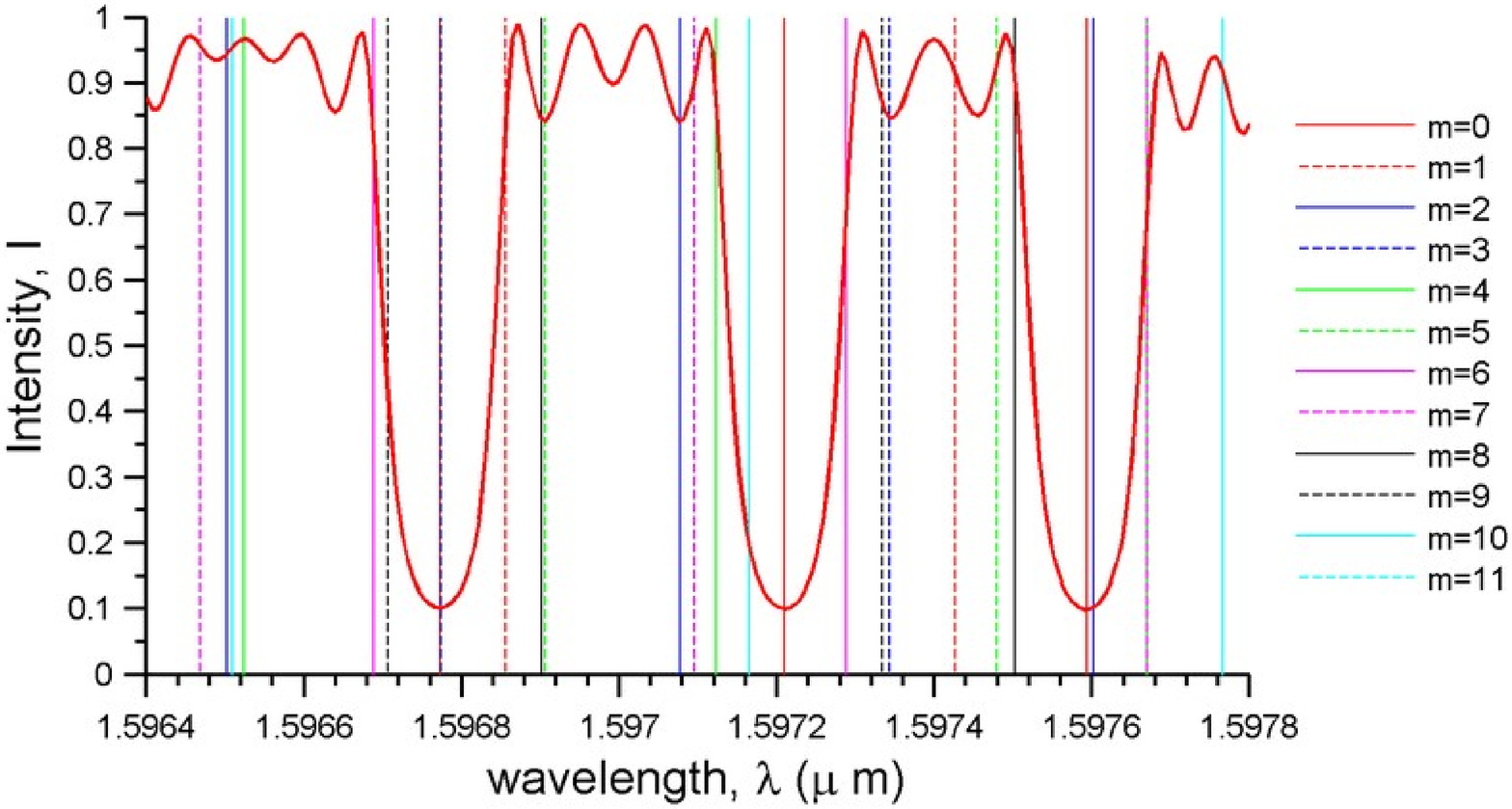}{The resonances of different families of modes, computed in accordance with~(\ref{eq_lam_res}). The peaks arising due to the coupling between the core mode and ${m = 0}$ family of modes, neglecting the coupling to higher family of modes. Here the grating length ${L = 10 mm}$ and the coupling constant ${C =  2 \cdot 10^{-4}}$.}

Using the two mode approximation and considering the dephasing expression~(\ref{eq_diphas_lam}) the resonances of a particular family of modes, let us say for~{$m=0$}, can be plotted along the wavelength of operation~$\lambda$, as shown in Figure~\ref{CMT_many_res}. 
For a single family of modes the resonance peaks have almost no overlap (Figure~\ref{CMT_many_res}), thus a resonance can be approximately computed with the two mode approximation independent of the other resonances. In other words, the energy coupling between the core mode and a particular cladding mode can be computed without considering other cladding resonances of the same family of modes.

However, in the case of TFBG, many resonances, corresponding to various families of modes, are present in close proximity to each other. In Figure~\ref{CMT_many_res}, resonances of the first $11$ families of modes are plotted along with the peaks computed for ${m = 0}$ family. 
Although the two mode analysis is useful to study basic properties of the problem, it can not be applied to our problem of interest due to the overlap between the resonances.

Going back to the initial coupled modes equation~(\ref{eq_coup6}) and, as previously, neglecting the rapidly oscillating terms, not contributing to the change of amplitudes, we obtain the following system of coupled equations:
\Eqaa{eq_many_modes_C}
{d_z A^{core}(z) &=& \sum_{k=1}^N i \frac{C_k}{2 \beta^{core}} e^{-i \delta_k z} A_k^{clad}(z),}
{d_z A_k^{clad}(z) &=& - i \frac{C_k}{2 \beta_k^{clad}} e^{+i \delta_k z} A^{core}(z).}
Here ${C_k = <\psi^{core}|\Delta|\psi_k^{clad}>}$ is the coupling constant between the core and the $k$-th cladding mode, as was discussed in the previous section; $N$ is the number of coupled modes.
The values of ${\delta_k(\lambda)}$, ${\beta_k^{clad}(\lambda)}$, ${\psi(\lambda)}$ and ${C_k(\lambda)}$ have to be computed at a particular operational wavelength $\lambda$.

The analytical solution, similar to~(\ref{eq_coup_sol}), can no longer be applied to the system of equations~(\ref{eq_many_modes_C}). Instead this system of coupled nonlinear differential equations can be solved numerically. 

In the case of co-directional coupling the initial value problem (IVP) is considered ${A^{core}(0)=1}$ and ${A_k^{clad}(0)= 0 }$, hence the system of equations can be directly integrated.
Unfortunately, in our case of contra-directional coupling the boundary value problem (BVP) ${A^{core}(0)=1}$ and ${A_k^{clad}(L)= 0 }$ has to be considered (as the boundary conditions are known at the opposite sides of the interval), and this significantly complicates the numerical routine.
Likely the problem~(\ref{eq_many_modes_C}) is linear with respect to the $z$ variable, hence we can start propagation from the opposite end ${z=L}$ assuming that ${A_k^{clad}(L)= 0}$ and setting amplitude of the core mode at ${z=L}$ to some arbitrary constant $C$ to be determined later: ${A^{core}(L)= C}$. Next the system of equations can be integrated as an initial value problem. Once the solution ${A^{core}(0)}$ at ${z = 0}$  is known we can renormalise the solution to ensure that ${A^{core}(0) = 1}$. The procedure is shown in Figure~\ref{Coupling_2modes}.
\Fig{Coupling_2modes}{1}
{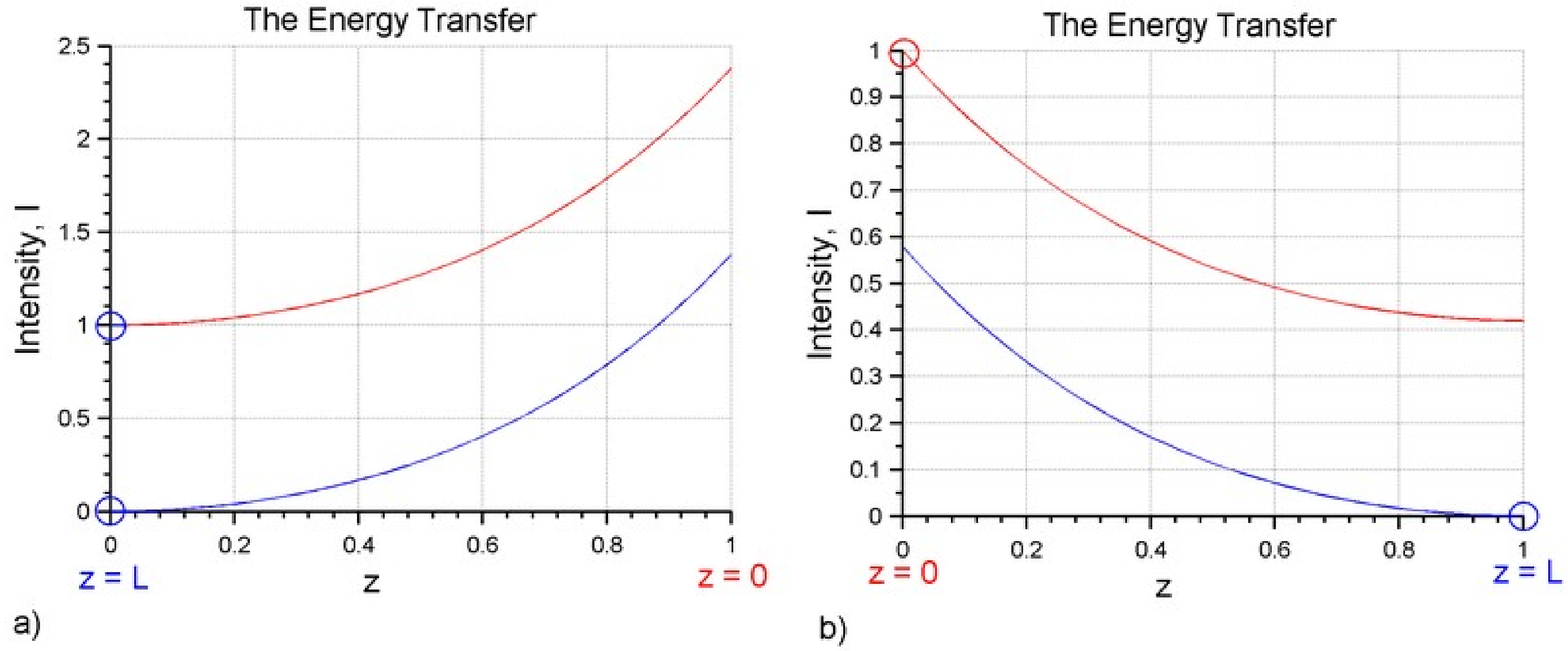}{Transforming the boundary value problem into the initial value problem. a) solving IVP by assuming ${A^{core}(L) = 1}$ and ${A^{clad}(L) = 0}$, b) renormalized the solution by setting ${A^{core}(0) = 1}.$}

The initial value problem is next solved with the help of the standard fourth order Runge-Kutta method. The result for the core mode coupled to the two closely positioned cladding mode resonances is shown in Figure~\ref{Coupling_3modes}.

\Fig{Coupling_3modes}{1}
{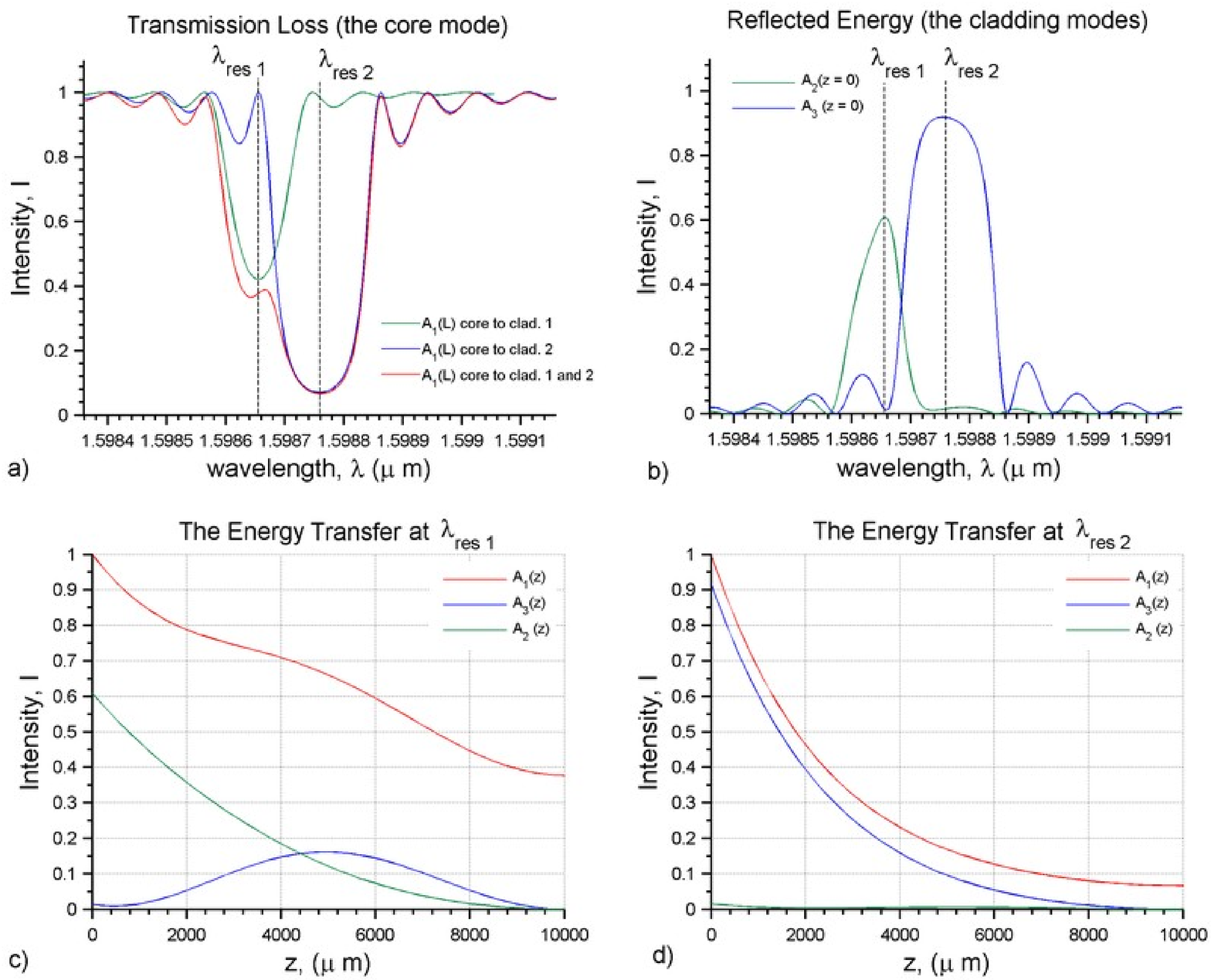}{Coupling between the core mode and two cladding modes. The grating length ${L = 10~mm}$ and  the coupling constant are ${C_1 = 2 \cdot 10^4}$ and ${C_2 = 1 \cdot 10^4}$ between the core and the two cladding modes. Here $A_1(z)$ is the amplitude of the forward propagating core mode (the red line), $A_2(z)$~and~$A_3(z)$ are amplitude of the backward propagating cladding modes (the blue and the green lines).}

In Figure~\ref{Coupling_3modes}~a) the difference between the two mode approximation, when the coupling to cladding modes is considered independently (the \C{green} and the blue curves), and the exact approach (the red curve) is clearly seen. It is also interesting to notice that the energy from the cladding mode Figure~\ref{Coupling_3modes}~c) (the blue curve) can be recoupled back to the core mode (the red curve).

Finally, the coupling problem for the TFBG structure of our interest can be solved as follows:

\begin{enumerate}
\item We introduce a grid vector $\lambda_j$ spanning the operation range of the sensor,~\textit{i.e.} ${\lambda_j \in [\lambda_{min}, \lambda_{max}]}$, where at each point $\lambda_j$ the system of coupled equations~(\ref{eq_many_modes_C}) has to be solved. We note that a typical peak width is about $0.2$~nm, where the operational range is about $100$ nm. Thus, at least $1000$ points $\lambda_j$ have to be considered to observe the resonances. Considering the necessity of solving the system of coupled differential equations~(\ref{eq_many_modes_C}) at each point, the described approach can be extremely time consuming.
We can overcome this difficulty by introducing a nonuniform grid. The positions of resonances is known~(\ref{eq_lam_res}):
\Eq{}
{\lambda_{res_j} = \frac{1}{2}\left(1+ \frac{N_j(\lambda)}{N_B}\right)\lambda_B,}
hence we can introduce a finer mesh in the vicinity of resonances, as shown in Figure~\ref{CMT_grid}.
\Fig{CMT_grid}{0.9}
{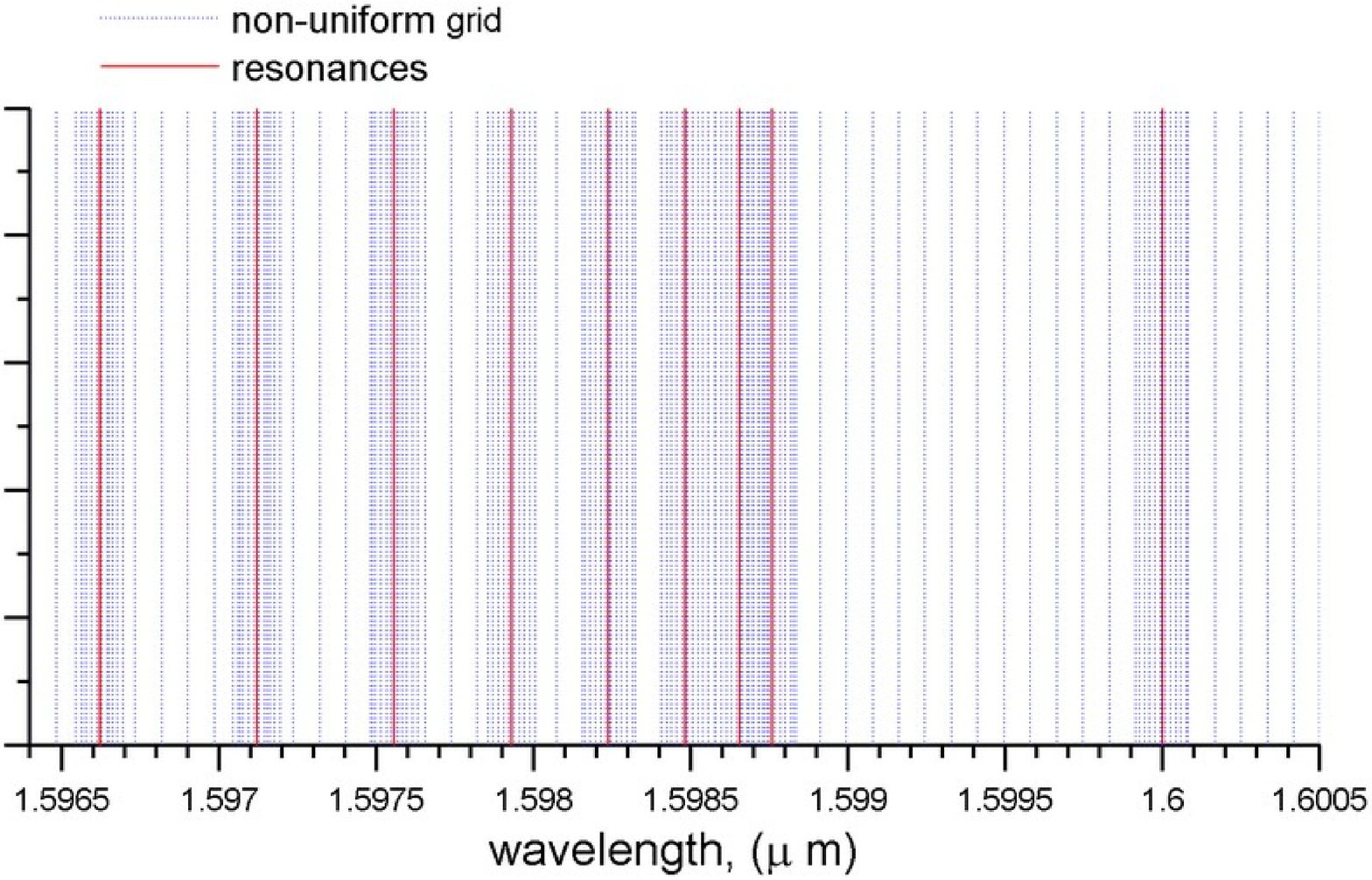}{The non-uniform grid with a finer mesh in the vicinity of resonances.}

\item Next the detuning parameter $\delta_j$ for each resonance can be computed in accordance with~(\ref{eq_diphas_lam}):  
\Eq{}
{\delta_j(\lambda) \sim 2\pi (N_B + N_j(\lambda))\left(\frac{\lambda - \lambda_j}{\lambda^2}\right).}

\item Unfortunately \C{the dispersion of modes} can not be neglected, as can be seen from Figure~\ref{Potentia_var_lam}. Hence, we have to find modes and coupling constants at each point $\lambda_j$ of the grid, or at least we can split the interval ${[\lambda_{min}, \lambda_{max}]}$ into a set of smaller subintervals and compute resonances and coupling coefficients for each of them, considering them to be constant within the subintervals.   
\Fig{Potentia_var_lam}{0.9}
{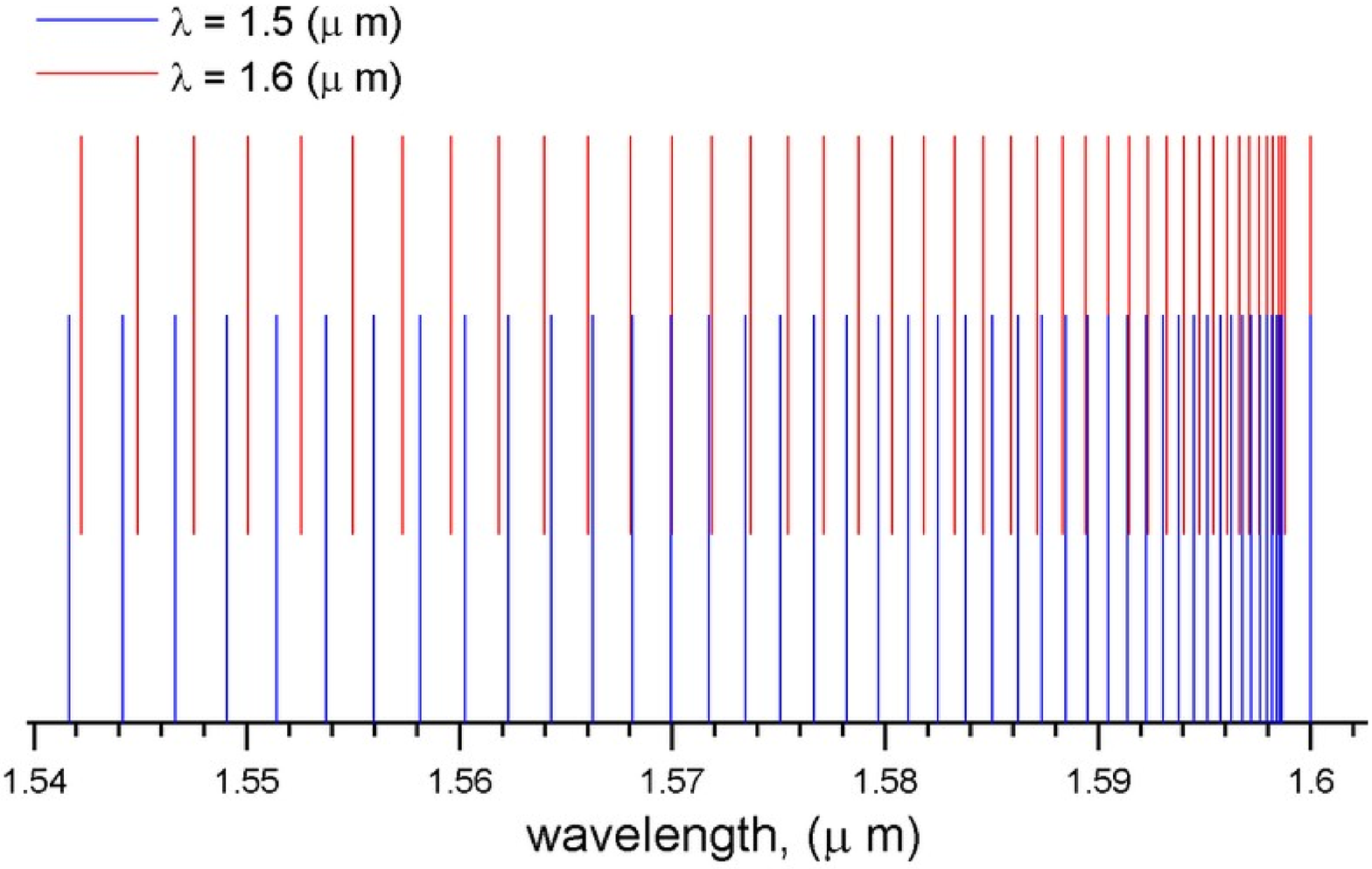}{\C{The dispersion of modes}. The resonances of ${m=0}$ family of modes are computed at ${\lambda_1 = 1.5}$~$\mu m$ and ${\lambda_2 = 1.6}$~$\mu m$ operational wavelengths.}

\item Finally the system of coupled nonlinear differential equations~(\ref{eq_many_modes_C}) can be solved independently at each point~$\lambda_j$ of the grid:
\Eqaa{}
{d_z A^{core}(z,\lambda_j) &=& \sum_{k=1}^N i \frac{C_k(\lambda_j)}{2 \beta^{core}(\lambda_j)} e^{-i \delta_k(\lambda_j) z} A_k^{clad}(z,\lambda_j),}
{d_z A_k^{clad}(z,\lambda_j) &=& - i \frac{C_k(\lambda_j)}{2 \beta_k^{clad}(\lambda_j)} e^{+i \delta_k(\lambda_j) z} A^{core}(z,\lambda_j).}

The number of coupled modes $N$ can be limited to $15$ modes by choosing only the modes with the smallest phase mismatch~${\delta_k(\lambda_j)}$. The remaining modes can be neglected due to a \C{weak coupling}. For each~$\lambda_j$ the choice of $N$ coupled modes should be reconsidered.

\end{enumerate}

The theoretically computed transmission spectra along with the experimentally measured spectra are shown in Figures~\ref{Done_2_exp} and~\ref{Done_2_Theor} for the $2^o$ degree grating, in 
Figures~\ref{Done_4_exp} and~\ref{Done_4_Theor} for the $4^o$ degree grating and in
Figures~\ref{Done_10_exp2} and~\ref{Done_10_pi2} for the $10^o$ degree grating.
\C{A more detailed information} can be obtained by \C{zooming in} to particular resonances, which are shown in 
Figures~\ref{Done_2_Zoom} for the $2^o$ degree grating, in 
Figures~\ref{Done_4_Fine_L},~\ref{Done_4_Fine_R} for the $4^o$ degree grating and in
Figures~\ref{Fine_10_L},~\ref{Fine_10_C},~\ref{Fine_10_R} for the $10^o$ degree grating. The coupling coefficients are plotted on the same figures. 

\C{We note} that the resonances have a similar structure along the whole operational range of the sensor. We also note the an interesting effect of peaks alternation. 
There are only two group of peaks: the peaks consisting of modes with odd azimuthal symmetry ${m = 1,3,5 ...}$, and modes with even azimuthal symmetry ${m = 0,2,6 ...}$ . In the spectral area where peaks are separate, these two groups of peaks are alternating. This is an important effect explaining some of the experimental data, to be discussed later.

\clearpage

\Fig{Done_2_exp}{1}
{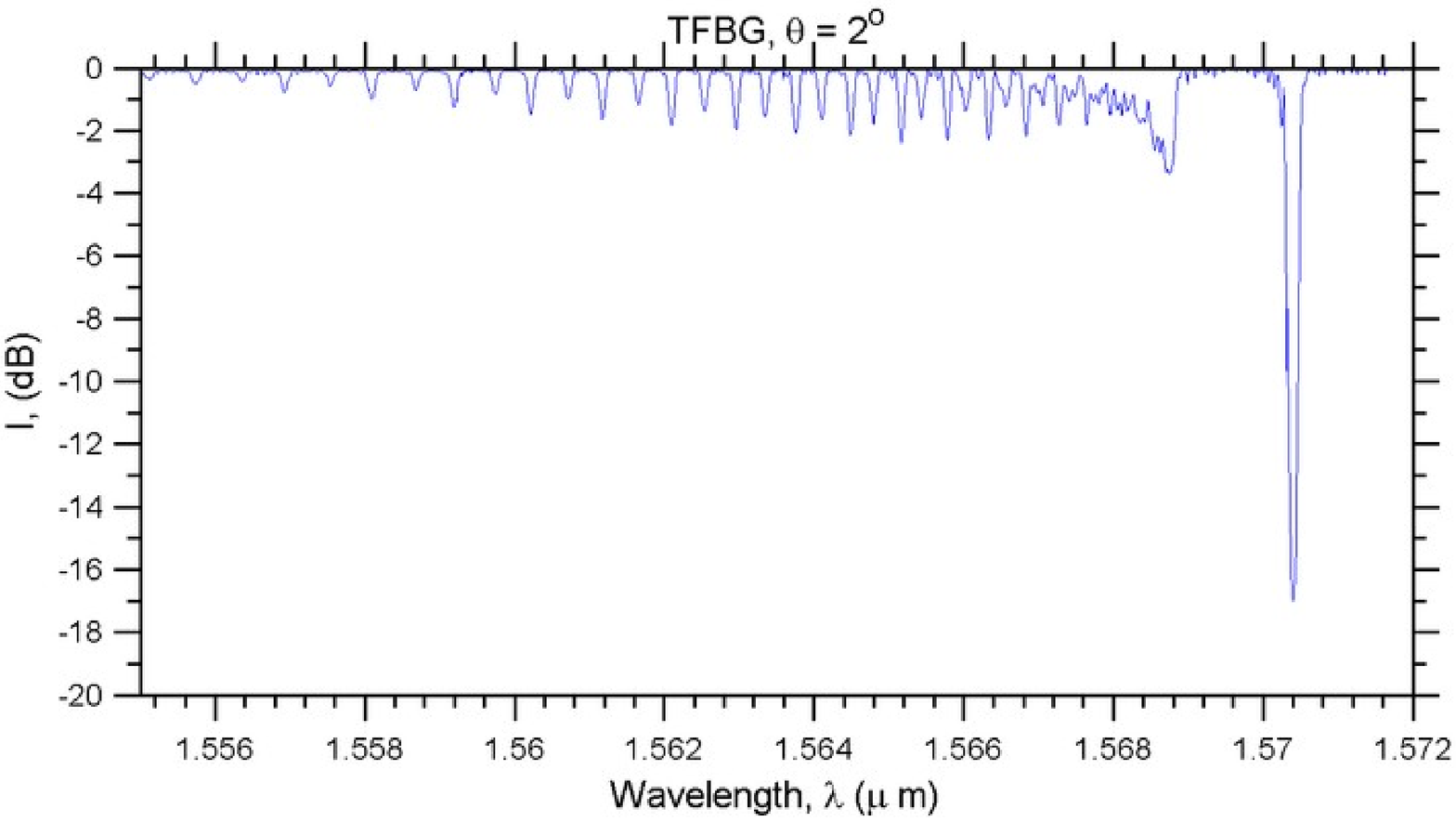}
{The experimentally measured spectra of the $2^o$ degree $1~cm$ long grating.}

\Fig{Done_2_Theor}{1}{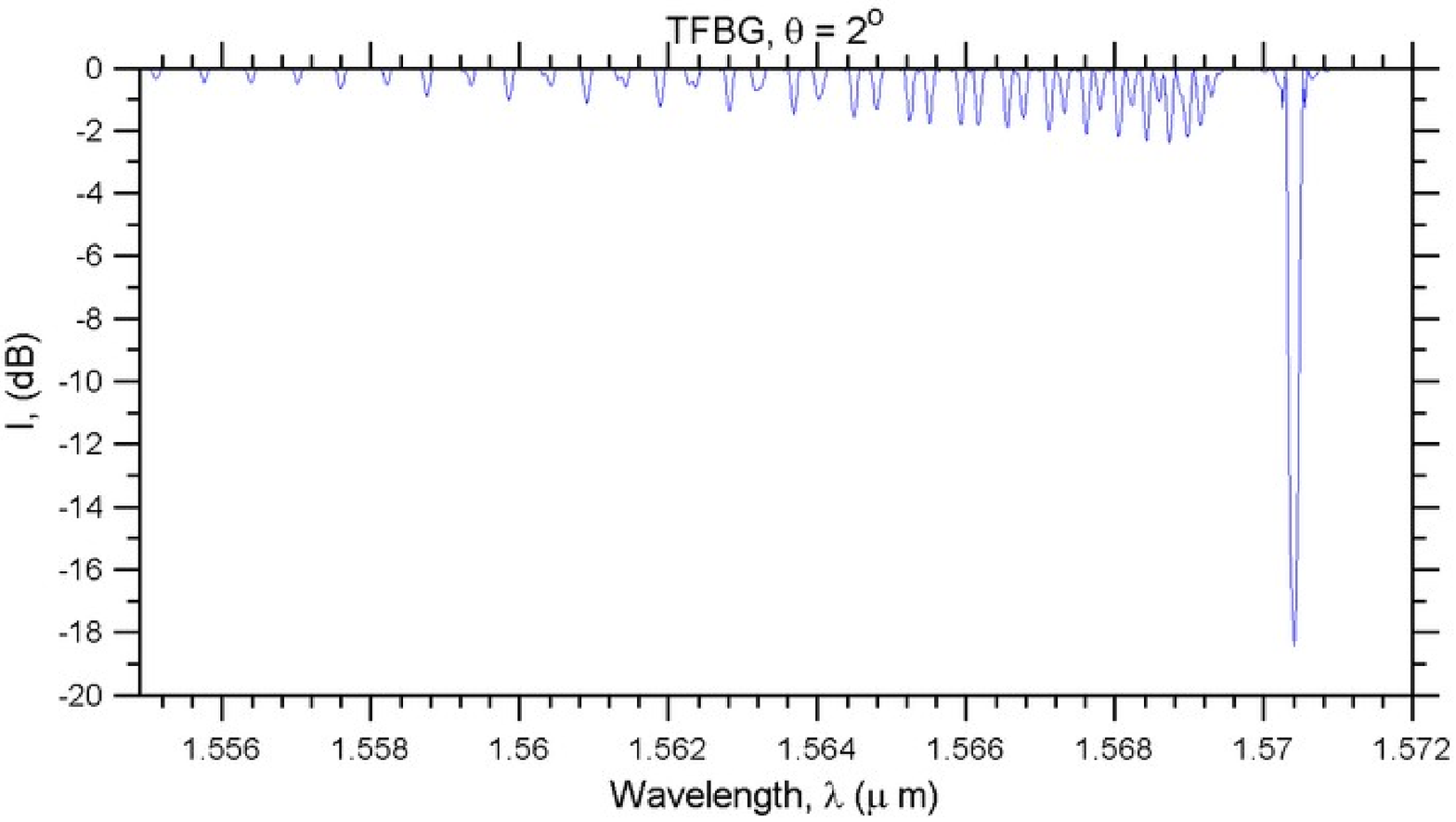}
{The theoretically computed transmission spectra of $2^o$ degree grating. ($L = 1~cm, \Delta n = 10^{-4}$)}

\Fig{Done_2_Zoom}{0.9}
{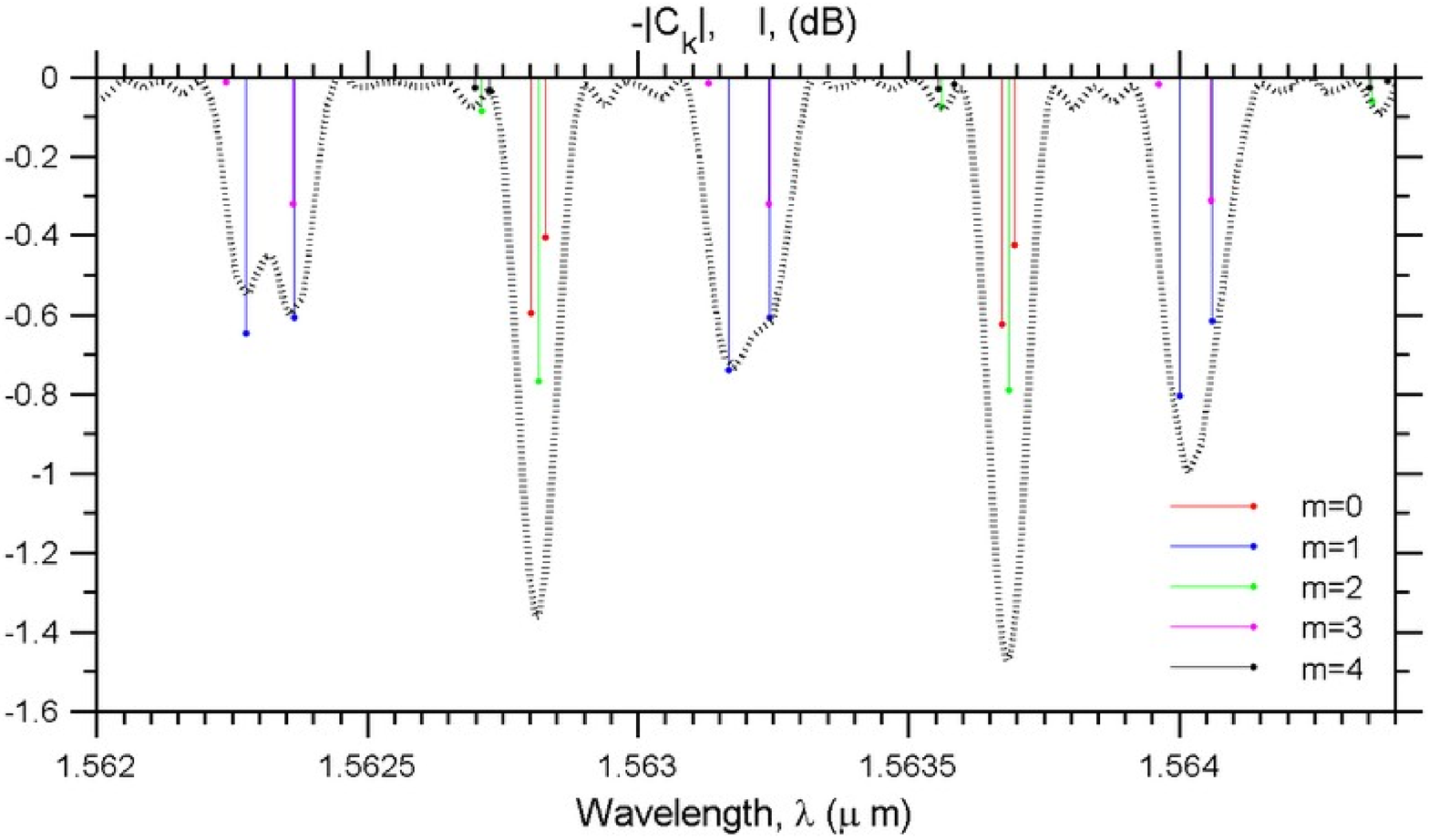}
{The fine structure of the particular resonances and corresponding coupling coefficients of $2^o$ degree grating.}

\clearpage
\Fig{Done_4_exp}{1}{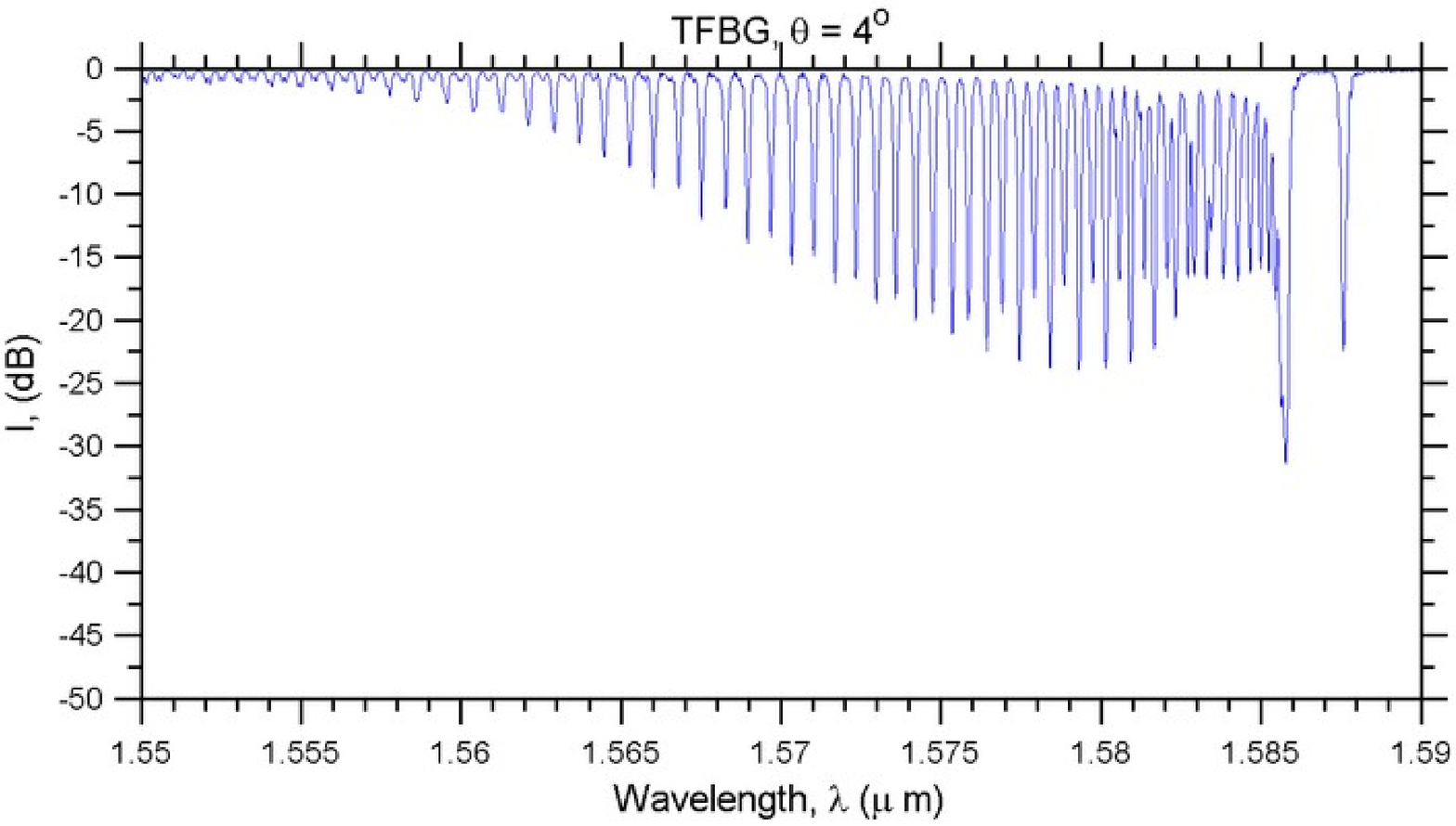}
{The experimentally measured spectra of the $4^o$ degree $1~cm$ long grating.}

\Fig{Done_4_Theor}{1}{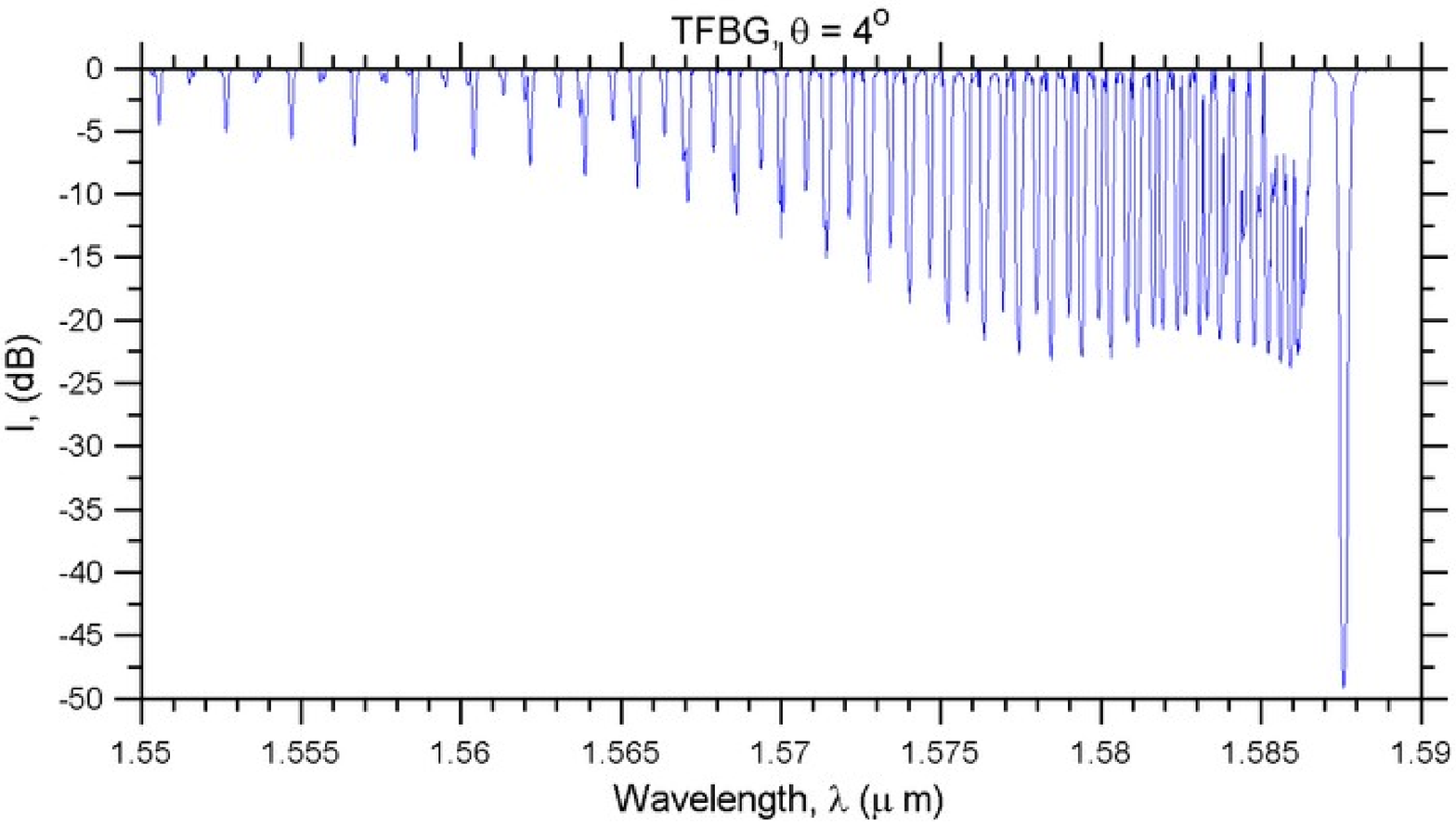}
{The theoretically computed transmission spectra of $4^o$ degree grating. ($L = 1~cm, \Delta n = 10^{-4}$)}

\Fig{Done_4_Fine_L}{0.9}{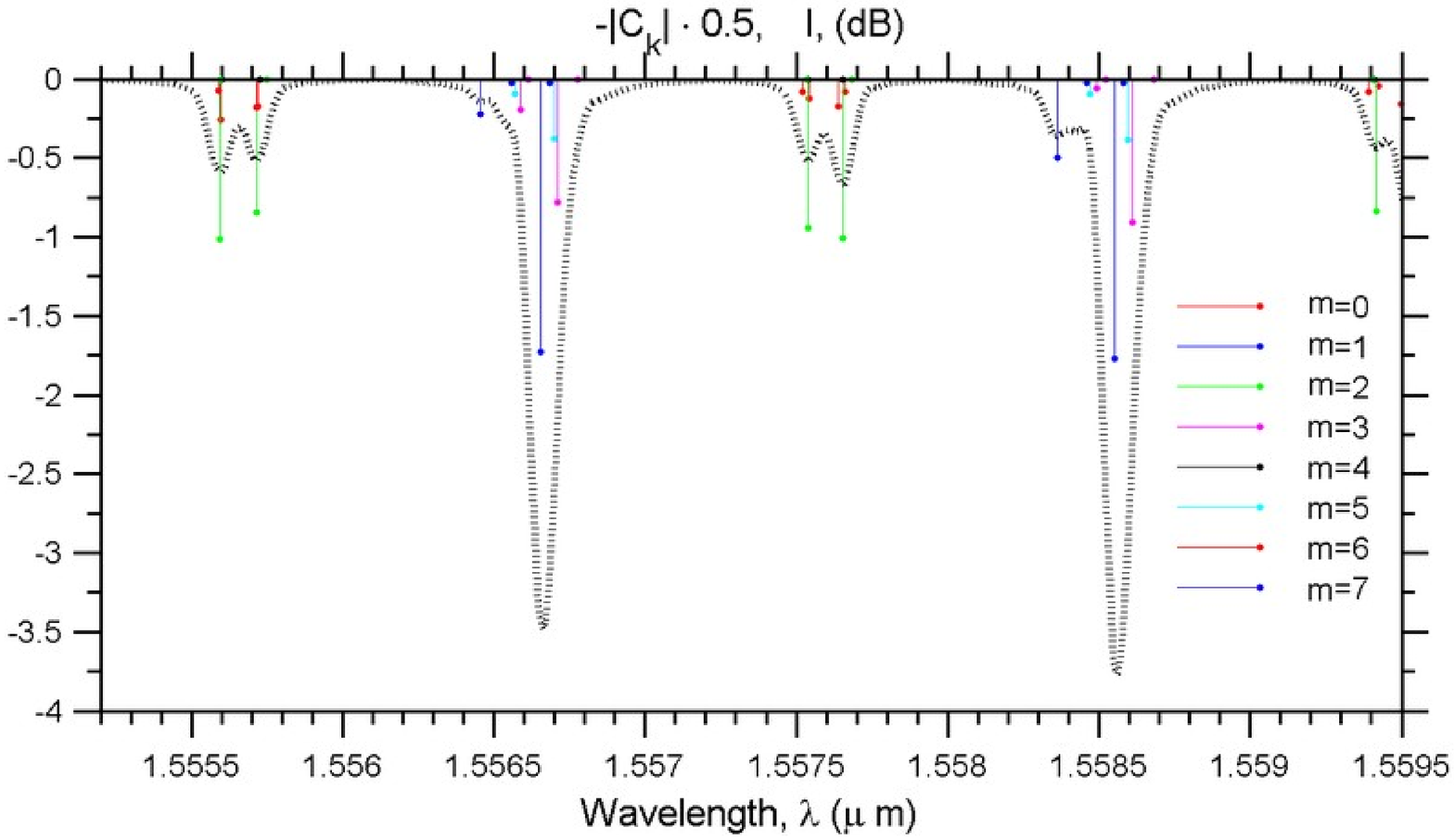}
{The fine structure of the particular resonances and corresponding coupling coefficients of $4^o$ degree grating.}

\Fig{Done_4_Fine_R}{0.9}{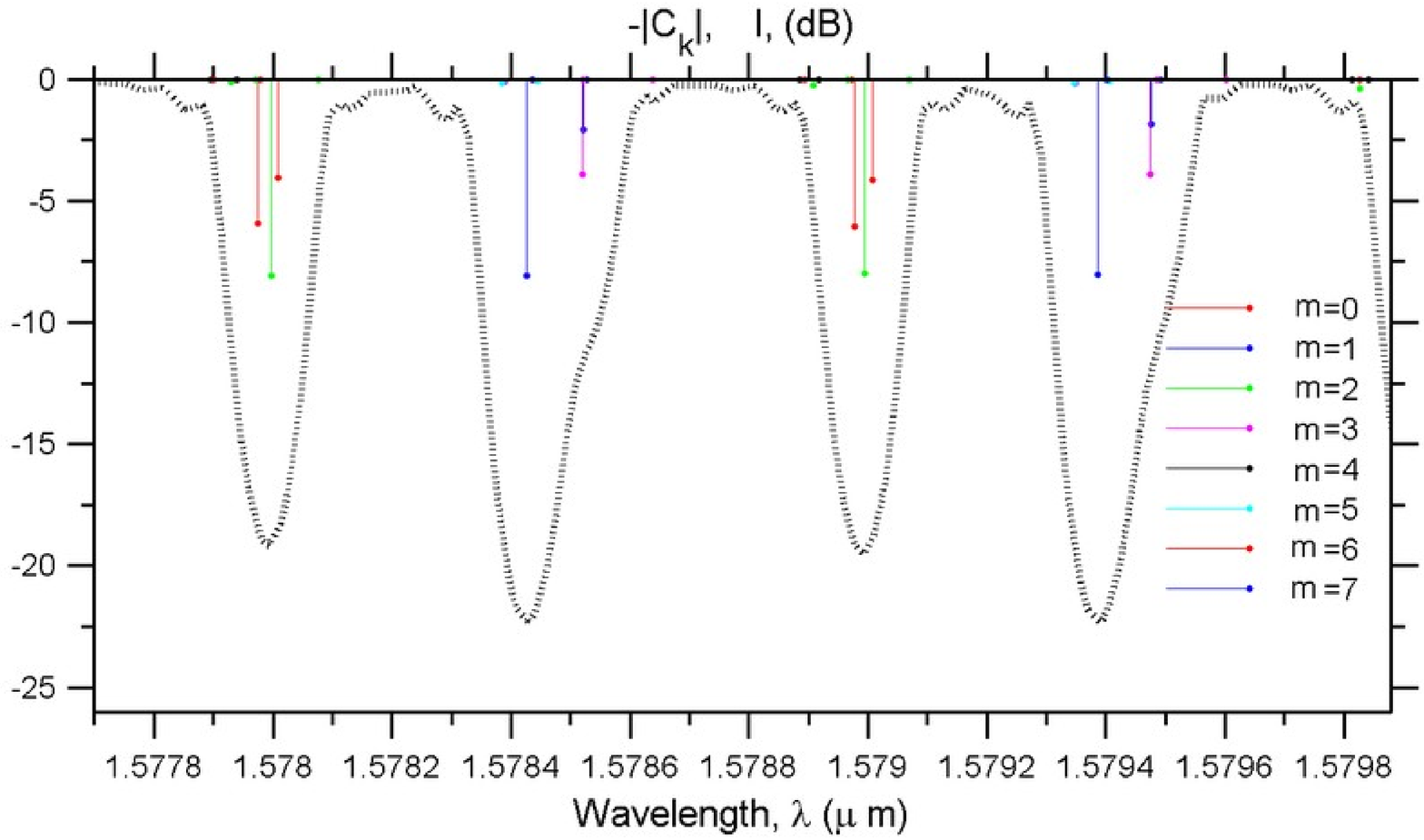}
{The fine structure of the particular resonances and corresponding coupling coefficients of $4^o$ degree grating.}

\clearpage
\Fig{Done_10_exp2}{1}{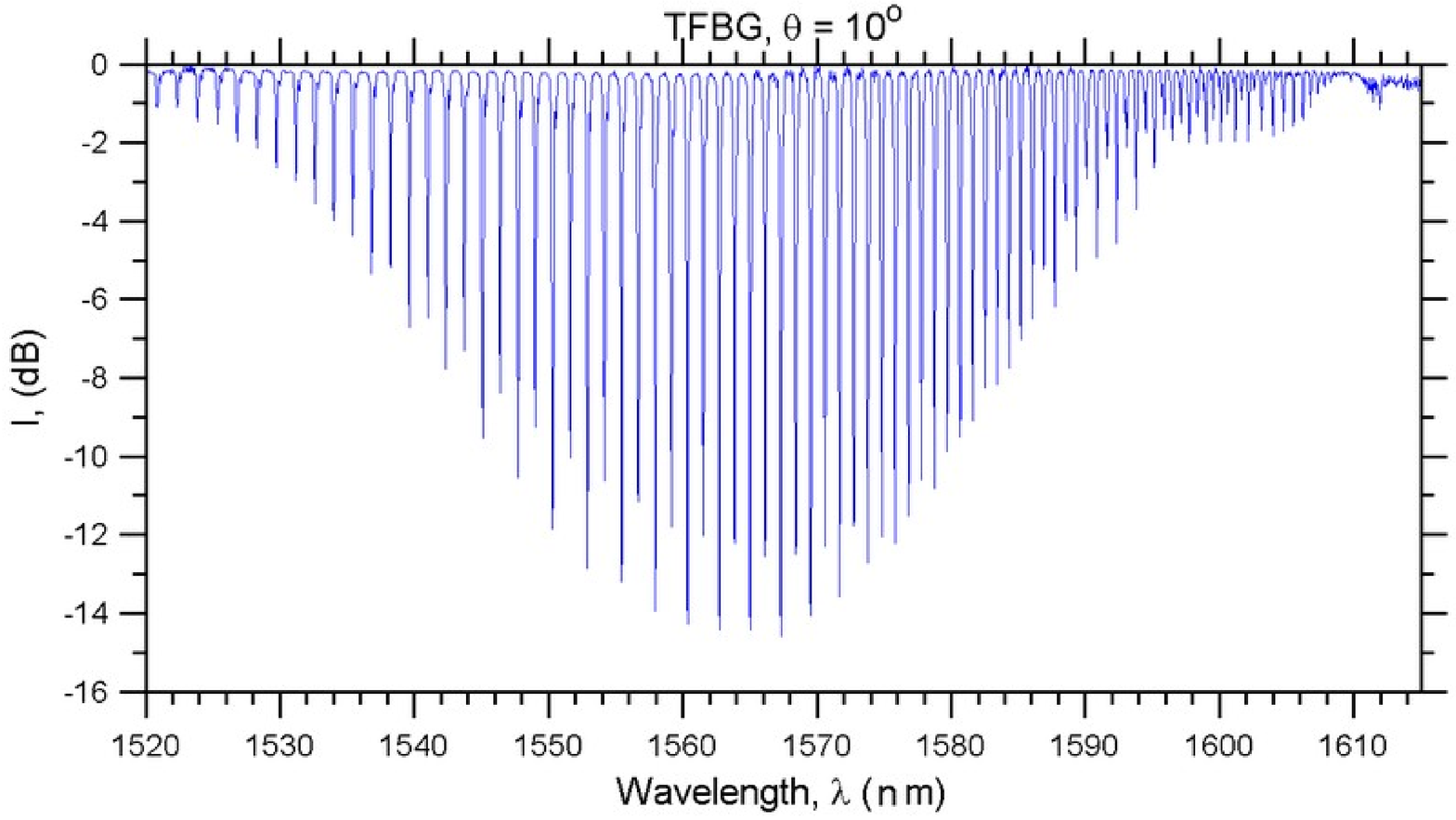}
{The experimentally measured spectra of the $10^o$ degree $1~cm$ long grating.}

\Fig{Done_10_pi2}{1}{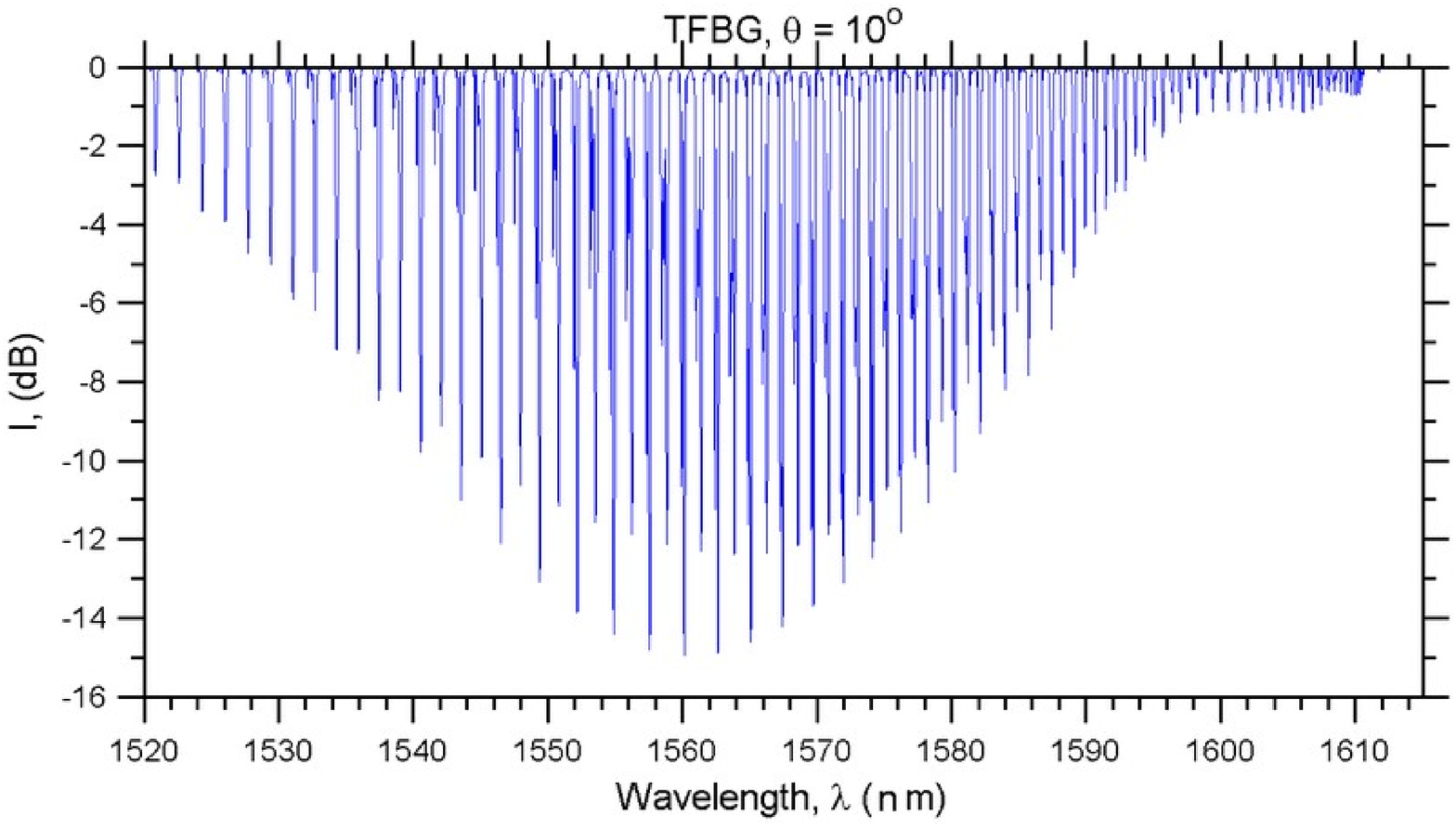}
{The theoretically computed transmission spectra of $10^o$ degree grating. ($L = 1~cm, \Delta n = 10^{-4}$)}

\Fig{Fine_10_L}{0.9}{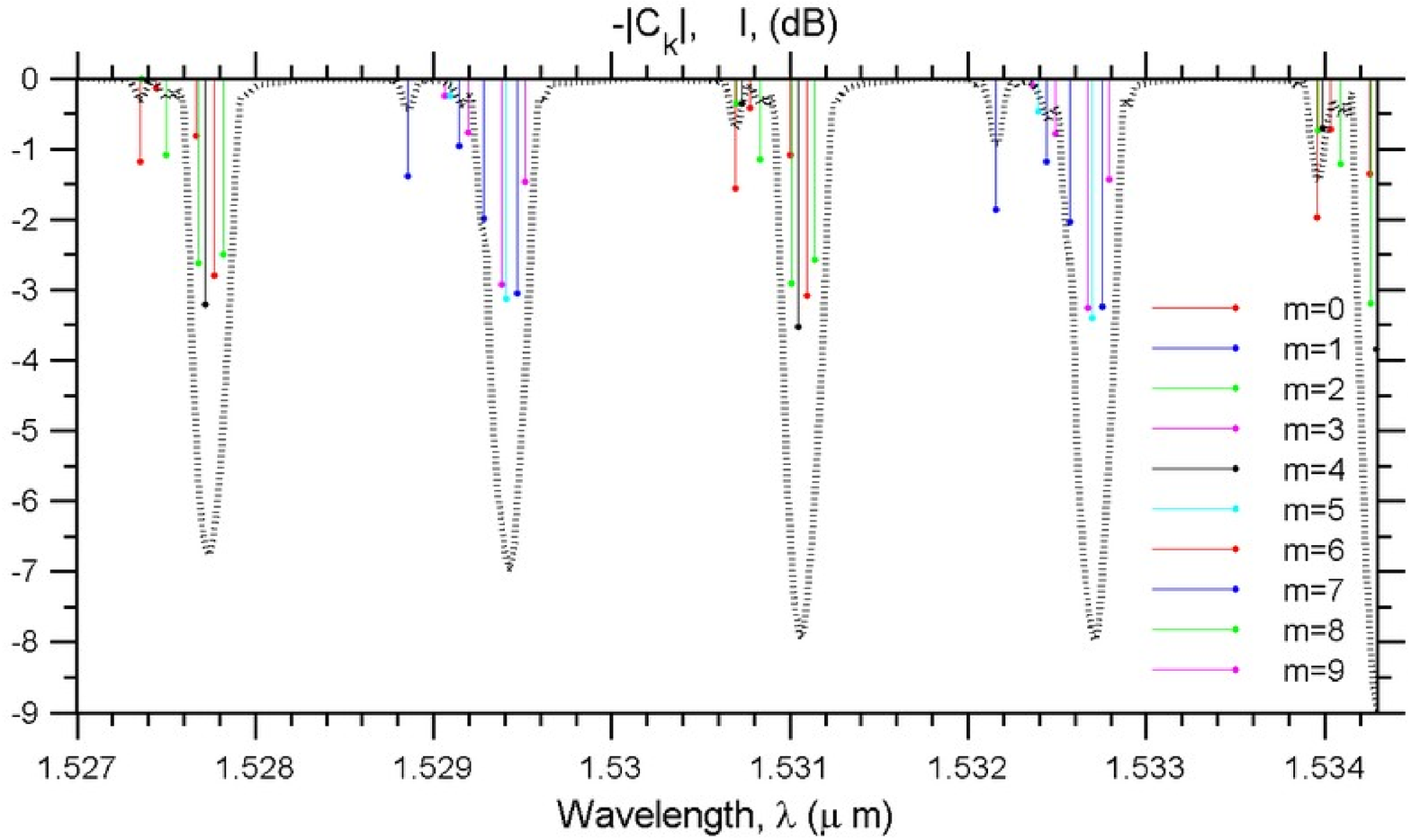}
{The fine structure of the particular resonances and corresponding coupling coefficients of $10^o$ degree grating.}

\Fig{Fine_10_C}{0.9}{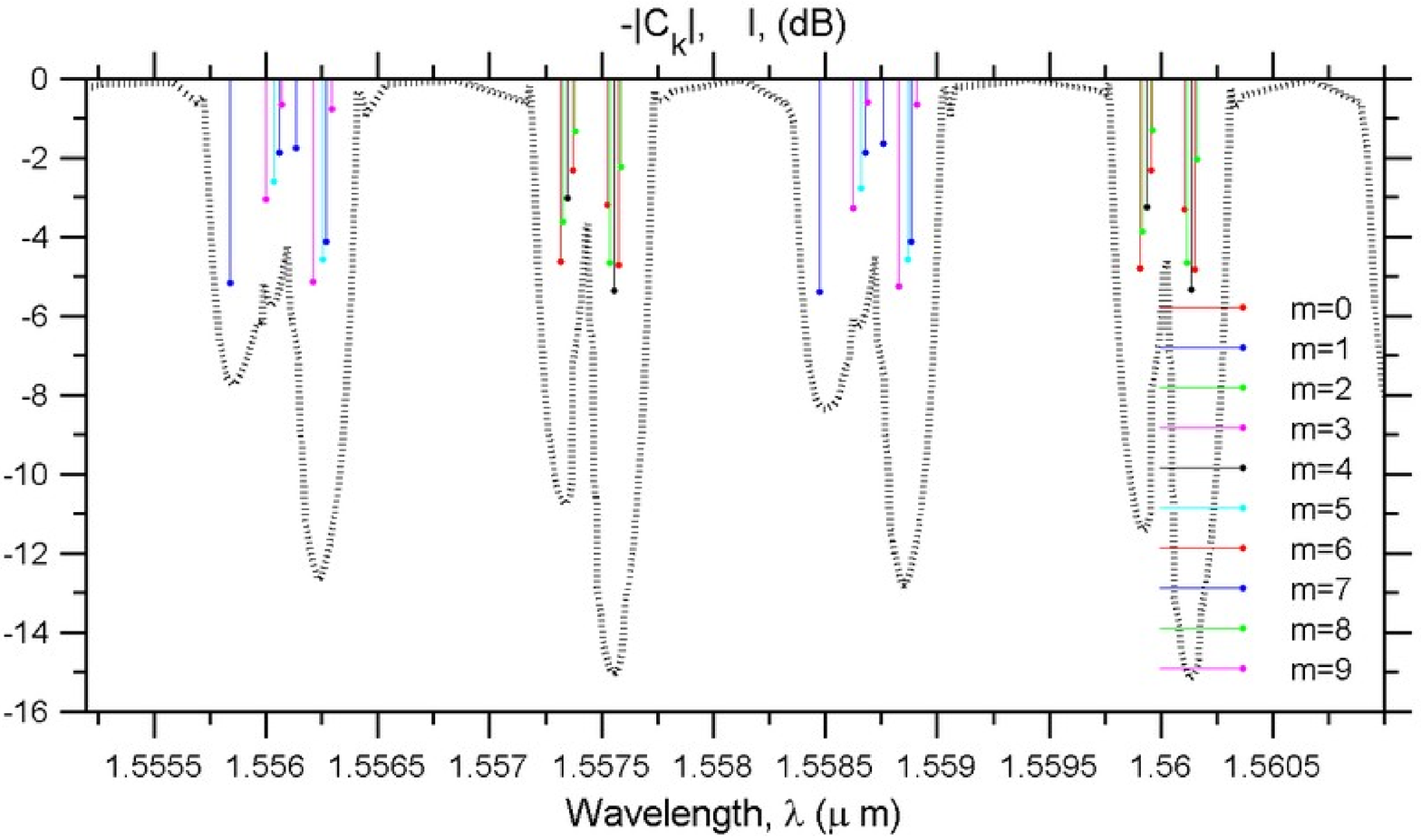}
{The fine structure of the particular resonances and corresponding coupling coefficients of $10^o$ degree grating.}

\Fig{Fine_10_R}{0.9}{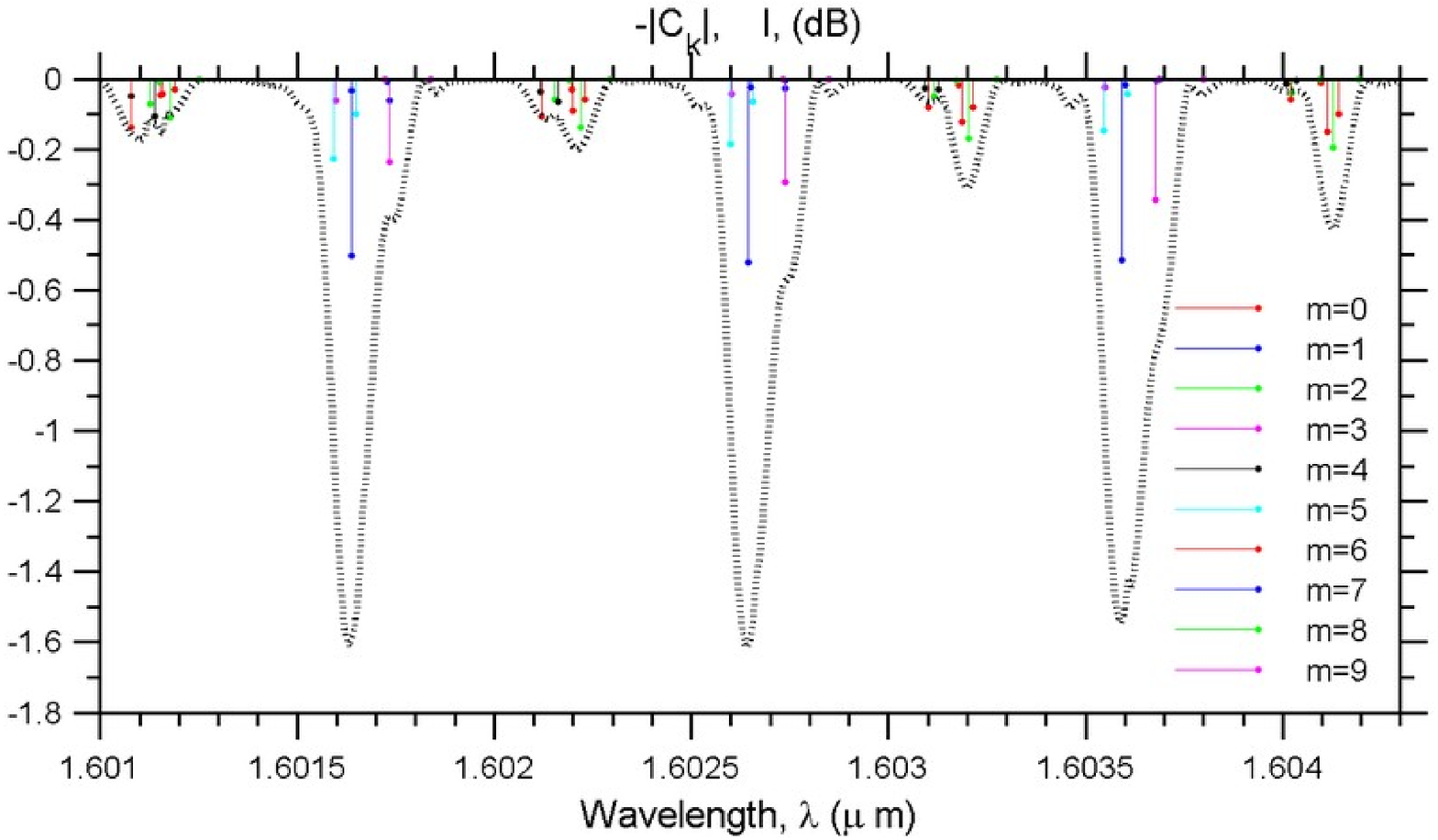}
{The fine structure of the particular resonances and corresponding coupling coefficients of $10^o$ degree grating.}

As can be seen from the presented figures the developed theoretical model provides results in good agreement with the experimental measurements. However, the small difference is present in the area of the so-called ghost modes~\cite{Chan:2007} for $4^o$ degree tilted grating. This area is highly populated with resonances, thus our approximation where only $15$ coupled modes are considered might not be sufficient. 
It also should be noted that we presume an ideal grating. In reality a physical grating might be subjected to various unaccounted-for effects, such as a non-uniform and one sided illumination by the UV beam in the process of grating inscription.
Nevertheless, from the point of view of practical application, we are interested in the spectral area with distinct resonances. This region is simulated in good agreement with the measurements.

%% file: Chap_TFBG_CMT_4_Polarization.tex
\section{Polarization-dependent coupling}
In this section we discuss polarization-dependent properties of the TFBG.
Considering that the core mode has ${m = 1}$ azimuthal symmetry it can be orientated differently in the plane transverse to the propagation axis.
The tilt of the grating planes breaks the cylindrical symmetry of the fibre, and defines the reference frame $x-y$ in which it is convenient to analyze the system. 
Considering the mutual orientation of the grating planes and the incident core mode, we can expect that the coupling coefficients might depend on the relative orientation between the polarization (or orientation) of the incident core mode and relative orientation of the grating planes, as shown in Figure~\ref{TFBG_polarization_intr}. 

\Fig{TFBG_polarization_intr}{0.7}{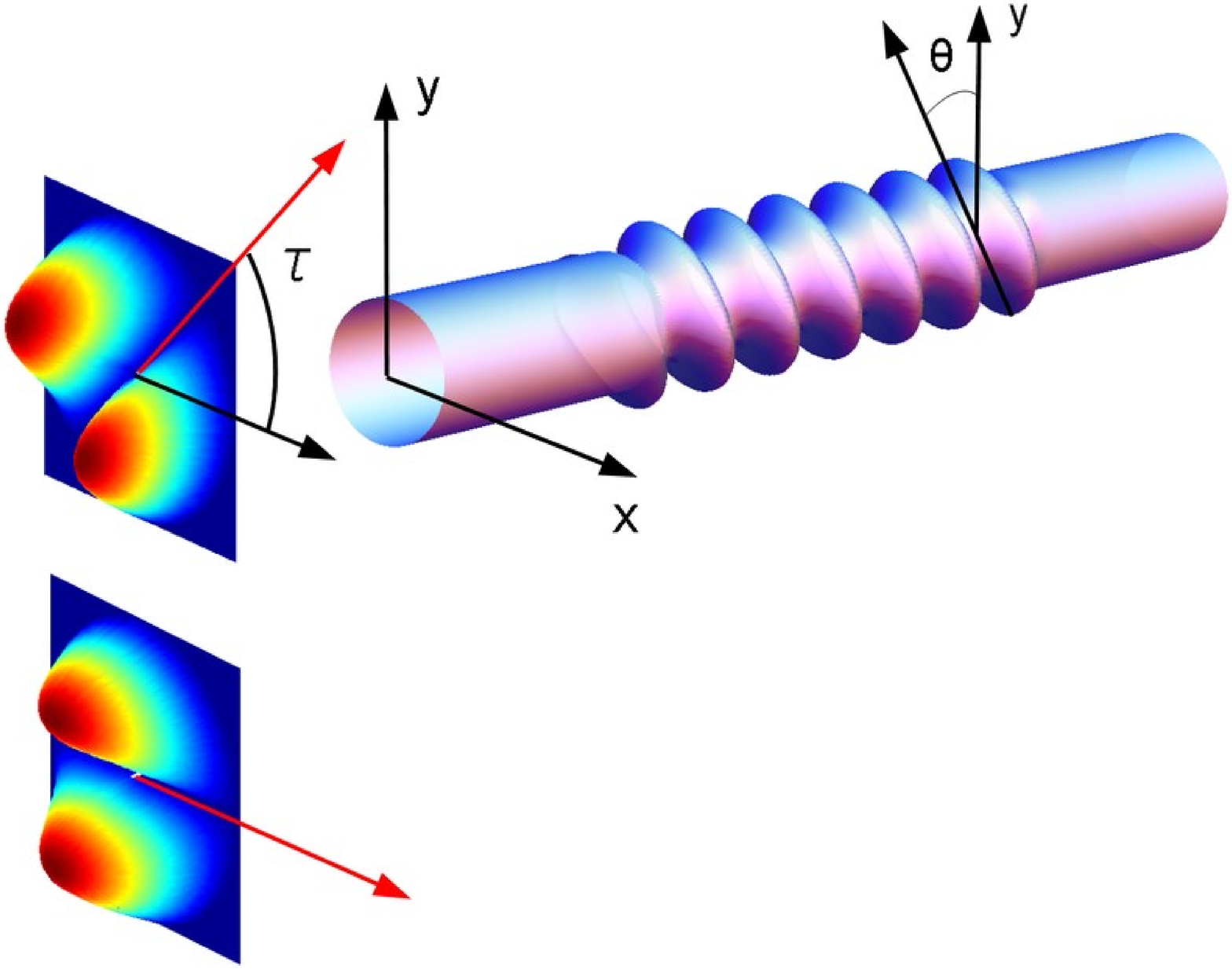}
{The schematic representation of the TFBG grating and the incident linearly polarized core mode (the $E_\rho$  component is shown). The core mode is rotated about the optical device axis by some angle~$\tau$. 
The grating is tilted by $\theta$ angle about the $x$ axis.}

The coupling coefficients~$C_{ik}^{nm}$ in accordance with~(\ref{coupl_coeff_0}) are proportional to
\Eq{}
{C_{ik}^{nm}(\tau) \sim \int_0^\infty \V R_i^n(\rho)\V R_k^m(\rho) n_o(\rho) \delta(\rho) \sigma^{mn}(\rho,\tau) d\rho  + c.c.,}
here $\tau$ is the angle at which linearly polarized light is incident on the TFBG structure and $\sigma^{mn}(\rho,\tau)$ is the weighted function, shown in Figure~\ref{Polarization}:
\Eq{}
{\sigma^{mn}(\rho,\tau) = \int_0^{2\pi} e^{j K_g \sin(\theta_g)\sin(\phi+\tau)\rho} e^{j(m-n)\phi}d\phi}

\Fig{Polarization}{1}{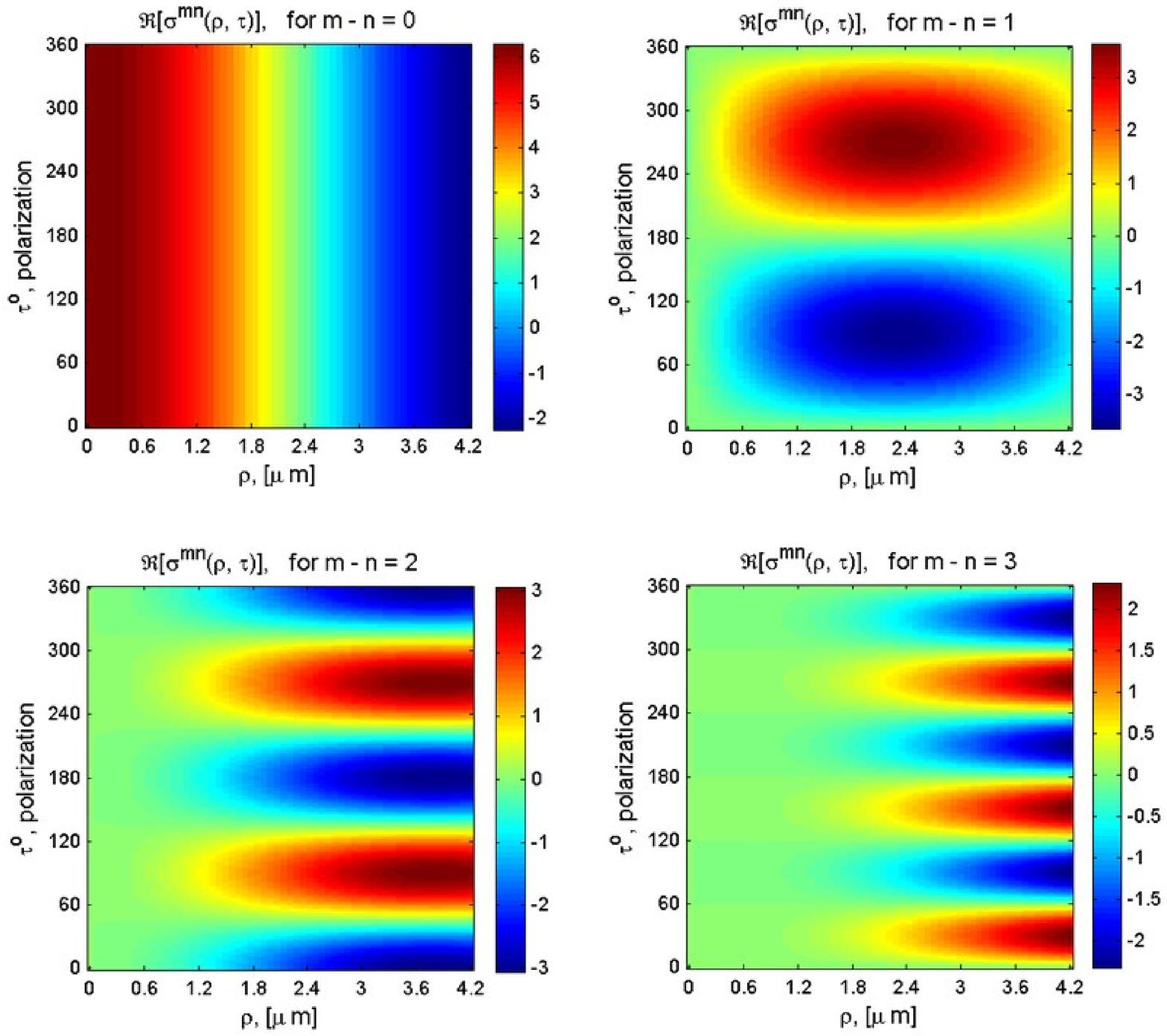}
{The weighted function ${\sigma^{mn}(\rho, \tau)}$ in the fibre core, for the $4^o$~degree tilted grating.}

As it shown in Figure~\ref{Polarization} the weighted function~$\sigma^{mn}(\rho,\tau)$ corresponding to coupling between the modes of the same family (in our case the core mode has the azimuthal number~${m = 1}$ is coupled to the~${m = 1}$ cladding modes) is not affected by the incident angle~$\tau$ of linearly polarized light, whereas coupling to different families of modes, with the~${m-n = 1,2,3}$, reveals a significant angle dependence. 

Hence, considering the weighted function~$\sigma^{mn}(\rho,\tau)$, the polarization-dependent coupling coefficients can be computed.
For instance for the $4^o$~degree tilted grating the coupling coefficients are shown in Figure~\ref{Polarization_4degree} for ${\tau = 0^o,45^o,90^o}$ polarization angles of the incident light.

\Fig{Polarization_4degree}{0.85}{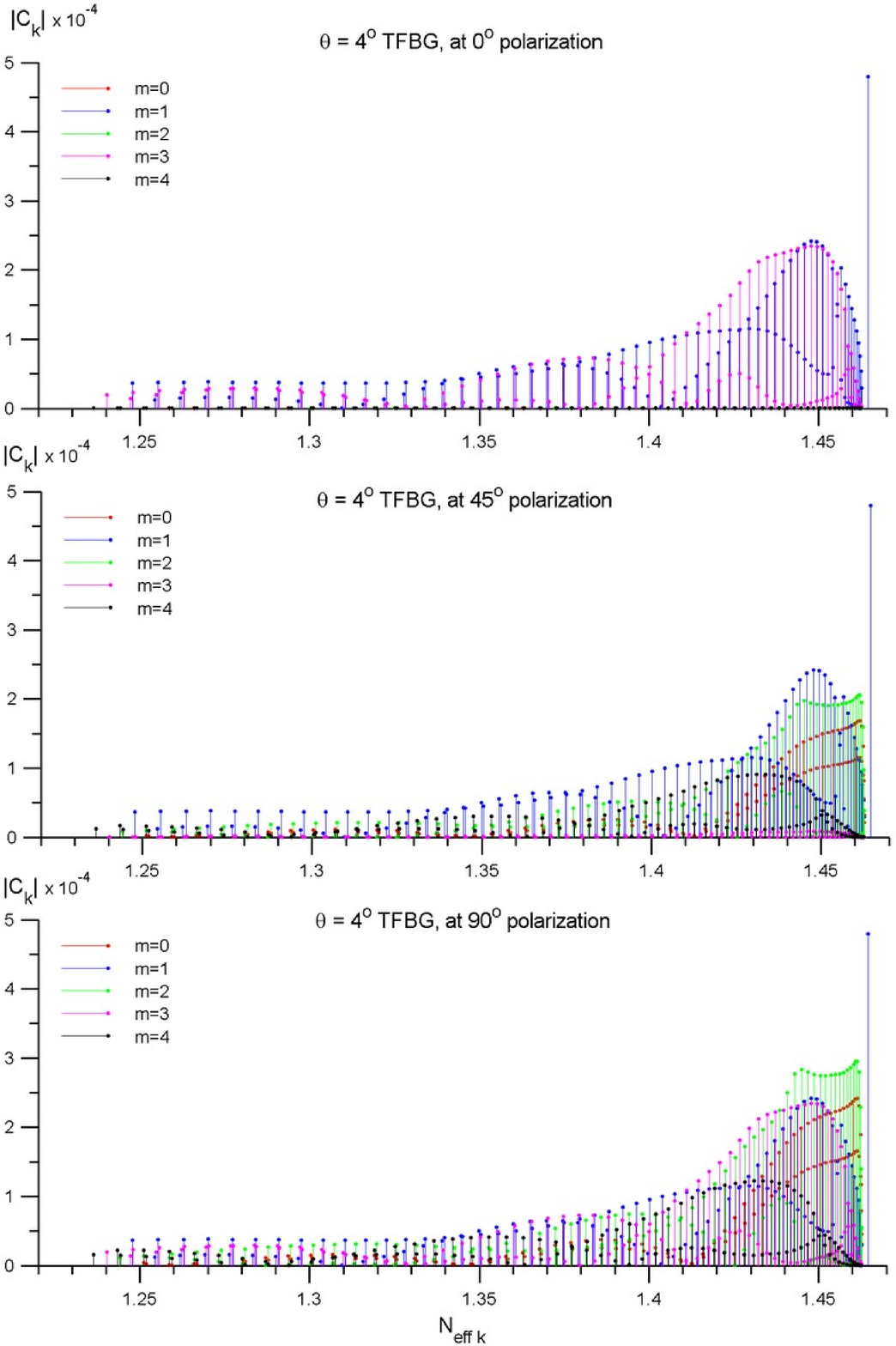}
{The coupling coefficients $C_k(\tau)$ for the $4^o$~degree TFBG computed for ${\tau = 0^o,45^o,90^o}$ polarization angles of the incident light.}

Now let us study the polarization dependence of particular resonances in more detail. As we mentioned in the previous section there are only two alternating groups of resonances with either odd or even azimuthal symmetry, thus it is sufficient to analyze two closely positioned resonances. 
The results for $4^o$ and $10^o$ degrees gratings are shown in Figure~\ref{Polarization_2Peaks_4} and  Figure~\ref{Polarization_2Peaks_10}, respectively. 

It is clearly seen that the resonances have a complicated inner structure. Several coupling coefficients of various polarization-dependent behaviors are bounded together by the system of coupled mode equations. 
We note that by changing the polarization of incident light the dominant coupling coefficient can be changed, thus the energy can be predominately coupled to a particular mode with a specific field distribution at the sensor surface.

\Fig{Polarization_2Peaks_4}{0.95}{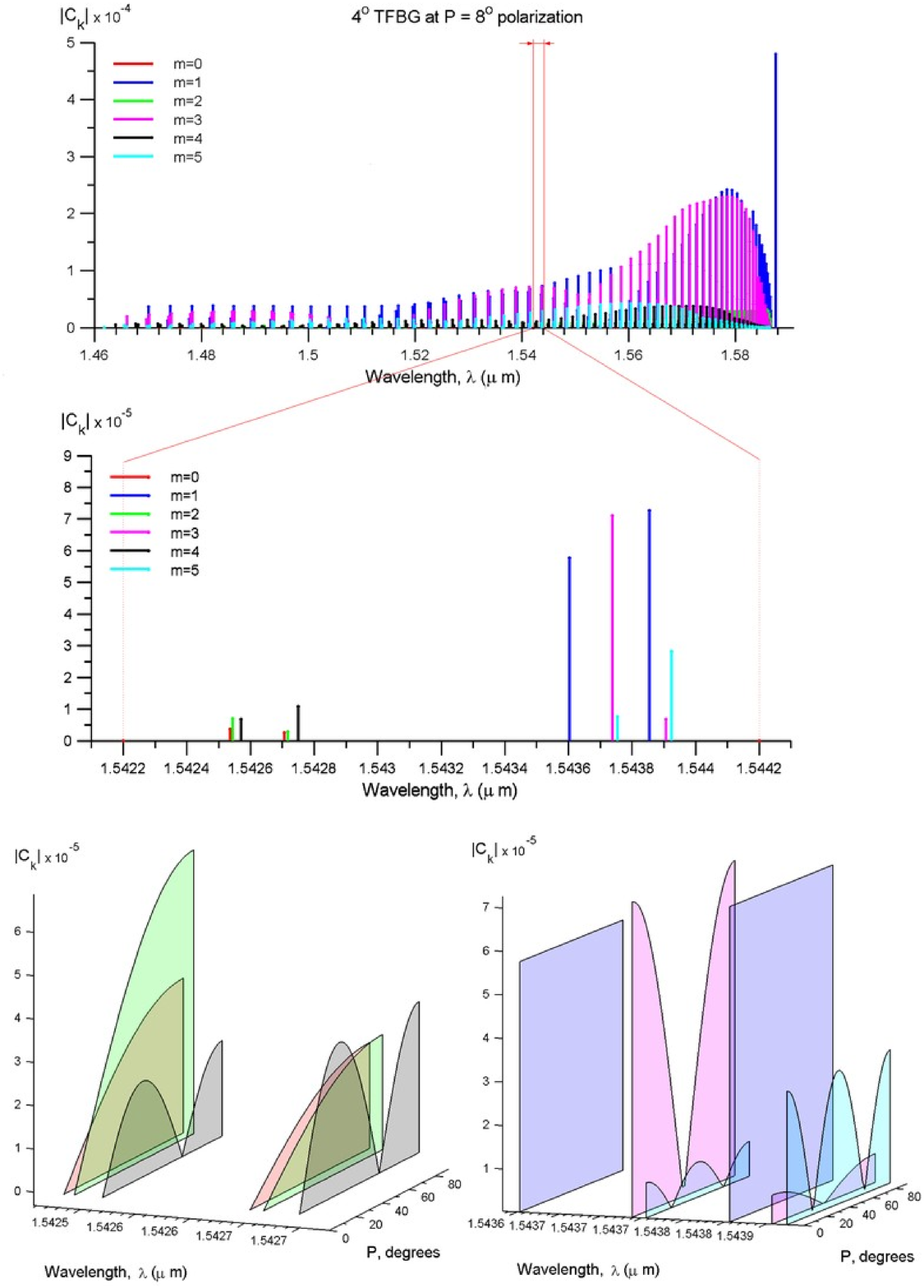}
{The polarization dependence of coupling coefficients corresponding to two resonances of the $4^o$ degree TFBG. }


\Fig{Polarization_2Peaks_10}{0.95}{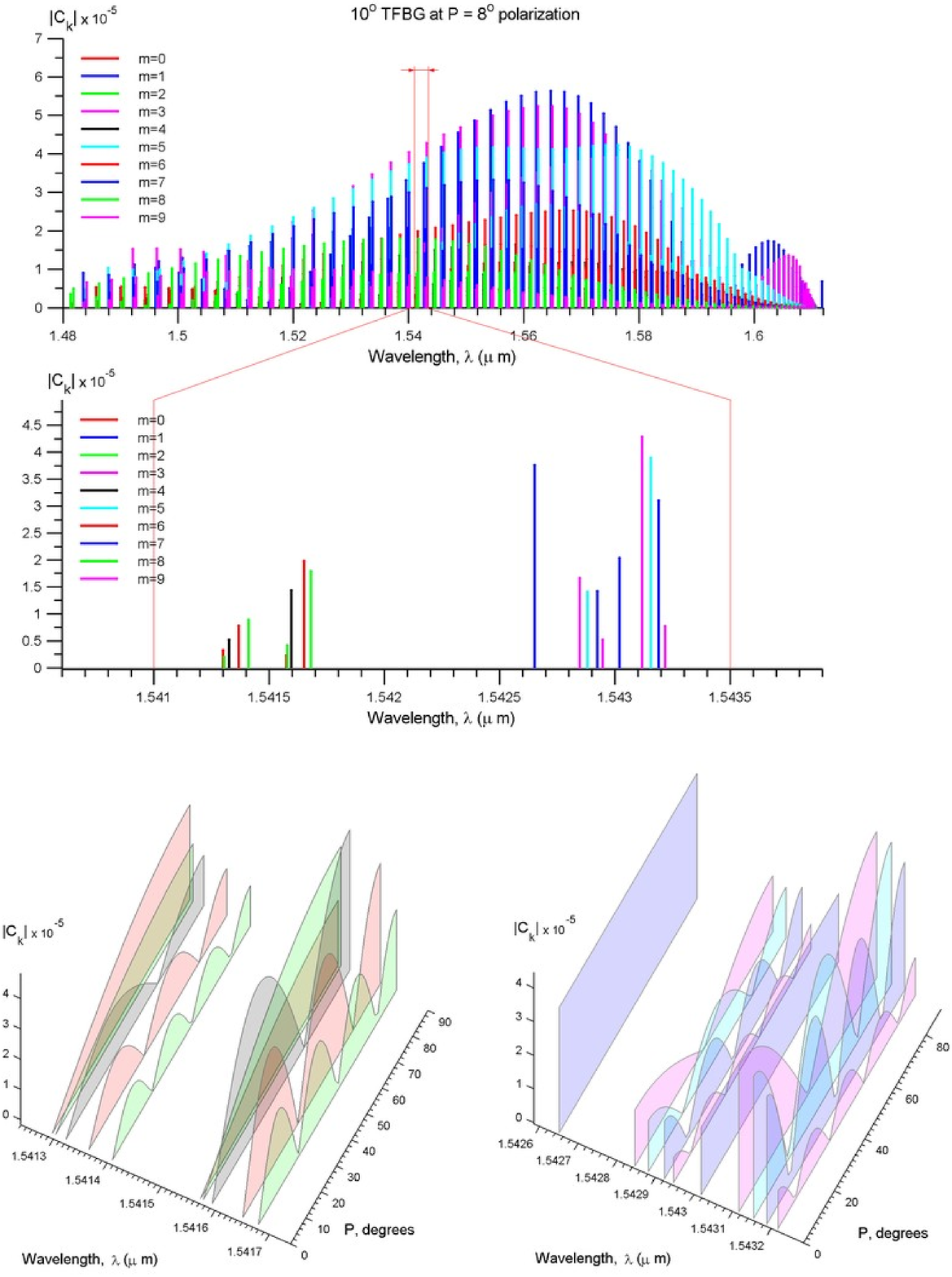}
{The polarization dependence of coupling coefficients corresponding to two resonances of the $10^o$ degree TFBG. }


\clearpage

Let us now assemble the density plot by computing the TFBG spectra at each angle of linear polarized light (the angle of rotation about the optical axis).
First we compute the coupling coefficients for cladding modes of ${m=0,1,2,3,4,5}$ azimuthal symmetry (as was previously shown the six first families of cladding modes are sufficient to accurately compute the spectrum of $4^o$~degree TFBG).


\Fig{Polarization_All_C15_sep}{1}{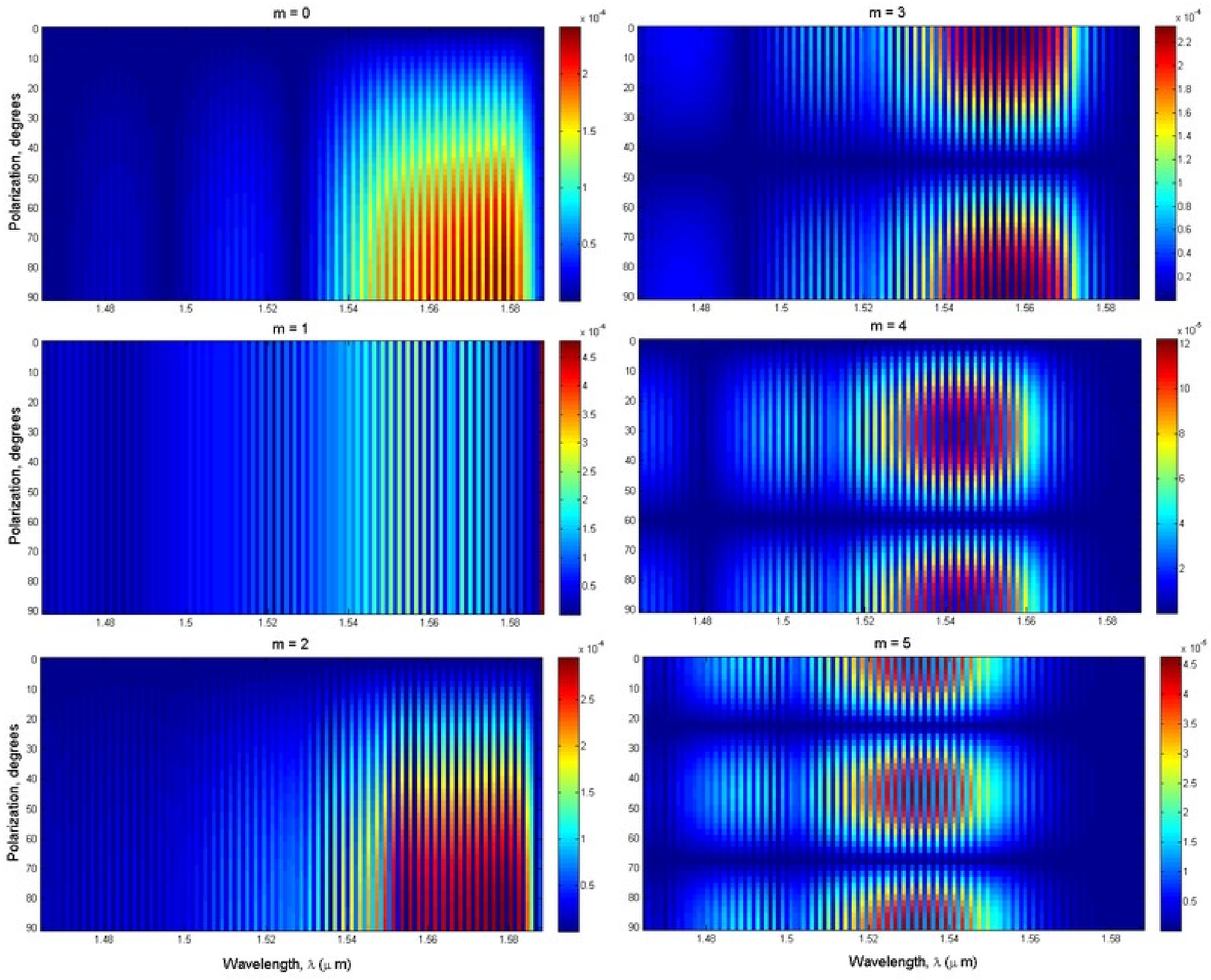}
{The coupling coefficients between the core and cladding modes of azimuthal order~${m=0,1,2,3,4,5}$ computed at various angles of linearly polarized light incident at the $4^o$~degree TFBG.}

The coupling coefficients of various $m$ azimuthal numbers plotted on the same figure are shown in Figure~\ref{Polarization_All_C15}.  

\Fig{Polarization_All_C15}{0.9}{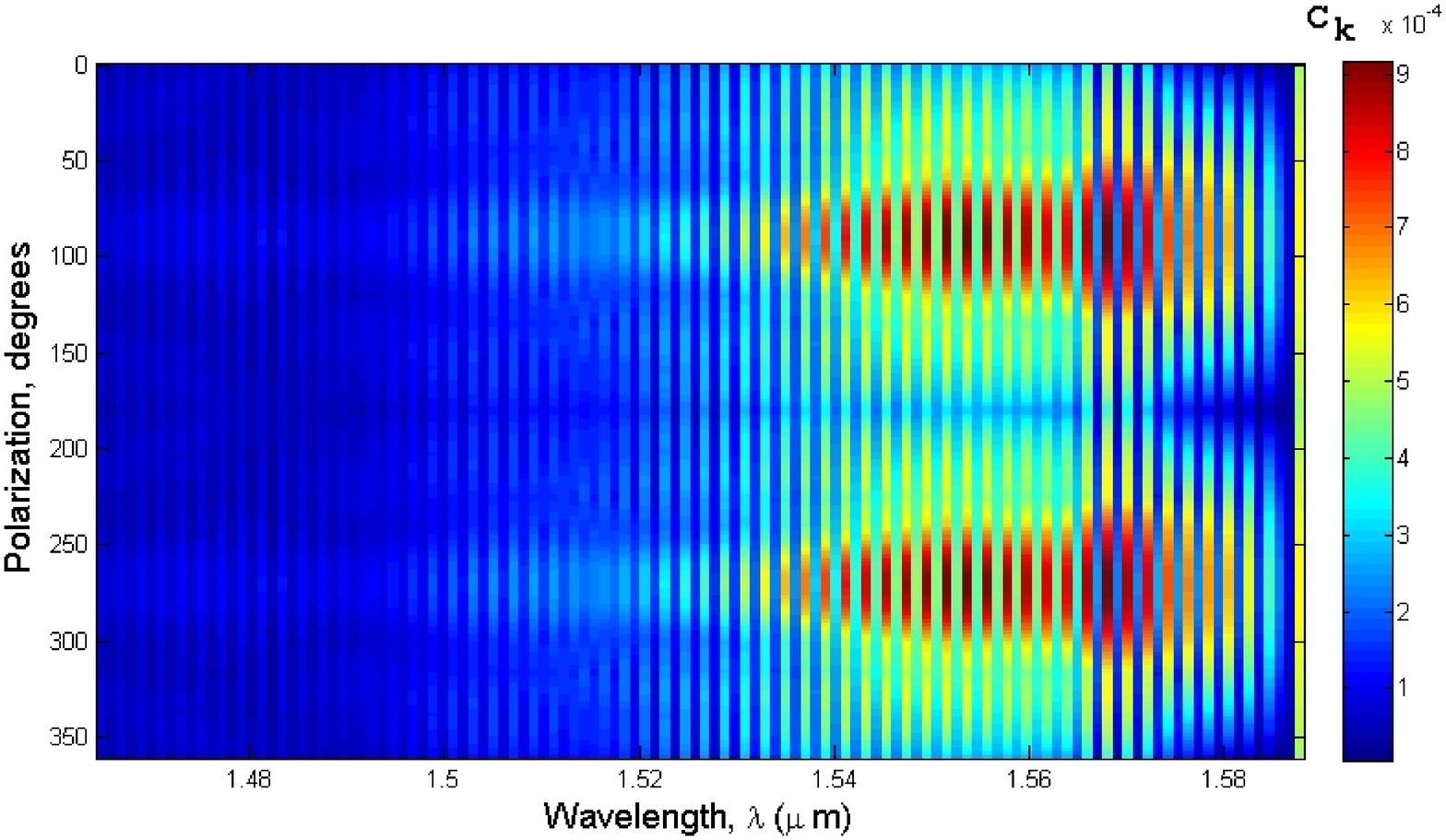}
{The coupling coefficients of the $4^o$~degree TFBG computed at various angles of linearly polarized light.}

Applying the developed in the previous chapter method we compute the corresponding transmission spectra for each angle of linearly polarized light incident at the TFBG structure. The result of the computation is shown in Figure~\ref{Polarization_All_TFBG_4}, the experimental measured spectra at various polarization angles are shown~Figure~\ref{Fig1}.

\Fig{Polarization_All_TFBG_4}{0.9}{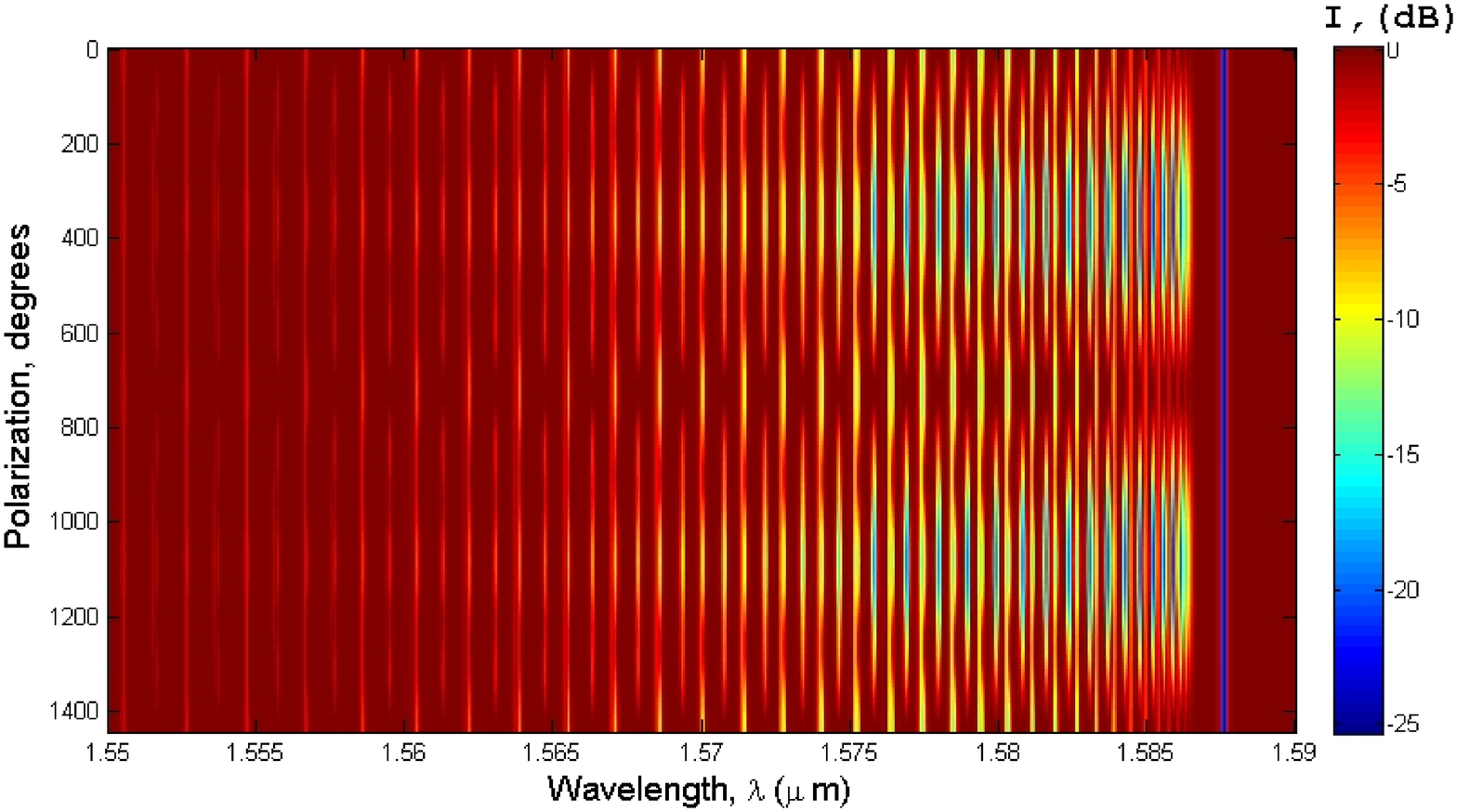}
{The transmission spectra of the $4^o$~degree TFBG, computed at various angles of linearly polarized light.}

\clearpage
\section{The electric field distribution at the fibre boundary}
In the following Chapters we discuss the possible ways of improving the TFBG sensor sensitivity by depositing nanoparticles on the sensor surface. The light particle interaction is dependent, among other parameters, on the incident electric field orientation.
In this section we discuss the electric field orientation at the fibre surface.

The developed vectorial mode solver allows us to determine $E_\rho$ and $E_\phi$ field components for each of the modes. 
In Figure~\ref{E_boundary_1} and Figure~\ref{E_boundary_2} the field components $E_\rho$ and $E_\phi$ are plotted along the effective refractive index of the modes with different azimuthal symmetries~${m=0,1,2,3,4,5}$. We note that by default the field of each mode is normalized to unity.

\Fig{E_boundary_1}{0.75}{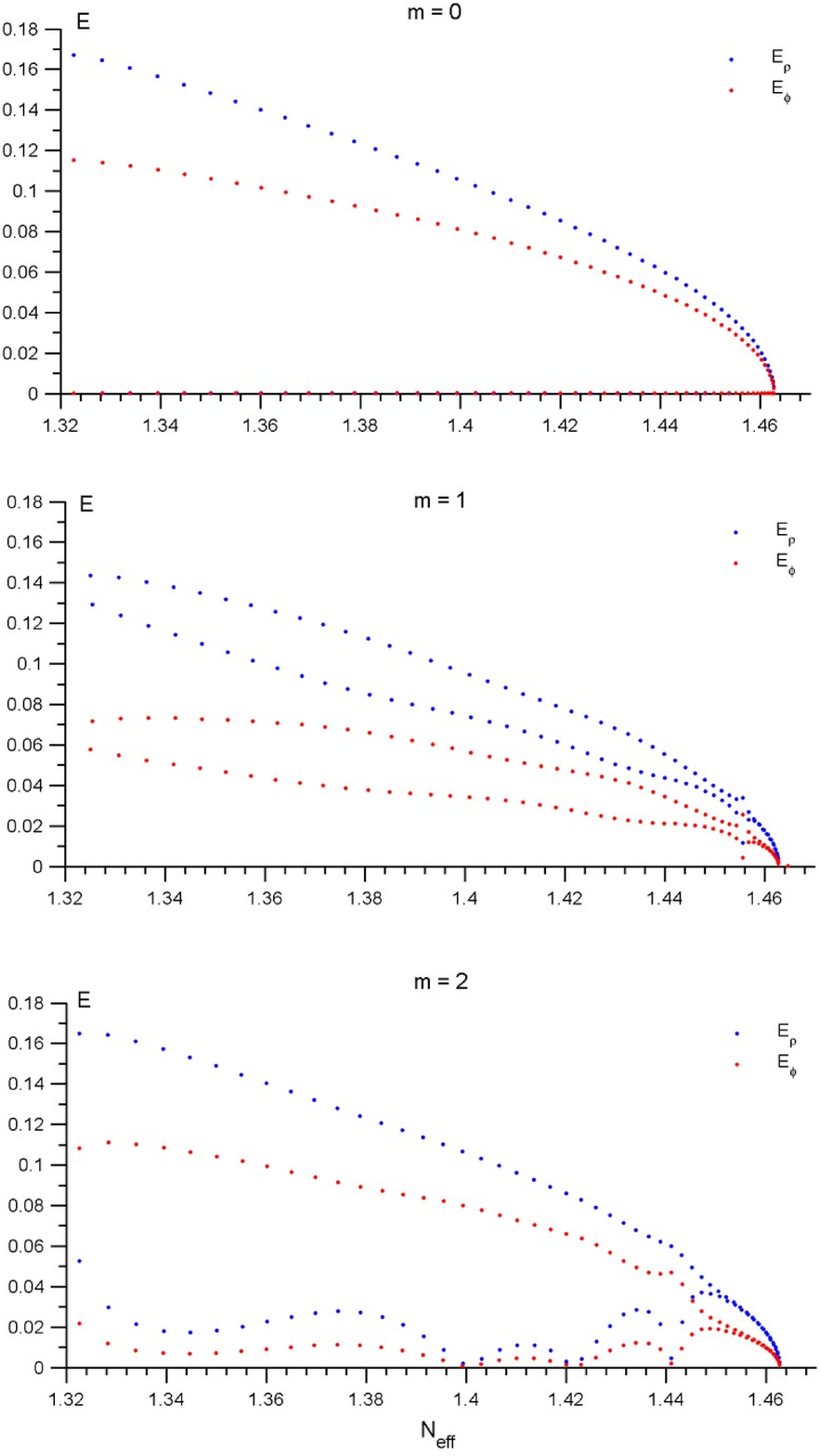}
{The field components $E_\rho$ and $E_\phi$ \C{computed} at the fibre boundary for modes with~${m=0,1,2}$ azimuthal symmetry.}

\Fig{E_boundary_2}{0.75}{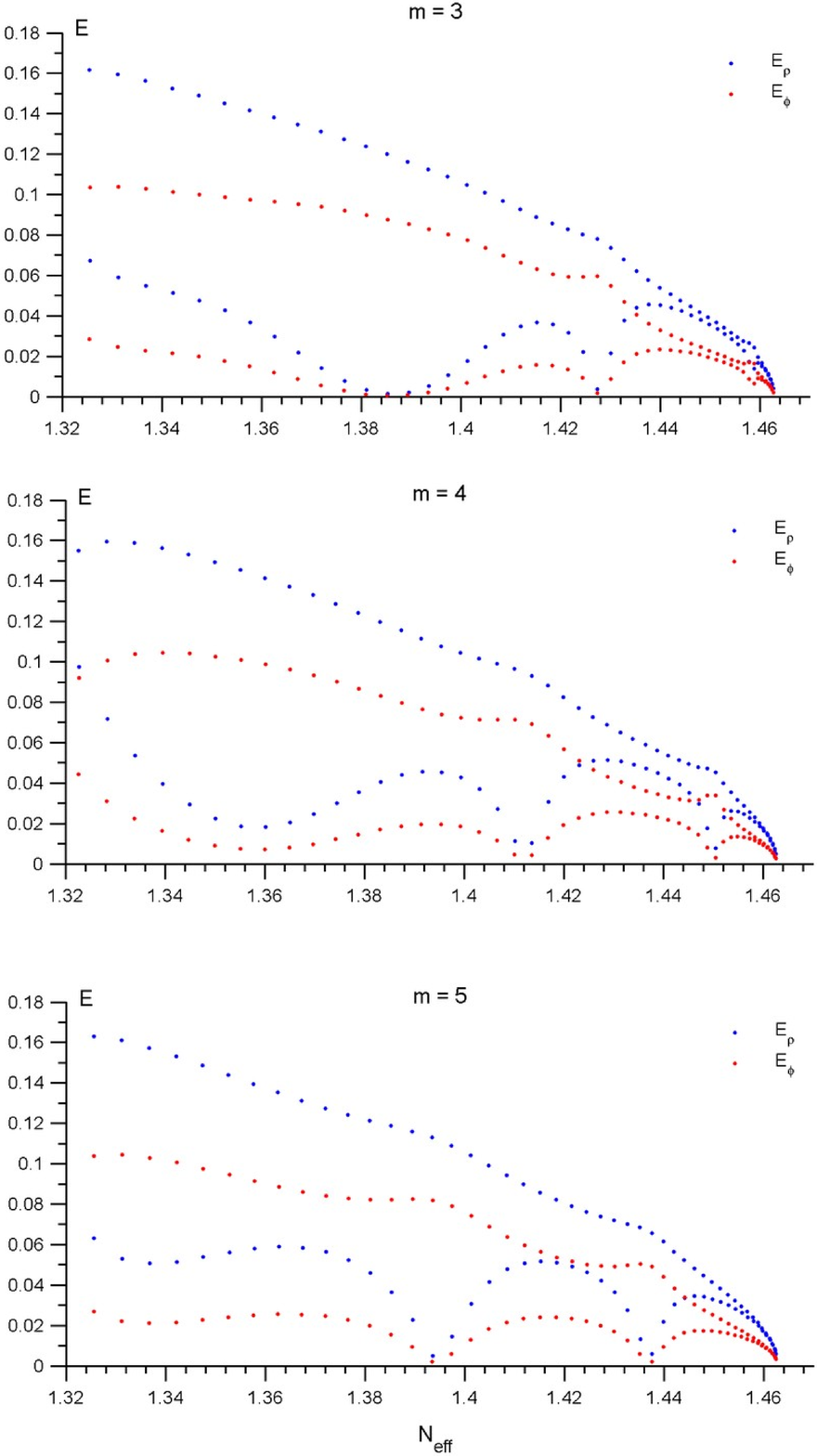}
{The field components $E_\rho$ and $E_\phi$ at the fibre boundary for modes with~${m=3,4,5}$ azimuthal symmetry.}

As can be seen from Figure~\ref{E_boundary_1} and Figure~\ref{E_boundary_2} the modes with a high effective refractive index are almost completely bounded inside the waveguide, with almost zero field at the fibre surface, whereas fields with low effective refractive index modes leak outside the waveguide.  
This effect can be seen in more detail if the field is computed at some distance from the fibre. In Figure~\ref{E_boundary_1nm} the field components $E_\rho$ and $E_\phi$ are computed at the distance of $1~nm$ away from the fibre boundary. We note that these modes with low effective refractive index have long exponential tails penetrating deep into the surrounding medium, these are the so-called evanescent waves. 
The long exponential tail of the modes is the reason for high sensitivity of the sensor in the region close to the cutoff.

\Fig{E_boundary_1nm}{0.8}{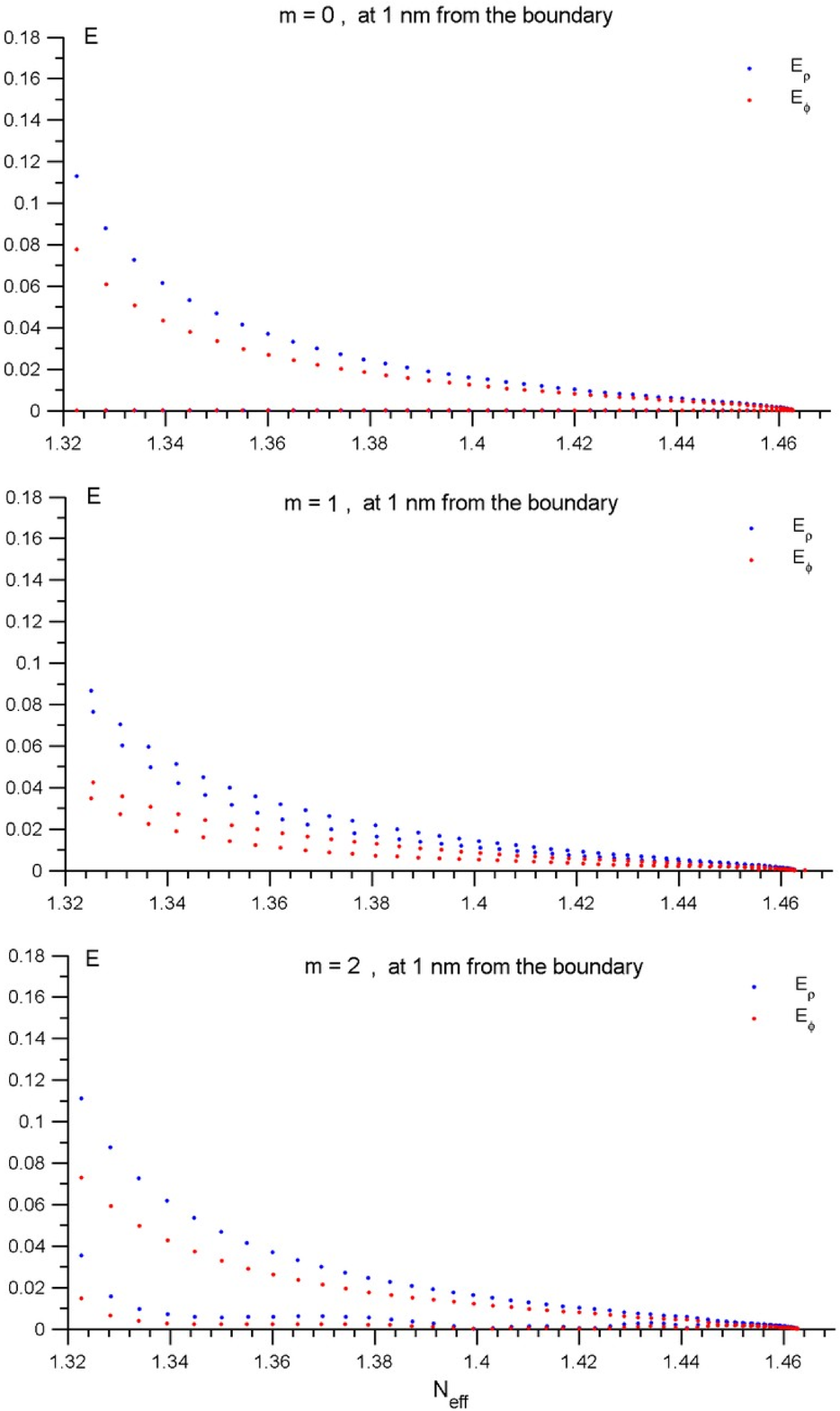}
{The field components $E_\rho$ and $E_\phi$ computed at $1~nm$ distance away from the fibre boundary.}

The Figures~\ref{E_boundary_1},~\ref{E_boundary_2} show components of the electric field normalized to unity. 
However, the field should depend on the energy coupled to a particular mode, as each mode is excited with different strength. The energy (or light) coupled to a particular $k$-th mode is proportional to the corresponding coupling coefficient $C_k$. Thus the resulting field at the sensor surface can be obtained by multiplying the electric field components~$E_\rho$ and $E_\phi$  of the $k$-th mode by the  $C_k$ coupling coefficient. The result is shown in Figure~\ref{E_boundary_All} for various states of linearly polarized light incident on the $4^o$ degree TFBG structure.

\Fig{E_boundary_All}{0.83}{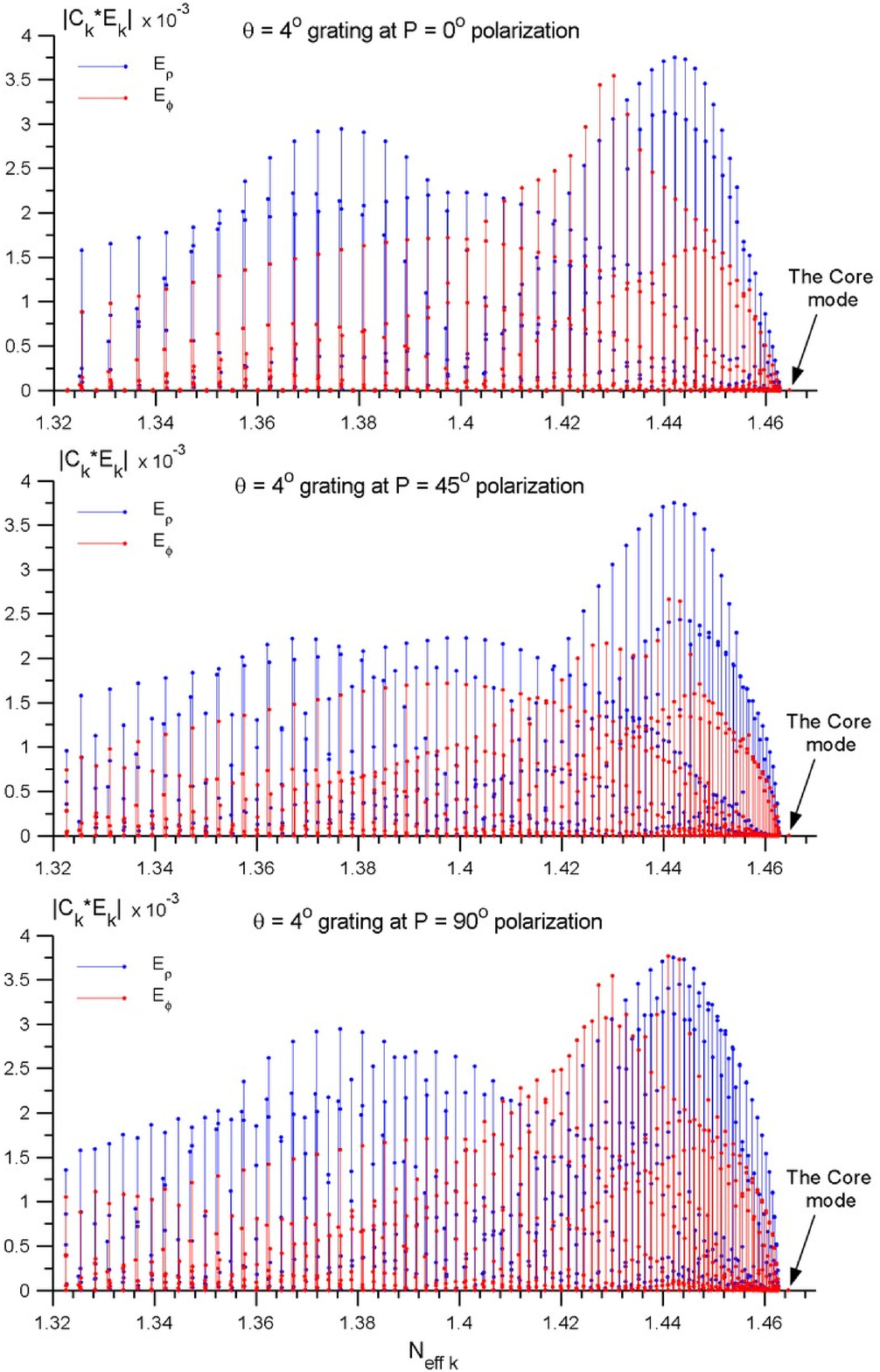}
{The intensity of electric field components~$E_\rho$ and~$E_\phi$ at the fibre boundary for $4^o$ degree TFBG, computed at various states of linearly polarized light ${P = 0^o, 45^o, 90^o}$.}
\clearpage

Let us zoom in on a few particular resonances. 
As mentioned previously each resonance consists of modes of either odd or even azimuthal symmetry, as shown in Figure~\ref{E_boundary_C}. 

\Fig{E_boundary_C}{0.8}{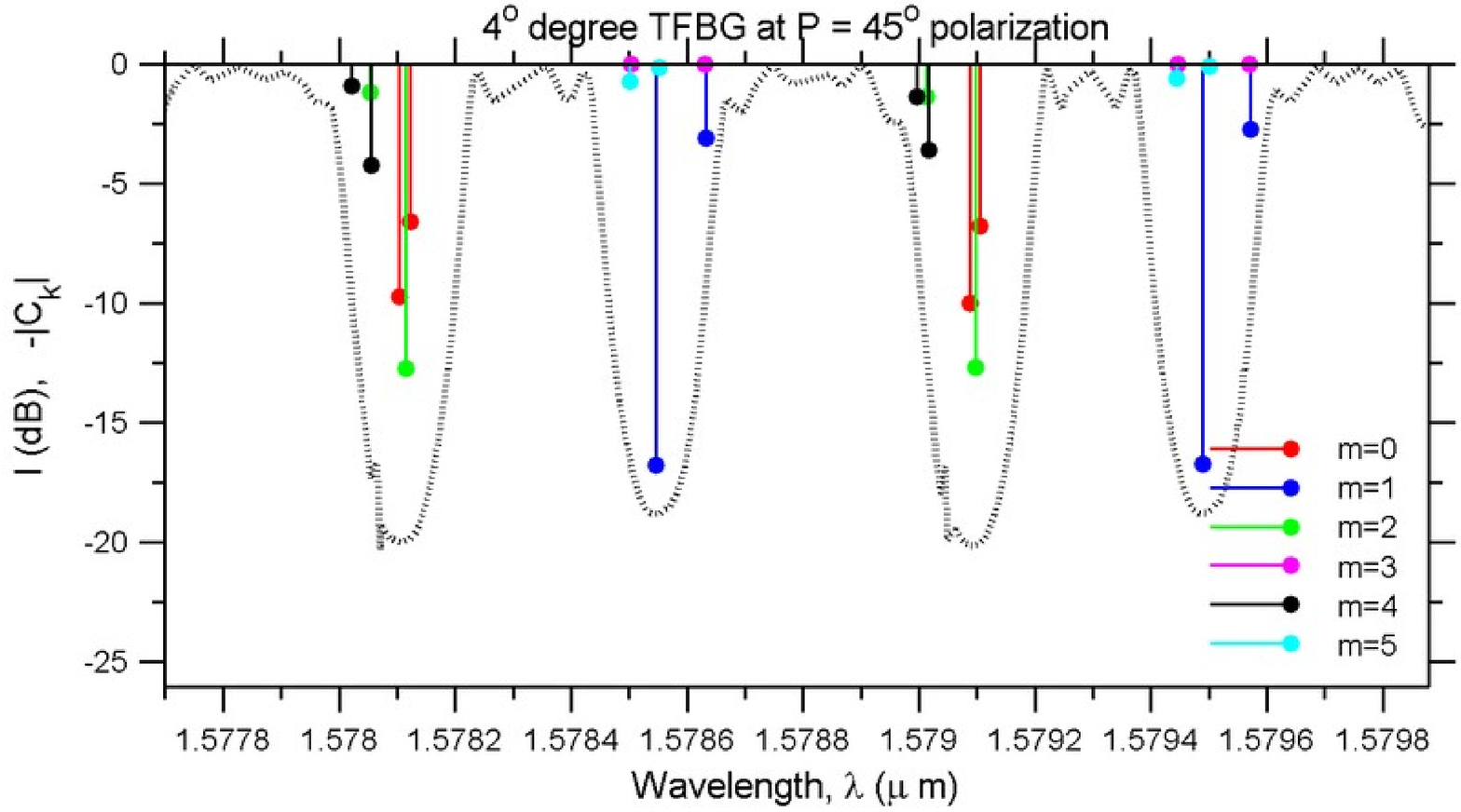}
{The structure of the particular resonances of $4^o$ degree TFBG.}

Multiplying the electric field components $E_\rho$ and $E_\phi$ at the fibre surface by the corresponding coupling coefficients we get the value of electric field present at the fibre surface, shown in Figure~\ref{E_boundary_zoom}.

\Fig{E_boundary_zoom}{0.8}{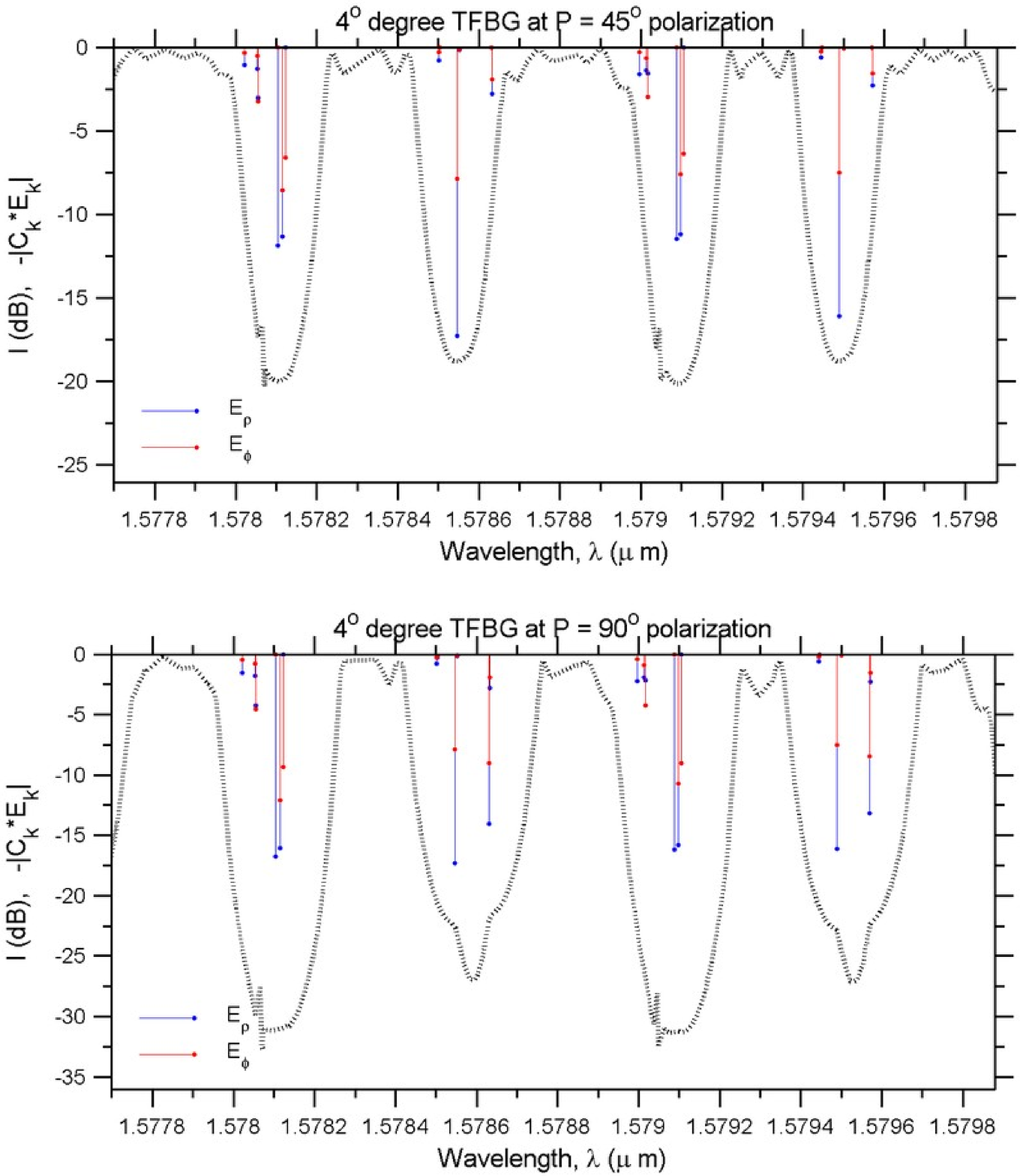}
{The electric field at the fibre surface, corresponding to particular resonances. The linear polorizes light incident at $P=45^o$ and $P=90^o$ angles at the $4^o$ TFBG.}

We conclude by stating that the electric field at the fibre surface is predominantly radially polarized (with electric field component normal to the fibre surface) in the wide spectral region, as can be inferred from Figure~\ref{E_boundary_All}. 
However, the composition of resonances is complex (Figure~\ref{E_boundary_zoom}) and consists of many modes with radial and tangential dominant polarization. 
Moreover, the composition depends on polarization of the incident light, as shown in Figure~\ref{E_boundary_All}.
Thus, by changing the core mode polarization, the coupling coefficient can be changed, and hence the energy couples to a particular mode with a specific field distribution at the sensor surface can be changed as well. 
The spectral response of the TFBG sensor is polarization-dependent, as shown in Figure~\ref{Polarization_All_TFBG_4}. 
The electric field at the sensor surface is polarized, with orientation of $\V{E}$ dependent on the core mode polarization. 
Hence the spectral response of the TFBG sensor can be further enhanced by coating its surface with a layer, the optical properties of which are polarization-dependent.

In the following Chapter~\ref{Chap_polarization} we discuss experimental measurement techniques taking advantage of the polarization-dependent response of the TFBG sensor. 
\C{The observed} polarization-dependence of the TFBG spectrum and electric field at the surface of the sensor will lead us to the sensitivity enhancement technique proposed in the following chapters.  




%% file: Chap_Exp_Opt_Setup_Polarization.tex
\chapter{Experimental polarization-based optical sensing with application to TFBG sensors}
\chaptermark{Experimental polarization-based sensing with TFBG}
\label{Chap_polarization}

In this Chapter we proposing a polarization-based sensing method developed for TFBG sensors. 

Polarization-based sensing is crucial for various types of optical sensors, in particular for stress analysis, plasmon-mediated sensing, and sensing of anisotropic media or other forms of perturbations~\cite{Zhang:12, Cranch:06, Kotov:13, Wang:13}. 
In the particular case of waveguide-type sensors (including optical fibres), it is possible to investigate the devices optical properties with polarized light but in general this requires very careful alignment and control of the input polarization. 

As we described in the previous sections, TFBG sensors reveal strong polarization-dependent properties~\cite{Caucheteur:2008, Bialiayeu:2012} due to the tilt of the grating planes which breaks the cylindrical symmetry of the fibre and strongly impacts the magnitude of the coupling coefficients between the incident core mode and the cladding modes excited by the grating~\cite{Lee_00}, as is schematically shown in Figure~\ref{TFBG_polarization_intr}. 

The grating tilt causes asymmetry in coupling between the cladding modes and the core mode, with respect to the grating tilt. 
In particular, for a grating tilted by an angle $\theta$ in the reference $y-z$~plane and a linearly polarized input core mode, whose polarization is rotated by an arbitrary angle~$\phi$ with the reference $x$ axis, the coupling to individual cladding modes will depend strongly on both $\theta$ and $\phi$.

\Fig{Fig1}{0.8}
{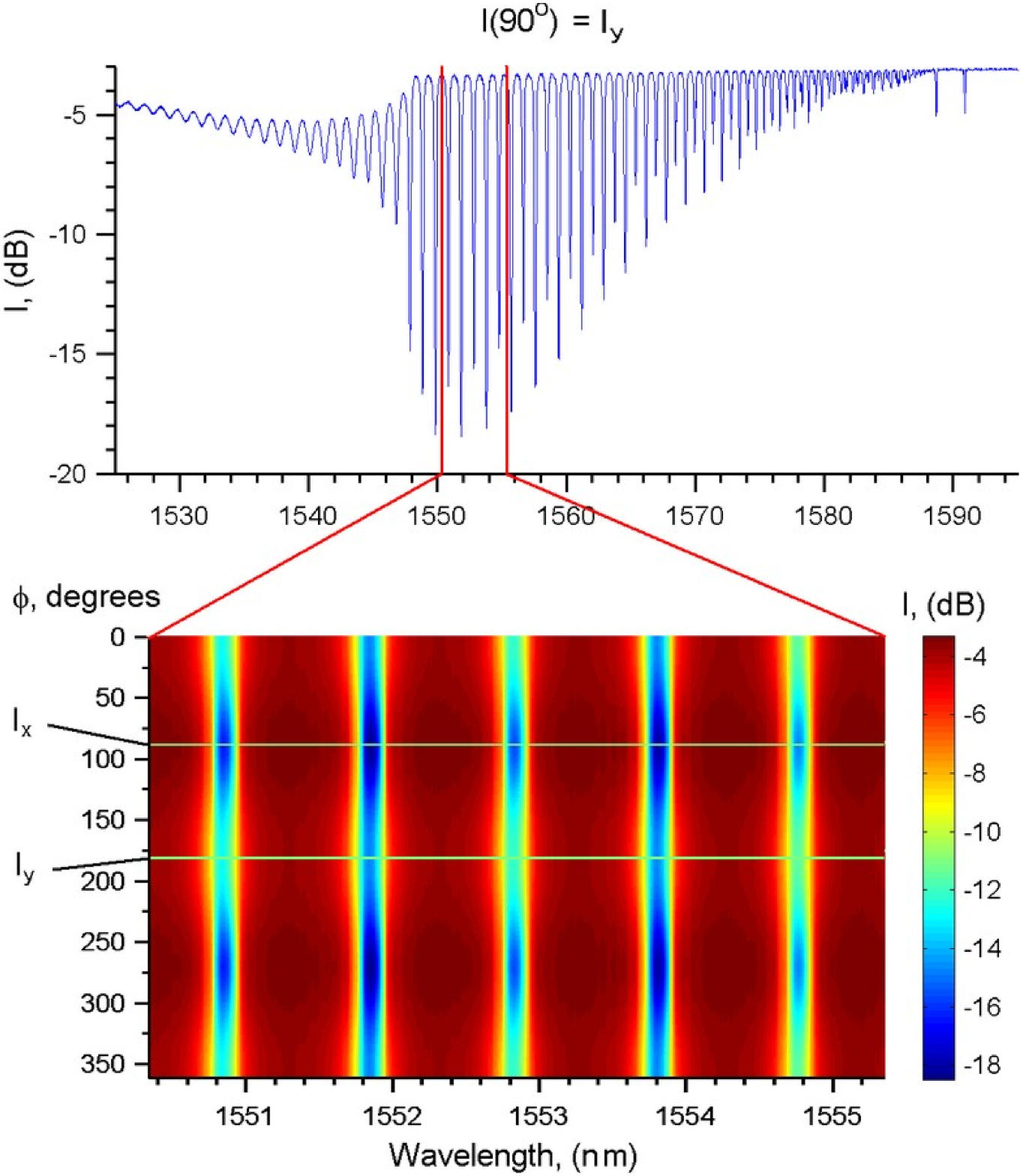}
{A typical TFBG transmission spectrum for linearly polarized light, and series of spectra obtained by rotating a linear polarizer about the optical axis are shown as the density plot.}

The asymmetry in sensing results from the different nature of interaction between the cladding TM-like and TE-like modes, with the predominant radial and tangential orientation of the electric field at the sensor surface, from one side, and the environment under test from another side. 
The asymmetry can be significantly enhanced if the sensor surface is coated with a thin metallic film\cite{Bialiayeu:2011} or metal nano rods~\cite{Bialiayeu:2012}, interacting differently with the radial and tangential oriented electric fields.

The polarization effects in optical fibres have been traditionally quantified with a polarization-dependent loss (PDL) parameter, provided by optical vector analyzer (OVA) instruments.
Alternatively, the polarization effects can be studied with regards to linearly polarized light aligned with a specific axis of the device under test~\cite{Bialiayeu:2011}. A typical series of spectra measured at various angles of linearly polarized light is shown if Figure~\ref{Fig1}.

The effect of the input mode polarization can be observed by measuring optical transmission spectra with a polarizer inserted between the light source and the grating~\cite{Bialiayeu:2011}. A typical series of spectra measured at various angles of linearly polarized incident light is shown in Figure~\ref{Fig1} for a $1~cm$ long TFBG immersed in water.

In the case of TFBG sensor modes with a different orientation of electric field at the sensor surface can be excited by rotating linear polarized light about the sensor axis~\cite{Lee_00}. To rotate the linearly polarized light an external polarizer is usually used, sometimes in combination with OVA instrument. 

In the presented work we compare the PDL based approach, which is based on measurements of transmission spectra along the orthogonal principal axes, with the measurements based on extracting linear states of polarization along a predefined axis either from the Jones matrix or the Stokes vector.

\section{The Optical Setup}
The measuring optical system was designed to acquire transmission spectra of tilted fibre Bragg sensor at various polarization states. 

Two alternative approaches were used. 
The first straightforward approach as shown in Figure~\ref{Exp_Set1}, is based on introduction of a linear polarizer in the optical path~\cite{Bialiayeu:2011}, thus allowing the collection of spectra at various states of linear polarization. 
The second approach was based on measuring the Stokes parameters or alternatively the Jones matrix elements, and will be discussed later.

\Fig{Exp_Set1}{0.8}
{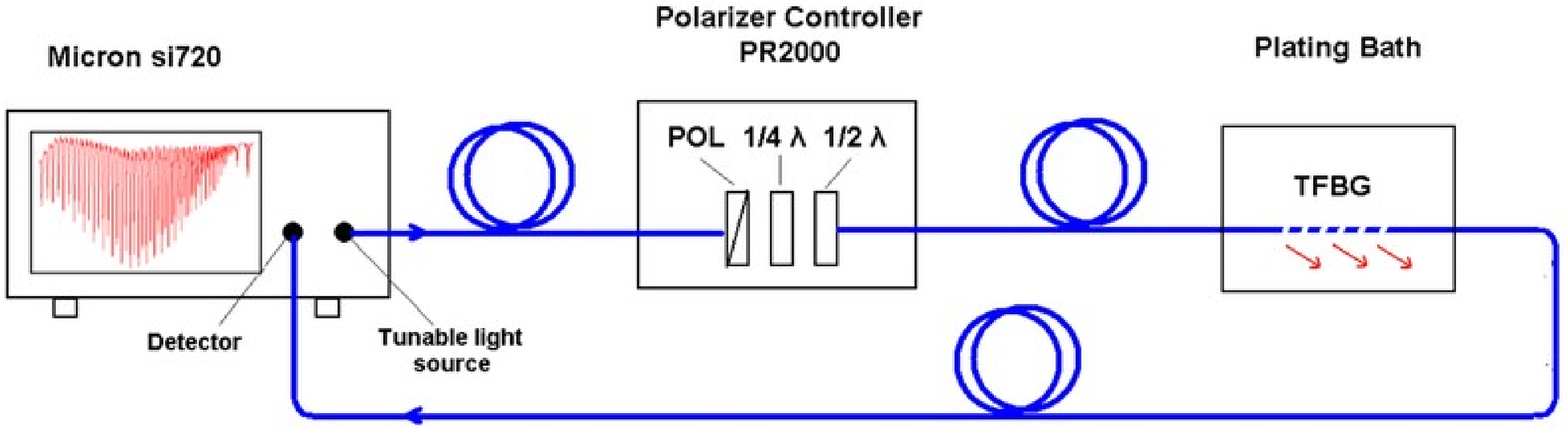}{The optical setup based on SI720 spectrophotometer and PR2000 polarization controller.}

The experimental setup is shown schematically in Figure~\ref{Exp_Set1} consists of a SI720 spectrophotometer (Micron Optics) and a polarization controller PR2000 (JDS Uniphase).
The SI720 spectrophotometer is based on the fast sweeping tunable laser and integrated photo detector. 
The polarization controller was set to continuously scan all linearly polarized light states in~$80$ seconds.

The spectra over the full operational range (from $1520$ to $1570~nm$) were taken continuously at a rate of one spectrum each $0.2$ seconds, thus the spectra were acquired with the resolution of less than $1^o$ degree of polarization angle. 
Such high resolution allowed us to collect a huge amount of data and, as a result, to precisely describe the spectral and polarization response of the TFBG-SPR sensor. 

An example of acquired spectra is shown in Figure~\ref{Fig1}. The spectra are represented in the form of a density plot. The spectra are stored in the matrix, with each row corresponding to a different angle of linearly polarized light.
Two orthogonal states of polarization $I_x$ and $I_y$ were extracted at the data processing stage, and correspond to the measurements along axes which the maximum and minimum transmission spectra.

Considering that light transmitted through a device under test is experiencing the insertion loss dependent on the state of incident light polarization, as a function of optical frequency, two alternative approaches to optical measurements are possible. The optical system can be fully characterized either by the Stokes parameters or the Jones matrix.
\Fig{Ex_Opt_Set_2}{0.7}{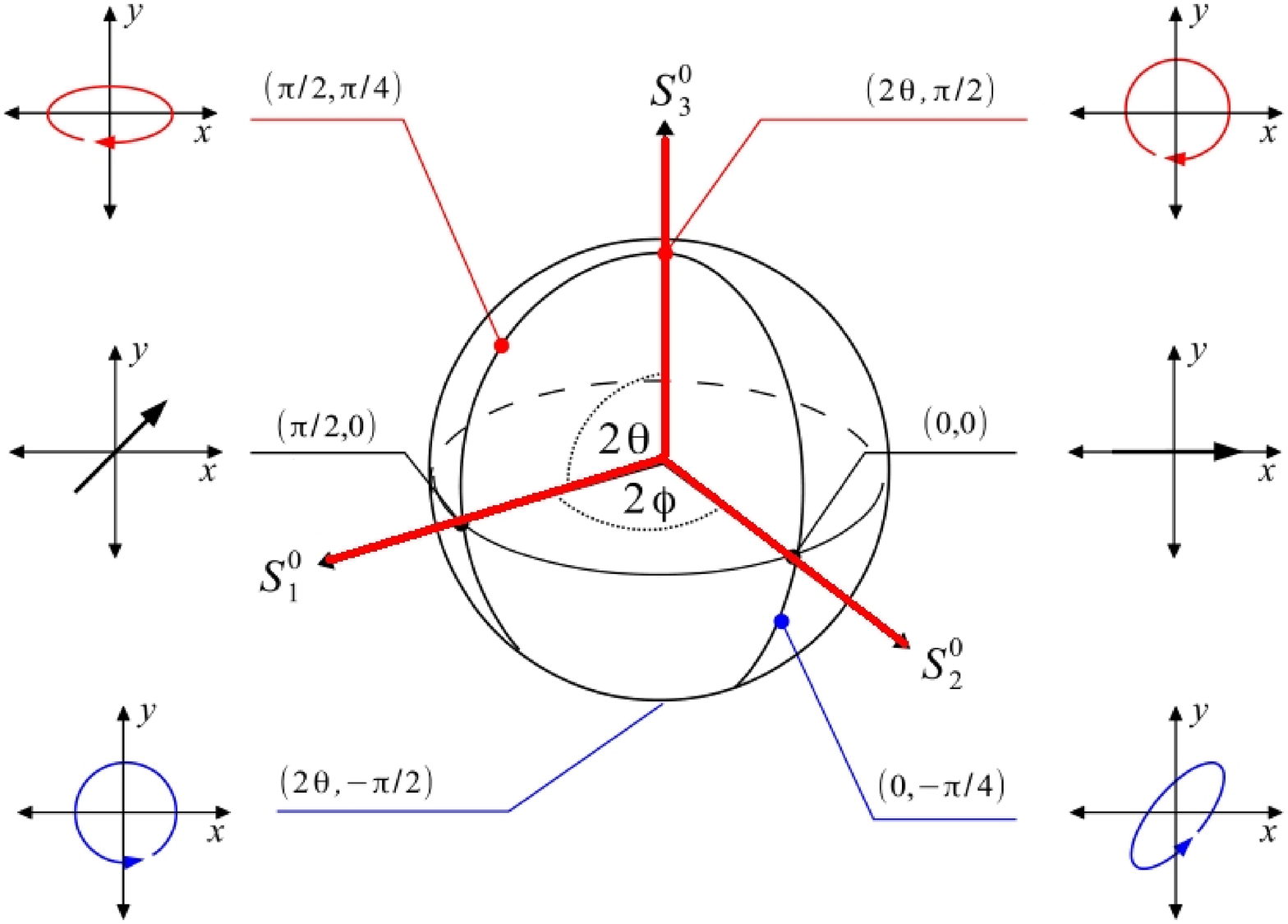}
{The Stokes vector representation.}

The Stokes parameters and Jones matrix were acquired with help of a JDS Uniphase SWS-OMNI-2 system and from an optical vector analyzer (OVA) 5000 from Luna Technologies, respectively. 
Both approaches are based on the same operational principle, except that the optical vector analyzer from Luna Technologies is also capable of measuring the phase delay in an optical device under test.

An optical vector analyzer usually consists of an optical source, a polarization controller capable of producing one of four known polarization states and a power meter measuring an insertion loss.
Only four measurements to determine Stokes parameters ${S_0, S_1, S_2, S_3}$ are required. 
The Stokes parameters can be visualized with help of a \C{Poincar\'{e}} sphere sphere as shown in Figure~\ref{Ex_Opt_Set_2}.
First, an optical signal is polarized to produce one of four known polarization states ${a,b,c}$ and $d$, described by the Stokes vectors ${S_{0,a},S_{0,b},S_{0,c}}$ and~$S_{0,d}$, and then transmitted through the device under test. The corresponding transmitted powers ${T_{0,a},T_{0,b},T_{0,c}}$ and~$T_{0,d}$ of the output signal are measured with the power meter~\cite{Favin:1994, Cyr:2008}.

The transmissivity of the optical component under test can be represented by a $4 \times 4$ Mueller matrix $[M]$:
\Eq{}
{\V S_{out} = [M] \V S_{in},}
where ${\V S_{in} = (S_0,S_1,S_2,S_3)}$ and ${\V S_{out} = (T_0,T_1,T_2,T_3)}$ are the input and the output states of polarization, respectively, represented by Stokes vectors. $T_0$ is the intensity of transmitted light, measured with the power meter. 
Hence, for each state of polarization ${a,b,c,d}$ the four transmitted intensities~${T_{0,a},T_{0,b},T_{0,c},T_{0,d}}$ can be measured, and four equations can be written by multiplying the first row of the Mueller matrix by the input Stokes vector: 

\Eq{}
{\begin{array}{ccc}
T_{o,a} &=& m_{00} S_{0,a} + m_{01} S_{1,a} + m_{02} s_{2,a} + m_{0k} S_{3,a}\\
T_{o,b} &=& m_{00} S_{0,b} + m_{01} S_{1,b} + m_{02} s_{2,b} + m_{0k} S_{3,b}\\
T_{o,c} &=& m_{00} S_{0,c} + m_{01} S_{1,c} + m_{02} s_{2,c} + m_{0k} S_{3,c}\\
T_{o,d} &=& m_{00} S_{0,d} + m_{01} S_{1,d} + m_{02} s_{2,d} + m_{0k} S_{3,d} 
\end{array}}
Solving the above system of equations the four matrix elements $m_{00}$, $m_{01}$, $m_{02}$ and $m_{03}$
 can be found, in terms of which other parameters, such as polarization-dependent loss~(PDL), can be expressed~\cite{Favin:1994}.

Unfortunately the described technique does not allow us to gather information about a phase delay in an optical system. The complete information about an optical device can only be obtained if the instrument used can measure phase as a function of frequency and polarization, thus providing the complete response of a system.
One of the possible realizations of such a device is schematically shown in Figure~\ref{Ex_Opt_Set}~\cite{Cyr:2008}.
\Fig{Ex_Opt_Set}{0.9}
{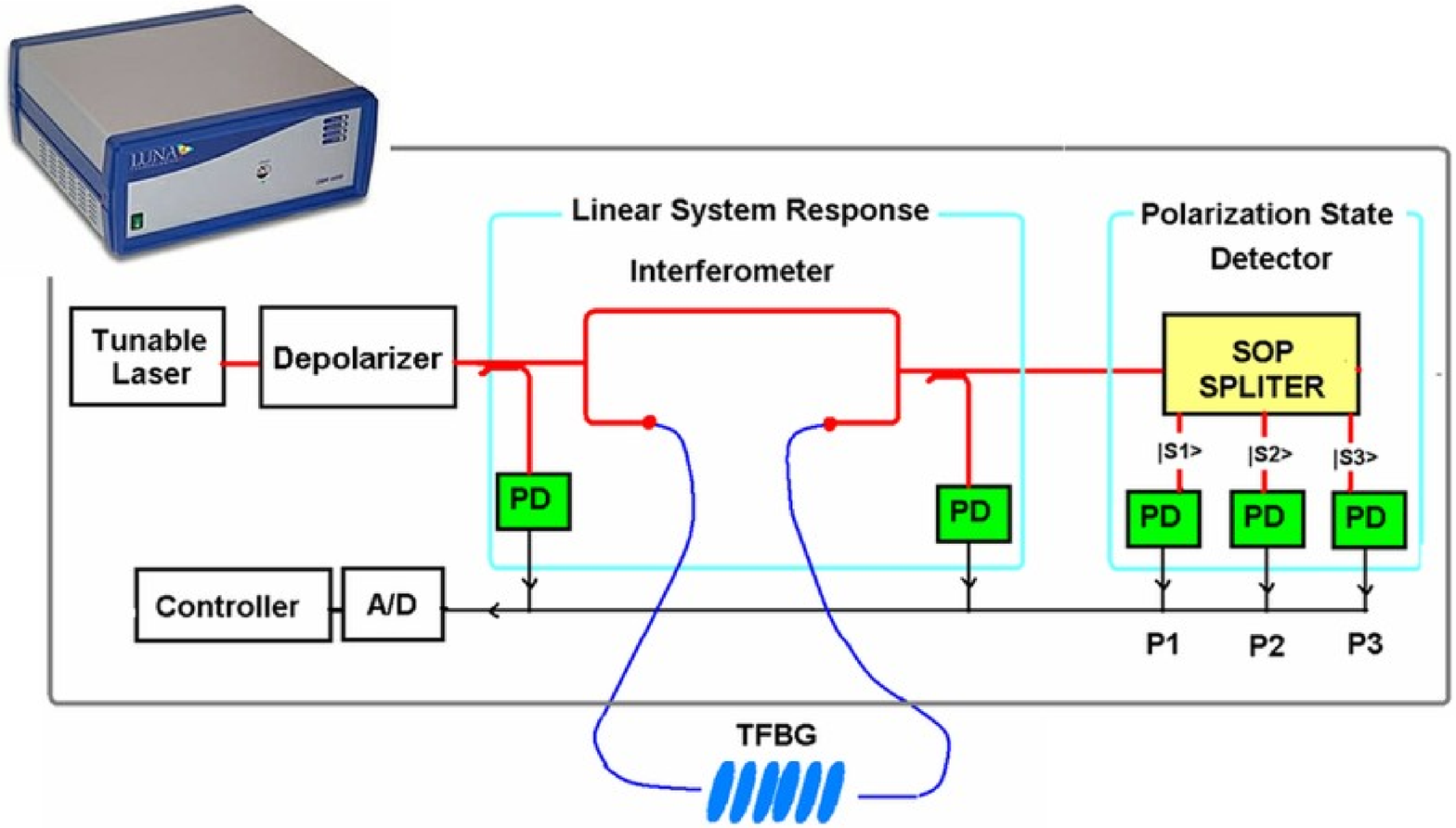}{The principle of operation of an optical vector analyzer .}

The principle of operation is based on the so-called swept-homodyne interferometry~\cite{Cyr:2008} performed for various states of polarization and fully characterizes an optical system by measuring the attenuation and phase delay as functions of optical frequency. 
As shown in Figure~\ref{Ex_Opt_Set} the device consists of an unpolarized light source, supplying unpolarized coherent light whose optical frequency is swept continuously as a function of time, which is first passed through the interferometer then through a three-way polarization splitter coupled to individual detectors. 
The detection polarization splitter is designed to measure the light intensities ${P_1, P_2, P_3}$ by projecting the light polarization state onto the three linearly independent preselected Stokes axes of polarization ${\V S_1, \V S_2, \V S_3}$, shown in Figure~\ref{Ex_Opt_Set_2}, as a function of optical frequency. 
The fourth measured $P_4$ is a total polarization-independent power. 
The optical delay is measured with a Michelson interferometer, where a device under test is coupled in one arm of the interferometer and the other arm of the interferometer is used for the reference.
As a result a $2 \times 2$ Jones matrix $[J]$ is computed at each optical frequency:
\Eq{}
{\V E_{out} = [J] \V E_{in}.}
The four elements of the Jones matrix are in general complex numbers, encoding the attenuation and phase delay of an optical system.

\section{The data processing technique}

The optical setup based on OVA implementation is shown in Figure~\ref{Exp_Set2}. In spite of the apparent simplicity, the data analysis is significantly more complicated than in the case where the state of polarization is rotated mechanically.
\Fig{Exp_Set2}{0.8}{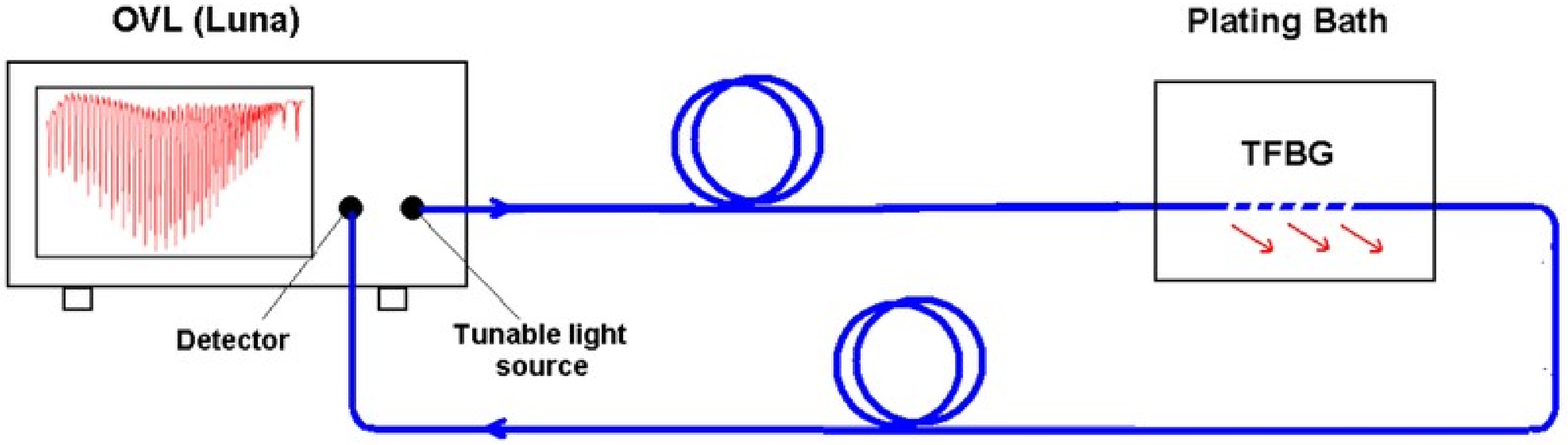}
{The Experimental Setup.}

In this section we describe the data processing technique which allowed us to extract polarization-based parameters, characterizing the sensor response, from the data provided by an optical vector analyzer.

\subsection{Measurements along principal axes of an optical system}
In this section we review polarization-dependent loss (PDL) technique and provide an approach to study the system transmission spectrum along its principal axes. 
 
A linear optical system can be represented in terms of the ${2 \times 2}$ Jones matrix $[J]$ connecting an incident and transmuted electric field vectors~\cite{JONES:41}: 
\begin{equation}
 \vec{E}_{out} = [J] \vec{E}_{in}
\end{equation}

The transmission spectrum of a device under test usually is characterized with two major parameters. 
The polarization independent parameter called the insertion loss ${I(\lambda)}$ and the polarization-dependent loss (PDL) parameter, both measured as a function of optical frequency $\lambda$.
To extract these parameters from the Jones matrix, a hermitian matrix $[H]$ has to be constructed first.
\begin{equation}
[H] = [J]^\dagger [J]
\end{equation}
Next, performing eigen decomposition of $[H]$ matrix
\begin{equation}
[H] = [U] [\Lambda] [U]^T
\end{equation}
we find diagonal matrix $[\Lambda]$ containing eigenvalues $\rho_1$ and $\rho_2$ of $[H]$ matrix.

Alternative singular value decomposition (SVD) of the Jones matrix can be computed~\cite{Heffner:1992}:
\begin{equation}
[J] = [U] [\Sigma][V]^T
\end{equation}
with the diagonal matrix $[\Sigma]$ containing two singular values $\sigma_1$ and $\sigma_2$ such that ${\rho_1 = \sigma_1^2}$ and ${\rho_2 = \sigma_2^2}$, due to the fact that ${[H] = [J]^\dagger [J]}$.

The matrix ${[U] = (\vec{u}_1, \vec{u_2})}$ contains two orthogonal eigenvectors (due to the properties of SVD decomposition), corresponding to $\rho_1$ and $\rho_2$ eigenvalues, align with principal axes of the system. 
The principal axes $\vec{u}_1$, $\vec{u_2}$ of TFBG sensor are showed schematically in Figure~\ref{Fig2}.

\Fig{Fig2}{0.6}{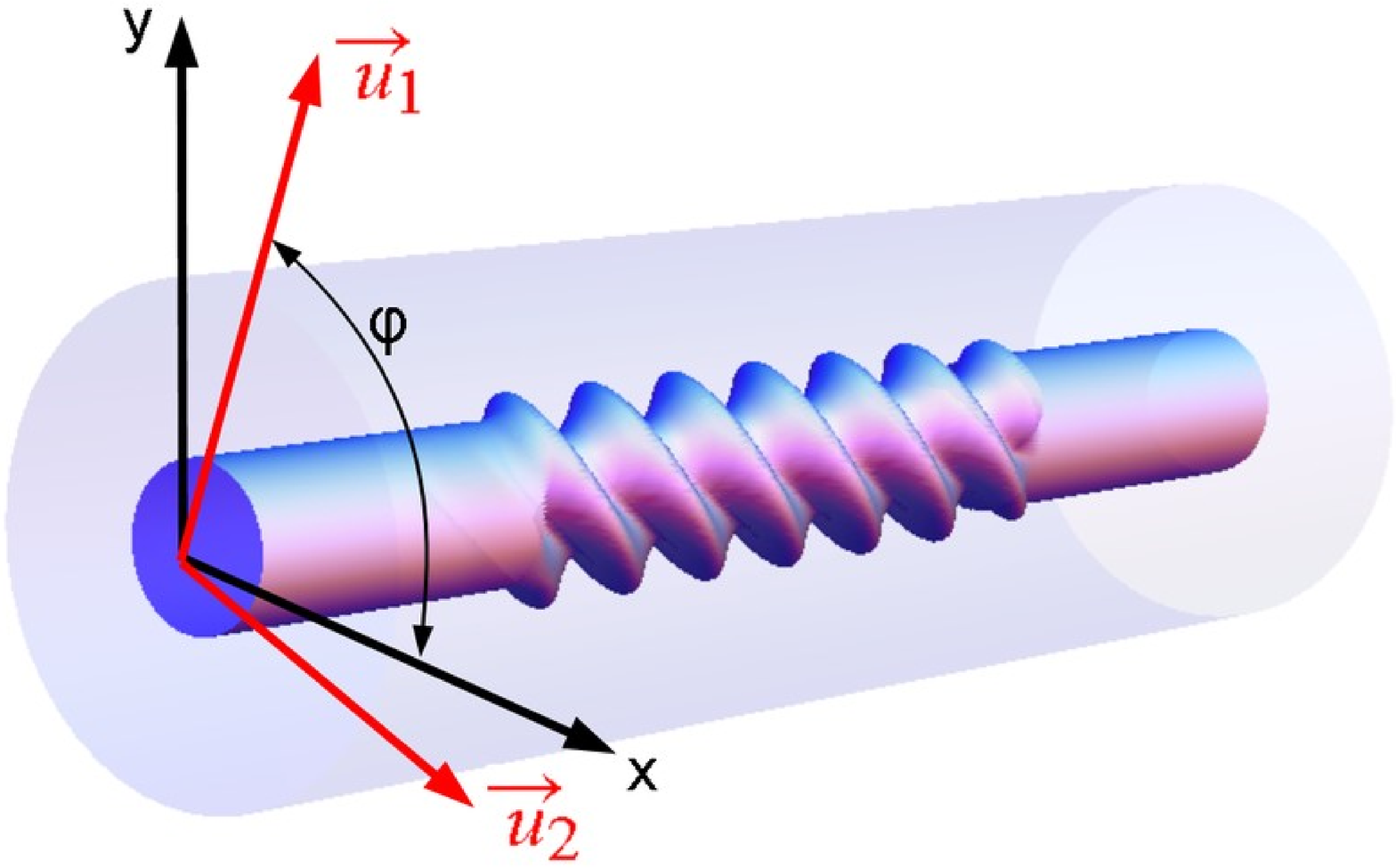}
{TFBG with physical $\hat{x}$, $\hat{y}$ axes and principal system axes $\vec{u}_1$ and $\vec{u}_2$ measured at some optical frequency $\lambda$.}

Having the singular values of the Jones matrix $[J]$ or eigenvalues of ${[H] = [J]^\dagger [J]}$ matrix the transmission loss can be computed as follows~\cite{Heffner:1992}.
\begin{eqnarray}
I_{out} & =& \langle \vec{E}_{out}|\vec{E}_{out} \rangle = \langle \vec{E}_{in}|[J]^\dagger [J]| \vec{E}_{in} \rangle = \nonumber \\
& =& \langle \vec{E'}_{in}|\Lambda| \vec{E'}_{in} \rangle = \frac{\rho_1 + \rho_2}{2} I_{in}
\end{eqnarray}
Which can be re-formulated in the more frequently used dB scale as:
\begin{equation}
\frac{I_{out}}{I_{in}} = 10 \log_{10}{\frac{\rho_1 + \rho_2}{2}}.
\end{equation}
Similarly, the polarization-dependent loss, which is the magnitude of the difference between the maximum and minimum device transmission over all possible input polarization actually corresponds to the difference in the loss measured along the system principal axes $\vec{u}_1$ and $\vec{u}_2$. It is defined as follows: 
\begin{equation}
PDL = 10 | \log_{10}{\frac{\rho_1}{\rho_2}|}
\end{equation}

Alternatively we can introduce a parameter called degree of polarization~\cite{Bialiayeu:2012}:
\Eq{}
{P = \frac{\rho_1 - \rho_2}{\rho_1 + \rho_2}}

The eigenvalues $\rho_k =\sigma_k^2$ can be interpreted as an observable transmission loss for the case when incident to an optical system electric vector $\vec{E}$ is align with the system principal axes,~\textit{i.e.} $\vec{E} \parallel \vec{u}_k$. 
The PDL parameter accounts for the difference in the loss measured along the system principal axes $\vec{u}_1$ and $\vec{u}_2$.

\Fig{Fig3}{1}{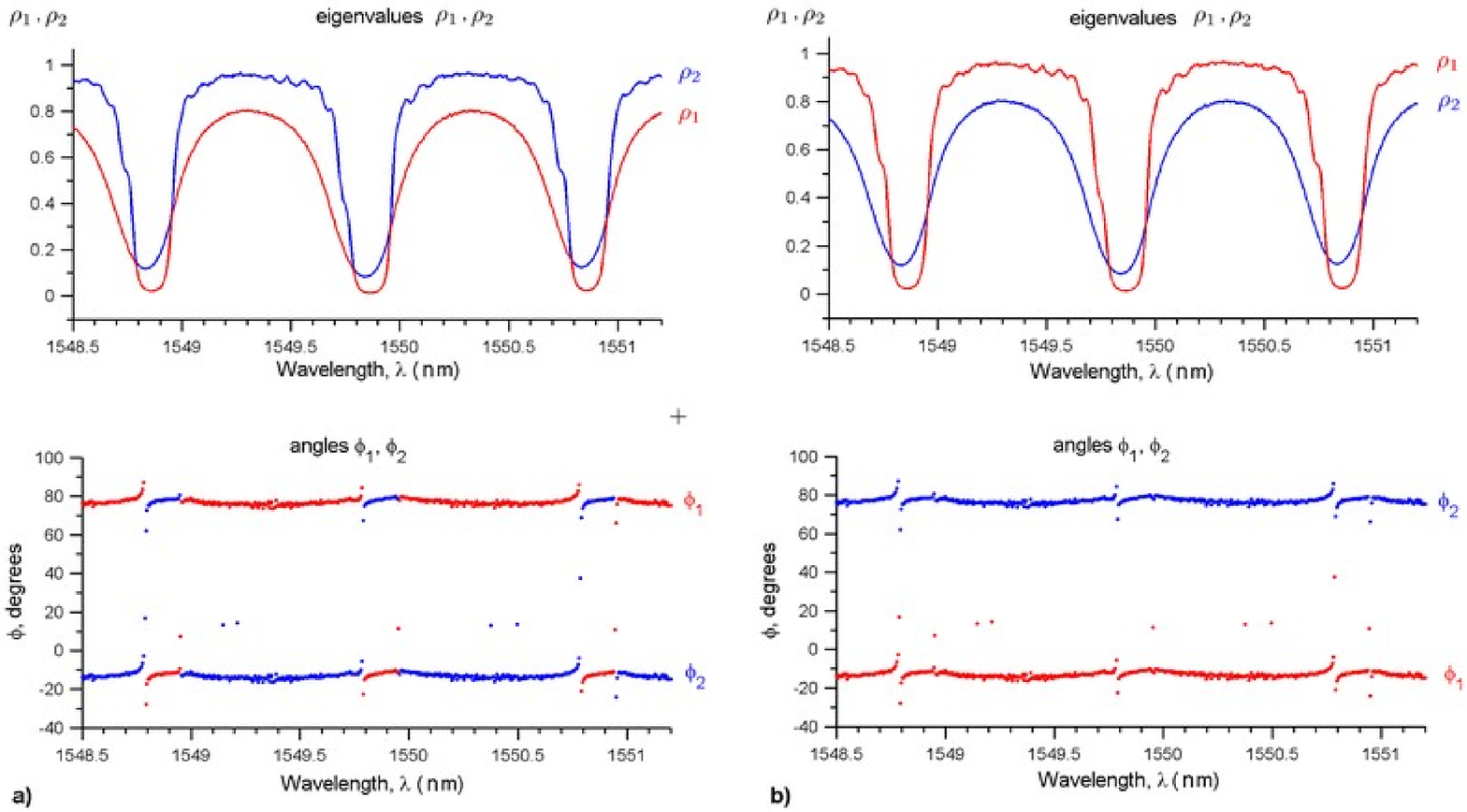}
{Eigenvalues $\rho_1$ and $\rho_2$ and the angle $\phi$ between the geometrical axes of the system $\hat x$, $\hat y$ and the coordinate system defined by the principal axes $\vec{u}_1$, $\vec{u}_2$, before a) and after b) the eigenvalues were reordered. (The data were obtained by means of the OVA 5000 Luna Technologies.)}

In addition to the eigenvalues of the system transmission matrix (plotted as a function of wavelength in Figure~\ref{Fig3}), we can also find the angle of rotation~($\phi$) of the principal axes about the device optical axis~($z$) and plot it as a function of wavelength (bottom frame of Figure~\ref{Fig3}). 
From the mathematical point of view, both eigenvalues are roots of a quadric polynomial and can be ordered arbitrary, usually in descending order: this is why $\rho_1$ is always larger than $\rho_2$ in Figure~\ref{Fig3}a, hence the eigenstates are interchanged when they cross. Because of this effect, the corresponding eigenvectors are interchanged as well, therefore the angle $\phi$ experiences ${\pi/2}$ shifts at the crossing points, as shown on the bottom panel of Figure~\ref{Fig3}a. Therefore, a strategy for restoring the individual transmission spectra along the principal axes is to interchange the eigenvalues and principal axes every time the ${\phi = \pi/2}$ jumps are detected. 

The result of such reordering is shown Figure~\ref{Fig3}b. Now the transmission spectrum corresponds to linearly polarized light with its electric field vector aligned with each principal axis of the system and the rotation angle jumps are eliminated.

In general, the principal axes of an optical system are not necessarily fixed with respect to a reference frame but may depend on the wavelength, as shown in Figure~\ref{Fig4} for a TFBG sensor. 
Indeed, by changing the optical frequency it is expected that an optical system would operate differently. The global behavior of the principal axes, along the whole operational range of the TFBG sensor, is shown in Figure~\ref{Fig4}b, and the corresponding insertion loss in Figure~\ref{Fig4}a.
To remove the noise only the points at which the difference between eigenvalues is noticeable (${|\rho_2 - \rho_1| > 0.05}$) are plotted. The noise arises from the fact that at the points of eigenvalues crossing the transmission matrix becomes degenerate, has the eigenvectors are not well defined, and oscillate rapidly with respect to the geometrical axis, as can be seen from Figure~\ref{Fig4}b at the points of crossing.

\Fig{Fig4}{0.8}{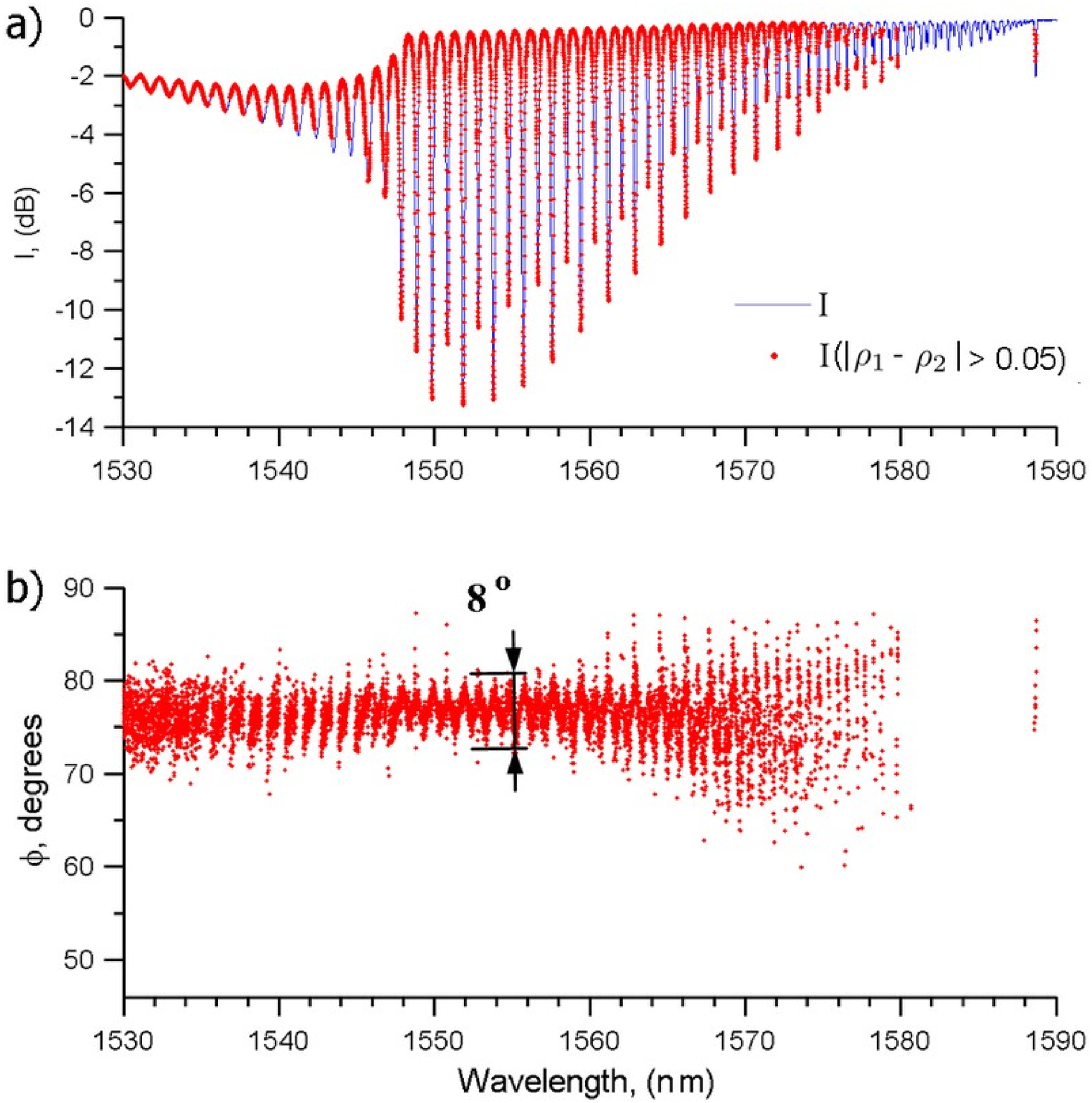}
{Rotation of the principal axes of TFBG device as a function of optical wavelength. (The data was obtained by means of the OVA 5000 Luna Technologies.)}

Figure~\ref{Fig4}b shows that the system possesses significant polarization asymmetry, or birefringence, in the wavelength range ${\lambda \in [1545-1575]}$. Although the principal axes are globally stable, near $75^o$ degrees relative to the reference frame of the LUNA interrogation system, locally they experience small oscillations, of about $8^o$ degrees about the optical axis. These oscillations are related to the resonances observed in the spectrum (Figure~\ref{Fig4}a), and reflect a wavelength dependent birefringence.

We also note the observed alternation between the peaks \C{(Figure~\ref{Fig4}a)}, which can be explained by the fact that the alternating peaks have different azimuthal symmetries and polarizations~\cite{Lee_00}.

\subsection{Measurements along geometrical axes of an optical system}
Instead of choosing principal axes as a reference coordinate system, geometrical axes can be chosen alternatively. In the case of TFBG it is convenient to choose orthogonal axes such that one of the axis is normal to the fibre axis and lay on the plane parallel to the grating blades, as shown in Figure~\ref{TFBG_polarization_intr}. Usually the spectra along geometrical axes are obtained by introducing an external linear polarizer, such that state $\vec{E}_{in}$ can be align along the required axis. 
In Figure~\ref{Fig1} such orthogonal states along $\hat{x}$ and $\hat{y}$ axes are shown on the density plot and denoted as $I_x$ and $I_y$, respectively.

\Fig{Fig5}{0.5}{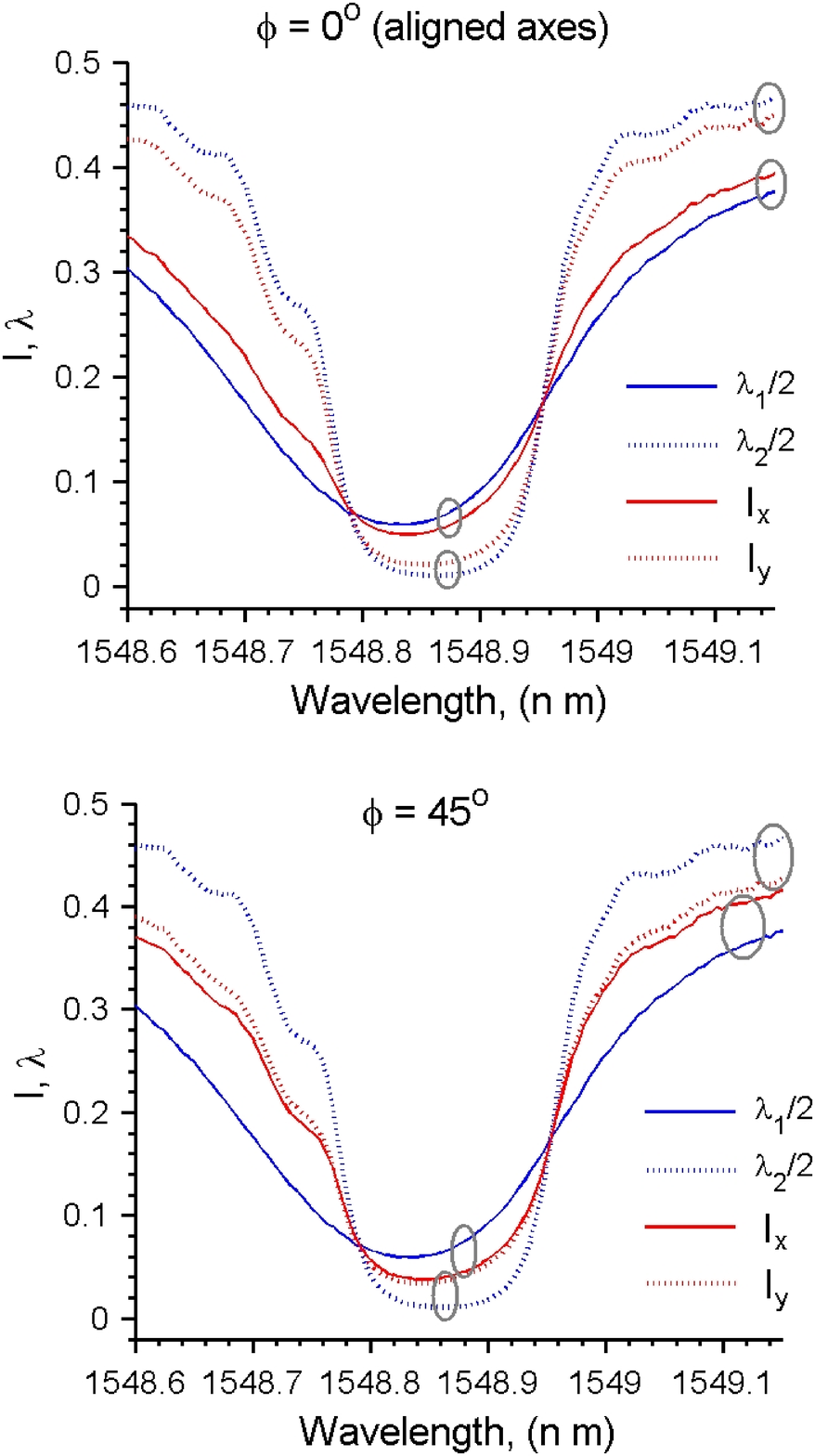}
{Transmission loss along the TFBG system principal axes and its geometrical axes for perfectly align coordinate systems ${\phi = 0^o}$ and rotated by $45^o$ degrees ${\phi = 45^o}$. The intersession loss $I_{x,y}$ and eigenvalues $\rho_{1,2}$ are given in linear scale.}

The difference in transmission loss along the principal axis of the TFBG sensor and its geometrical axes is shown in Figure~\ref{Fig5}. 
It can be clearly seen that in the case when geometrical axes are align with the principal axes of an optical system both approaches yield almost identical result. 
The small difference comes from the aforementioned wavelength dependent oscillation of the principal axes (which cannot be compensated for in the direct measurement), and to a possible slight change in polarization state between polarizer and the grating (since a non-polarization maintaining fibre is used). The following section will demonstrate that in spite of this small inaccuracy, the parameters extracted from the Jones matrix data provide excellent spectral sensitivity results for refractometric sensing.

Nevertheless, although the singular value decomposition, or egen decomposition, provide an elegant way to introduce coordinate axes along which the system can be studied, it might be desirable to fix the reference axis permanently to an optical system geometry, for example in the cases of TFBG sensor, when it is known that the system behaves physically different along particular axes. 

\subsection{Extracting transmission spectra measured along geometrical axes from the Jones matrix or the Stokes vector data}
In this section we describe an approach allowing to extract transmission loss spectra along the given geometrical axes from the Jones matrix or the Stokes vector without introduction of external polarizer.
Thus a single measurement by OVA can provide a complete information about transmission loss along all possible geometrical axes.

\subsubsection{The Jones matrix analysis}

Transmission spectrum of linear polarized light, with the electric field vector $\vec{E}$ align along a given geometrical axis can be extracted directly from the Jones matrix, measured for example by means of OVA 5000 from Luna Technologies.

The instrument was designed to provide a full characterization of the polarization-dependent transmission properties of optical fibre devices as a function of wavelength in the form of the Jones matrix elements. The spectral accuracy of the OVA is $+/- 1.5 pm$ and its insertion loss accuracy is $+/-0.1$ dB, over a wavelength range from $1525$ to $1610$ nm.

The OVA 5000 Luna is capable of capturing four complex transfer functions (amplitude and phase) of a fibre as a function of wavelength and polarization.
The incident and transmitted light can be represented in terms of the electric field vector, written as a column vector: ${\V E = \begin{pmatrix} E_x\\ E_y\\ \end{pmatrix}}$, known as a Jones vector~\cite{JONES:41}, where the field components $E_x$, $E_y$ are projections of the electric field vector $\V E$ on the $x$, $y$ axes. 
The optical TFBG sensor then can be represented as a four port device with two input and output states of polarization. 
These four transfer functions can be assembled into a Jones matrix~\cite{JONES:41}, giving the complete characterization of a device under test. 
\Eq{}
{\V E_{out} = [J] \V E_{in}}

\Eq{}
{[J(\lambda)] = \begin{pmatrix} a(\lambda)&b(\lambda)\\ c(\lambda)&d(\lambda) \end{pmatrix}}

Here, $J(\lambda)$ is the Jones matrix consisting from the four complex transfer functions $a(\lambda), b(\lambda), c(\lambda), d(\lambda)$ of the optical frequency $\lambda$.

The Jones matrix of an optical device under test connected to a linear polarizer is give by: 
\begin{equation}
[J_{dev + pol}] = [J_{pol}][J_{dev}].
\end{equation}
Here ${[J_{dev}]}$ is the measured by an optical vector analyzer Jones matrix of a bare device, and $[J_{pol}]$ is the Jones matrix of linear polarizer known from theory.

The Jones matrix of a linear horizontal polarizer ${[J_{dev}(\theta)]}$, rotated by $\theta$ angle with respect to the optical device axes, is given by: 
\begin{eqnarray}
[J_{pol}(\theta)] &=& [R(\theta)] \begin{bmatrix} 1 & 0 \\ 0 & 0 \end{bmatrix} [R(-\theta)] = \nonumber \\
&=& \begin{bmatrix} \cos^2(\theta) & \cos(\theta)\sin(\theta) \\ \sin(\theta)\cos(\theta) & \sin^2(\theta) \end{bmatrix},
\end{eqnarray}
where $[R(\theta)]$ is the rotation matrix: 
\begin{equation}
[R(\theta)] = \begin{bmatrix} \cos(\theta) & -\sin(\theta) \\ \sin(\theta) & \cos(\theta) \end{bmatrix},
\end{equation}
rotating the coordinates system so that the horizontal polarizing element has the simplest representation and then rotating the system back down into the original system of coordinates. 

Thus, on optical device with a known Jones matrix ${[J_{dev}]}$, connected to a linear polarizer rotated by $\theta$ angle, is described by the Jones matrix:
\begin{equation}
[J_{dev + pol}(\theta)] = \begin{bmatrix} \cos^2(\theta) & \cos(\theta)\sin(\theta) \\ \sin(\theta)\cos(\theta) & \sin^2(\theta) \end{bmatrix}[J_{dev}].
\end{equation}

Finally the transmission spectrum ${I(\theta,\theta)}$, for a given angle~$\theta$ of the linear polarizer, can be computed as 
\begin{equation}
I(\theta,\lambda) = 10 \log_{10}{\frac{\rho_1(\theta,\lambda) + \rho_2(\theta,\lambda)}{2}},
\end{equation}
where ${\rho_1(\theta,\lambda)}$ and ${\rho_2(\theta,\lambda)}$ are the eigenvalues of ${H(\theta,\lambda) = [J_{dev + pol}(\theta,\lambda)]^\dagger [J_{dev + pol}(\theta,\lambda)]}$ matrix.

\subsubsection{The Stokes vector analysis}
In a similar way we can use the Stokes vector. The Stokes vector can be measured with a polarizer controller or a special instrument, such as JDS Uniphase SWS-OMNI-2 system.

A beam of light can be completely described by the four parameters~\cite{Berry:77, Goldstein:1992}, represented in the form of the Stokes vector:
\begin{equation}
\vec{S} = \begin{pmatrix} S_0 \\ S_1\\ S_2\\ S_3 \end{pmatrix} = \begin{pmatrix} I(0^o)+I(90^o) \\ I(0^o)-I(90^o) \\ I(45^o)-I(135^o) \\ I_{RHS} - I_{LHS} \end{pmatrix},
\end{equation}
here $I(\theta)$ is the intensity of light polarized in the direction defined by the angle $\theta$ in the plane perpendicular to the direction of light propagation, and $I_{RHS}$, $I_{LHC}$ are the intensities of right- and left-handed polarized light, respectively.

Since the Stokes parameters are dependent upon the choice of axes, they can be transformed into a different coordinate system with a rotation matrix $[R]$.
Considering that a second coordinate system is obtained by rotating the original coordinate system about the direction of light propagation on the angle $\theta$, we can write~\cite{McMaster:1954} 

\begin{equation}
\vec{S'}(\theta) = [R(\theta)] \vec{S},
\end{equation}
or
\begin{eqnarray}
\begin{pmatrix} S_0' \\ S_1'\\ S_2'\\ S_3' \end{pmatrix} &=& \begin{bmatrix} 1 & 0 & 0 & 0 \\ 0 & \cos(2\theta) & \sin(2\theta) & 0 \\ 0& -\sin(2\theta) & \cos(2\theta) & 0 \\ 0 & 0 & 0& 1 \end{bmatrix} \begin{pmatrix} S_0 \\ S_1 \\ S_2 \\ S_3 \end{pmatrix} = \nonumber \\
 &=& \begin{pmatrix} S_0 \\ S_1 \cos(2\theta) + S_2 \sin(2\theta) \\ -S_1 \sin(2\theta) + S_2 \cos(2\theta) \\ S_3 \end{pmatrix},
\end{eqnarray}
here ${S_0, S_1, S_2}$ and $S_3$ are the Stokes parameters.

Now considering that
\begin{eqnarray}
I(0^o)+I(90^o) &=& S_o, \nonumber \\
I(0^o)-I(90^o) &=& S_1 \cos(2 \theta) + S_2 \sin(2 \theta),
\end{eqnarray}
we conclude that the transmission loss spectra along the two orthogonal axis, rotated by the angle $\theta$ with respect to the original system of measurements, are given by the following expressions:
\begin{eqnarray}
I_x(\theta,\lambda) = \frac{1}{2}(S_o(\lambda) + S_1(\lambda) \cos(2 \theta) + S_2(\lambda) \sin(2 \theta)), \nonumber \\
I_y(\theta,\lambda) = \frac{1}{2}(S_o(\lambda) - S_1(\lambda) \cos(2 \theta) - S_2(\lambda) \sin(2 \theta)). \nonumber \\
\end{eqnarray}
Hence, if the Stokes vector is known in one coordinate system, the transmission spectrum of linearly polarized light with the electric field $\vec{E}$ align along a given geometrical axis, defined by the angle $\theta$, can be recovered.

We should also note that the Jones matrix and Mueller matrix representations are connected, and can be expressed as~\cite{Hiroyuki:2007}:
\Eq{}
{M = U(J \otimes J)U^{-1},}
where $J \otimes J$ is the direct product of Jones matrices, and matrix $U$ is given by
\Eq{}
{U = \begin{bmatrix}
	1&0&0&1 \\
	1&0&0&-1 \\
	0&1&1&0 \\
	0&i&-i&0
 \end{bmatrix}}

Although the two approaches are related the phase information is available only in the case when the Jones matrix is known. The phase information allows to determine additional phase related parameters, such as Group Delay (GD).

\section{Polarization-based detection of small refractive index changes with TFBG sensors}
In this section we investigate which polarization-based measurement techniques provide the best signal-to-noise ratio when TFBG sensors are used to detect small refractive index changes, and use the special case of a TFBG coated with gold nanorods (as described in~\cite{Bialiayeu:2012}). This choice is made because the polarization dependence of waveguide-type sensors is much enhanced when metal coatings are used. 
As indicated earlier, the interaction of guided waves with metal interfaces depends strongly on whether the electric fields of the waves are tangential or normal to the metal boundary. 
It was further mentioned that when the core-guided input light of a TFBG is linearly polarized along the principal axes, the electric fields of high order cladding modes are either tangential or radial (hence normal) to the cladding boundary. To be precise, $y$-polarized input light (corresponding to light polarized parallel to the plane of incidence on the tilted grating fringes, as shown in Figure~\ref{TFBG_polarization_intr},~\textit{i.e.} P-polarized) couples to radially polarized cladding modes while $x$-polarized light (perpendicular to the tilt plane, or S-polarized) couples to azimuthally polarized cladding modes (tangential to the boundary). Further polarization effects arise with non-uniform metal coatings, since they have boundaries that are both tangential and radially oriented relative to the cylindrical geometry of the fibre~\cite{Zhou:13}.
It is therefore desirable to carry out two transmission measurements along the principal axis to observe directly such polarization effects on the strengths and positions of the cladding mode resonances~\cite{Berini:2011}. 
On the other hand, it has been shown here that a Jones matrix measurement can provide this information as well, in addition to other parameters of interest, such as PDL.
While the PDL spectrum \C{``hides''} the physical effects responsible for difference in transmission due to the different polarization states, it has been shown in the past to yield excellent limits of detection for surface plasmon resonance based TFBG sensors~\cite{Caucheteur:11}. We now proceed to compare the signal noise for refractive index measurements by the various polarization-dependent data extraction techniques.

As shown in Figure~\ref{Fig6} for a TFBG coated with a sparse layer of gold nanorods and immersed in water, the PDL parameter provides the absolute value of the difference between resonances observed in the transmission spectra measured along the principal axes. 
The relative position of the peaks, their amplitude, and their width become convoluted in the PDL parameter, which provides instead a maximum located somewhat in between the individual resonance maxima and a zero on either side corresponding to the wavelengths where the spectra cross each other.

\Fig{Fig6}{0.6}{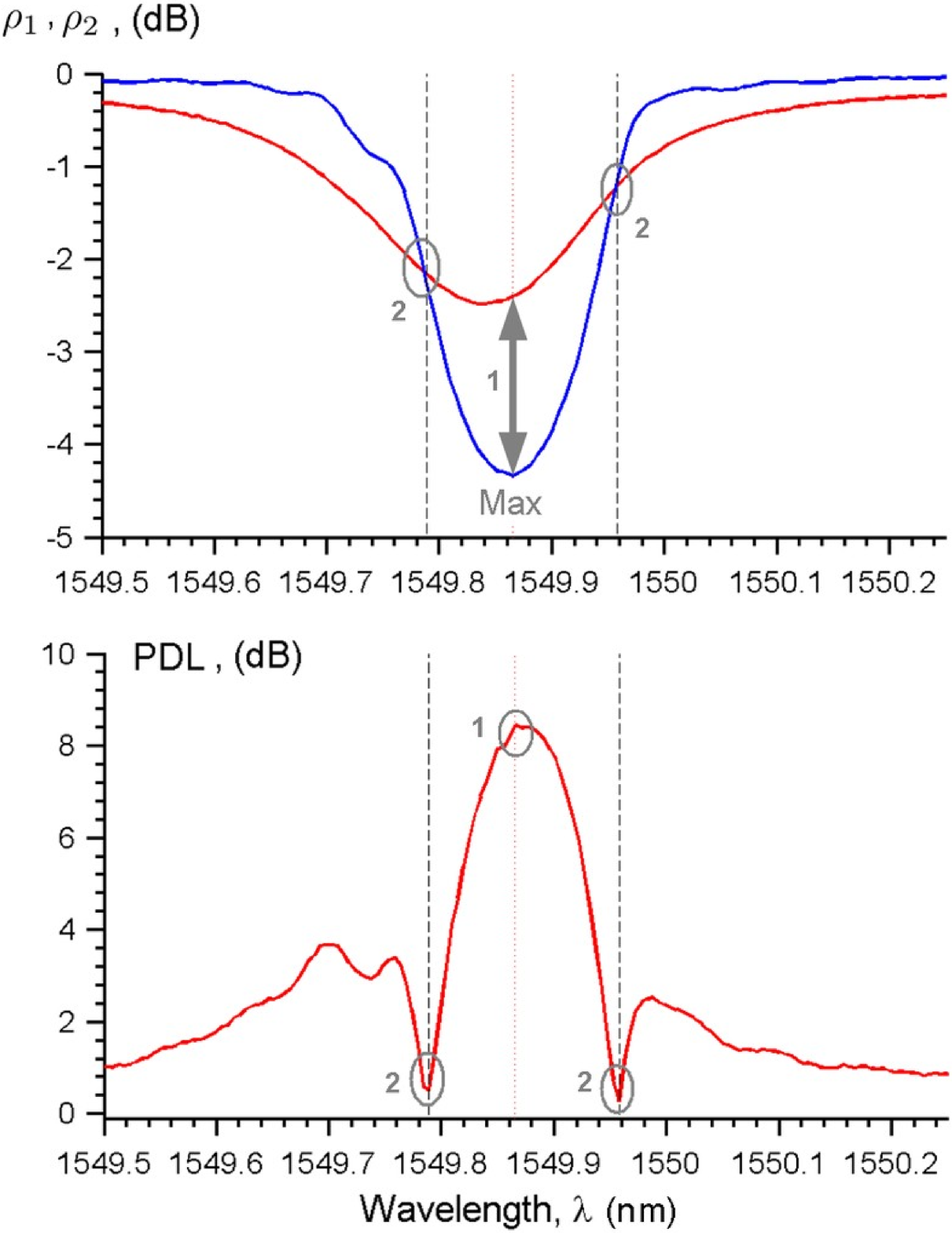}
{Two eigenvalue spectra (individual transmission along the principal axes) and corresponding polarization-dependent loss (PDL) parameter. The peak in PDL spectrum is denoted by ``1'' and zeros by ``2''. Here the more common dB scale is used.}

When the refractive index of the medium surrounding such TFBG changes the waveguiding characteristics of the cladding are modified and the resonances observed in the transmission spectrum change accordingly. 
Therefore, to detect changes in refractive index we can either follow the amplitudes and positions of individual resonances~\cite{Berini:2011} or of the PDL features. Here, the sensor was immersed in water and the refractive index was incrementally increased in steps of ${\Delta n =1.517 \times 10^{-4}}$ by adding ${10~\mu l}$ of ethylene glycol ( ${C_2 H_4 (OH)_2}$ ) to ${5~m l}$ to the water. The impact of each increase in refractive index on all parameters of interest is shown in~Figure~\ref{Fig7} for a typical slice of the spectrum (it was noted in~\cite{Bialiayeu:2012} that there was little difference in wavelength shifts across the TFBG spectrum for this device).

\Fig{Fig7}{0.9}{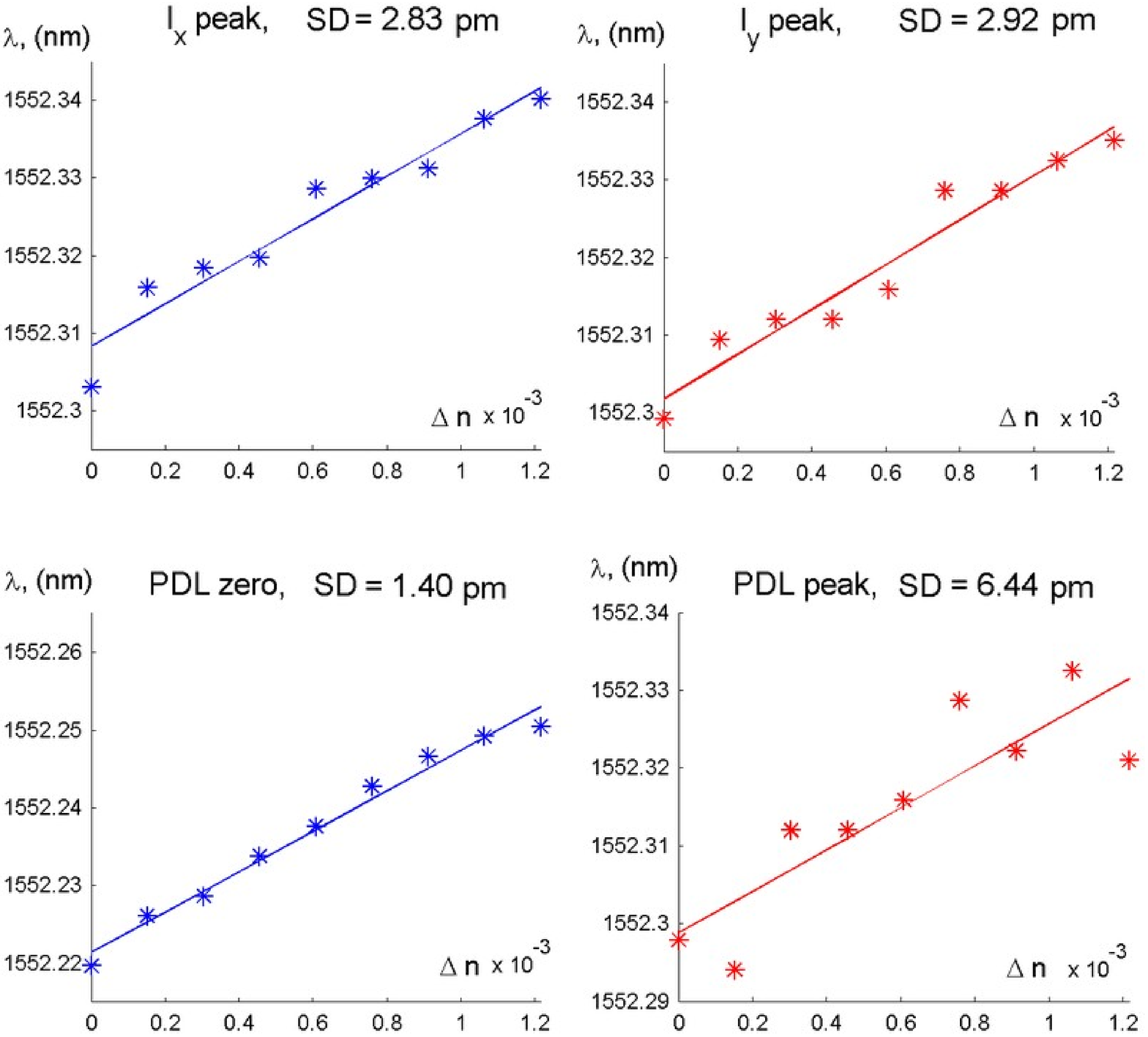}
{ Position of the resonance a) in the transmission spectra $I_x$ and $I_y$ of light polarized along ${\hat x}$ and ${\hat y}$ geometrical axes here aligned with the principal axes, and b) in PDL spectra (the maximum and zero values of PDL are detected), as a function of refractive index change. 
The continuous lines represent the least square approximation to the measured data, and $SD$ is the standard deviation from the linear approximation}

The refractive index change was chosen to have a relatively small value of ( $\Delta n =1.517 \times 10^{-4}$) to test the sensor detection limits and a linear fitting was used because the sensor response for such small changes is expected to be linear. The standard deviation of the errors from the linear fit was calculated in the usual manner by: 

\begin{equation}
SD = \sqrt{\frac{1}{N} \sum_{i=1}^N (x_i - \mu)^2},
\end{equation}
here $\mu$ is the expected value of $\vec{x}$, and ${\vec{x} = \vec{y}_{appr} - \vec{y}_{data}}$ is the difference between the measured data~$\vec{y}_{data}$ and its linear approximation~$\vec{y}_{appr}$.

The results of the fits show that the most accurate detection of the refractive index change is achieved with the zeros in the PDL spectrum, as this measurement provides the smallest standard deviation of~${SD = 1.40~pm}$. 
The worst result were obtained with the PDL peak (${SD = 6.44~pm}$), while the detection of individual polarized resonances provides an intermediate value of the standard deviation (essentially equal to $3$~pm). It is not surprising that zeros of PDL should provide the most accurate results as they consist essentially of a differential measurement between two spectra with well-defined crossing points that occur on the sides of the individual resonances, where the spectral slope is highest. On the other hand, at the expense of an increase in noise by a factor of approximately 2, other effects such as differences in the change in the resonance amplitude for the two polarizations (which can be linked to differential loss or scattering) can be studied when the principal axis spectra are used.

\section{Conclusion}

In this Chapter we investigated the use of Jones matrix and Stokes vector based techniques and polarization analysis to extract information from optical fibre sensors in non-polarization maintaining fibres. 
In particular we showed how to calculate transmission spectra with electric fields $\vec{E}$ aligned with the system principal axes or along any other system axes. 
We also proposed the method of extraction spectra of light polarization along a given geometrical axis either from the Jones Matrix or the Stokes vector data. 
In the case of a TFBG inscribed in a non-polarization maintaining fibre for instance, the principal axes of the system are determined by the direction of the tilt of the grating planes (and its perpendicular). 
The transmission spectra for light polarized in the tilt plane (P-polarized) or out of the tilt plane (S-polarized) can thus be extracted without having to separately align a linear polarizer upstream from the grating and hoping that the polarization remains linear between the polarizer and the TFBG.
Polarization-resolved spectra can also be obtained at a faster rate, for applications in chemical deposition process monitoring for instance, without the need to line up or rotate a polarizer between each measurement. 

The transmission spectra of P- and S-polarized light were compared with the TFBG spectral response along the principal axes. A small oscillation of about 8 degrees in the orientation of the principal axes as a function of wavelength was observed.

Furthermore, it was determined that the best TFBG sensor detection limits are achieved when zeros in the PDL spectrum are followed, mainly because of the sharpness of crossing points occurring on the steep sides of the individual resonances, but at the expense of the direct observation of the sensor response to polarized light (which are also available from the Jones matrix data). In the latter case, while the standard deviation of the spectral sensitivity is doubled relative to PDL measurements, additional information becomes available regarding the influence of the measured medium on cladding mode loss, allowing further uses from the TFBG data sets~\cite{Wenjun:2014}.

%% file: Chap_Material_Metals.tex
\chapter{Optical properties of materials}
\label{opt_prop_metals}

In the presented work we investigate methods of increasing TFBG sensor sensitivity by coating its surface with various types of coatings, such as metal films and films consisting of nanoparticles.

Before running numerical simulations, the optical properties of materials should be defined. 
In this chapter we will focus our attention on optical properties of metals since it is known the that sensitivity of an optical sensor can be increased by excitation of surface plasmon resonances (SPR) observed in metal films and nanoparticles of a proper geometrical dimension~\cite{Bottomley:2011, Ahamad:2012, Anker:2008, Mayer:2011}. 
We also consider various surface morphologies and review methods that allow us to account for the dipole-dipole radiative interaction between elements of a film or between closely deposited nanoparticles.

\section{The methods of measurement of optical constants}

It is not a trivial task to choose the correct optical parameters for thin metal films or nanoparticles. Various sources of information do not always provide consistent results. The theoretically computed permittivity of metals is not always consistent with the experimental measurements, and experimental results sometimes are limited by a particular method. 

In this section we provide a brief review of various methods available for measuring of optical constants and discuss limitations imposed by a particular method, so that we can choose a reliable data source for further computation. Next we will consider models for mixed media needed to describe rough films resulting from nanoparticle-based coatings.

The optical constants $n$ and $k$ of a medium at frequency $\omega$ can be determined either by~(1) measuring $n$ and $k$ independently at a given $\omega$ or (2)~one of the parameters ($n$ or $k$) should be measured over the entire frequency range, then the remaining parameter can be deduced from the measured data.
 
Several alternative methods were proposed as well, including the Drude ellipsometry method introduced in 1889~\cite{Drude:1902} in which the relative amplitude and phase shift are measured simultaneously, however, the method is significantly affected by the sample surface quality~\cite{Collins:1998}.

\subsection{The methods based on single parameter measurements}
\label{subsec_KK}

Historically, this approach was based on Bode's electric network theory introduced in 1945, where it had been shown that the attenuation and phase at the output of an electric network are not independent of one another. 
In Robinson's paper from 1952~\cite{Robinson:1952} the absorption spectrum was derived from a normal incidence reflection spectrum. The phase change of a reflected wave was deduced from the curve of the reflection attenuation as a function of frequency, and the optical constants $n$ and $k$ were then determined with the aid of a Smith chart. 

A modern treatment of the problem can be found in~\cite{Ahrenkiel:1971, Rubloff:1971, Schmidt:1971}.
The usual optical procedure is to measure the reflectance at normal incidence over as wide range as possible, and then use, for example, the free-electron extrapolation for the reflectance outside the range. 
Once the reflectance is approximated over the range from zero to infinity the Kramers-Kronig (KK) integration might be implemented:

\Eqaa{eq_KK}
{\chi_1(\omega) & = & \frac{2}{\pi} P \int_0^\infty \frac{\Omega \cdot \chi_2(\Omega)}{\Omega^2-\omega^2} d\Omega,}
{\chi_2(\omega) & = & - \frac{2 \omega}{\pi} P \int_0^\infty \frac{\chi_1(\Omega)}{\Omega^2-\omega^2} d\Omega.}
Here the integration is done in the sense of the Cauchy principal value, defined as the $P \int$ and $\chi(\omega)$ is a complex analytical function, vanishing faster than $\frac{1}{|\omega|}$ as $|\omega| \to \infty $, $\chi_1(\omega): = Re[\chi(\omega)]$ and $\chi_2(\omega): = Im[\chi(\omega)]$. 

Thus the optical constants $n$ and $k$ can be connected through the Kramers-Kronig relations:
\Eqaa{}
{\chi(\omega) = \tilde{n}(\omega) &=& n(\omega) + ik(\omega) =}
{&=& \chi_1(\omega) + i\chi_2(\omega)}

However, the results of such calculations are often quite sensitive to the shape of the extrapolated spectrum, therefore the extrapolation of the reflectance at both ends of the spectrum is crucial.
While a reasonable low-energy extrapolation is possible by simply assuming that $R = 100\%$, additional experimental information is needed to make a reasonable high-energy extrapolation. 
The measurements have to be extended further into the vacuum ultraviolet region in order to determine the $n$ and $k$ in the visible band. The necessity of such measurements is the major disadvantage of this method~\cite{Schmidt:1971, Nestell:1972}. 


\subsection{The methods based on simultaneous measurement of both parameters}

Simultaneous measurement of both parameters is usually difficult in practice, especially in the region of strong absorption. 
To overcome this problem several new techniques were proposed and successfully implemented in~\cite{Nestell:1972, Johnson:1972}.
Instead of a single normal incidence reflectivity measurement, the parameters for oblique incidence at several angles and various polarizations were measured. The right choice of the incident angle and polarization state can significantly improve the results in the region where normal incidence measurements are inaccurate~\cite{Nestell:1972}. 
The proposed method is also less sensitive to the surface roughness, unlike the ellipsometry method.

First the reflectance $R$ and transmittance $T$ of a thin film deposited on a transparent substrate were calculated for the normal and oblique incidence, for $s$ and $p$ polarization, and different film thicknesses. 
The calculation involves solution of Maxwell's equation boundary-value problem from which the reflectance $R$ and transmittance $T$ can be expressed in terms of the material constants $n$ and $k$.

Next the obtained function $R(k,n)$ and $T(k,n)$ were inverted to obtain $n$ and $k$ expressed in terms of the measured $R$ and $T$. 
Since the optical constants $n$ and $k$ are independent of the boundary conditions, the $n$ and $k$ values were expected to depend only on the incident light wavelength (which excites the electronic transition), but not on polarization and angle of incidence (if material is isotropic) or on the film thickness.
Therefore at each wavelength, in the range of interest, reflectance $R$ and transmittance $T$ at specified incident angles and polarizations were measured.
Then the film thickness was measured by means that were independent of the $R$ and $T$ measurements and compared with the result deduced directly from $R$ and $T$ measurements~\cite{Nestell:1972}. 
The process of inversion requires solution of several nonlinear equations, and a number of iterations till the convergence is achieved~\cite{Nestell:1972}.

\section{Optical constants of metals}

In our calculation we are going to use optical constants obtained in~\cite{Johnson:1972} by the method described in the previous section~\cite{Nestell:1972}, where the optical constants $n$ and $k$ were obtained for copper, silver, and gold from reflection and transmission measurements. 
The films were created by vacuum-evaporation with film-thickness in the range of $185-500$~$\mathring{A}$, and the measurements were conducted in the spectral range of $0.5-6.5$~eV. 

It was observed that the results were independent of film thickness only above a certain critical thickness of about $250$~$\mathring{A}$, and were unchanged after vacuum annealing or aging in air. 
The optical properties of evaporated thin films have been found to be the same as for bulk materials for the thickness of the films greater than about~$300$~$\mathring{A}$~\cite{Johnson:1972}.


\subsection{Free electron approximation, the Drude-Sommerfeld model}
The obtained experimental data can be approximated with the free-electron model.
Assuming that the motion of electrons is confined to a region much smaller than the wavelength we can implement the Drude-Sommerfeld model~\cite{Sommerfeld:1933, Drude:1902}. 
The model does not include a restoring force, assuming free electrons, thus an equation of motion of a single electron can be written in the following form:
\Eq{eq_eps1}{m_o \cdot d_t^2 x(t) + m_o \gamma \cdot d_t x(t) = F_{ext}(t), }
where $m_o$ is the effective optical mass of electron, $\gamma = \frac{1}{\tau}$~is the macroscopic damping constant due to the dispersion of the electrons caused by the crystalline structure, $\tau$~is the relaxation time and ${F_{ext}(t)}$ is the external force applied to the electron.

Assuming harmonic excitation: $ F_{ext}(\omega) = - e E(\omega)$ and taking the Laplace transform of equation~(\ref{eq_eps1}) we have:

\Eq{eq_eps2}{x(\omega) = \frac{e}{m_o} \frac{1}{\omega^2 + i \omega \gamma} E(\omega).}

The macroscopic polarization can be written in the following form:

\Eqaaa{eq_eps3}
{P(\omega) &=& N \cdot p(\omega) = -N e \cdot x(\omega) = - \frac{N e^2}{m_o} \frac{1}{\omega^2 + i \gamma \omega} E(\omega) =}
{&=& \epsilon_o \chi (\omega) E(\omega) =}
{&=& \epsilon_o \left(\epsilon(\omega) - 1\right) E(\omega),}
where: $\chi$ is the electric susceptibility, $N$ is the number of conducting electrons per unit volume (density), $\epsilon_0$ is the electric permittivity of free space and $\tau$~is the relaxation time.

Excluding $E(\omega)$ from equation~(\ref{eq_eps3}), the expression of $\epsilon(\omega)$ can be obtained in terms of effective optical mass and macroscopic damping of a free electron. We denote it as $\epsilon^f(\omega)$.

\Eq{eq_eps4}{\epsilon^f(\omega) = 1 - \frac{\omega_p^2}{\omega^2 + i \gamma \omega},}
here ${\omega_p = \sqrt{\frac{N e^2}{\epsilon_o m_o}}}$ is the plasma frequency.

The complex dielectric constant
 \Eq{eq_eps6}{\hat{\epsilon}= \epsilon_1 + i \epsilon_2}
and the complex index of refraction, (measured in experiments)
 \Eq{eq_eps7}{\hat{n} = n + ik}
 are connected by the relation $\hat{\epsilon} =\hat{n}^2$, so that
\Eqaa{eq_eps8}
{\epsilon_1 &=& n^2 - k^2,}
{\epsilon_2 &=& 2nk.}

Considering (\ref{eq_eps4}) it is possible to separate $ \hat{\epsilon}^f$ into its real and imaginary parts:
\Eqaa{eq_eps9}
{\epsilon_1^f &=& 1 - \frac{\omega_p^2 \tau^2}{1+ \omega^2 \tau^2},}
{\epsilon_2^f &=& \frac{\omega_p^2 \tau}{\omega(1 + \omega^2 \tau^2)}.}

We note that free-electron approximation to the dielectric function is determined by the relaxation time $\tau$ and the the plasma frequency $\omega_p$ defined by the electron optical mass~$m_o$.

\subsection{The near infrared band}

For metals at near-infrared frequencies we can assume $ \omega \gg \frac{1}{\tau}$.
Considering equations (\ref{eq_eps4}) and (\ref{eq_eps9}) we can write: 
\Eqaa{eq_eps10}
{\epsilon_1^f \sim 1 - \frac{\omega_p^2}{\omega^2} &=& 1 - \frac{\lambda^2}{\lambda_p^2},}
{\epsilon_2^f \sim \frac{\omega_p^2}{\omega^3 \tau} &=& \frac{\lambda^3}{\lambda_p^2} \tau',}
where 
\Eqaa{eq_eps11}
{\frac{1}{\lambda_p^2} &=& \frac{Ne^2}{\pi m_o c^2},}
{\tau' &=& 2 \pi c \tau.}

The values of $n(\omega)$ and $k(\omega)$ can be measured experimentally, hence the values of $\epsilon_1$ and $\epsilon_2$ can be calculated using (\ref{eq_eps8}). 
Assuming that the experimental results at near-infrared bands can be satisfactorily approximated with the free-electron model we set
\Eq{eq_eps12}{\epsilon = \epsilon^f .}
Then the optical mass $m_o$ can be determined from the experimental results for $\epsilon_1$ from the slope of a plot of $-\epsilon$ vs $\lambda^2$. 
Next, using the expression for $\epsilon_2$ and plotting $\epsilon_2/\lambda$ vs $\lambda^2$ we can determine $\tau$ from the slope of the graph in the infrared band. 
Such derivation was conducted in~\cite{Johnson:1972} where the effective optical mass and relaxation time were obtained. 
It was shown that the effective optical mass and relaxation time can be considered to be constant for the whole near-infrared band.


The results for copper, silver, and gold are shown in Table~\ref{tab_1}~\cite{Johnson:1972}.
The results for $m_o$ are relatively accurate since in the infrared $k \gg n$ and can be measured precisely. The error in the $\tau$ measurements is larger due to the error in the $n$ measurements in the infrared band~\cite{Johnson:1972}.

\begin{table}[!htb]
\caption{Optical masses and the relaxation times for copper, silver and gold.~\cite{Johnson:1972}} \label{tab_1} 
\begin{center}
 \begin{tabular}{ l | c | c }
	Metal & $m_0$ (electron masses) & $\tau \times 10^{-15}$ (sec) \\
 \hline 
	Copper & $1.49 \pm 0.06$ & $6.9 \pm 0.7$ \\
	Silver & $ 0.96 \pm 0.04$ & $31 \pm 12$ \\
	Gold & $ 0.99 \pm 0.04$ & $9.3 \pm 0.9$ \\
\end{tabular}
\end{center}
\end{table}

We can conclude that the properties of metals at the near-infrared band can be approximated with the free-electron model, defined by the parameters $m_o$ and $\tau$, which can be obtained from the experimental measurements of the optical constants $n$ and $k$.

\subsection{The visible band. The interband absorption}

In the visible and near-ultraviolet regions Drude's free-electron theory fails and the interband absorption should be taken into account.
The absorption in the visible and ultraviolet band is significantly influenced by the transitions from the completely occupied $d$ bands to an empty state above the Fermi level in the conduction band.
Moreover, these transitions depend on the nanoparticle size. 
The interband absorption peak becomes sharp and its peak position shifts towards the lower energy side~\cite{Balamurugan:2007}. 

We are going to review the size-induces changes in the $d$ band in the following chapter, here we simply state the significance of the interband absorption.
The interband contribution to the imaginary part of the dielectric constant can be obtained by subtracting the free-electron contribution value $\epsilon_2^f$ from the experimentally determined value of $\epsilon_2$. 
The comparison between the experimental data and theoretical prediction, based on the classical free-electron Drude theory, was conducted in~\cite{Johnson:1972}.

The interband contribution was also compared with the theoretical prediction based on the band-structure model, where the joint density of states and the transition-probability matrix elements throughout the Brillouin zone were taken into account~\cite{Christensen:1971, Williams:1972, Fong:1970}.


The discussion of optical constants of Ag and Cu in terms of free-electron effects, interband transitions and collective oscillations can be found in Ehrenreich and Philipp \cite{Ehrenreich:1962}.

We conclude that the free-electron expression of $\hat{\epsilon}^f $ is useful only for photon energies below a threshold energy and can be applied for the near-infrared band only.
Above this threshold energy the form of the $\epsilon_2$ curve depends on the specific material band structure, and also exhibits size-dependent properties \cite{Balamurugan:2007}.

Assuming that interband contribution in known either from experiments or theory we can write 
\Eq{eq_eps13}{\epsilon(\omega) = \epsilon_{intra}(\omega) + \epsilon^f(\omega)}
where $\epsilon(\omega)$ is the dielectric function of a metal, $\epsilon^f(\omega)$ is the free-electron contribution and $\epsilon_{intra}(\omega)$ is the interband contribution.

Although the interband contribution can be satisfactory calculated only in the framework of quantum mechanics, or obtained from the experimental measurements, the classical approximation is still possible. 
The electrons in the $d$ band are not free as valence band electrons, but rather bounded by the lattice ions. 
Correcting the classical free electron model~(\ref{eq_eps1}), by introducing a linear restoring force we come to the Lorentz-Drude oscillator model, based on the damped harmonic oscillator approximation:

\Eq{eq_eps14}{m_d \cdot d_t^2 x(t) + m_d \gamma \cdot d_t x(t) + m_d \omega_d^2 x(t) = F_{ext}(t),}
thus the single induced dipole moment is: 
\Eq{eq_eps15}{p(\omega) = e x(\omega) = - \frac{e^2}{m} \frac{1}{-\omega_d^2 + \omega^2 + i \omega \gamma} E(\omega).}
The macroscopic polarization $P(\omega)$ depends on the model describing dipole-dipole interaction. The discussion on this subject can be found in~\cite{Stenzel:2005}. 

In practice it is not always possible to approximate interband absorption with one specific type of electron (which has an effective mass $m_d$ and resonates at some eigenfrequency $\omega_d$). 
Usually several different types of electrons with different effective masses bounded by different restoring forces should be taken into consideration. 

Thus for each resonance in a given absorption spectrum the restoring forces and the fraction of each type of electrons can be chosen for accurate approximation. 
Such investigation was conducted in~\cite{Rakic:98} were the six-oscillator Lorentz-Drude model was used to fit optical functions of eleven widely-used metals in optoelectronic. 
The parameters were chosen for the best fit of experimental data, obtained by the method described in the previous section~\cite{Johnson:1972}.

In accordance with Lorentz-Drude oscillator model, a complex dielectric function $\epsilon(\omega)$ can be expressed \cite{POWELL:70, Rakic:98}:
\Eq{eq_eps16}{\epsilon(\omega) = \epsilon^f(\omega) + \epsilon^b(\omega),}
where the free-electron or Drude model is represented by the first term:
\Eq{eq_eps17}{\epsilon^f(\omega) = 1 - \frac{f_o \omega_p^2}{\omega^2 - i \omega \Gamma_0},}
and the interband part of the dielectric function is represented by the bounded Lorentz electron model, similar to the model used for insulators:
\Eq{eq_eps18}{\epsilon^b(\omega) = - \sum_{j=1}^N \frac{f_j \omega_p^2}{\omega^2 - \omega_j^2 - i \omega \Gamma_j}.}
Here $\omega_p$ is the plasma frequency, $N$ is the number of oscillators with resonant frequency $\omega_j$, and oscillator strength $f_j$, and $\Gamma_j = \frac{1}{\tau_j}$ is the macroscopic damping constant of the $j$th oscillator ($\Gamma_0$ corresponds to the plasma damping constant ).

The obtained results can be summarized in Table~\ref{tab_2}~\cite{Rakic:98} (where $\omega_j$ is given in electron volts, and $\Gamma_j$ in $sec^{-1}$).

\begin{table}[!htb]
\caption{Values of the Lorentz-Drude Model Parameters~\cite{Rakic:98}.} \label{tab_2}

{\centering
\vskip6pt

\begin{tabu} to \textwidth {|X|X|X|X|X|X|X|X|X|X|X|X|}
 \hline
	 & Ag & Au & Cu & Al & Be & Cr & Ni & Pd & Pt & Ti & W\\
 \hline 
	$\omega_p$ & 9.01 & 9.03 & 10.83 & 14.98 & 18.51 & 10.75 & 15.92 & 9.72 & 9.59 & 7.29 & 13.22\\ 
	$\omega_1$ & 0.816 & 0.415 & 0.291 & 0.162 & 0.100 & 0.121 & 0.174 & 0.336 & 0.780 & 0.777 & 1.004\\
	$\omega_2$ & 4.481 & 0.830 & 2.957 & 1.544 & 1.032 & 0.543 & 0.582 & 0.501 & 1.314 & 1.545 & 1.917\\
	$\omega_3$ & 8.185 & 2.969 & 5.300 & 1.808 & 3.183 & 1.970 & 1.597 & 1.659 & 3.141 & 2.509 & 3.580\\
	$\omega_4$ & 9.083 & 4.304 & 11.18 & 3.473 & 4.604 & 8.775 & 6.089 & 5.715 & 9.249 & 19.43 & 7.498\\
	$\omega_5$ & 20.29 & 13.32 & 0 & 0 & 0 & 0 & 0 & 0 & 0 & 0 & 0\\

	$\Gamma_o$ & 0.048 & 0.053 & 0.030 & 0.047 & 0.035 & 0.047 & 0.048 & 0.008 & 0.080 & 0.082 & 0.064\\
	$\Gamma_1$ & 3.886 & 0.241 & 0.378 & 0.333 & 1.664 & 3.175 & 4.511 & 2.950 & 0.517 & 2.276 & 0.530\\
	$\Gamma_2$ & 0.452 & 0.345 & 1.056 & 0.312 & 3.395 & 1.305 & 1.334 & 0.555 & 1.838 & 2.518 & 1.281\\
	$\Gamma_3$ & 0.065 & 0.870 & 3.213 & 1.351 & 4.454 & 2.676 & 2.178 & 4.621 & 3.668 & 1.663 & 3.332\\
	$\Gamma_4$ & 0.916 & 2.494 & 4.305 & 3.382 & 1.802 & 1.335 & 6.292 & 3.236 & 8.517 & 1.762 & 5.836\\
	$\Gamma_5$ & 2.419 & 2.214 & 0 & 0 & 0 & 0 & 0 & 0 & 0 & 0 & 0\\
	
	$f_o$ & 0.845 & 0.760 & 0.575 & 0.523 & 0.084 & 0.168 & 0.096 & 0.330 & 0.333 & 0.148 & 0.206\\
	$f_1$ & 0.065 & 0.024 & 0.061 & 0.227 & 0.031 & 0.151 & 0.100 & 0.649 & 0.191 & 0.899 & 0.054\\
	$f_2$ & 0.124 & 0.010 & 0.104 & 0.050 & 0.140 & 0.150 & 0.135 & 0.121 & 0.659 & 0.393 & 0.166\\
	$f_3$ & 0.011 & 0.071 & 0.723 & 0.166 & 0.530 & 1.149 & 0.106 & 0.638 & 0.547 & 0.187 & 0.706\\
	$f_4$ & 0.840 & 0.601 & 0.638 & 0.030 & 0.130 & 0.825 & 0.729 & 0.453 & 3.576 & 0.001 & 2.590\\
	$f_5$ & 5.646 & 4.384 & 0 & 0 & 0 & 0 & 0 & 0 & 0 & 0 & 0\\ 
\end{tabu}

\vskip6pt
}
\end{table}

\clearpage
\subsection{Dispersion curves}
In this section, as an example, we plot the real and imaginary parts of the refractive index ${\hat{n}(\omega) = n(\omega) + ik(\omega)}$ of Au, Ag, Al and Cu metals.
The optical properties of metals were described phenomenologically using the Lorentz--Drude model, in which parameters of six oscillators were fitted for the best consistency with experiments~\cite{Rakic:98}. The results are shown in Figure~\ref{disperssion_metals}.

\Fig{disperssion_metals}{1}{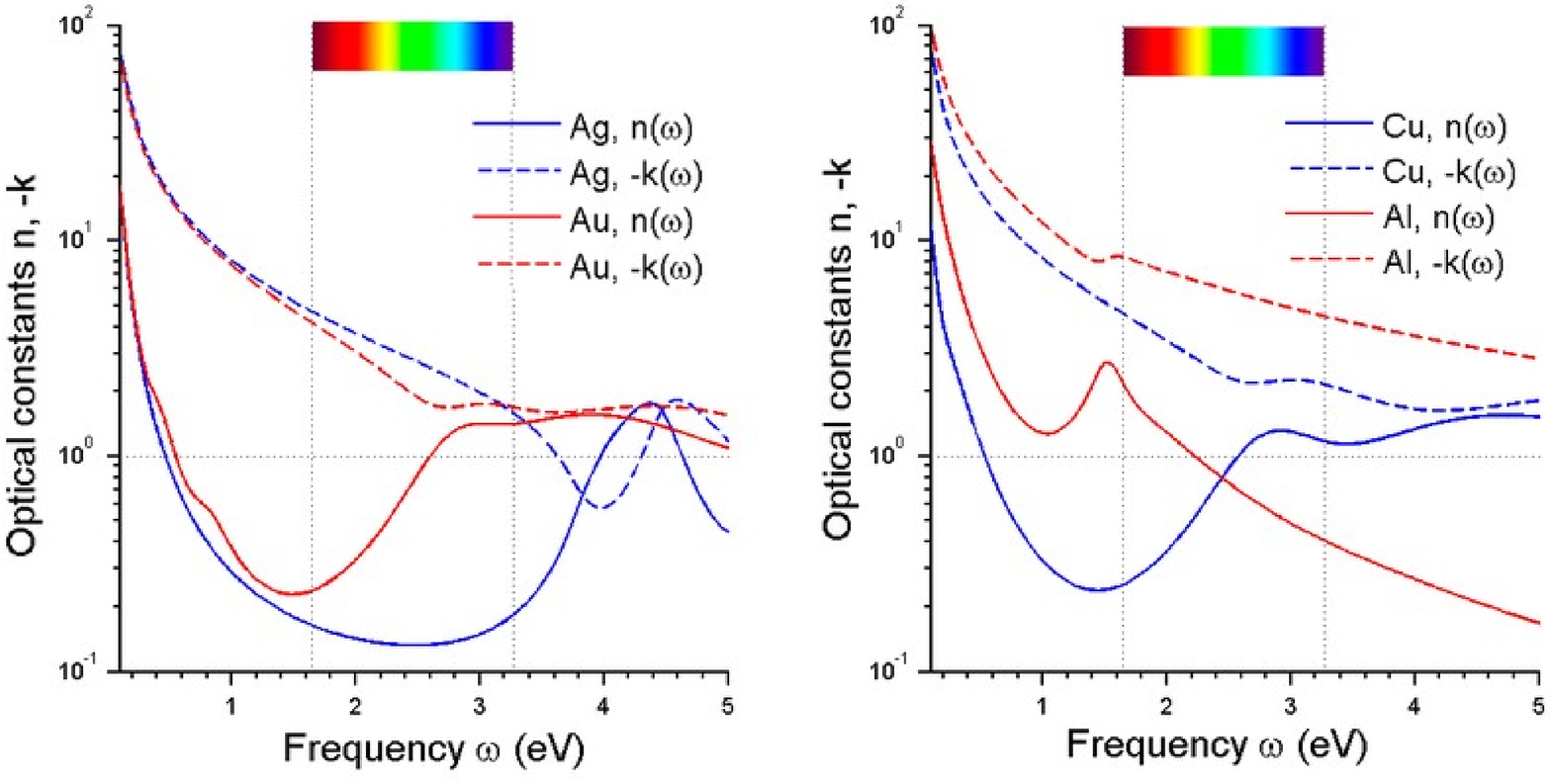}
{Optical constants $n$ and $k$ of Au, Ag, Cu and Al as a function of photon energy.}

It can be seen from Figure~\ref{disperssion_metals} that the simple free electron Drude model can be applied only below the visible band for Ag and Au, where is in the case of Al and Cu the interband absorption resonances are present even in the IR band.

Once the dispersion curves of various metals are known, we can proceed with simulation of various metal-based coatings.

%% file: Chap_Material_Mixtures.tex
\section{Optical properties of mixtures and rough surfaces}
\label{Optical_properties_of_mixtures} 

In this section we review a simple yet efficient approach allowing us to describe a nonuniform metal coating, rough surfaces and surfaces coated with nanoparticles.

\subsection{Local field effects and effective medium theory}

Probably the simplest way to calculate dielectric permittivity of a given material with inclusions of another material is to apply the effective medium theory. 
The effective medium theory describes macroscopic properties of a medium based on the properties and the relative fractions of its components. 

In 1909 Lorentz~\cite{Lorentz:1909} had shown that the dielectric properties of a substance can be related to the polarizabilities $\alpha$ with the Clausius-Mossotti (or Lorentz-Lorenz) relation~\cite{Lorenz:1880, Mossotti:1850, Jackson:1998}: 

\Eq{eq:EMT1}{ \alpha = \frac{3}{N} \frac{\epsilon - 1}{\epsilon + 2},}
where $N$ is the number of dipole particles per unit volume and $\epsilon$ is the dielectric permittivity. 
The derivation goes as follows~\cite{BornWolf:1999, Jackson:1998}: an external field $E_{ext}$ induces local molecular dipole moments $p_j$ at each $j$th site, proportional to the local field $E_{local}$, with the proportionality constant $\alpha$ - called the atomic polarizability:
\Eq{eq:polariz}{ p = \alpha E_{local}.}
In the general case $\alpha$ is a complex number, meaning that the polarization may be shifted in phase with the external electric field, dependent on the external field frequency. 
The polarization can be a tensor for a non isotropic material.
The local field $E_{local}$ is not equivalent to the external field $E_{ext}$ and should be corrected 1) by taking into account the field from all other dipoles (excluding the one under consideration) and 2) in some cases by including the dipole radiation, if the external field oscillates at a sufficiently high frequency.

The first effect is considered by subtracting a microscopic sphere from the continuous medium and finding the electric field inside the formed imaginary cavity. The sphere diameter has to be chosen much smaller than the wavelength, hence the average field may be assumed to be spatially homogeneous.
The resulting field is the local field $E_{local}$ (which corrects the $E_{ext}$ field) and is directly connected to the induced local dipole moment $p$. 
The macroscopic polarization $P$ is simply a sum of all the local microscopic dipole moments. 
Then macroscopic observables, such as the dielectric function of a medium, can be easily deduced from the known connection between an external field $E_{ext}$ and the correspondent macroscopic polarization $P$.

The second effect (the dipole radiated field) can be neglected in many cases, and the result can be viewed as a zero-frequency limit~\cite{Jackson:1998}. If the effect of the dipole radiative interaction is significant it can be accounted with the discrete dipole approximation method.

In the presented work we are interested in the calculation of the dielectric permittivity of particles made of the same material (the inclusions) which are embedded in another material (the host). The derivation of the general optical mixing formula can be found in~\cite{Stenzel:2005, Jackson:1998, Aspnes:1982} and written in the following form:

\begin{equation} \label{eq_MixMed}
\frac{\epsilon - \epsilon_h}{\epsilon_h + (\epsilon-\epsilon_h)L} = \sum_j p_j \frac{\epsilon_j - \epsilon_h}{\epsilon_h + (\epsilon-\epsilon_h)L},
\end{equation}
where\\
$\epsilon$ - is the effective dielectric function.\\
$\epsilon_j$ - the dielectric functions of the constituents.\\
$\epsilon_h$ - is the dielectric function of the host material.\\
$L$ - is the depolarisation factor, defined by the morphology. For spherical inclusions $L = 1/3$.\\
$p_j = V_j/V$ - is the filling factor of the constituents, also called the volume fraction.\\
$V$ - is the full volume occupied by mixtures,\\
$V_j$ - is the volume fraction occupied by $j$th constitute.

By making several assumptions the general form of equation~(\ref{eq_MixMed}) can be reduced to a set of particular cases, best suited for certain types of media, such as aerosols, porous media and metamaterials.

\subsubsection{Maxwell Garnett (MG) approach}

If one of the constituents is regarded as the host material, and the others as inclusions, the equation~(\ref{eq_MixMed}) can be rewritten as Maxwell--Garnett (MG) equation~\cite{Maxwell:1906, Mallet:2005}:

\begin{equation} \label{eq:eqMG}
\frac{\epsilon - \epsilon_m}{\epsilon_m + (\epsilon-\epsilon_m)L} = \sum_{j \ne m} p_j \frac{\epsilon_j - \epsilon_m}{\epsilon_m + (\epsilon-\epsilon_m)L},
\end{equation}
where $\epsilon_m$ is the dielectric functions of the host material.

It should be noted that the result depends on whether the first material is embedded into the second or the second material is considered to be embedded in the first material.

\subsubsection{Lorentz-Lorenz (LL) approach}

The Lorentz--Lorenz approach is based on the assumption that the host material is a vacuum (i.e. $\epsilon_h = 1$)~\cite{Lorenz:1880, Lorentz:1909}:
 
\begin{equation} \label{eq:eqLL}
\frac{\epsilon - 1}{1 + (\epsilon-1)L} = \sum_{j\ne m} p_j \frac{\epsilon_j - 1}{1 + (\epsilon-1)L} .
\end{equation}

\subsubsection{Effective medium approximation (EMA) or Bruggeman approach}

Assuming that the effective dielectric function acts as a host medium for inclusions,~\textit{i.e.} $\epsilon = \epsilon_h$, the resulting equation~(\ref{eq_MixMed}) can be rewritten with the left hand side equated to zero~\cite{Bruggeman:1935}:

\begin{equation} \label{}
0 = \sum_j p_j \frac{\epsilon_j - \epsilon}{\epsilon + (\epsilon-\epsilon)L} .
\end{equation}

\subsubsection{Discussion}

One of the approaches is most effective depending on the particular type of medium~\cite{Stenzel:2005}.
The MG theory should be applied when constituents clearly may be subdivided into the inclusions and the matrix material.
The EMA theory works best in the presence of molecular mixtures, where a clear subdivision into inclusions and the host material is not possible.
The LL approach is best suited for porous materials.
The effective medium theory can be viewed as the first order approximation, allowing us to obtain the effective permittivity of the host material with the inclusions.

One of the objectives of the presented work is to obtain optical constants for various metallic films deposited on the fibre surface with either the chemical or CVD technique. Two examples of the film morphology are shown in Figure~\ref{fig_films}.

\Fig{fig_films}{0.9}{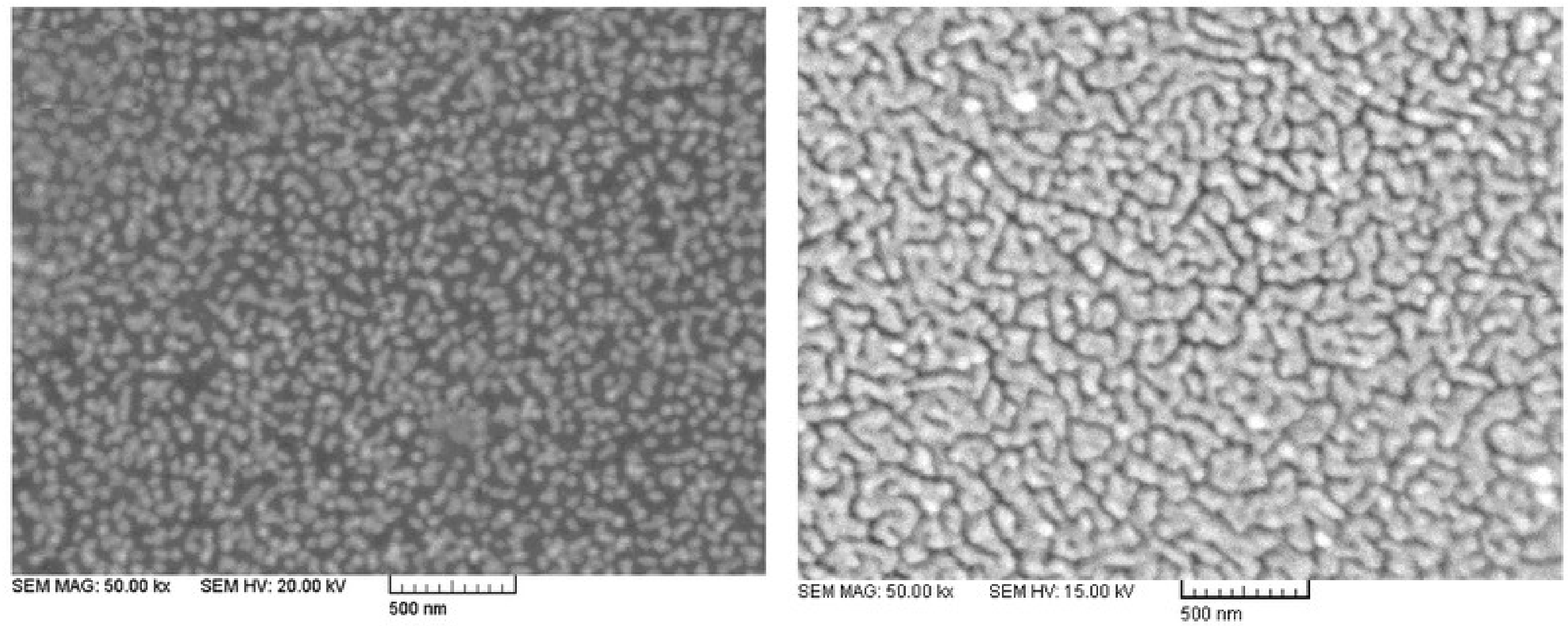}
{SEM images of gold film surface morphology, taken at different stages of chemical deposition~\cite{Bialiayeu:2011}.}

The effective optical constant can be obtained in fairly straightforward way using the Bruggeman effective medium theory (EMT). 
An example of EMT application to a thin film nucleation and growth on a silica wafer can be found, for example, in~\cite{Collins:1998}.




\subsection{The exact numerical method accounting for the interaction between film elements.}
\label{NP_FDTD_sim} 

The full numerical simulation of the optical properties of a film can be used as an alternative to the simple mixtures models described in the previous section. 

The absorption, and thus the imaginary part of the refractive index, can be obtained if the incident and scattered intensity are known. 
Next, the real part of the refractive index can be found by implementing the Kramers-Kronig integration method, as was described in Section~\ref{subsec_KK}.
The idea here was to use the Finite-Difference Time Domain (FDTD) method to compute the film absorption properties by illuminating it with a short light pulse. The absorbed power can be found by taking the difference between the incident and scattered intensities.
In order to separate the incident and reflected pulses in time domain, either a large simulation area in the space domain has to be chosen or the pulse should also be sufficiently short in the time domain. 
However, the pulse should be sufficiently long so that its spectral image is sufficiently narrow, allowing the required spectral resolution to be achieved.
Therefore, several contradictory conditions have to be met.

We chose \C{``FDTD-method Maxwell solver''} from Lumerical Solutions for our simulations.
The software package was designed specifically for electromagnetic field simulations and is used extensively. Examples of its applications to the simulation of nanopoarticles can be found, for example, in~\cite{Chowdhury:2008, Tanev:2009}.

In our research we simulated the rough metal film shown in Figure~\ref{fig_films}. The film was approximated by a metallic film substrate in which spherical NP inclusions were immersed, as shown in Figure~\ref{My_FDTD_Lumerical}. The metal film and immersed nanoparticles were chosen to have an identical refractive index.
\Fig{My_FDTD_Lumerical}{0.8}{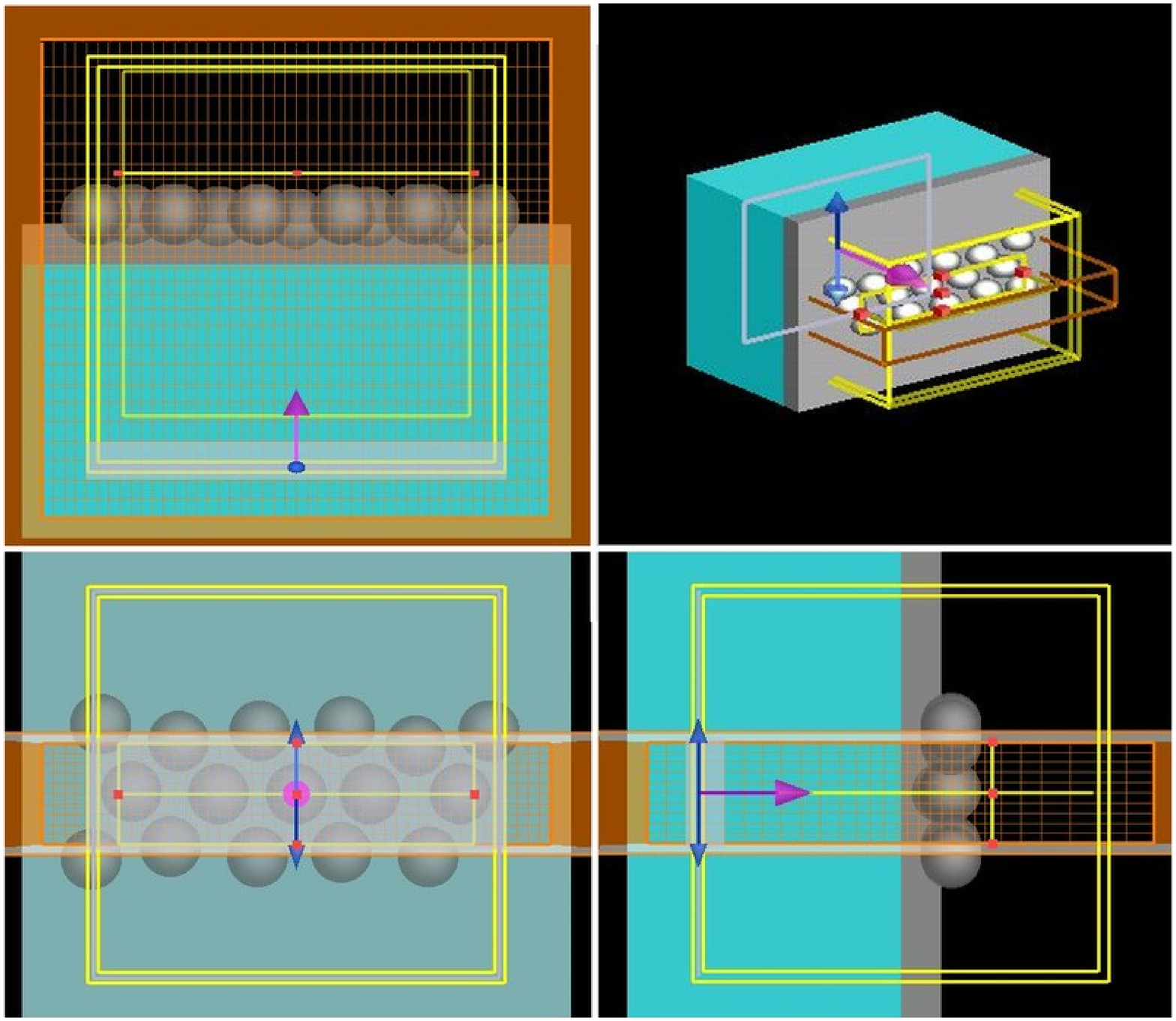}
{Simulation of light scattering by a rough silver film deposited on the glass substrate.}
Considering the enormous amount of time required for FDTD simulation, only a small area with randomly assign roughness was simulated. 
The sample was assumed to be infinite. This condition was satisfied by surrounding the sample with four planes, transverse to its surface, which enforced the periodic boundary condition. 
The perfect matching layers were placed above and beyond the sample, in order to cancel reflection from the  computational domain boundaries, as shown in Figure~\ref{My_FDTD_Lumerical}.
The pulse was set to incident the film surface from within the glass substrate, which is the case for fibre optical sensors. 
The incident and scattered fields were measured with twelve power sensors (2D rectangular plates) detecting the incident field. 
The first six sensors were arranged in a 3D cube configuration including the simulation region but excluding the emitting antenna.
The last six sensors were arranged into the outer cubic shell confining all the simulated objects including the antenna.
The inner and the outer shells were set to detect only ingoing (antenna) and outgoing (scattering) radiation energies, respectively.
The absorbed energy was computed by taking the difference between the incident and scattered energies.
Alternatively, the system geometry can be chosen such that the incident and the scattered pulses can be separated in time domain. 
Repeating the measurements for a number of frequencies, in the bandwidth of interest, the absorption spectrum was obtained. 

Although the proper 3D film simulation takes considerable amount of time, more than $24~h$, some useful conclusions can be drawn from a simple 2D simulation. 
The result of light interaction with an array of spherical nanospheres is shown in Figure~\ref{My_FDTD_Lumerical2}. The obtained simulation results helped to explain the observed anomalies in transmission spectra of a metal coated TFBG sensor~\cite{Bialiayeu:2013}, \C{which} was my contribution to that work.

\Fig{My_FDTD_Lumerical2}{0.6}{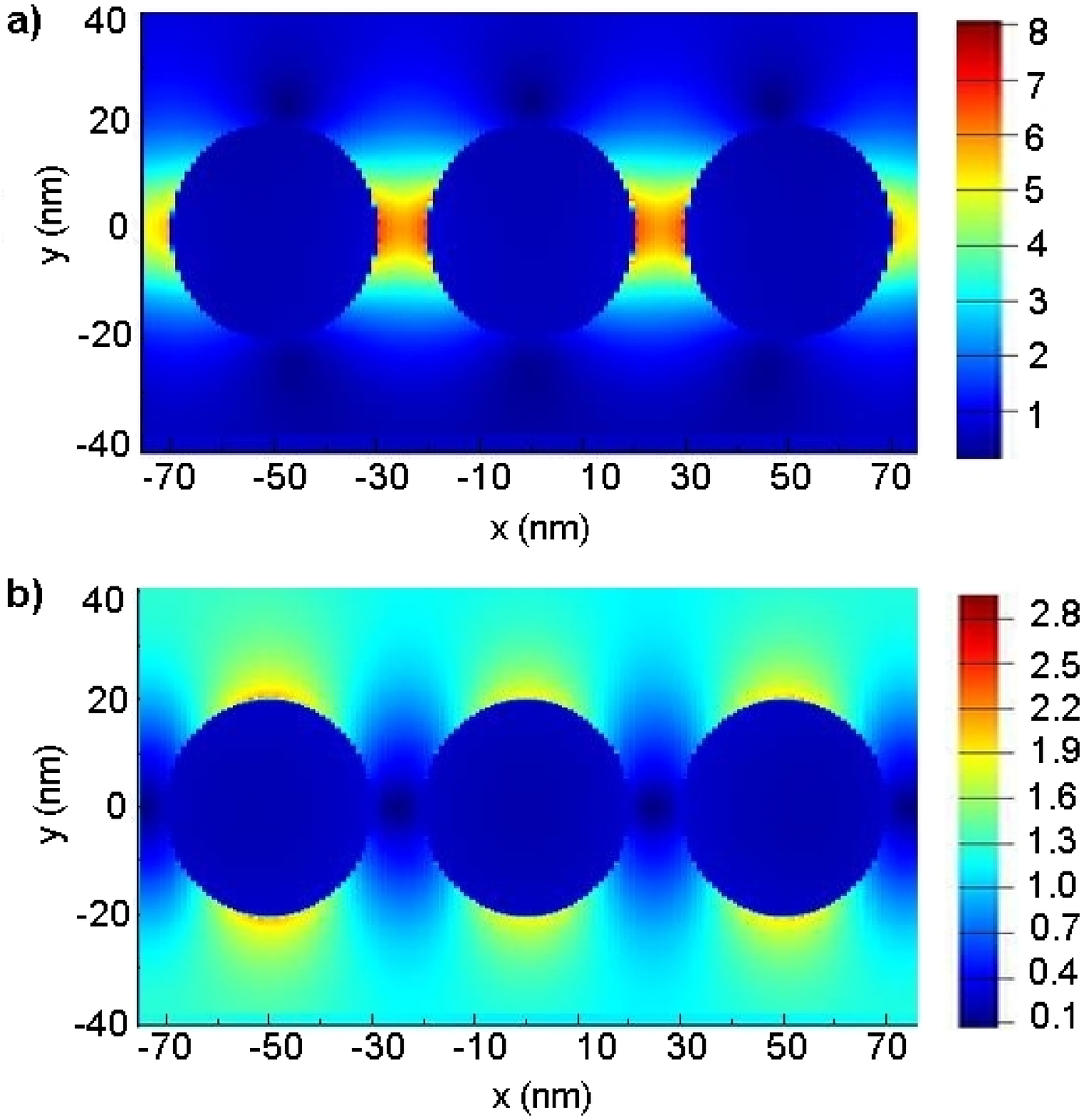}{Electric field of S (a) and P (b) polarized light interacting with an array of metallic spheres~\cite{Bialiayeu:2013}.}

It should be further noted that the 2D simulation took less than $1~h$, wheres a full 3D simulation requires more than $24~h$. Furthermore, if the goal is to establish polarization or angular dependency, the parametric space should be scanned (by changing values of the incident angle and polarization state) which makes this approach prohibitively time expensive. In such a case it would be reasonable to choose a different software package capable of running on a supercomputer, or to use a different simulation technique, for example the discrete dipole approximation, described in the following section.

%% file: Chap_Material_DDA.tex
\section{Discrete dipole approximation (DDA) or coupled dipole approximation (CDA) method }

The discrete dipole approximation (DDA) is an extremely flexible and powerful numerical method used for computing the scattering and absorption of electromagnetic waves by a target of arbitrary geometry, whose dimensions are comparable or larger than the incident wavelength.

The basic idea of the DDA was introduced in 1964 by DeVoe with application to the optical properties of molecular aggregates~\cite{DeVoe:1964, DeVoe:1965}. At first the retardation effects were not included, and treatment was limited to aggregates that were small compared with the wavelength. The retardation effects were added in 1973 by Purcell and Pennypacker~\cite{Purcell:1973}. 
The accuracy of the DDA method had been confirmed by comparing it against the analytical results and experimental measurements~\cite{Penttila:2007, Draine:2008}.

The DDA method was popularized by Draine and Flatau. In 1993 they distributed discrete dipole approximation open source code DDSCAT~\cite{Draine:1994, Draine:2008}, which is now freely available. 
DDSCAT is an open-source Fortran-90 software package applying the discrete dipole approximation (DDA) to calculate the scattering and absorption of electromagnetic waves by targets with arbitrary geometries and a complex refractive index.
The target may be an isolated object as well as 1-d or 2-d periodic arrays of target unit cells. The method can be used to study absorption, scattering, as well as electric fields.
The description of the latest version DDSCAT 7.2 can be found in~\cite{Draine:2012}.
At the present day a few other DDA implementations are available.
The highly optimized computational framework OpenDDA was written in C language by J.~M.~Donald at University of Ireland~\cite{Donald:2009}. Another implementation of DDA, called ADDA has been developed through over a period of ten years at University of Amsterdam by M.~A.~Yurkin and A.~G.~Hoekstra. Their software package is capable of running on multiple processors and clusters of computers under Linux Operating system~\cite{Yurkin:2007}.



The DDA method implies an approximation of a continuum target (scatterer) by an ensemble of polarizable points in a cubic lattice. 
The polarizable points acquire dipole moments in response to the local electric field.
Each of the dipoles is driven by an incident electric field and by contributions from the other dipoles, hence the method is also known as the coupled dipole approximation (CDA) method.
These interacting dipoles give rise to a system of linear equations, from which a self-consistent solution for the oscillating dipole moments is found, and next the absorption and scattering cross sections are computed.
If DDA solutions are obtained for two linearly independent states of polarization of the incident wave, then the complete amplitude scattering matrix for a given scatterer can be determined as well.
An extensive review of the DDA method and its applications was given in~\cite{Draine:1994, Mishchenko:2000}.
The theory behind the DDA method was discussed in~\cite{Tsang:2000, Kahnert:2002, Yurkin:2007, Chiappetta:1997, Smith:2006} here we only show the idea behind the derivation. 

Let us assume an array of $N$ polarizable point dipoles located at $ \vec r_j $ with each one characterized by a polarizability $ \alpha_j $.
The system is excited by a monochromatic incident plane wave: 

\begin{equation} \label{eq:DDA1}
	\vec E_{loc}(\vec r,t) = \vec E_o e^{i (\vec k \vec r - \omega t)},
\end{equation}
inducing a dipole moment in each polarizable point:

\begin{equation} \label{eq:DDA2}
	\vec P_j = a_j \vec E_{j,loc},
\end{equation}
where index $j$ denotes position $\vec r_j$ of the $j$th element and $\vec P_j$ is the dipole moment of the $j$th element.

Each dipole of the system is subjected to an electric field that can be split in two contributions: 
the incident radiation field $\vec E_{in}$, plus 
the field radiated by all the other induced dipoles $\vec E_{dp}$. 
The sum of both fields is the so-called local field at each dipole:

\begin{equation} \label{eq:DDA3}
	\vec E_{j,loc} = \vec E_{j,in} + \vec E_{j,dp}.
\end{equation}

Now excluding $\vec E_{j,loc}$ from the consideration by substituting (\ref{eq:DDA2}) into (\ref{eq:DDA3}) we obtain

\begin{equation} \label{eq:DDA4}
	\frac{1}{a_j} \vec P_j - \vec E_{j,dp} = \vec E_o e^{i \vec k \vec r_j}.
\end{equation}

Now let's consider the field radiated by all the other dipoles $\vec E_{j,dp}$. 
First we use the well-known expression for radiation of a dipole located at position $\vec r_k$ evaluated at point $\vec r_j \neq \vec r_k $~\cite{Jackson:1998}: 

\begin{equation} \label{eq:DDA5}
\vec E_{jk} = \frac{e^{i \vec k \vec R_{jk}}}{R_{jk}} \left (k^2 \vec n_{jk} \times (\vec n_{jk} \times \vec P_k) + \left (\frac{1}{R_{jk}^2} -\frac{ik}{R_{jk}}\right)(3 \vec n_{jk} \cdot (\vec n_{jk} \cdot \vec P_k) - \vec P_k) \right )
\end{equation}
where $\vec R_{jk}= \vec r_j - \vec r_k$, and $\vec n_{jk}$ is a unit vector pointing from $\vec r_k$ point to $\vec r_j$ point.

Taking a close look at equation~(\ref{eq:DDA5}) we note that at the given $j$ and $k$ this equation simply expresses components of the $\vec E_{jk}$ vector in terms of linear combination of the $\vec P_j$ vector components. Thus we can rewrite this equation in a matrix form:

\begin{equation} \label{eq:DDA13}
\vec E_{jk} = [A_{jk}] \vec P_k,
\end{equation}
where $[A_{jk}]$ is a $3 \times 3$ matrix.

Now summing all the fields from all $k \neq j$ dipoles we obtain:

\begin{equation} \label{eq:DDA6}
	\vec E_{j,dp} = \sum_{j \neq k} [A_{jk}] \vec P_k .
\end{equation}
Basically we obtained a representation of $\vec E_{j,dp}$ vector in terms of all the $3\cdot(N-1)$ components of all the $\vec P_j$ vectors where $j \neq k$. 

Finally, (\ref{eq:DDA4}) can be rewritten as:

\begin{equation} \label{eq:DDA7}
	\frac{1}{a_j} \vec P_j - \sum_{j \neq k} [A_{jk}] \vec P_k = \vec E_o e^{i \vec k \vec r_j} .
\end{equation}
This is the final equation that has to be solved, consisting of $N$ coupled vectorial equations from which $N$ vectors $\vec P_j$ can be found, thus providing a solution to the scattering problem. It is more convenient to formulate the problem as a set of scalar equations, therefore we can simply rewrite the above equations as:
\begin{equation} \label{eq:DDA8}
	[\hat A] \hat P = \hat E_{in} .
\end{equation}
For $N$ dipole elements the $[\hat A]$ is a $3N \times 3N$ symmetric matrix, $\hat P~=~(\vec P_1, ..., \vec P_N)$ and $\hat E_{in} = (\vec E_o e^{i \vec k \vec r_1}, ..., \vec E_o e^{i \vec k \vec r_N})$ are $3N$ vectors. 

Once the polarizations $\vec P_j $ are found, the extinction, absorption and scattering cross section can be expressed as follows~\cite{Draine:1988}:

\begin{equation} \label{eq:DDA9}
	C_{ext} = \frac{4 \pi k}{ |E_{in}|^2} \sum_j^N Im[\vec E_{j, in}^* \cdot \vec P_j],
\end{equation}
the extinction cross section is computed from the forward-scattering amplitude using the optical theorem~\cite{Draine:1988};

\begin{equation} \label{eq:DDA10}
	C_{abs} = \frac{4 \pi k}{ |E_{in}|^2} Im[\vec E_{j, in}^* \cdot \alpha \vec E_in],
\end{equation}
the absorption cross section is obtain by summing rate of energy dissipation by each of the dipoles;

\begin{equation} \label{eq:DDA11}
	C_{sca} = \frac{4 \pi k}{ |E_{in}|^2} \int \left | \sum_j^n \left [\vec P_j - \vec n \cdot (\vec n \cdot \vec P_j)\right] e^{- i k (\vec n \vec r_j)} \right |^2 d \Omega ,
\end{equation}
where $\vec n (\Omega)$ is a unit vector in direction of scattering and $d \Omega$ is the solid angle element.

%% file: Chap_Material_Conclusion.tex
\section{Conclusion}
In this chapter we have reviewed the optical properties of materials, briefly discussed experimental methods which are used to determine the optical constants $n$ and $k$. We have chosen parameters for eleven metals commonly used in optics and plotted their dispersion curves, which are extensively used in the following simulations.

The morphology of various films can be modeled with the simplest mixtures model. We are going to use the Maxwell--Garnett model to compute the effective refractive index of rough films, and the Lorentz--Lorenz model to determine the refractive indices of liquid mixtures.  

The properties of nanoparticles are discussed in the next Chapter, where we will obtain absorption coefficients and deduce the effective optical constants of nanoparticles. 
The obtained effective permittivity of nanoparticles can be considered as the permittivity of inclusions in the Maxwell--Garnett model. 
Thus it should be possible to characterize metallic films of various roughnesses as well as films consisting of nanoparticles. 

A more accurate result can be obtained if the radiative dipole-dipole interactions between the film elements are considered. Unfortunately the FDTD method is prohibitively time expensive, although we did \C{reveal} some useful results. 

Alternatively we might consider the DDA method. For the sake of simplicity, properties of a particular particle can be found with other methods, such as the Mie theory discussed in the following Chapter, and then used to define the starting points of the DDA method. Therefore all the elements of nanoparticle-based coating become intrinsically interconnected, thus allowing us to account for a collective effect. 
However, for sparse coatings the analysis can be solely based on optical properties of one single nanoparticle.

%% file: Chap_NP_Lit.tex
\chapter{Optical properties of nanoparticles}
\label{NP_lit_rev}

In this chapter we discuss optical properties of metal nanoparticles. 
Next we present results of our simulations.
The results of simulation would allow us to deduce parameters for an optimal nanoparticle based coating enchaining the TFBG sensor sensitivity.

\section{Review}
\subsection{Characterization of light scattering by particles}
We start by introducing parameters which are usually used for particles characterization.
Let us consider a single particle illuminated by a plane wave with intensity $I_{inc}$, than the total power scattered by this particle is $W_{sca}$, defined as: 
\Eq{}{W_{sca} = C_{sca} I_{inc},}
here $C_{sca}$ is the scattering cross section, with dimensions of area.

Particles can also absorb electromagnetic radiation. 
The rate of absorption $W_{abs}$ should also be proportional to the incident intensity: 
\Eq{}{W_{abs} = C_{abs} I_{inc},}
where $C_{abs}$ is the absorption cross section.
 
The sum of the scattering and absorption sections is called the extinction cross section: 
\Eq{}{C_{ext} = C_{sca} + C_{abs}.}


In practice, the cross sections are usually normalized by the particle’s area projected onto a plane perpendicular to the incident beam, and defined as $Q_{ext}$, $Q_{sca}$, $Q_{abs}$. 

\subsubsection{polarization-dependent properties}
Particles can be thought as miniature polarizers and retarders. Assuming that the incident field is polarized, the polarization of the scattered field would depend on the nanoparticle shape, size and the direction of scattering. Even unpolarized light can become partially polarized upon particular direction of scattering.

The polarization properties of a particle can be described, for example, with Jones matrix. 
Assuming that the incident wave, with the wavenumber $k$, propagates along the $z$ axis, the general relation between incident and scattered fields can be written in the following from~\cite{Huffman:1983}:
\Eq{}{\V E_{sca} = \frac{e^{i \V k(\V r - \V z)}}{-i k r} \lbrack S \rbrack \V E_{inc}.}
For the known direction of incident light and a particular chosen direction of the scattered light, the scattering plane can be introduced, spanned by these two directions. 
Now fields can be decomposed into two orthogonal components, one parallel, the other perpendicular to the scattering plane: 
\Eq{}{ \begin{pmatrix} E_{\parallel,sca} \\E_{\bot,sca} \end{pmatrix} = \frac{e^{i \V k(\V r - \V z)}}{-i k r} \begin{bmatrix} S_{11} & S_{12} \\ S_{21} & S_{22} \end{bmatrix} \begin{pmatrix} E_{\parallel,inc} \\ E_{\bot,inc} \end{pmatrix},}
The elements $S_{jk}$ of the scattering matrix $\lbrack S \rbrack$ (or Jones matrix) are complex-valued functions of the scattering direction defined, for example, by $\theta$ and $\phi$ in polar coordinate system.

The scattering Mueller matrix can also be introduced, connecting the Stokes parameters, which can be measured directly~\cite{Huffman:1983}: 
\Eq{}{ \begin{pmatrix} I_{sca} \\Q_{sca} \\U_{sca} \\V_{sca} \end{pmatrix} = \frac{1}{k^2 r^2} \begin{bmatrix} S_{11} & S_{12} & S_{13} & S_{14} \\ S_{21} & S_{22} & S_{23} & S_{24} \\ S_{31} & S_{32} & S_{33} & S_{34} \\ S_{41} & S_{42} & S_{43} & S_{44} \end{bmatrix} \begin{pmatrix} I_{inc} \\Q_{inc} \\U_{inc} \\V_{inc} \end{pmatrix},}
here
\Eqaa{}
{I = \frac{k}{2\omega\mu} \langle E_{\parallel} E_{\parallel}^* + E_{\bot} E_{\bot}^* \rangle,} 
{Q = \frac{k}{2\omega\mu} \langle E_{\parallel} E_{\parallel}^* - E_{\bot} E_{\bot}^* \rangle,} 

\Eqaa{}
{U = \frac{k}{2\omega\mu} \langle E_{\parallel} E_{\bot}^* + E_{\bot} E_{\parallel}^* \rangle,} 
{V = \frac{k}{2\omega\mu} \langle E_{\parallel} E_{\bot}^* - E_{\bot} E_{\parallel}^* \rangle .} 
It can be noted that all the parameters $I, Q, U, V$ are real numbers and can be measured by optical means with quadric detector, without the necessity to measure the phase of the signal, unlike in the case of characterization by the Jones matrix. This subject was discussed in more detail in Chapter~\ref{Chap_polarization}.

An example calculation can be found in reference~\cite{1996:Hovenier}, where polarization properties of light scattered by four different homogeneous particles: a prolate spheroid, an oblate spheroid, a finite cylinder and a bisphere with touching components were considered. 
The exact analytical solution exists for such scattering problems, which made it possible to verify T-matrix and DDA methods techniques~\cite{1996:Hovenier}. 
The particle was illuminated by a plane harmonic wave propagating along the positive z-axis. 
The scattering matrix ${[S] = S_{ij}}$ was computed as a function of the scattering angle. 
It was shown for all scattering angles the elements $S_{13}$, $S_{14}$, $S_{23}$, $S_{24}$, $S_{31}$, $S_{32}$, $S_{41}$, $S_{42}$ vanish and $S_{11} = S_{22}$, $S_{12}= S_{21}$, $S_{34}= - S_{43}$, $S_{33}= S_{44}$. 
Therefore only the remaining elements $S_{11}$, $S_{21}$, $S_{33}$, $S_{43}$ were considered.
A strong polarization dependence was revealed, especially in the case of prolate spheroid.
Another example, revealing a strong polarization dependence of optical absorption can be found in~\cite{Noguez:2007}, where a prolate spheroid with an aspect ratio of $1.6$ and a major axis of $8$ nm embedded in silica was illuminated with linear polarized plane wave.

The extensive study of polarization properties of light scattered by spherical particles can be found in the original Mie's paper~\cite{Mie:1908} as well as in~\cite{Wriedt:2012, Horvath:2009}, where it was shown that for a particle smaller than 100nm the scattered light behaves according to the Rayleigh scattering law and has the maximum polarization at scattering angle ${\theta = 90^o}$. Even if particles are illuminated by unpolarized light, the scattered light becomes partly polarized.
The scattering patterns of a spherical particles were also considered in~\cite{Wriedt:2012}, where computation method was based on numerical evaluation of Mie's coefficients. It was shown that the maximum of polarization moves to ${\theta =120^o}$ with the increase in the size of a particle. For particles with sizes larger than 100nm the degree of polarization diminishes rapidly. Light scattered sideways is always linearly polarized, regardless of the particle size~\cite{Wriedt:2012, Horvath:2009}.

We conclude, that even spherically symmetric particles have nontrivial polarization properties which depend on the scattering angle, particle size and material properties from which the particle is made. Even a small variation in material absorption is causing significant change in the polarization properties of scattered light. 

\subsubsection{Size dependent effects}
The absorption and scattering phenomena can be viewed as a size-dependent process. 
The first extensive review of this subject was given in Mie's original paper~\cite{Mie:1908}, where absorption and scattering spectra were calculated for gold nanoparticles with various sizes. 

The classical Mie's approach can be applied with no limitations to nanoparticles larger than $20$ nm~\cite{Soonnichsen:2001}. For smaller particle the material properties would differ significantly from measured bulk material properties. 
The imaginary part of the dielectric function, which is responsible for the absorption, increases by a factor of $10$ compared to the bulk value for particles of $2.5~nm$ whereas for $20~nm$ particles the difference is only of a factor $1.5-2$, depending on wavelength~\cite{Soonnichsen:2001}. 

For a particle of a few nanometers in size the mean free path of conduction electrons is limited by the particle boundary, which interferes with the electron relaxation time and leads to the increase in absorption at a narrow wavelength range. 
This effect can be taken into account by introducing an extra damping term into the Drude-Sommerfeld free-electron model.

The experimental investigation performed by Soonnichsen~\cite{Soonnichsen:2001} confirmed that the quantum treatment is not necessary for metallic particles larger than $20$nm. 
It was also shown that the collective oscillation dephasing is adequately described by single electron dephasing, following from the classical solution. 
The surface scattering effect and chemical damping by the outside medium can be negligible as well. 
Thus the bulk material properties can be successfully applied to simulation of particles with sizes larger then $20~nm$.



\subsubsection{Shape and substrate dependent effects}

As was mentioned previously, the exact analytic solutions to the scattering problems is known only for a few simple geometries.
Numerical methods, such as discrete dipole approximation, have to be applied in the general case. 

The size dependent scattering and absorption properties of non-spherical particles were studied, for example, in~\cite{Yurkin:2010}. 
Discrete dipole approximation (DDA) method was chosen for simulations of absorption and scattering spectra of gold spheres, cubes and rods, ranging in size from $10$ to $100$ nm.
The problem of light scattering by silver particles with various geometries (spheres, cubes, truncated cubes) was also reviewed in~\cite{Noguez:2007}. First the polarizability was calculated using the Clausius--Mossotti relation with dielectric function of bulk silver. It was observed that the optical response below $325$ nm is independent of a particle shape and is defined by intra-band transitions in silver. 

The optical properties of particles are also strongly dependent on the substrate properties. 
For example, if a nanoparticle is deposited on a substrate, the substrate can be considered as an array of dipoles interacting with the nanoparticle~\cite{Albella:2011}. Alternatively, the model can be simplified by replacing the substrate with an induced image of the nanoparticle, thus dipolar interaction between the particle and its image can be considered instead~\cite{Noguez:2007}. 
However such dipolar interaction leads to the strongly inhomogeneous field, especially when the particle is close to the substrate. In such a case multipolar interactions \C{have} to be considered~\cite{Ruppin:1983}.

%% file: Chap_NP_Scat_Analit.tex

\section{Simulation of light scattering by small particles}
\label{Light_scattering_by_small_particles}

The problem of light scattering by a spherical particle made of an arbitrary material, including absorbing materials such as metals, can be solved exactly by analytical methods. The exact solutions are also known for spheroids, or infinite cylinders.

The solution is usually obtained by expanding the incident, scattered, and internal fields in a series of vector spherical harmonics. The expansion coefficients are found from the boundary condition ensuring continuity of tangential components of the electric and magnetic fields across the surface of a particle. Observable quantities are expressed in terms of the coefficients.

An interesting overview of the early work was given in~\cite{Logan:1966, Horvath:2009, Kerker:1969}, an extensive Mie's theory review can be found in~\cite{Wriedt:2012, Kerker:1969}.
As early as in 1863 Alfred Clebsch (1833--1872) published a general solution of the elastic wave equation in terms of the vector wave functions~\cite{Clebsch:1863} and Ludvig Lorenz (1829--1891) completely solved the problem in terms of the ether theory.
Lord Rayleigh (John Strutt) (1842--1904) introduced his theory of elastic light scattering by small dielectric particles in 1871~\cite{Rayleigh:1877, Rayleigh:1871}. The particles were considered to be much smaller than the wavelength of the light so that the general electromagnetic problem can be reduced to electrostatic problem of an isotropic, homogeneous, dielectric sphere in a uniform field.

The interaction of electromagnetic waves with a perfectly conducting sphere was reviewed by Thomson~\cite{Thomson:1893}.
The solution was generalized by Hasenorl~\cite{Hasenorl:1902} who introduced a non-zero conductance for metals. His paper actually already gives the full solution contained in Mie's later publication~\cite{Horvath:2009}. 
Ehrenhaft~\cite{Ehrenhaft:1905} gave a rigorous treatment of the scattering of light by small absorbing spheres, which is an even more elegant than Mie's work~\cite{Horvath:2009}.
A similar treatise was given by Debye~\cite{Debye:1909}, who provided an alternative approach by utilizing two scalar potential functions.

Finally in 1908 Gustav Mie presented his solution to the problem in recognizably modern notation, and presented a comprehensive comparison of experimental results with theoretical findings, providing many numerical examples~\cite{Mie:1908}. 
Although in his calculation Mie was using only three terms in the expansion series it was sufficient to completely explain all the optical phenomena observed until then.
All Mie's conclusions published in his paper from 1908 remain valid.
Today the problem of light scattering by a homogeneous isotropic sphere is firmly attached to the name of Gustav Mie, although Mie was rather the last one who solved this problem.

Mie had to limit his results to particles smaller than $200~nm$ due to the limits imposed by calculation. The calculations were performed by hand, and only three first terms in the infinite series were considered.
He also limits his finding to spherical particles, although he knew that ellipsoidal particles can be treated in the similar way~\cite{Horvath:2009}.
Today, with availability of computers, these limitations can be easily removed.


The exact solutions to Maxwell's equations are known only for special geometries such as spheres, spheroids, or infinite cylinders, so approximate methods are required for the general case.
An excellent review of various methods can be found in~\cite{Kahnert:2003, Kahnert:2010, Wriedt:1998, Jones:1999, Chiappetta:1997, Tsang:2000}. Mie's theory is usually used as reference, to validate other methods.


\subsection{Analytical solution to the problem of electromagnetic wave scattering on spherical particles}
\label{Mie_analit}
In this section we briefly outline the steps taken to solve the problem of electromagnetic wave scattering on small spherical particles. An excellent review of the subject can be found in~\cite{Mishchenko:2000, Tsang:2000, Kahnert:2002, Huffman:1983}.

We start with formulation of the problem. 
Let us assume that an incident electromagnetic field $\V E_{inc}$, $\V H_{inc}$ interacts with a particle that occupies a bounded region.
The total fields $\V E_{tot}$, $\V H_{tot}$ in the surrounding medium are equal to the sum of the incident and the scattered fields:
\Eqaa{}
{\V E_{tot} &=& \V E_{inc} + \V E_{sca}}
{\V H_{tot} &=& \V H_{inc} + \V H_{sca}.}

We are interested in a free space solution outside the particle boundaries, therefore we assume that there are no sources in space. 
Hence, Maxwell's equations can be reduced to the vectorial Helmholtz equations, as was shown in Chapter~\ref{Mode_Solver}:
\Eqaa{eq_Hlmz}
{(\nabla^2 + k^2(\V r)) \V E(\V r) & =& 0,}
{(\nabla^2 + k^2(\V r)) \V H(\V r) &=& 0,}
where 
\Eq{}{k^2(\V r) = \epsilon(\V r) \mu(\V r) \frac{\omega^2}{c^2} = n^2(\V r) k_o^2.}

Now let us consider the boundary condition: (i) the tangential components of the total fields have to be continuous across the particle surface; (ii) the radiation condition, requiring that the tangential components of the electric and magnetic fields have to approach zero at rate $1/r$ as the distance $r$ from the origin approaches infinity.

The solution to the vector Helmholtz equations~(\ref{eq_Hlmz}) can be searched in terms of scalar basis functions, obtained as a solution to the scalar Helmholtz equation:
\Eq{eq_Hlmz_sc}{(\nabla^2 + k^2(\V r)) \psi (\V r) = 0.}
In spherical coordinates equation~(\ref{eq_Hlmz_sc}) becomes:
\Eq{}{\frac{1}{r}\frac{\partial^2}{\partial r^2}(r \psi) + \frac{1}{r^2 \sin(\theta)}\frac{\partial}{\partial \theta} \left(\sin(\theta)\frac{\partial \psi}{\partial \theta}\right) + \frac{1}{r^2 \sin^2(\theta)}\frac{\partial^2 \psi}{\partial \phi^2} + k^2 \psi = 0 .}
Separating the variables with \textit{ansatz}
\Eq{eq_spher_ansatz}{\psi(r,\theta,\phi) = R(r)\Theta(\theta)\Phi(\phi),}
the three ordinary differential equations are obtained:
\Eqaaa{eq_spher3}{0 &=& \left( \frac{d^2}{d r^2} + k^2 - \frac{l(l+1)}{r^2}\right) (r R), }
{0 &=& \frac{1}{\sin(\theta)} \frac{d}{d \theta} \left(\sin(\theta) \frac{d \Theta}{d \theta }\right) + \left(l(l+1) - \frac{m^2}{\sin^2(\theta)} \right)\Theta, }
{0 &=& \left( \frac{d^2}{d \phi^2} + m^2 \right)\Phi. }

The solution to the radial equation in~(\ref{eq_spher3}) can be expressed in terms of special functions $u_l^{(k)}$:
\Eq{}{R(r) = u_l^{(k)},} 
where\\
$u_l^{(1)} = J_l(r)$ are spherical Bessel functions,\\ 
$u_l^{(2)} = N_l(r)$ are spherical Neumann functions,\\ 
$u_l^{(3)} = H_l^{(1)}(r) = J_l(r) + i N_l(r)$ are spherical Hankel functions of the first kind and \\
$u_l^{(4)} = H_l^{(2)}(r) = J_l(r) - i N_l(r)$ are spherical Hankel functions of the second kind. 

The solutions to the polar equation in~(\ref{eq_spher3}) are the associated Legendre functions: 
\Eq{}{\Theta(\theta) = P_l^m(\theta),} 
where $ m = 0, \pm 1, .... \pm l $. 

Finally the solutions to the azimuthal equation are simply harmonic functions: 
\Eq{}{\Phi(\phi) = e^{i m\phi}.}

Thus the solution to the scalar Helmholtz equation in spherical coordinates can be written in the following form:
\Eq{eq_HlmzSol}{\psi_{l,m}^{(k)}(r,\theta,\phi) = u_l^{(k)} P_l^m(\theta)e^{i m\phi}.}
Using the above scalar basis function we can create vectorial basis functions:
\Eqaaa{eq_SperWave}{\V L_{l,m}^{(k)}(\V a) &=& \nabla \cdot ( \psi_{l,m}^{(k)} \cdot \V a),} 
{\V M_{l,m}^{(k)}(\V a) &=& \nabla \times (\psi_{l,m}^{(k)} \cdot \V a),}
{\V N_{l,m}^{(k)}(\V a) &=& \frac{1}{k} \nabla \times \V M_{l,m}^{(k)}.}
Here $\V a$ is some constant vector or the position vector $\V a = \V r $, in such a case functions~(\ref{eq_SperWave}) satisfy the vector Helmholtz equation in spherical coordinates~(\ref{eq_Hlmz})~\cite{Tsang:2000, Huffman:1983}. 

However, we assumed that Maxwell's equations are divergence-free ,~\textit{i.e.} the particle is immersed in a charge free medium, but the functions $\V L_{l,m}^{(k)}(\V a)$ are not divergence-free and therefore should be excluded from consideration.
Hence, the incident, scattered and internal electric fields can now be represented in terms of spherical vector wave functions:
\Eqaaa{}{\V E_{inc}(k_o \V r) &=& \sum_{l=0}^L \sum_{m=-l}^l \left(\alpha_{l,m}^M M_{l,m}^{(1)}(k_o \V r) + \alpha_{l,m}^N N_{l,m}^{(1)}(k_o \V r) \right),}
{\V E_{sca}(k_o \V r) &=& \sum_{l=0}^L \sum_{m=-l}^l \left(\beta_{l,m}^M M_{l,m}^{(3)}(k_o \V r) + \beta_{l,m}^N N_{l,m}^{(3)}(k_o \V r) \right),}
{\V E_{int}(k \V r) &=& \sum_{l=0}^L \sum_{m=-l}^l \left(c_{l,m}^M M_{l,m}^{(1)}(k \V r) + c_{l,m}^N N_{l,m}^{(1)}(k \V r) \right).}
Here $\V E_{int}$ defines the field inside the particle and $k$ is the wavenumber inside the particle. 

To satisfy the boundary condition at the origin a non-singular at the the origin spherical vector wave basis functions of the first kind have to be used. 
The scattered field has to satisfy the radiation condition, therefore it can be expanded in spherical vector wave functions of the third kind.

Considering the continuity of the tangential field components across the surface of the particle, the boundary condition can be set:
\Eqaa{}{\hat n_{+} \times (\V E_{inc}(k_o \V r) + \V E_{sca}(k_o \V r) - \V E_{int}(k\V r)) = \V 0,}
{\hat n_{+} \times (\V H_{inc}(k_o \V r) + \V H_{sca}(k_o \V r) - \V H_{int}(k\V r)) = \V 0.}
Here $\hat n_{+}$ denotes the outward pointing vector, normal to the boundary surface of the particle.

The boundary conditions yield a system of linear equations, from which the unknown expansion coefficients $\beta_{l,m}^M, \beta_{l,m}^N $ and $c_{l,m}^M, c_{l,m}^N $ can be expressed in terms of the known expansion coefficients $\alpha_{l,m}^M, \alpha_{l,m}^N$ , characterizing the known incident field.

For spherical particles this system of linear equations can be inverted analytically, and leads to the well-known Mie's solution~\cite{Mie:1908, Debye:1909}.

Observable quantities such as scattering $C_{sca}$, absorption $C_{abs}$ and extinction $C_{ext} = C_{sca} +C_{abs} $ cross sections can be expressed as follows~\cite{Huffman:1983}:

\Eqaa{}{C_{ext} &=& \frac{2 \pi}{k^2} \sum_{l=1}^\infty (2l+1) Re(a_l + b_l),}
{C_{sca} &=& \frac{2 \pi}{k^2} \sum_{l=1}^\infty (2l+1)(|a_l|^2 +|b_l|^2).}
Here
\Eqaa{}{a_l &=& \frac{n \psi_l(nx)\psi_l'(x) - \psi_l(x)\psi_l'(n x) }{n \psi_l(nx)\xi_l'(x) - \xi_l(x)\psi_l'(n x)},}
{b_l &=& \frac{\psi_l(nx)\psi_l'(x) - n \psi_l(x)\psi_l'(n x) }{\psi_l(nx)\xi_l'(x) - n \xi_l(x)\psi_l'(n x)},}
with size parameter $x = k a$, where $a$ is the radius of the sphere and $k$ is the wavenumber of the incident light in the surrounding medium, $\psi_l$ and $\xi_l$ are the Riccati-Bessel functions, and $n = \sqrt{\epsilon}$ is particle refractive index. 

Alternatively the Mie solution can be derived with help of Debye potentials~\cite{BornWolf:1999}.

The described procedure can be applied in any other coordinate system in which the scalar Helmholtz equation becomes separable. Hence various particles with non-spherical symmetry can be considered as well~\cite{Sinha:1977, Asano:1975}. However, the basis functions might no longer be orthogonal and a resulting system of linear equations can be only inverted numerically.

\subsection{The quasi-static approximation}
In this section we review the quasi-static approximation, which can significantly simply analysis of non spherical particles in the following section.

For a particle with size significantly smaller than the wavelength of the incident light, the electric field of the light can be viewed as an uniform and static. 
Thus the interaction between a small particle and light can be described in terms of electrostatics rather than electrodynamics. The electron cloud of the particle can be viewed as displaced by the electric field, hence a restoring force arises from Coulomb attraction between electrons and nuclei, resulting in electron cloud oscillation.

If a homogeneous, isotropic sphere is placed in medium with a different permittivity, subjected to a uniform static electric field, a charge will be induced on the surface of the sphere.
The electric fields inside and outside the sphere (of radius $R$) can be described by Laplace's equation in terms of scalar potentials $\phi_1(r,\theta)$ and $\phi_2(r,\theta)$ written in spherical coordinates~\cite{Huffman:1983, Jackson:1998}:

\Eqaa{np_eq_1}
{\nabla^2 \phi_1 (r < R ,\theta) &= & 0,}
{\nabla^2 \phi_2 (r > R ,\theta) &= & 0,}

At the boundary between sphere and medium the potentials must satisfy continuity condition:

\Eqaa{np_eq_2}
{\phi_1(R,\theta) &= & \phi_2(R,\theta), }
{\frac{1}{\epsilon_1}\partial_r \phi_1(R,\theta) &= & \frac{1}{\epsilon_2} \partial_r \phi_2(R,\theta).}

Thus we have the partial differential equation and two boundary conditions. 
In addition, by requiring the field at infinity to be equal to the external homogeneous field $E_o$, the solution can be written in the following form: 

\Eqaa{np_eq_3}
{\phi_1(r,\theta) &= & - \frac{3 \epsilon_2}{\epsilon_1 + 2 \epsilon_2} E_o r \cos(\theta) ,}
{\phi_2(r,\theta) &= & - E_o r \cos(\theta) + R^3 E_o \frac{ \epsilon_1 - \epsilon_2 }{\epsilon_1 + 2 \epsilon_2} \frac{\cos(\theta)}{r^2}.}

Now superposing the above two potentials and comparing the result with the potential of an ideal dipole:

\Eq{np_eq_4}{\phi (r,\theta) = \frac{1}{4 \pi \epsilon_2} \frac{\V{p} \cdot \V{r}}{ r^3} = \frac{1}{4 \pi \epsilon_2} \frac{p \cos{\theta}}{ r^2}, }
The solution (\ref{np_eq_3}) can rewritten in the following form:

\Eqaa{np_eq_5}
{&& \V{p} = \epsilon_2 \alpha \V{E_o},}
{&& \alpha = 4\pi R^3 \frac{ \epsilon_1 - \epsilon_2 }{\epsilon_1 + 2 \epsilon_2},}
where $\alpha(R, \epsilon_1, \epsilon_2)$ is the polarizability, which depends on the particle size and material permittivity, $\epsilon_1$ is the material permittivity, from which the particle is made and $\epsilon_2$ is the medium permittivity.

The above result states that a sphere in an electrostatic field is equivalent to an ideal dipole, with momentum $\V{p} = \epsilon_2 \alpha \V{E_o}$. 
Therefore the problem of electromagnetic wave scattering by a sphere can be reduced to a well studied problem of dipole radiation.
The cross section for absorption and scattering of field by an ideal dipole exposed to a plane wave source is given by the following relations~\cite{Huffman:1983, Jackson:1998}: 

\Eqaa{np_eq_6}
{C_{abs} &=& k \Im(\alpha), }
{C_{sca} &=& \frac{k^4}{6 \pi}|\alpha|^2,}
here $k = n_2 \frac{\omega}{c}$ is the wavenumber of incident wave in the medium outside the particle. 
However it is more convenient to use in practice unitless parameters called efficiency:
\Eqaa{np_eq_7}
{Q_{abs} &=&\frac{C_{abs}}{\pi R^2}, }
{Q_{sca} &=&\frac{C_{sca}}{\pi R^2},}

\subsection{The ellipsoidal shape particles}
The analytical solution for ellipsoidal shape particle can be found in a similar way as in the case of spherical particle by introducing ellipsoidal coordinates and exploiting particle symmetries to solve Laplace's equation. The derivation can be found for example in~\cite{Huffman:1983, Jackson:1998}. Here we review the quasi-static approximation instead.

As before we solve the problem of electric charge distribution on a particle of a given shape subjected to a uniform static electric field.
The result can be written in the similar form as in the previous section. 
For a cigar-shaped ellipsoid with two equal length principle axes, defined by $b$ and $a > b$ (such spheroid is called prolate) we have: 
\Eqaa{np_eq_13}
{&& \V{p} = \epsilon_2 \alpha \V{E_o},}
{&& \alpha_x =\frac{ 4\pi}{3} a b^2 \frac{ \epsilon_1 - \epsilon_2 }{\epsilon_2 + L_x (\epsilon_1 - \epsilon_2)},}
The material permittivities for the particle and the surrounding medium are defined as $\epsilon_1$ and $\epsilon_2$, respectively.
The parameter $L_x$ is called depolarization factor and depends on the relative orientation of the long axis $a$ of prolate spheroid and the polarization direction of incident light, which is incident transversely to the plane on which particles are deposited. 
For incident light with the polarization direction parallel to the long axis of the prolate ellipsoid $ x = a $ the depolarization factor is defined as:
\Eq{np_eq_14}
{L_a = \frac{1 - e^2}{e^2}\left(\frac{1}{2e} \ln \left(\frac{1+e}{1-e}\right) -1 \right),}
and for the polarization direction perpendicular to the long axis $ x = b $ the depolarization factor $L_b$ can be defined in terms of $L_a$~\cite{Huffman:1983}:
\Eq{np_eq_15}
{L_b = \frac{1}{2}\left(1 - L_a \right),}
here parameter $e = \sqrt{1- \eta^2}$ is the eccentricity, with the aspect ratio $\eta = \frac{b}{a}$. 

The cross sections $C_{abs}, C_{sca}, C_{ext}$ defined in the previous section by equation (\ref{np_eq_6}) can be applied without modification, but the efficiencies in~(\ref{np_eq_7}) should be modified by replacing the particle radius with the product of its principle axes:
\Eq{}{Q_j =\frac{C_j}{\pi (a b^2)^\frac{2}{3}}.}

\section{Conclusion}
At this point we know how to define optical properties of metals, how to simulate properties of separate nanoparticles, either exactly or with help of the quasi-static approximation, and how to model films with different morphology. In the next Chapter we apply this knowledge to design the optimal coating for TFBG based refractive index sensors.  


%% file: Chap_NP_Results.tex
\chapter{The optimal parameters for a nanoparticle-based coating}
\label{NP_optimal_coat}

In this chapter we study optical properties of metal nanoparticles embedded in a homogeneous ambient medium, with particular attention to the absorption properties of nanoparticles. The absorption properties allow us to compute the effective refractive index of a medium by applying the Kramers-Kronig relation and connecting the imaginary and real part of the refractive index. Next we discuss the choice of an optimal nanoparticle-based coating.

\Fig{NP_enhenc2}{0.5}{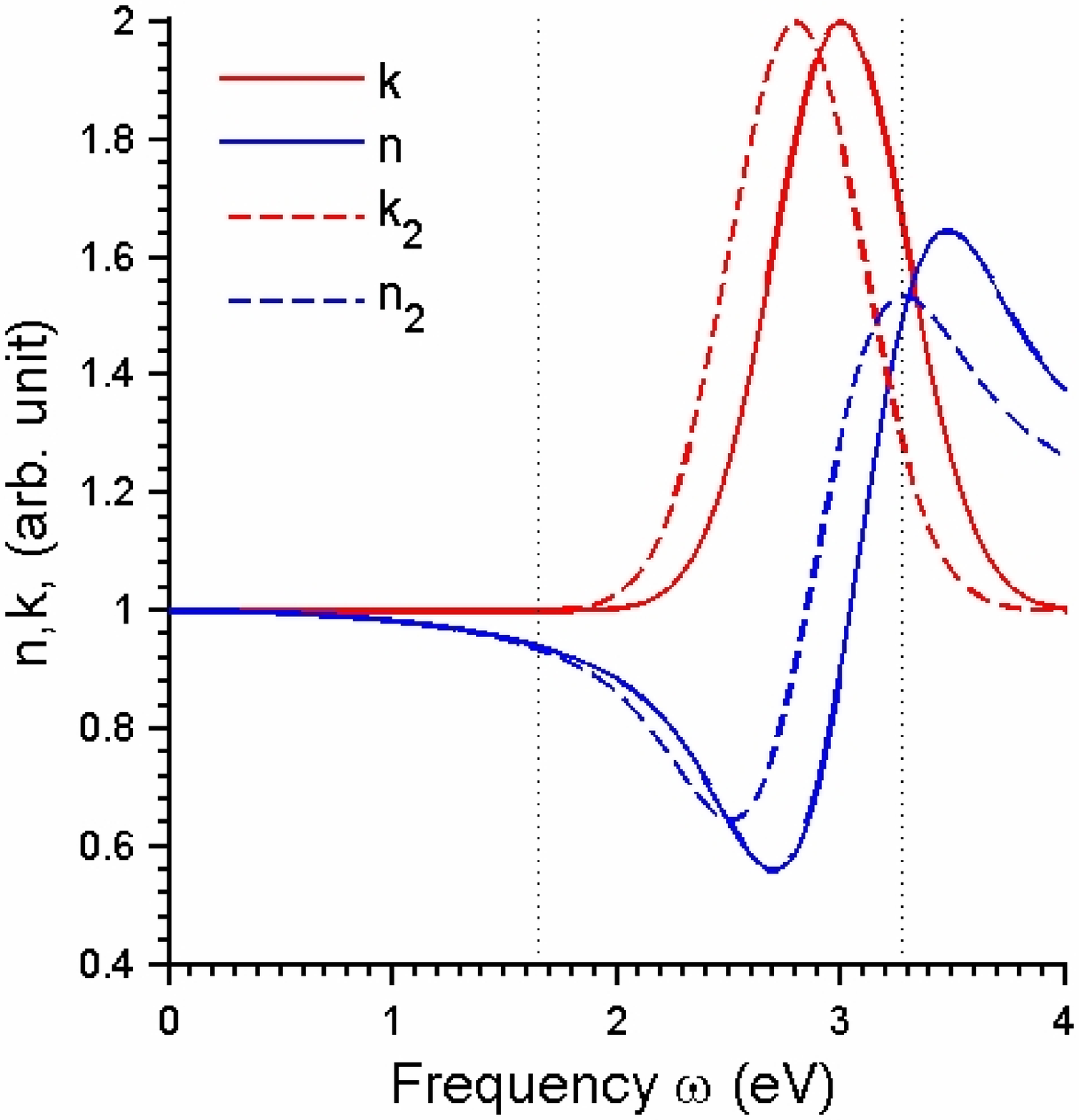}
{The shift in absorption peak (red) and the corresponding shift in real refractive index \C{of a medium} consisting from nanoparticles.}

\section{The idea behind sensitivity enhancement}
Although the optical properties of bulk materials are known, the optical properties of a particle can not be directly deduced. The particle size and its shape introduce additional resonances and hence additional absorption. 
The absorption peaks are not only influenced by the shape and size of the particle but also by the external mediums refractive index. 
Hence, if the refractive index of the external medium is changed the particles resonances will change their locations as well. 
Thus a coating layer can be created with optical properties defined by the refractive index of an external medium.
In such a material not only the imaginary part of the refractive index becomes sensitive to the environmental changes, but also the real part through the Kramers-Kronig connection, as shown in Figure~\ref{NP_enhenc2}. 

We are mainly interested in the real part of a materials effective refractive index, as it defines the positions of resonances, which can be observed in the TFBG spectrum, and hence the change in external medium can be easily detected. 

The connection between high quality factor resonances of TFBG structure and low quality factor resonance of nanoparticles is schematically shown in Figure~\ref{NP_enhenc}.  

\Fig{NP_enhenc}{0.7}{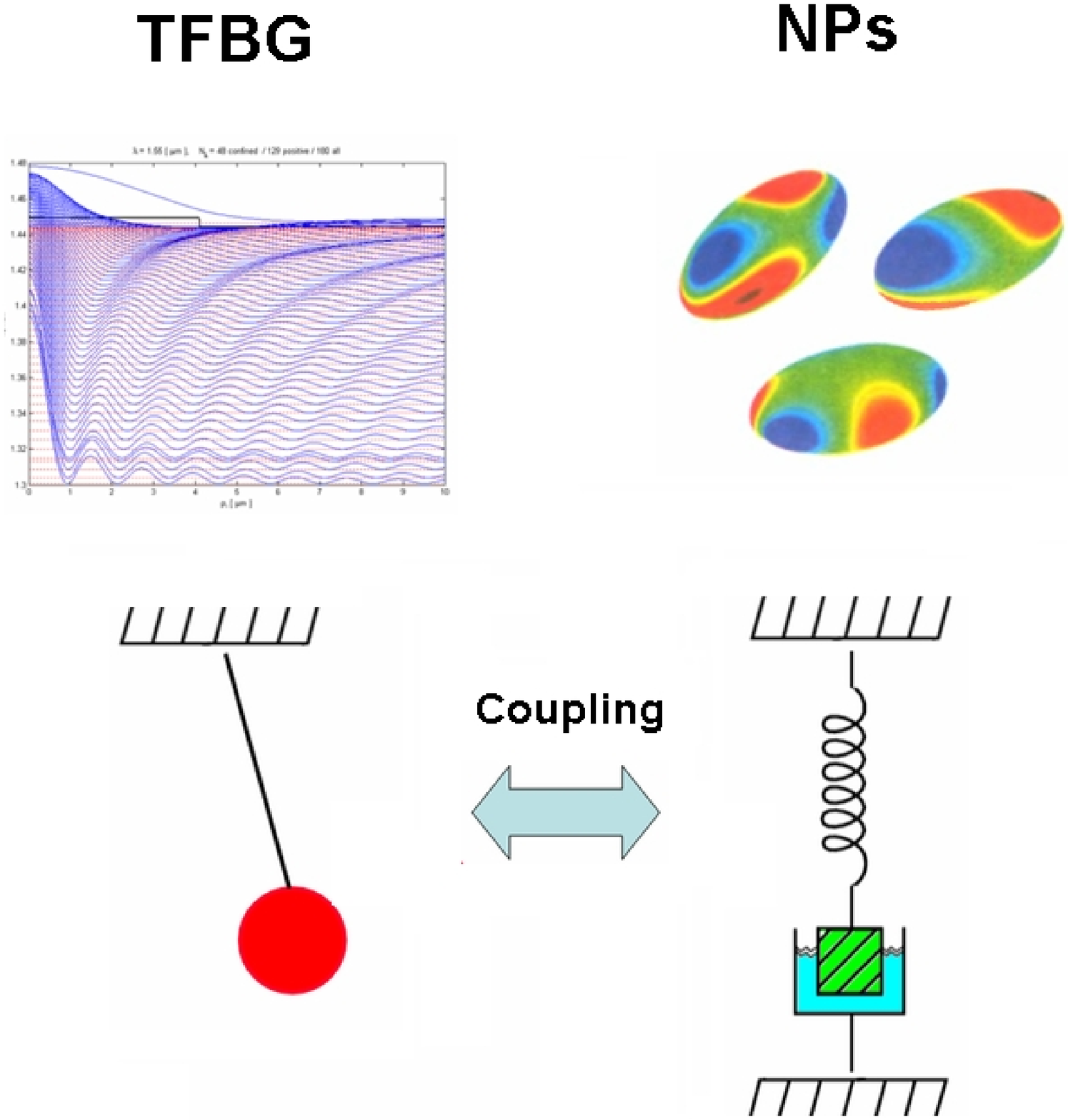}
{The schematic representation of connection between high Q resonances of TFBG and low Q resonances of nanoparticles.}

\section{Simulation of optical properties of metal nanoparticles}
\subsection{The quasi-static approximation}

Let us first simulate optical properties of small spherical nanoparticles made of various metals.

Following the results from Chapter~\ref{opt_prop_metals} we obtain complete characterization of optical properties of $11$ metals in the wide range of frequencies (from $\omega =0$ up to $10-20$ eV ).
The metals were described phenomenologically by Lorentz--Drude model, in which parameters of six oscillators were fitted for the best consistency with experiments~\cite{Rakic:98}.

We start with the quasi-static approximation, which is valid for the problem of light scattering on spherical particles with the particle size significantly smaller then the wavelength. The wavelength of interest is ${\lambda = 1.5~\mu m}$, hence the the approximation should yield the correct result for particles about ${100~ nm}$ in diameter.

Considering the Lorentz--Drude model from Chapter~\ref{opt_prop_metals} and using equations~(\ref{np_eq_5})-(\ref{np_eq_7}) the absorption and scattering efficiencies can be plotted for various metals as shown in Figure~\ref{Qabs}.

The next~Figure~\ref{Qabs_Qsca_varn} shows the absorption and scattering efficiencies of gold nanoparticle immersed in solutions with various refractive indexes.

\Fig{Qabs}{0.63}{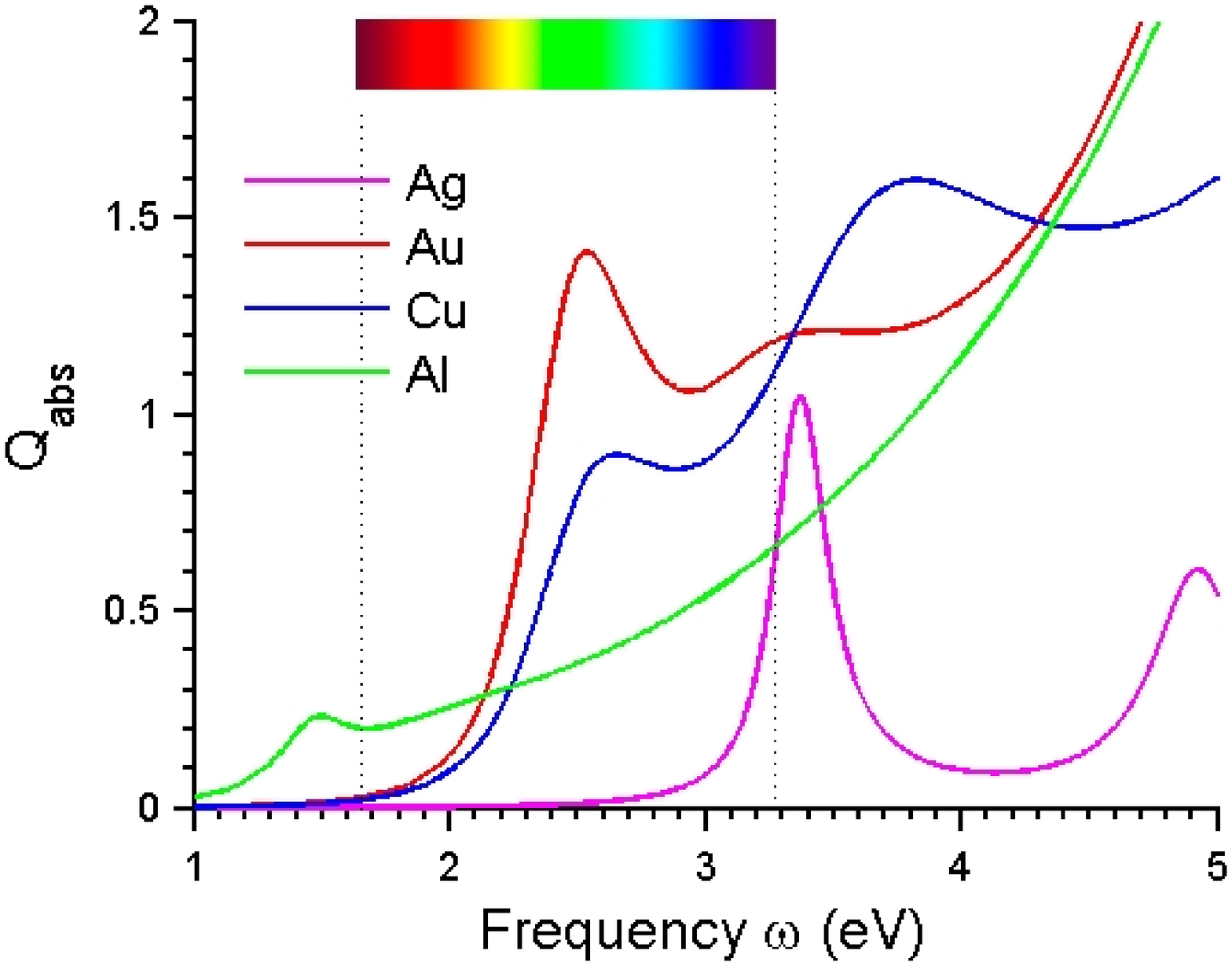}
{\C{Simulated} absorption efficiency of 30 nm spherical nanoparticle made of Au, Ag, Cu and Al metals, as a function of photon energy (the graphs for Ag and Al were divided and multiplied by 10, respectively). }

\Fig{Qabs_Qsca_varn}{0.93}{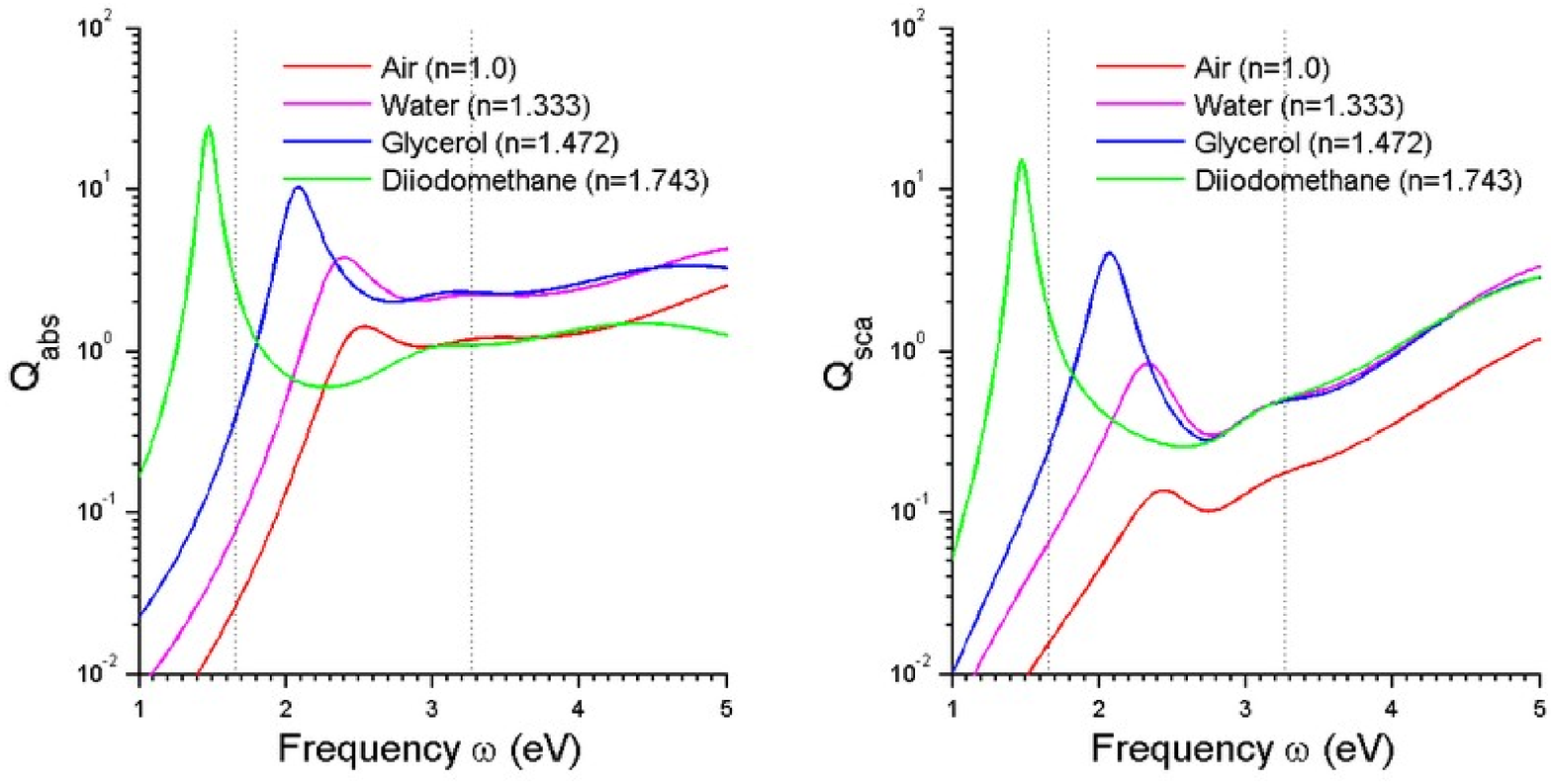}
{\C{Simulated} absorption and scattering efficiency of 30 nm gold spherical NP, immersed in media with various refractive indexes. }

From the above results we can draw the following conclusions:
\begin{enumerate}

\item From equations (\ref{np_eq_5}) and (\ref{np_eq_7}) we note that the absorption $Q_{abs}(\omega)$ and scattering $Q_{sca}(\omega)$ curves are independent of the particle size, hence the position of plasmon resonance is determined only by the dielectric functions of the particle $\epsilon_1$ and the surrounding media $\epsilon_2$.
The position of the resonance is given by the pole of the polarizability in equation (\ref{np_eq_5}),~\textit{i.e.} when $\epsilon_1 + 2 \epsilon_2$ is approaches to zero. 
This pole is commonly referred as a polarization mode or plasmon resonance.

\item The position of the absorption resonance is different for different metals. As can be seen from Figure~\ref{Qabs} for Au, Ag and Cu metals the absorption resonance is located in the visible band, wheres for Al it is located at IR band, not far away from the interband absorption peak. 
We can also conclude that the Al coating might look promising with regards to sensing application in IR band.

\item Although we used a first order approximation, the quasi-static approximation, valid only for a particle with size significantly smaller than the wavelength, we came to the correct qualitative as results reported in the literature. The obtained absorption spectra shown in~Figure~\ref{Qabs_Qsca_varn} are in a good agreement with experimental measurements, predicting the red colors of light transmitted through a suspension of fine gold particles. This effect happens due to the strong interband absorption of the shorter wavelength by the gold. 

\item With the increase of a particle size the electronic cloud displacement can exhibit high-multipolar charge oscillations, thus several poles in the polarizability function can appear and significantly affect the absorption spectrum. The quasi-static approximation can no longer be applied, hence the exact Mie's theory has to be used, as shown in the next section.
\end{enumerate}

\clearpage
\subsection{The exact solution}

As was discussed in Section~\ref{Mie_analit} the problem of light scattering by a spherical particle can be solved exactly.
The absorption, scattering and extinction efficiencies can be written in the form of infinite series~\cite{Huffman:1983}:

\Eqaaa{na_eq_8}
{Q_{ext}(x) &=& \frac{2}{ (R k)^2 } \sum_{l=1}^N (2l+1) \Re(a_l(x) + b_l(x)),}
{Q_{sca}(x) &=& \frac{2 \pi}{(R k)^2} \sum_{l=1}^N (2l+1)(|a_l(x)|^2 +|b_l(x)|^2),}
{Q_{abs}(x) &=& C_{ext}(x) - C_{sca}(x),}
with $a_l(x)$ and $b_l(x)$ given by:

\Eqaa{na_eq_9}
{a_l(x) &=& \frac{m \psi_l(m x) d_x \psi_l(x) - \psi_l(x) d_x \psi_l(m x) }{m \psi_l(m x) d_x \xi_l(x) - \xi_l(x) d_x \psi_l(m x)},}
{b_l(x) &=& \frac{\psi_l(m x) d_x \psi_l(x) - m \psi_l(x) d_x \psi_l(m x) }{\psi_l(mx) d_x \xi_l(x) - m \xi_l(x) d_x \psi_l(m x)}.}

Here $x = k_o R$ is the size parameter, dependent on the radius $R$ of the particle, the wavenumber $k_o$ of the incident light ( $k_o = \frac{\omega}{c}$ ) and the relative complex refractive index of the sphere 
 ${m = n_{particle}/n_{medium}}$. The wavenumber $k$ corresponds to the wave in the surrounding medium. 

The Riccati-Bessel functions $\psi_l(x)$ and $\xi_l(x)$ are defined as follows:
\Eqaa{na_eq_10}
{ \psi_l(x) &=& \sqrt{\frac{\pi x}{2}} J_l(x),}
{ \xi_l(x) &=& \sqrt{\frac{\pi x}{2}} H_l^1(x),}
where $H_l^1(x) = J_l(x) + i Y_l(x)$ is the Hankel function of the first kind, and $J_l(x)$ and $Y_l(x)$ are Bessel functions of the first and the second kind.

The number $N$ in equation~(\ref{na_eq_8}) is the number of terms in Mie series. 
In order to apply these equation in practice the series must be limited to a finite number of terms $N_{max}$. 
Analysis of the convergence behavior of these series reveals that the sufficient number of terms, dependent only on the size parameter $x$,  is determined by the following equation~\cite{Wiscombe:80}: 
\Eq{na_eq_11}
{N_{max} = x + 4 x^\frac{1}{3} + 1}
The relation~(\ref{na_eq_11}) applies to all refractive indices since the series~(\ref{na_eq_8}) convergence is determined entirely by Bessel functions of $x$ alone, as can be seen from~(\ref{na_eq_9}). 
Mie himself considered only three terms as all calculations were done by hand. At the present moment it's possible to compute hundreds of terms in several minutes on a typical PC.

The terms in the Mie series were computed with the help of Mathematica software from Wolfram Research.
The code is presented in Appendix~\ref{Mathematica_code}, and results for the absorption, scattering and extinction efficiencies are shown in Figure~\ref{Append_Mie_fig}.

\Fig{Append_Mie_fig}{0.7}{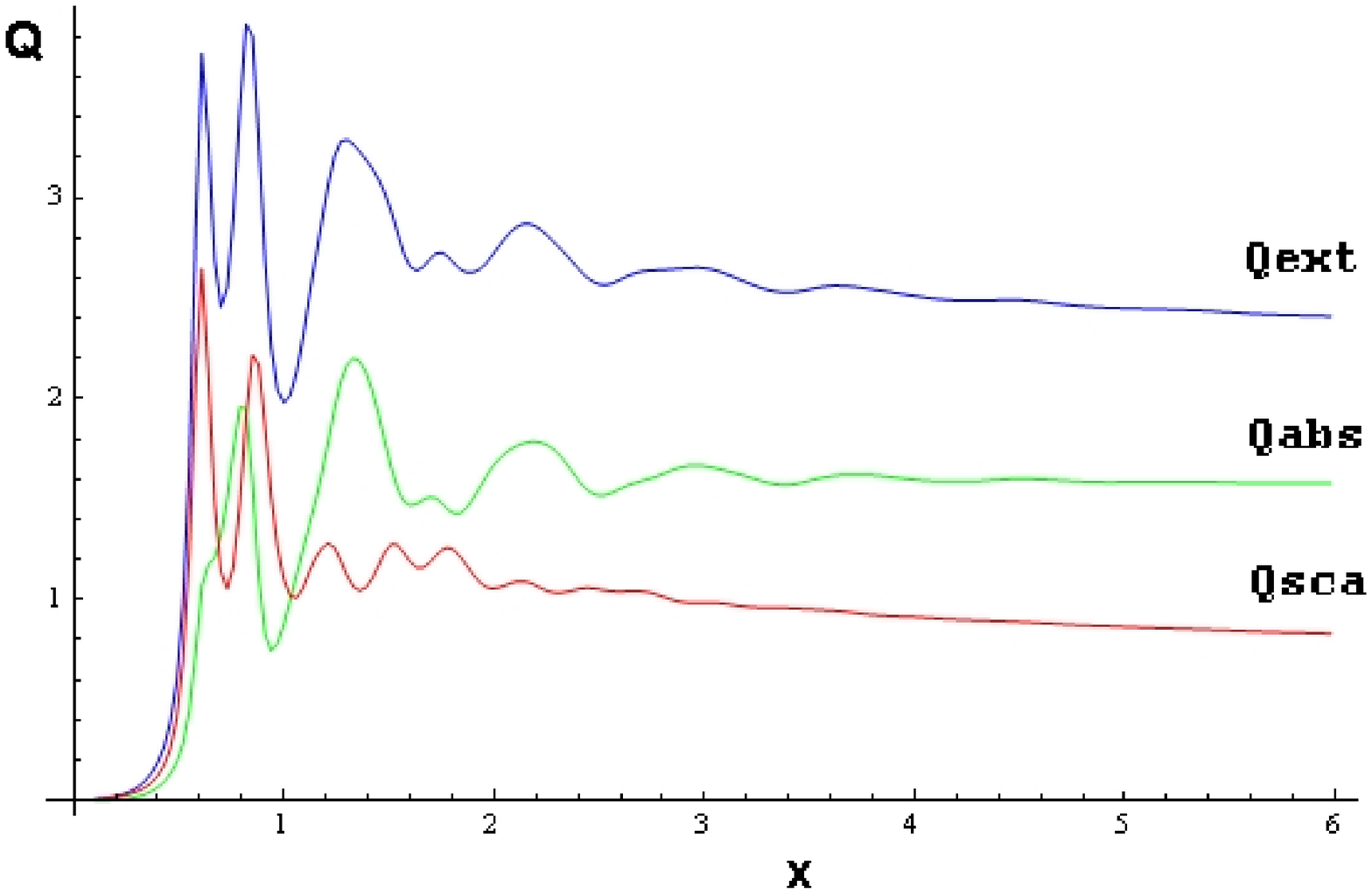}
{The absorption, scattering and extinction efficiencies as functions of the size parameter $x$, for a particle with the relative refractive index $m = 5 + j 0.4$.}

Next, considering optical properties of metals we can compare absorption efficiency for $30$ nm spherical nanoparticle made of various metals, as shown in Figure~\ref{Qabs_Mie_vs_Dip}.
The figure also shows the results based on the quasi-static approximation. 

\Fig{Qabs_Mie_vs_Dip}{0.9}{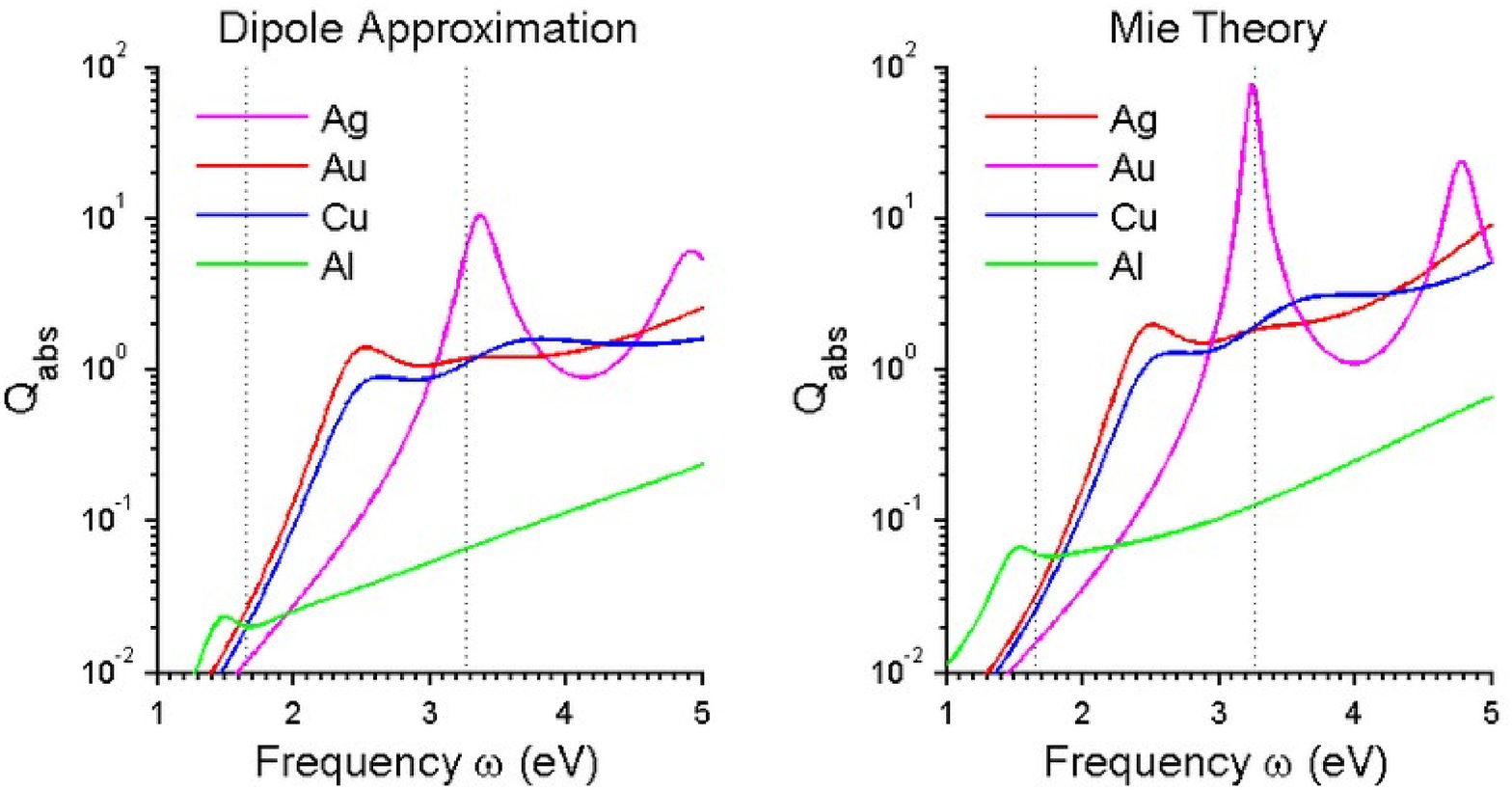}
{The absorption efficiency of $30$ nm spherical nanoparticle made of various metals ( Au, Ag, Cu and Al ) as a function of photon energy. The result were obtained with use of the exact Mie's theory and with the quasi-static approximation theory. }

It can be noted that the proximate and the exact results are in good mutual qualitative correspondence, although quantitative differences are also noticeable, especially for nanoparticles made from silver.
This confirms the well known fact, which states that the optical properties of small metal particles are predominately defined by the bulk material properties from which the nanoparticles are made. 

Next we will consider the influence of the refractive index of the surrounding medium on the absorption and scattering spectra of gold nanoparticles. The results are shown in Figure~\ref{Qabs_Qsca_Mie_varn}.
\Fig{Qabs_Qsca_Mie_varn}{0.9}{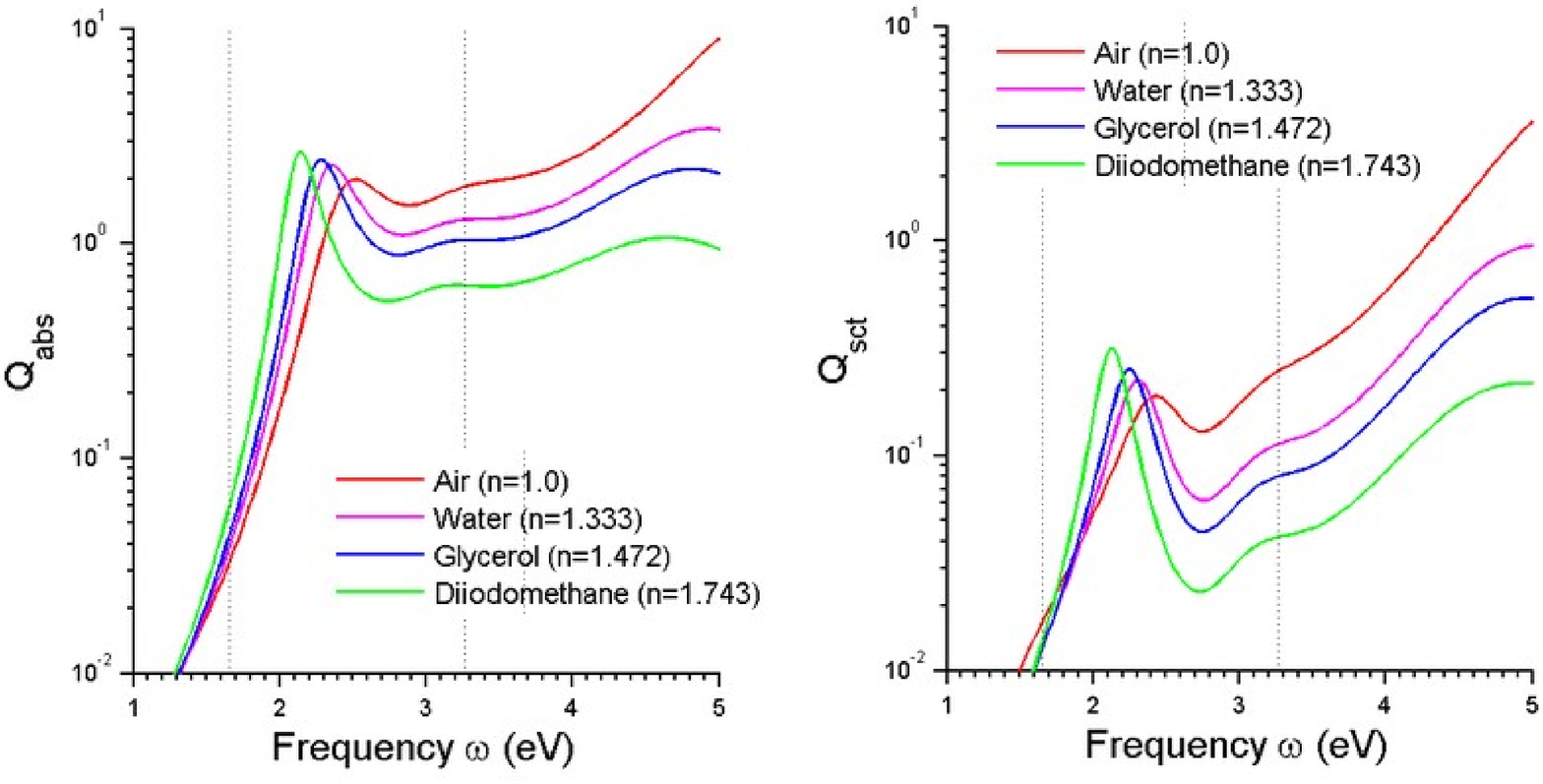}
{Absorption and scattering efficiency of $30$ nm gold spherical nanoparticle, immersed in media with various refractive indexes.}
Comparing these results with the quasi-static approximation Figure~\ref{Qabs_Qsca_varn} we can conclude that the exact Mie solution predicts a significantly different response, with a smaller correlation between the change in the external refractive index and the absorption and scattering spectra.

The most striking difference is observed when the absorption and scattering efficiencies are plotted against the particle radius, as shown in Figure~\ref{Q_vs_r_Au}. 
In the case of the exact solution a number of local resonances is observed, where in the case of the quasi-static approximation only a single resonance is predicted. 
The single resonance is defined by the approximation equation~(\ref{np_eq_6}) where $Q_{abs} \sim R$ and $Q_{abs} \sim R^4$ , in  accordance with the Rayleigh scattering theory~\cite{Rayleigh:1877, Rayleigh:1871}.

\Fig{Q_vs_r_Au}{0.6}{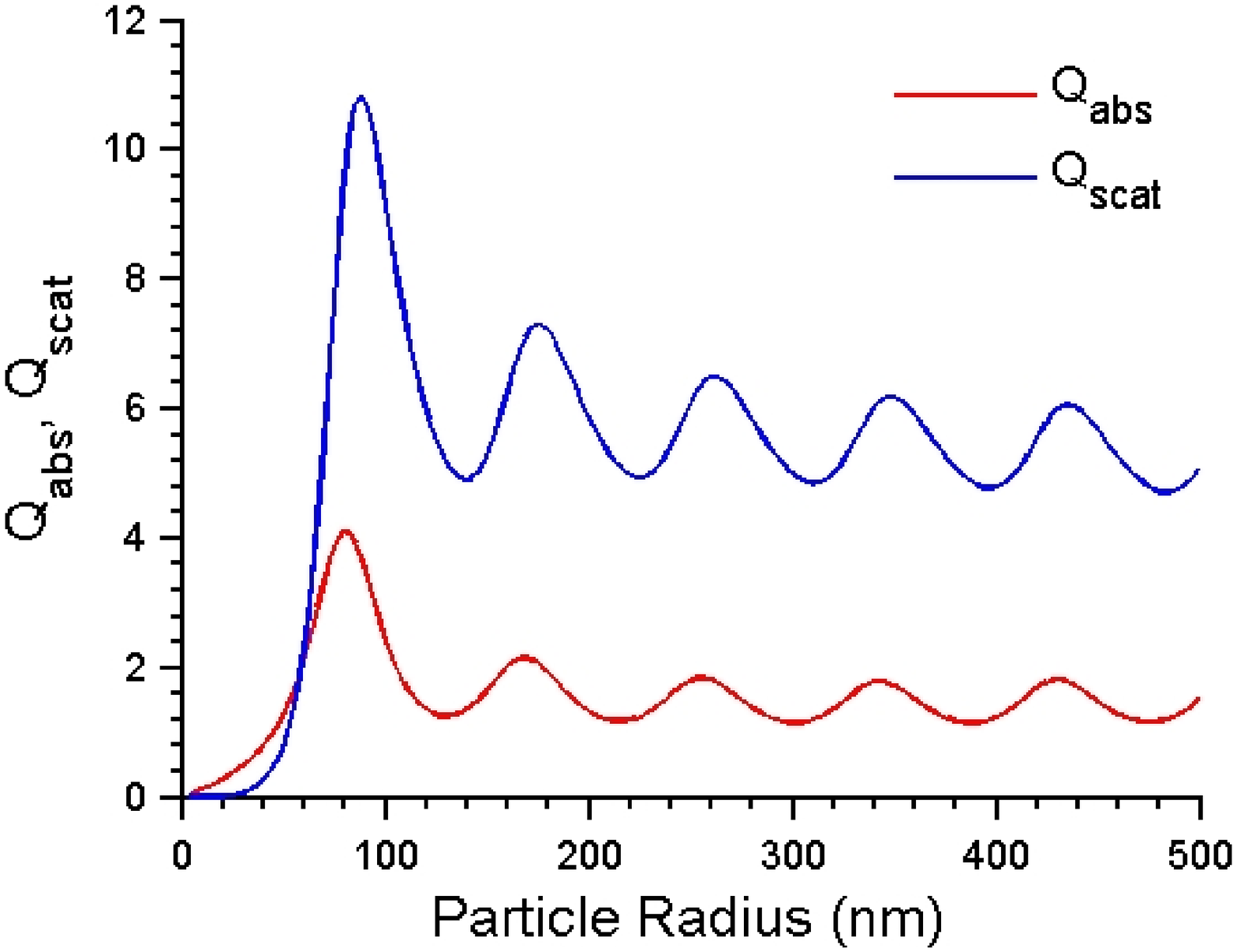}
{The size dependence of the scattering and absorption efficiencies of gold NP illuminated by electromagnetic wave at $\lambda= 560 nm$. The particle is assumed to be in the medium with refractive index $n =1$.}

Our goal here is to find the optimal set of system parameters, such that the local plasmon resonances of nanoparticles can be effectively coupled to the cladding modes excited in the TFBG sensor. 
We have a particular interest in the absorption efficiency $Q_{abs}$ of nanoparticles as it affects the effective refractive index of the surrounding medium detected by the sensor.
With this in mind we can plot $Q_{abs}$ as a function of various parameters, and thus study the parametric space. 

First, let us study the dependence on the particle radius and the material optical properties from which particle is made. Figure~\ref{Qsca_All_2D} shows the absorption efficiency $Q_{abs}$ as a function of particle radius and incident photon energy. The distinct location of the plasmon resonances is intrinsically connected with the material from which the particle is made. 
It is also should be noted that as the particle size increases higher modes of plasmon oscillation can occur, such as quadrupole modes and higher order modes.
However such higher order modes are not always excited, for example if the particle is illuminated by a plane wave, therefore the energy absorption might not be observed in the spectrum. Such modes are often referred as dark plasmon modes.

\Fig{Qsca_All_2D}{1}{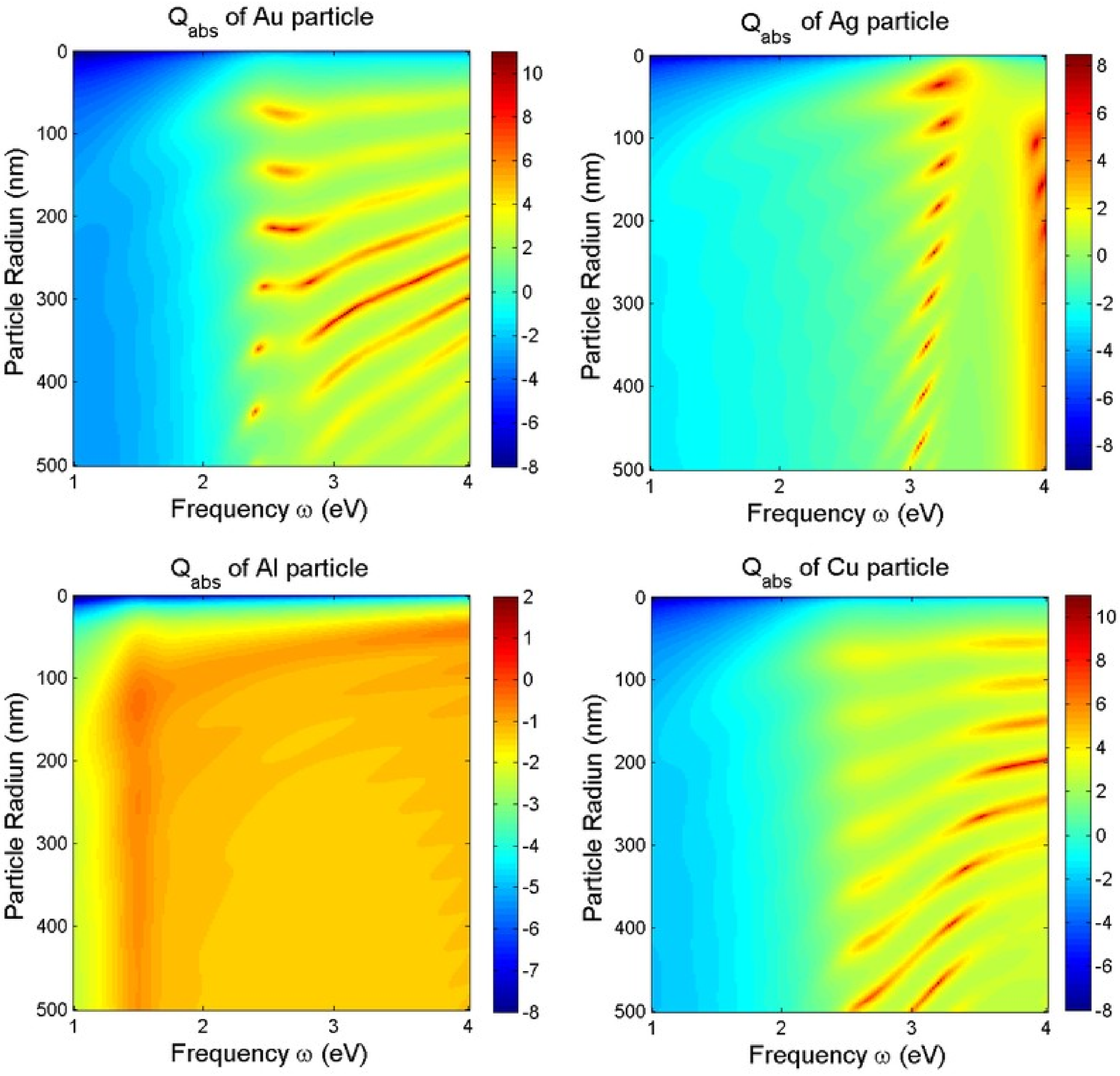}
{The absorption efficiency of gold, silver, aluminum and copper particles as a function of particle size and incident photon energy (plotted in logarithmic scale).}

We can summarize our findings as follows:
\begin{enumerate}
\item Both theories: the approximation quasi-static theory and the exact Mie theory are predict similar qualitative results, stressing the importance of the metal dielectric function from which the particle is made.
However the predictions vary significantly for large particles. 

\item From Figure~\ref{Qsca_All_2D} it can be seen that there is an optimal particle size for each given wavelength, at which the absorption peak has the greatest value. This result was rather unexpected, but a similar conclusion can be drawn from Figure~\ref{Q_vs_r_Au}, obtained in several other papers as well.

\item The material choice is extremely important. 
If we seek to excite plasmons in the visible band, a nanoparticle made of gold is a good option. However, as is seen from Figure~\ref{Qsca_All_2D}, copper can also be a good choice for a relatively large nanoparticle.
If the objective is to obtain a resonance in the IR band aluminum might be a reasonable choice, although the resonance is extremity weak.
\end{enumerate}

\clearpage
\subsection{The shape effect. Ellipsoidal nanoparticles.}
\label{compute_ellipsoidal}
In the previous section we have established that the quasi-static approximation theory can provide satisfactory results in the case of small particles. 
This conclusion was drawn by comparing the exact scattering theory with the quasi-static approximation.
Therefore we can use the quasi-static approximation to investigate the influence of a particle geometrical shape on its optical properties.

We have a particular interest in randomly orientated silver nanorod particles, shown in Figure~\ref{Depos_nanorod}, as some interesting experimental result were reported~\cite{Bialiayeu:2012}.

\Fig{Depos_nanorod}{0.6}{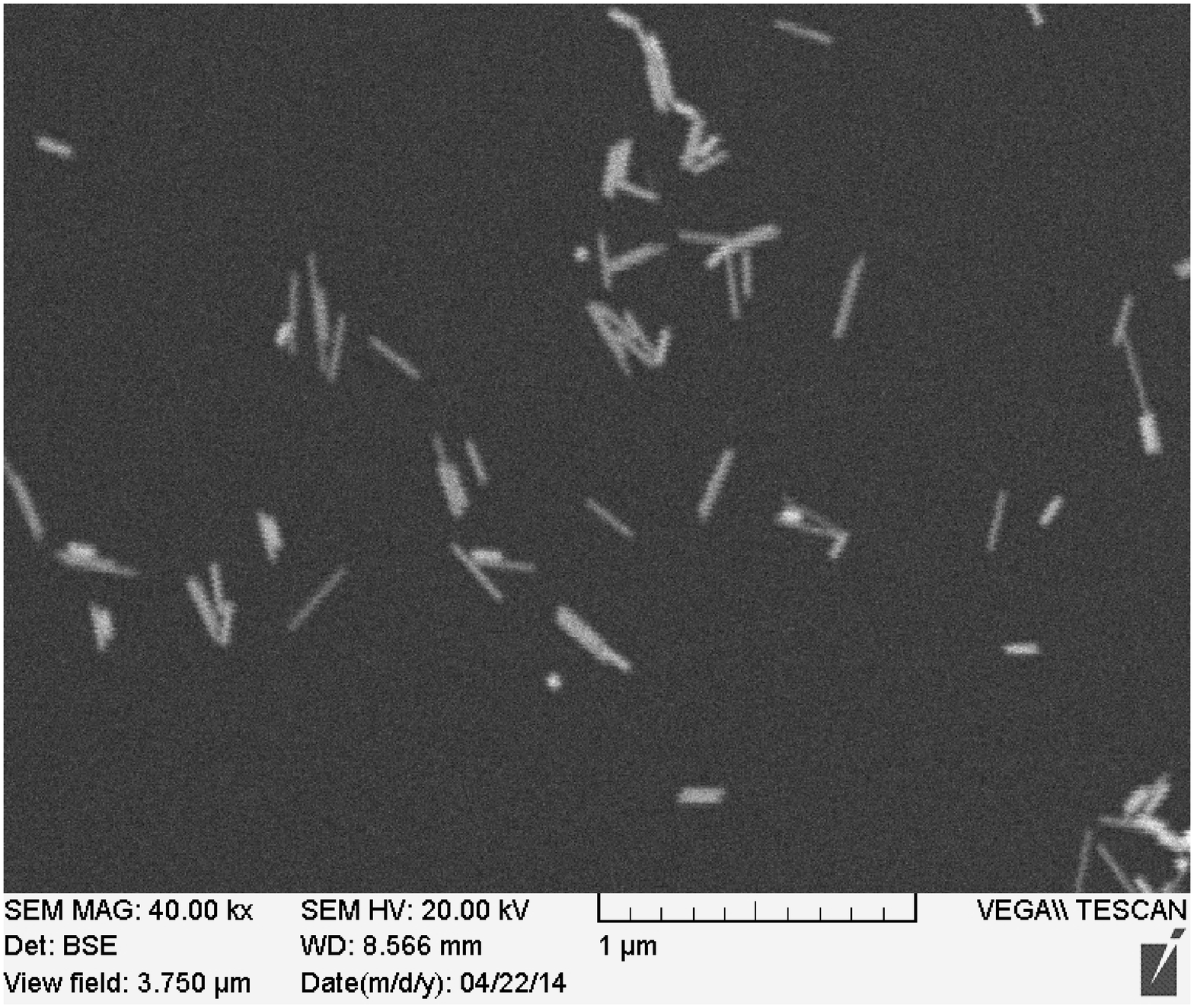}
{SEM image of the fibre surface coated with gold nanorod particles}

Such nanorod particles can be treated as prolate spheroids, with a significant ratio between its principle axes $a$ and $b$. 
Random orientation can be considered simply by taking the average polarizability~\cite{Kooij:06}:
\Eq{}
{\alpha =\frac{1}{3}\alpha_a + \frac{2}{3}\alpha_b,}
here $\alpha_a$ and $\alpha_b$ are polarizabilities along $a$ and $b$ principle axes, respectively.
The cross sections for absorption and scattering processes can be found with help of equations~\ref{np_eq_6} and~\ref{np_eq_7}, as was discussed previously.

The simulation result is shown in Figure~\ref{Qabs_Qsca_dip_prolate_varn}. 
It can be observed that silver and gold particles have distinct resonances in the visible band and look promising for the further investigation. 
\Fig{Qabs_Qsca_dip_prolate_varn}{1}{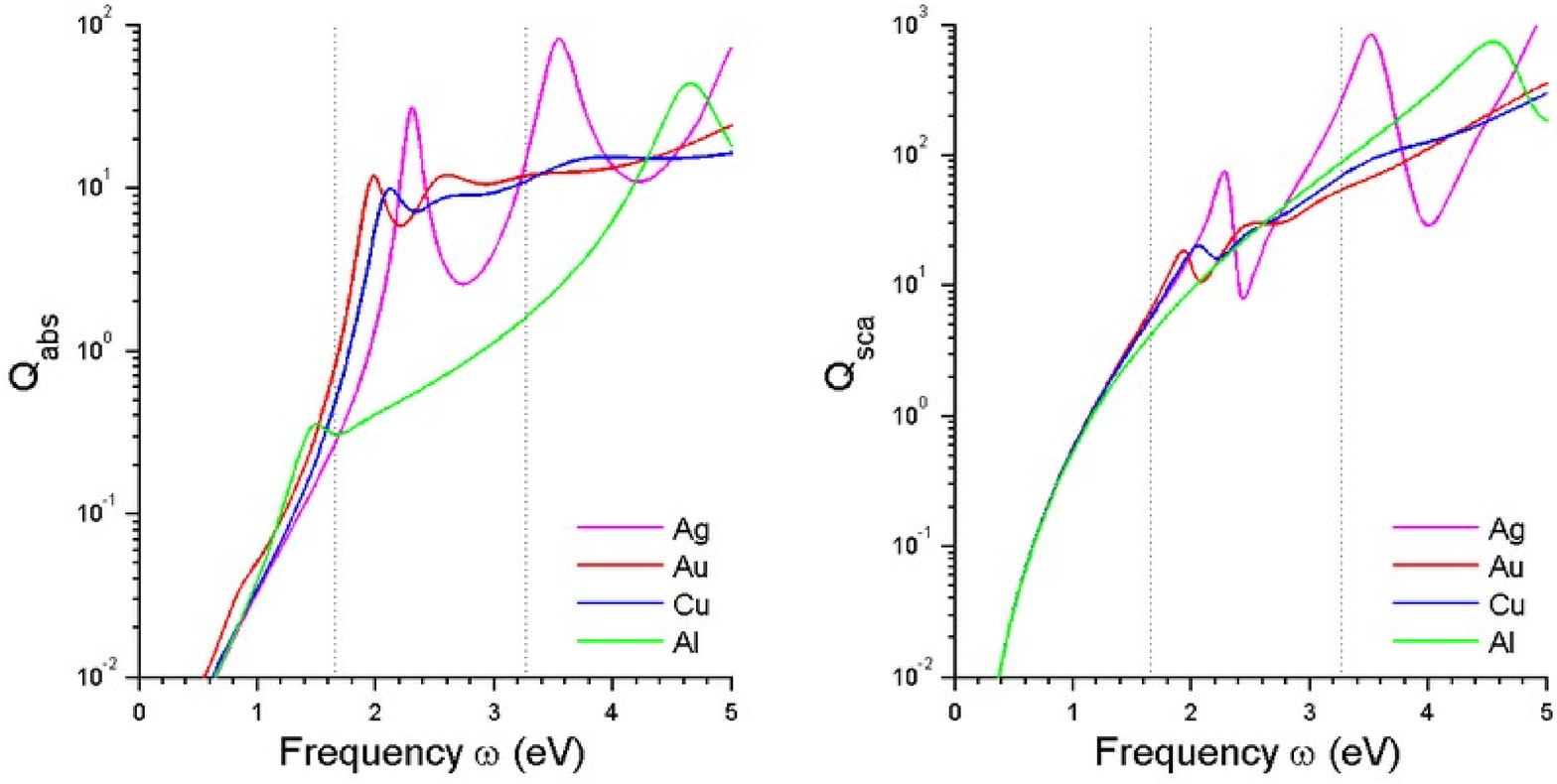}
{Comparison of the optical absorption and extinction efficiencies of prolate ellipsoids made from various metals with the principal axes $b=30, a=100$ nm.}

We next show that by changing the aspect ratios between principal axes of an ellipsoid particle we can tune the absorption resonance into the operational range of TFBG sensor.
The figure~\ref{Qabs_Qsca_dip_prolate_varn} shows the absorption efficiency of randomly orientated prolate silver particles with various aspect ratios.

\Fig{prolate_Ag_sif_asp_res}{0.7}{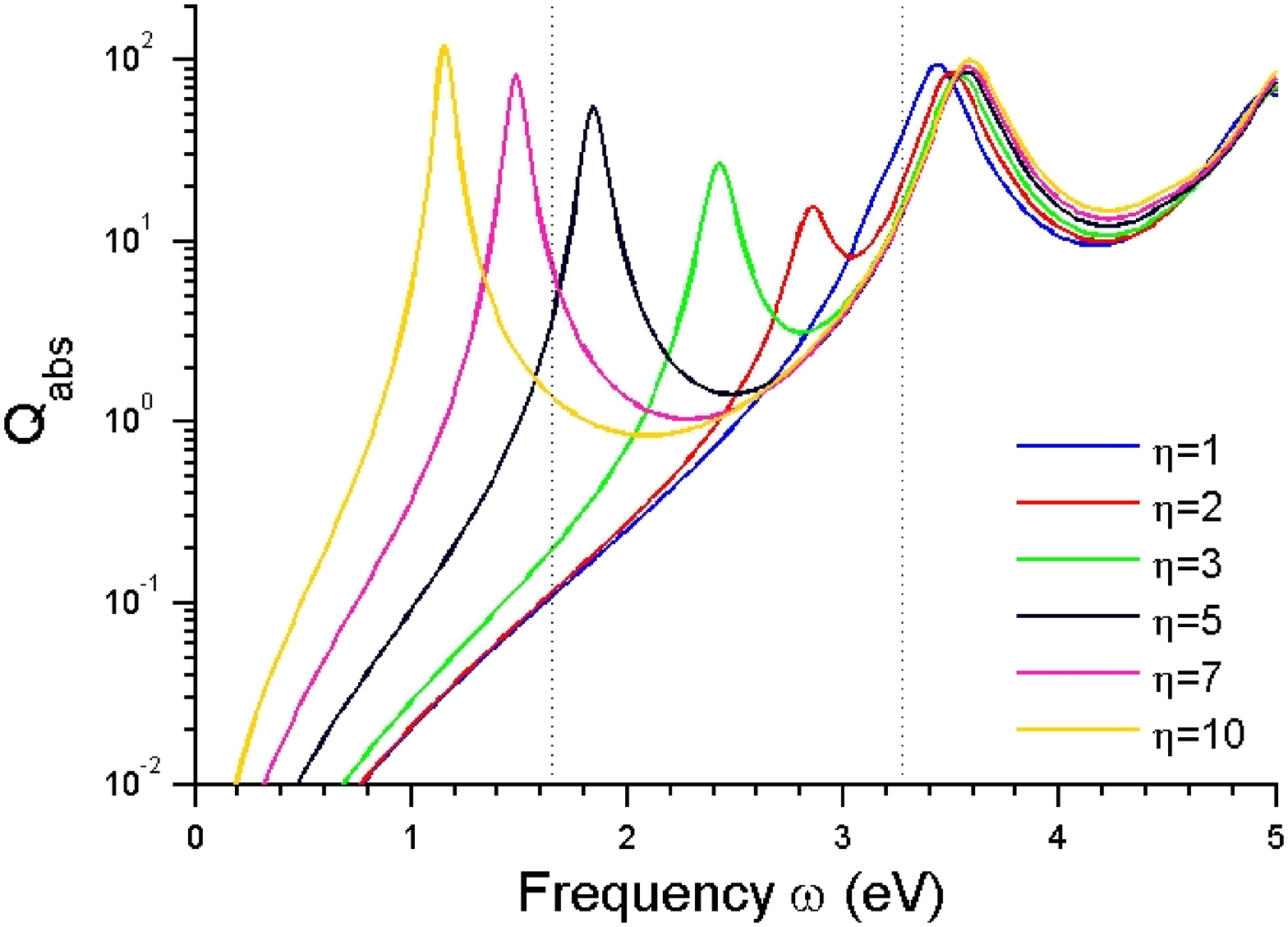}
{Comparison of the absorption efficiency for prolate silver nanoparticles with different aspect ratios between the principal axes $\eta = \frac{b}{a}$}


The properties of the TFBG sensor coated with elongated nanoparticles might be simulated in the usual manner, except that we have to account for the film anisotropy, particularly for the difference in optical properties of the film seen by the tangential and transverse polarized electric field. 
The difference arises from the fact that the position of the absorption resonances of an elongated nanoparticle is a polarization-dependent function, hence the real part of the effective dielectric permittivity function is also polarization-dependent.
It is convenient to introduce effective dielectric permittivities along the tangential and transverse axes associated with the film planes, as shown in Figure~\ref{NP_rod_assymetry}.  

\Fig{NP_rod_assymetry}{0.9}{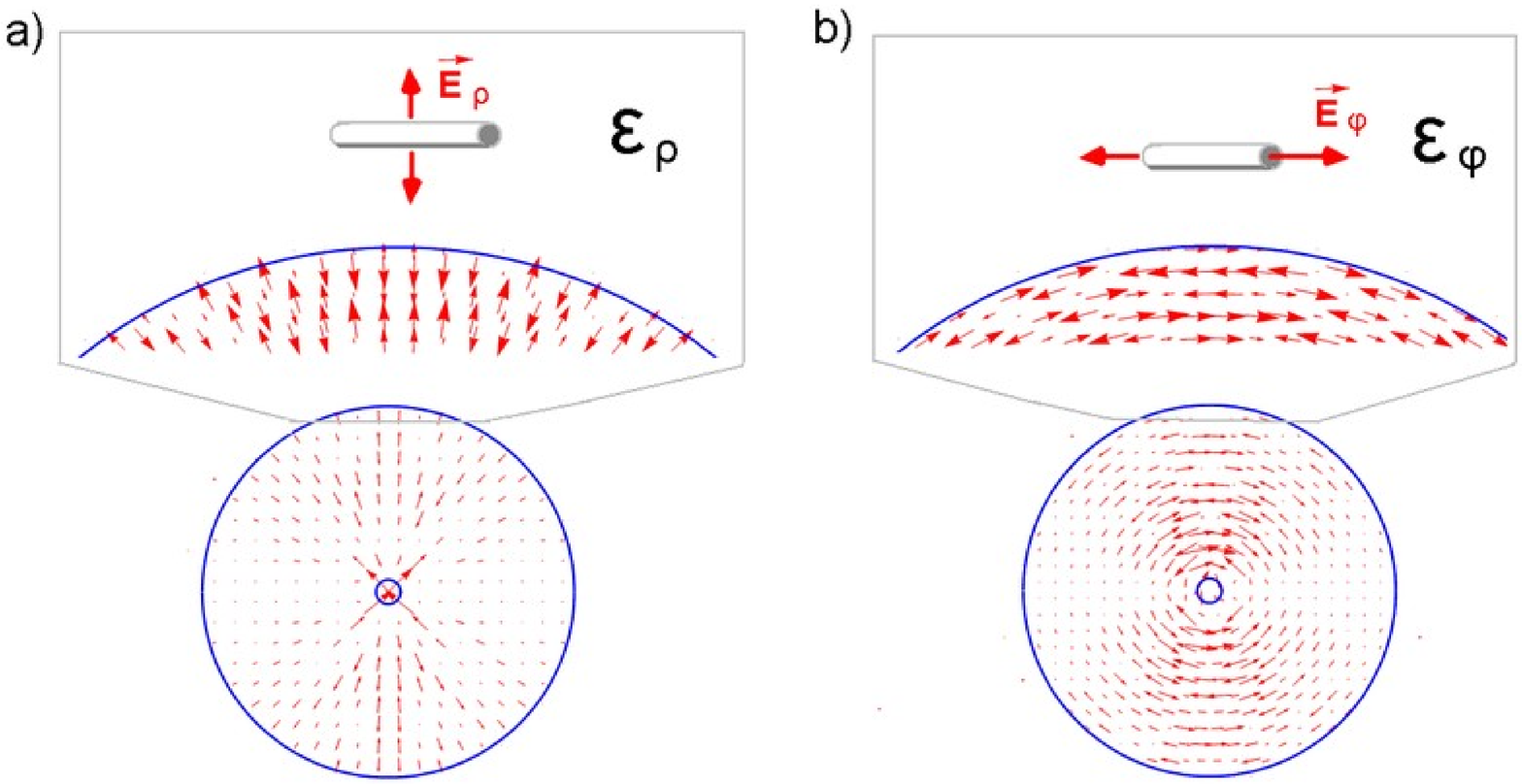}
{The asymmetry of the coating made of elongated nanoparticles. 
a) The transversely polarized electric field $\V{E}_\rho$ encounters $\epsilon_\rho$ dielectric permittivity and b) the tangentially polarized electric field $\V{E}_\phi$ sees the $\epsilon_\phi$ dielectric permittivity.}

The basis functions of a cylindrical waveguide coated with elongated nanoparticles can be found from the modified equation~(\ref{eq_cyl_vec_eigexact}):
\Eq{eq_cyl_vec_rods}
{\Scale[0.8]{
\begin{bmatrix}
d_\rho^2 + \frac{1}{\rho} d_\rho + (\ln \epsilon_\rho )' d_\rho - \frac{m^2+1}{\rho^2} + \epsilon_\rho k_o^2 + (\ln \epsilon_\rho )'' - \beta^2 & - \frac{j2m}{\rho^2}\\
\frac{j2m}{\rho^2} + \frac{jm}{\rho}(\ln \epsilon_\rho )' & d_\rho^2 + \frac{1}{\rho} d_\rho - \frac{m^2+1}{\rho^2} + \epsilon_\phi k_o^2 - \beta^2
\end{bmatrix}\begin{pmatrix}
E_\rho\\
E_\phi\\
\end{pmatrix}
= \V 0,
}
}
here $\epsilon_\rho$ and $\epsilon_\phi$ are the dielectric permittivities as seen by the transversely and tangentially polarized electric fields $\V{E}_\rho$ and $\V{E}_\phi$, respectively.

Our main interest is to increase the TFBG sensor sensitivity operating the in the IR band. 
We can expect that by coating the sensor with silver or gold prolate particles, with appropriate principal axes ratio of about ${\eta \sim 7..10}$, the sharp longitudinal resonance of nanoparticles can be tuned to the operational range of cladding modes resonances, thus making the TFBG sensor more sensitive to perturbations in the surrounding media. 
The sensor can even be coated with silver nanorods of various lengths, we still should expect that a fraction of these particle would have an appropriate aspect ratio and thus would contribute to an increase of the TFBG sensor sensitivity. 
We also should note that in the case of prolate particles with high aspect ratio the quasi-static approximation might not provide a satisfactory prediction, as was also the case with large spherical nanoparticles.


\section{The optimal parameters choice for coatings based on spherical nanoparticles}

Our goal is to design a nanoparticle coating with a sharp absorption resonance, located in the operational range of the TFBG sensor with location sensitive to the refractive index of surrounding media.
From the previews section we know that silver or gold prolate nanoparticles might be an excellent choice, as the  position of the plasmon resonance can be tuned at the exact location by choosing particles with the proper aspect ratio between its axes~($\eta \sim 7..10$).

In this section we discuss choice of parameters for coatings based on spherical nanoparticles.
The sensor coated with spherical nanoparticles is schematically shown in Figure~\ref{Depos_spheres}.

\Fig{Depos_spheres}{0.6}{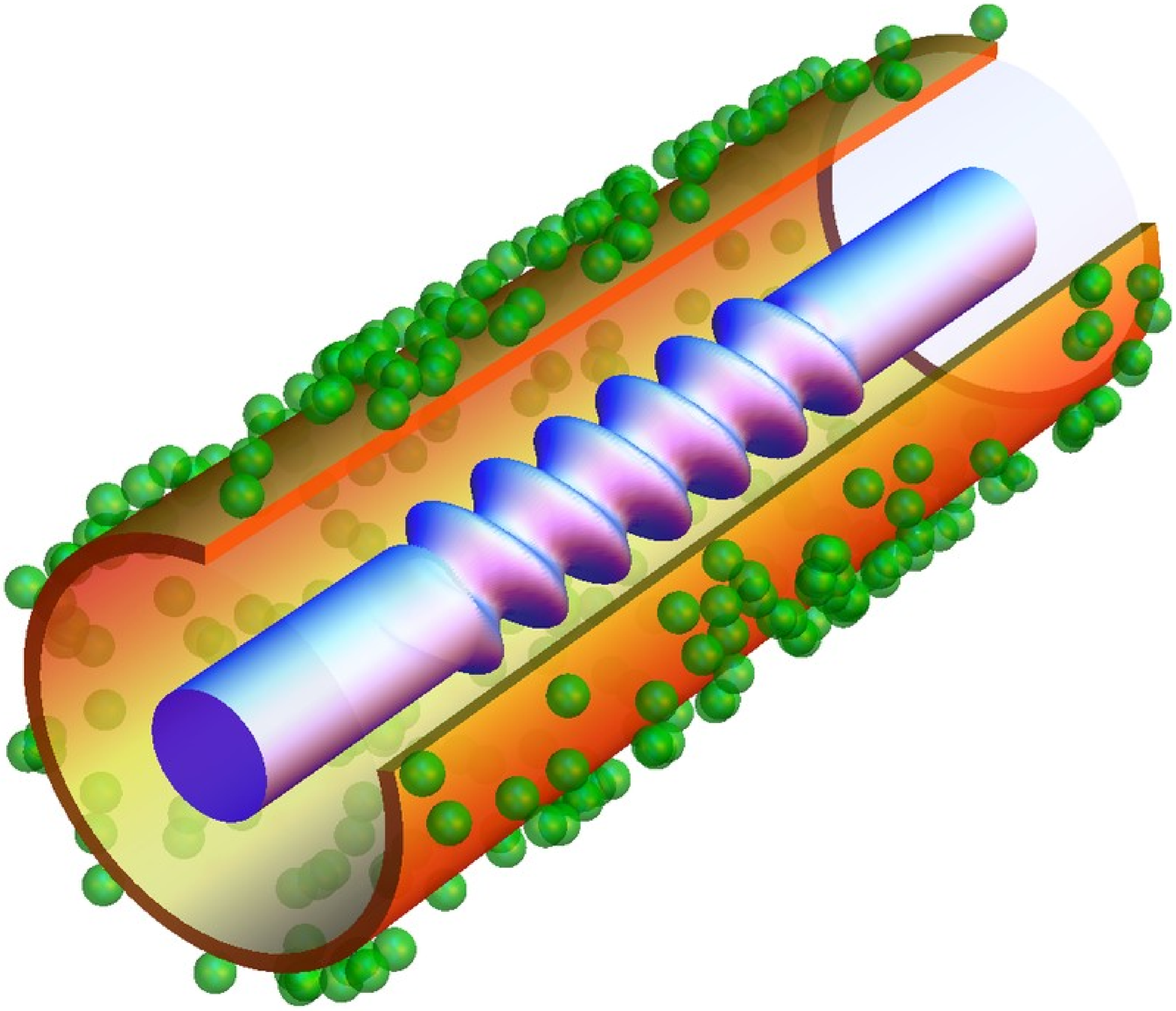}
{A schematic representation of a TFBG sensor coated with spherical nanoparticles.}

Let us plot absorption efficiencies for various particle sizes and materials. 
As a first step we will fix the the imaginary part of the particle permittivity, and compute $Q_{abs}$ as a function or real part of the relative complex refractive index ${\Re[m] =}$ ${ = \frac{\Re[n_{particle}]}{n_{medium}}}$ and the particle size parameter ${x = \frac{2\pi}{\lambda} R}$, as shown in Figures~\ref{Mie_choose_Re} and~\ref{Mie_choose_Re2}.

\Fig{Mie_choose_Re}{0.9}{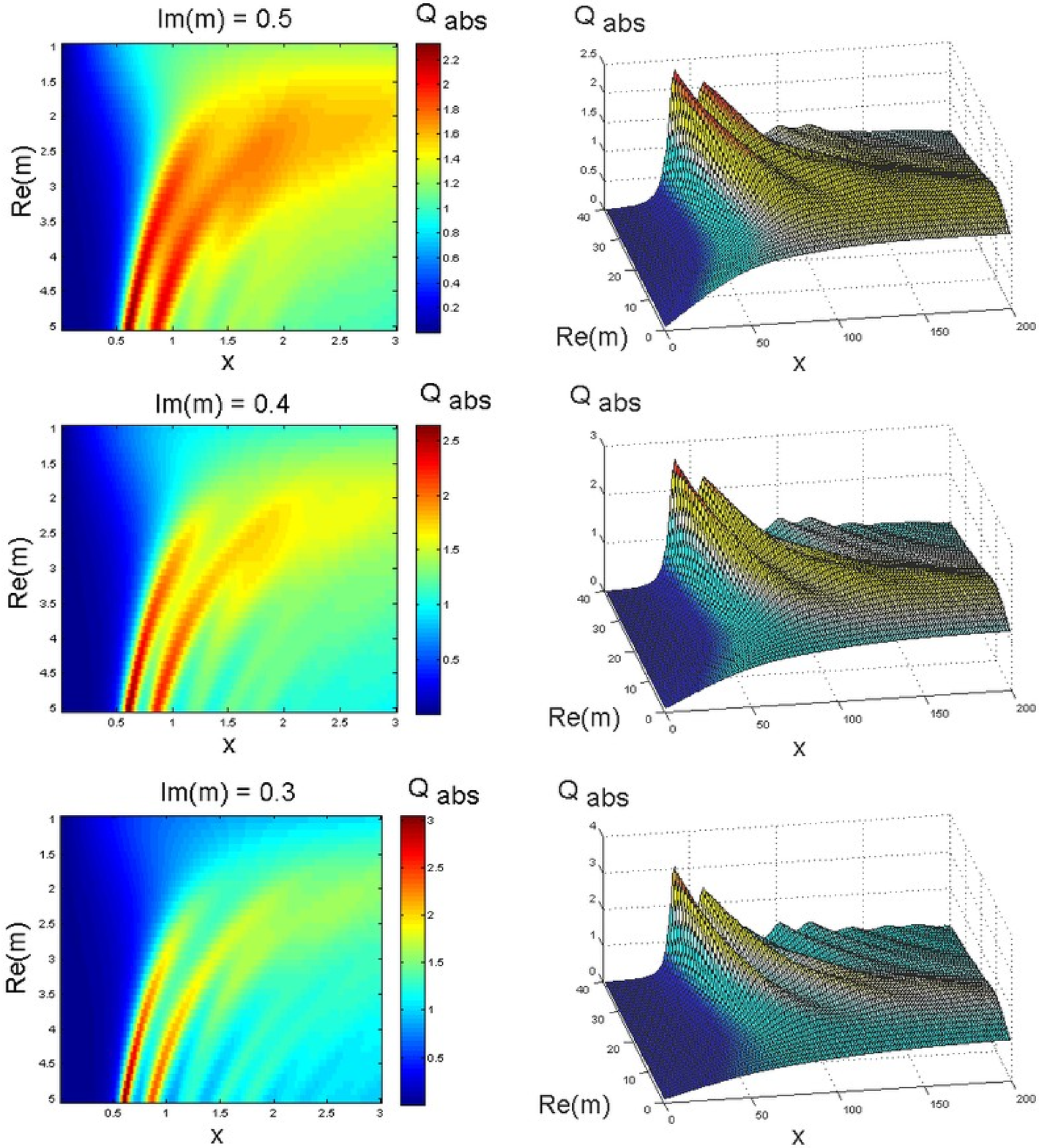}
{The absorption efficiency as a function of the particle size parameter $x$ and the real part of the relative complex refractive index $\Re[m]$ at various fixed values of $\Im[m]$.}

\Fig{Mie_choose_Re2}{0.9}{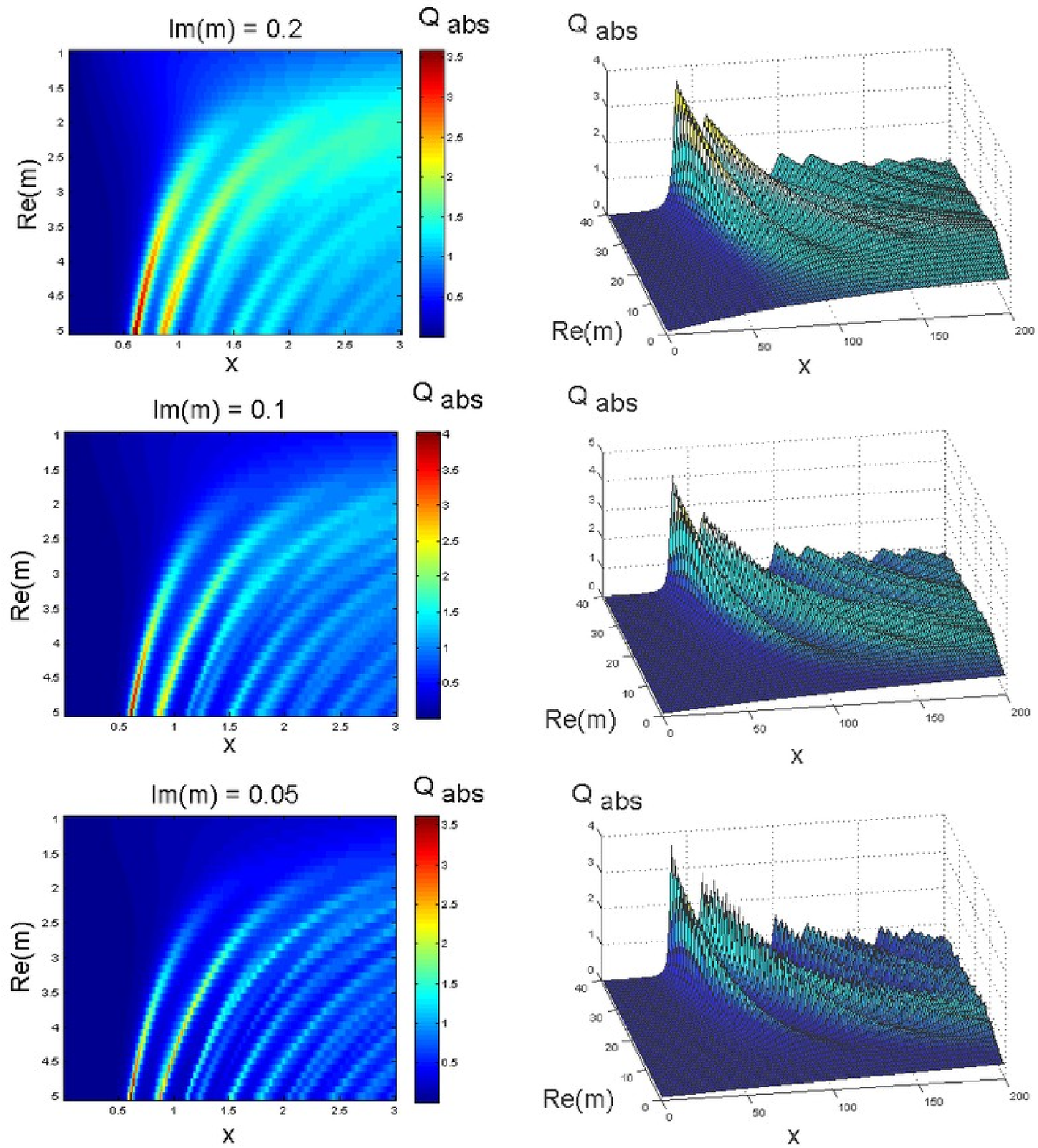}
{The absorption efficiency as a function of the particle size parameter~$x$ and the real part of the relative complex refractive index $\Re[m]$ at various fixed values of $\Im[m]$.}

Our goal here is to choose an optimal material and particle size. 
There are two main parameters to be considered: first we need a sharp prominent resonance, so that the real part of the effective refractive index can be significantly altered, and the second is the resonance steepness. 
We have a particular interest in the steepness of resonances. 
The steeper is the resonance the more its location is affected by the refractive index of the external medium, hence the higher is the sensor sensitivity.

We can conclude that it is desirable to have particles made of material with a small imaginary part of the dielectric permittivity $\Im[m] \sim 0.05$ and choose a particle size parameter $x$ such that ${\Re[m] \sim 2-3.5}$. We note that the smaller is the $\Re[m]$ parameter, the steeper is the resonance, unfortunately for small $\Re[m]$ we can no longer obtain a prominent resonance.  
Once the real part of dielectric permittivity was chosen, we can study the effect of material absorption in more detail, as shown in Figure~\ref{Mie_choose_Im}. We note that for $\Im[m]< 0.05$ the resonance amplitude starts to decrease, hence a material with small but noticeable loss is required. 

\Fig{Mie_choose_Im}{0.9}{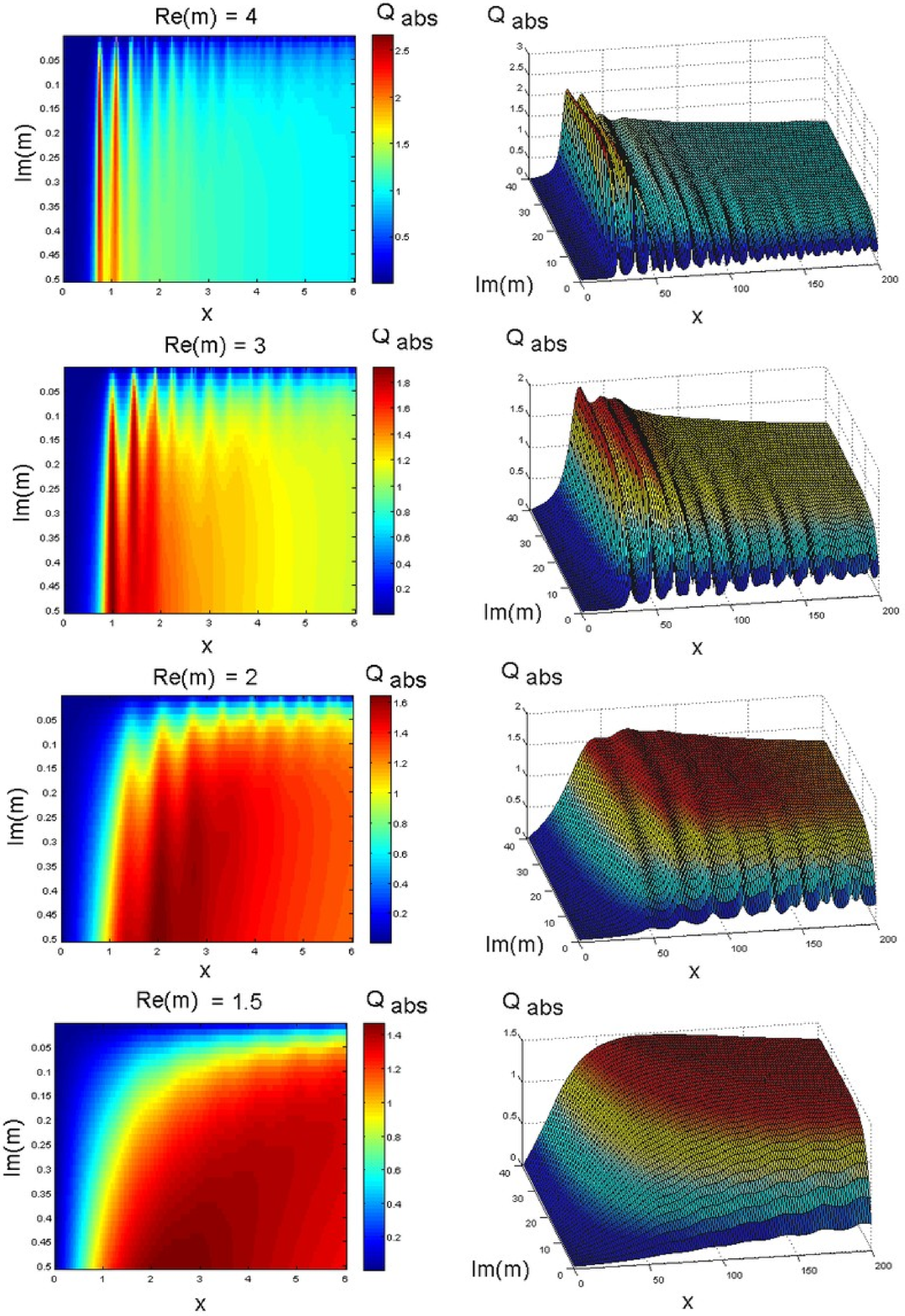}
{The absorption efficiency as a function of the particle size parameter $x$ and the imaginary part of the relative complex refractive index $\Im[m]$ at various fixed values of $\Re[m]$.}
\clearpage

Considering the large value of the required relative refractive index $m$ our choice is limited to only a few materials with high refractive indices, such us Zirconium Silicate and Titanium Dioxide.

The refractive index of Titanium Dioxide is defined by equation~(\ref{TiO2}) and is shown as a function of frequency in Figure~\ref{TiO}.
\Eq{TiO2}
{n = 5.913 + \frac{0.2441}{\lambda^2 - 0.0803}}

\Fig{TiO}{0.7}{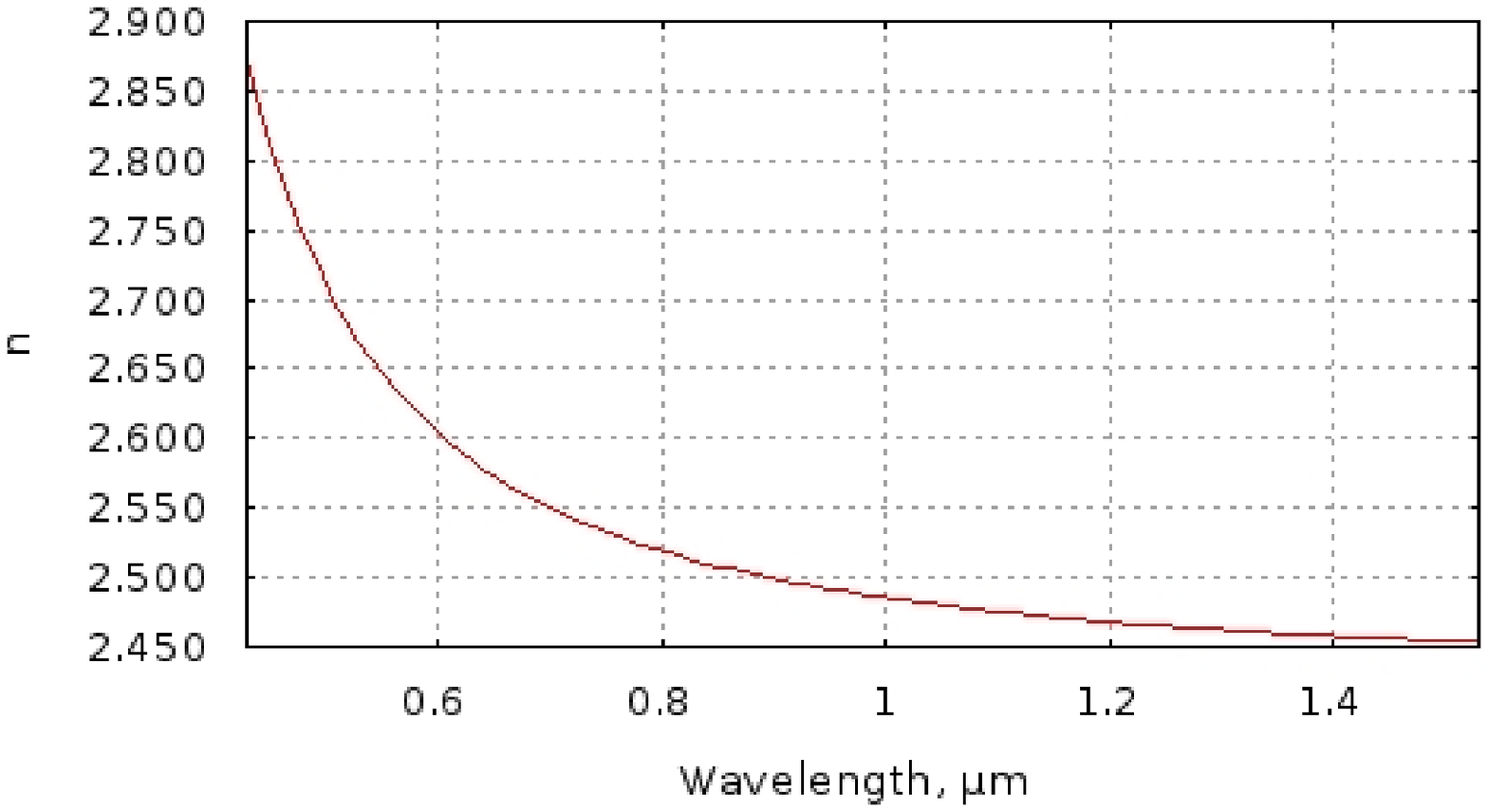}
{Refractive index of titanium dioxide, $Ti O_2$.}
 
Considering the operational wavelength ${\lambda = 1.55 \mu m}$ and the refractive index of the external medium (${n_{medium} = 1.318}$ for water) we find that the relative refractive index of $Ti O_2$ particle immersed in water is ${m \sim 1.9}$. The horizontal line corresponding to ${m \sim 1.9}$, as shown in Figure~\ref{Mie_Choose}, crosses several Mie resonances, hence a proper particle size can be chosen at one of such crossing points. Let us choose the particle size parameter ${x = 2.1}$ , corresponding to the second resonance. Then the particle radius can be found from the relation ${x = k_o n_m R}$ and for the above mentioned parameter is~${R = 393~nm}$.

We note that we only barely met the required conditions. With the available refractive index for particles we reached the very beginning of the resonance peak, as shown in Figure~\ref{Mie_Choose}. A material with a higher refractive index in the IR band is highly desirable, unfortunately our choice is limited.

\Fig{Mie_Choose}{0.65}{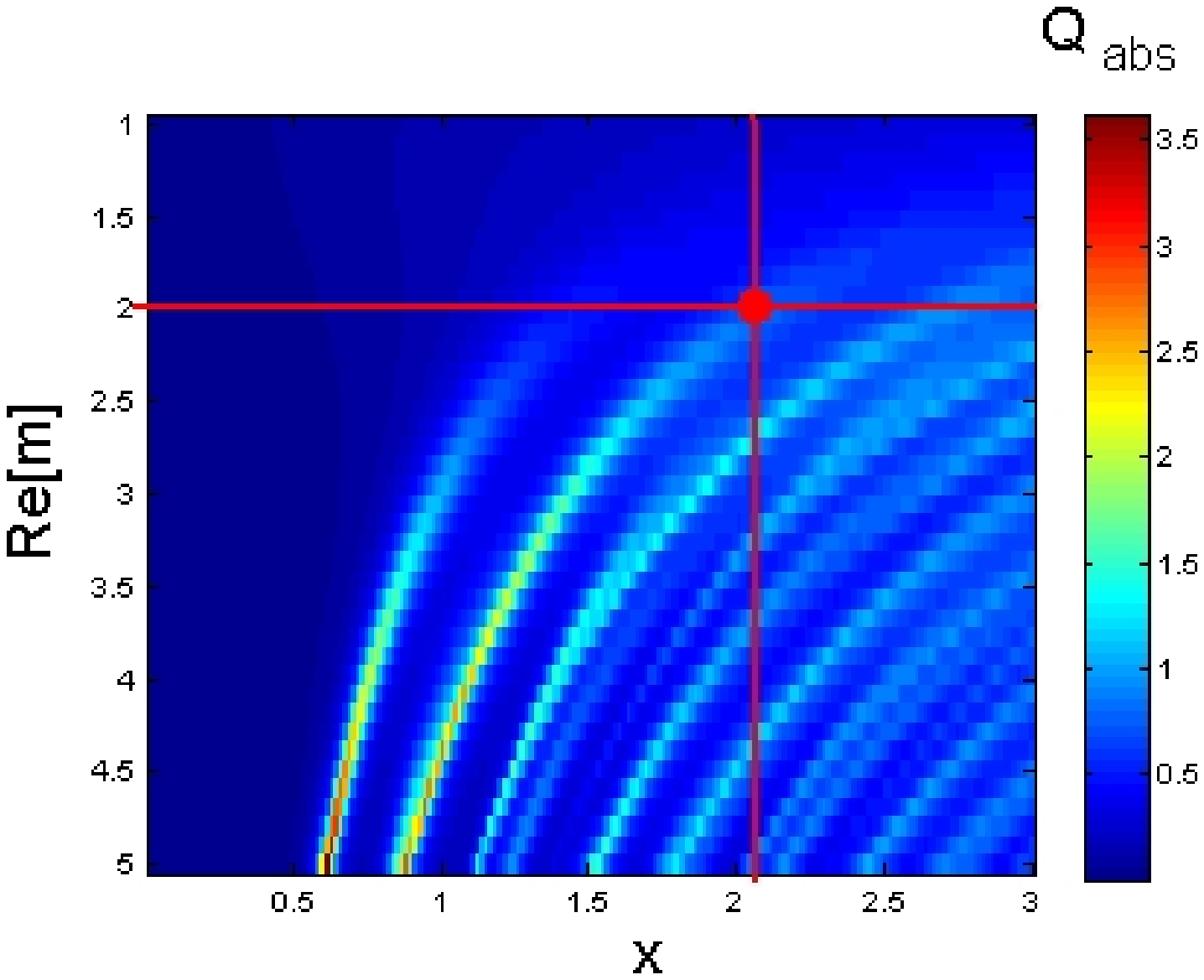}
{The optimal size of a particle made of Titanium Dioxide.}

The properties of the TFBG sensor coated with spherical nanoparticles can be modeled with equation~(\ref{eq_cyl_vec_rods}) as previously.
In the case of sparsely deposited spherical particles the dipole-dipole radiative interaction between the particles can be neglected. Thus the effective dielectric permittivity function can be assumed to be an isotropic function $\epsilon_\rho = \epsilon_\phi$. Hence, once the particles absorption is known the refractive index can be computed with help of the Kramers-Kronig relation~(\ref{eq_KK}) and the Maxwell--Garnett model, used to compute the effective refractive index of the particles embedded in the surrounding medium.

\section{Conclusion}
In this chapter we presented results of our simulations of spherical and ellipsoidal nanoparticles made of various materials. We presented the method for increasing TFBG sensor sensitivity based on resonant coupling between the modes of TFBG sensor and the modes of particles deposited on the fibre surface. 
The particle are chosen to have a sharp resonance with its location sensitive to the external refractive index.

It was shown that the metallic \C{nanorod-like} particles made of silver or gold look promising for the sensitivity-enhancing coating.
The resonance of \C{nanorod} particles can be tuned exactly to the operational range of the sensor by choosing particles with ${\eta \sim 7..10}$ aspect ratio between its principal axes.

Alternatively, dielectric spherical particles with high refractive index can be chosen. We found that particles made of Titanium Dioxide with the diameter ${D \sim 800~nm}$ look promising for the sensitivity enhancing coating.

We should also note that the particle-based coatings should be sufficiently sparse, so that the interaction between particles, and hence the broadening of the resonance peak can be neglected. However, the sensitivity enhancement effect is proportional to the number of particles deposited on the fiber the surface.
The effective refractive index of nanoparticle-based coatings can be computed with the Maxwell--Garnett model (see Section~\ref{Optical_properties_of_mixtures}), if the particles are assumed to be mutual independent and non-interacting. 

In the next chapter we present our experimental findings.

%% file: Chap_Chem_Deposition.tex
\chapter{Modification of the sensor surface with various types of nano-scale coatings}
\chaptermark{Modification of the sensor surface}

In this chapter we present our experimental results on enhancing the TFBG sensor sensitivity by coating its surface with various types of nano-scale coatings. 
It is already known that the TFBG sensor performance can be improved by coating its surface with a nano-scale metal film~\cite{Shevchenko:2007,Caucheteur:11}, thus coupling the TFBG resonances to a surface plasmon resonance~(SPR) excited in the metal film. This technique was thoroughly investigated in~\cite{Shevchenko:2007,Caucheteur:11}. In the presented work we are mainly interested on a possible sensitivity improvement with the introduction of nanoparticle-based coatings.

As was shown in Chapter~\ref{NP_optimal_coat}, we propose using an external resonant system, such as small nanoparticles coupled to the TFBG resonances, to boost the sensor performance.
It is well known that the position of a particle resonance is strongly affected by the surrounding medium, as shown in Figures~\ref{Mie_Choose} and~\ref{Qabs_Qsca_Mie_varn}, thus a thin coating layer created out of such particles, with optical properties strongly dependent on the refractive index of the external environment, can be engineered to boost the TFBG sensor sensitivity. 
For example, such effect was observed in~\cite{Bialiayeu:2012} where the sensor was coated with silver nanowires. 
A $3.5$-fold increase in the TFBG sensor sensitivity was reported.
In this context we have considered various types of particles. However, spherical and ellipsoidal particles are particularly interesting as the exact analytical solution exists for such cases.


During the course of our research we investigated several coatings, such as metallic films of various morphologies and thicknesses, as shown in Figures~\ref{film_my_gold} and coatings with nanoparticles of various shapes and materials. In particular we have tested coatings with nanocubes of various sizes ($40-80~nm$) synthesized from silver and gold metals, as shown in Figure~\ref{Films_NC}, coatings with elongated gold and silver nanoparticles, as shown in Figures~\ref{film_wires2} and~\ref{Depos_nanorod_2}, and coatings with nanospheres, shown in Figure~\ref{Depos_spheres_2}. The particles were deposited at various densities creating films of different morphologies. 
The surface images were obtained by means of a scanning electronic microscope (SEM) and an atomic force microscope (AFM).

\Fig{film_my_gold}{0.9}{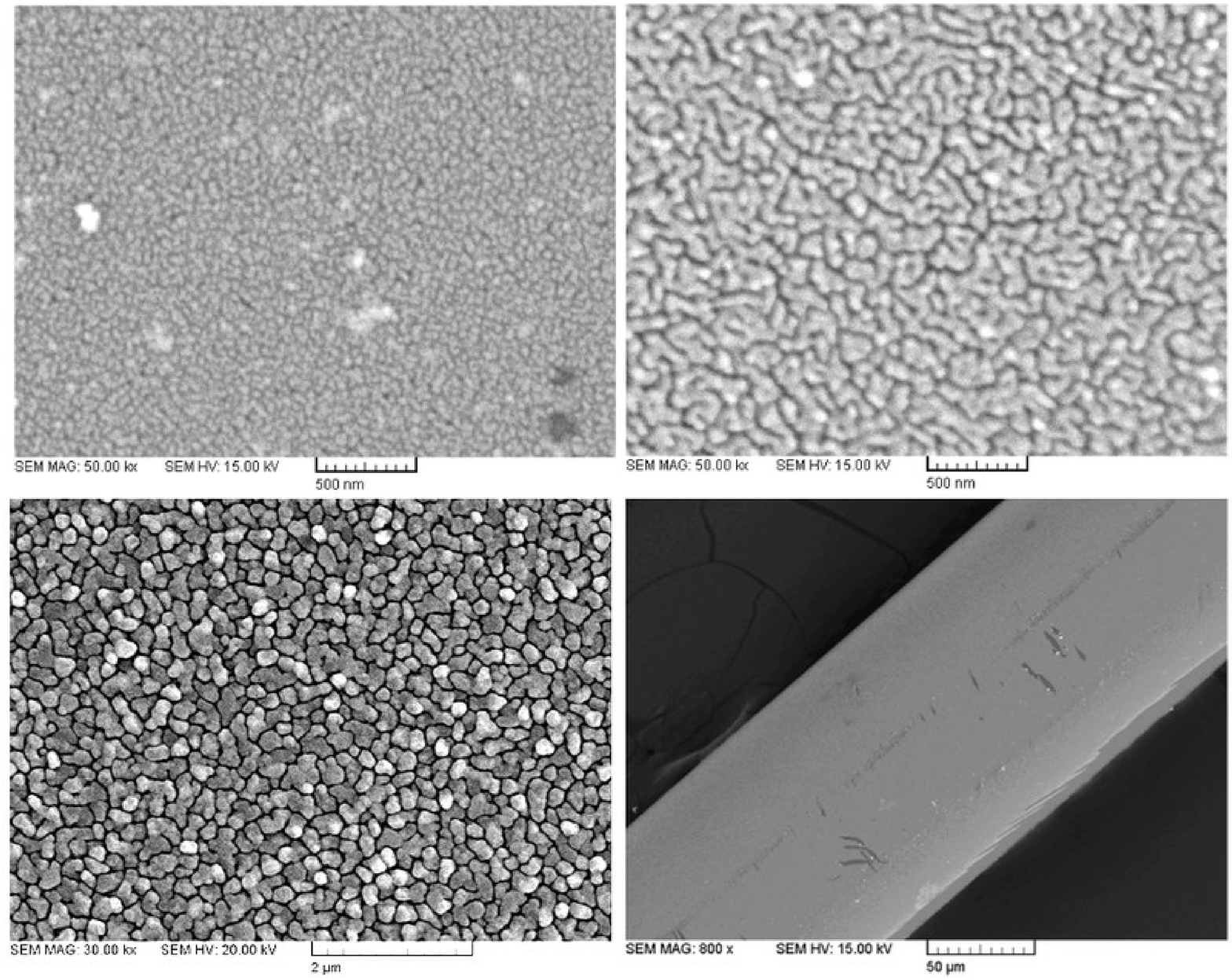}
{The SEM mages of the fibre surface coated with gold (top) and copper (bottom) films of different morphologies and thickness~\cite{Bialiayeu:2011}.}

\Fig{film_wires2}{0.95}{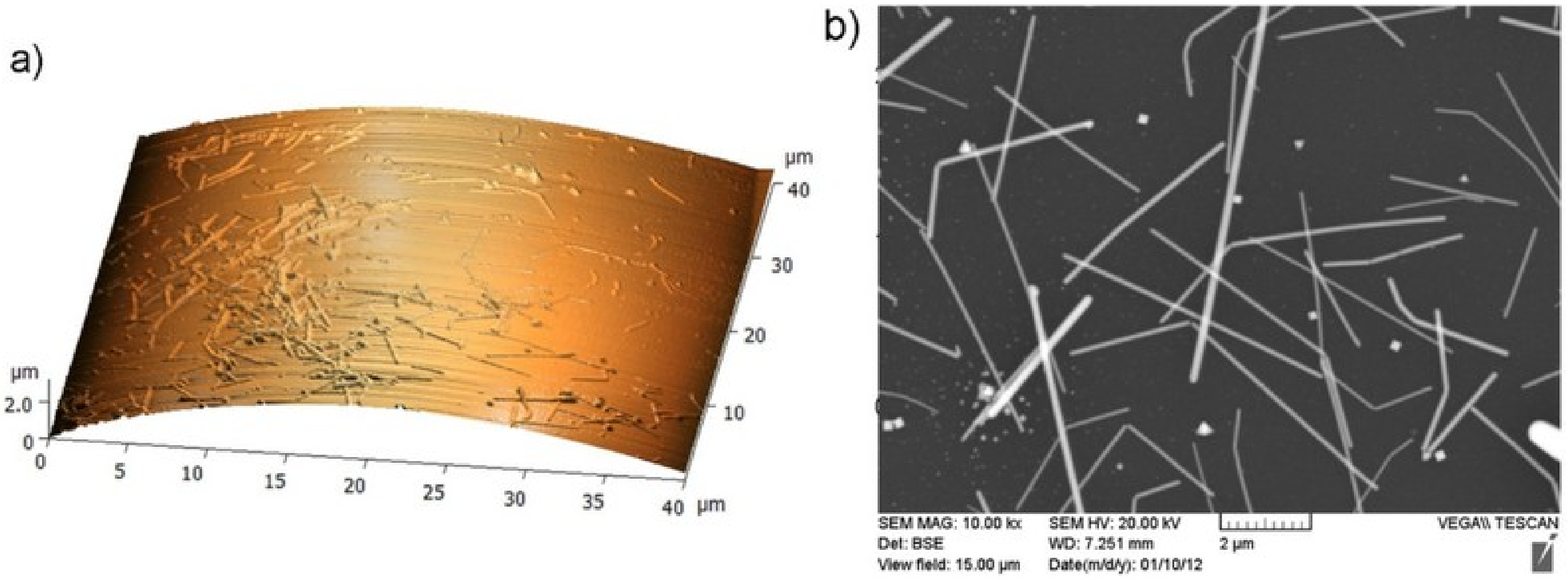}
{The AFM (a) and SEM (b) images of the fibre surface coated with silver nanowires~\cite{Bialiayeu:2012}.}

\Fig{Depos_nanorod_2}{0.95}{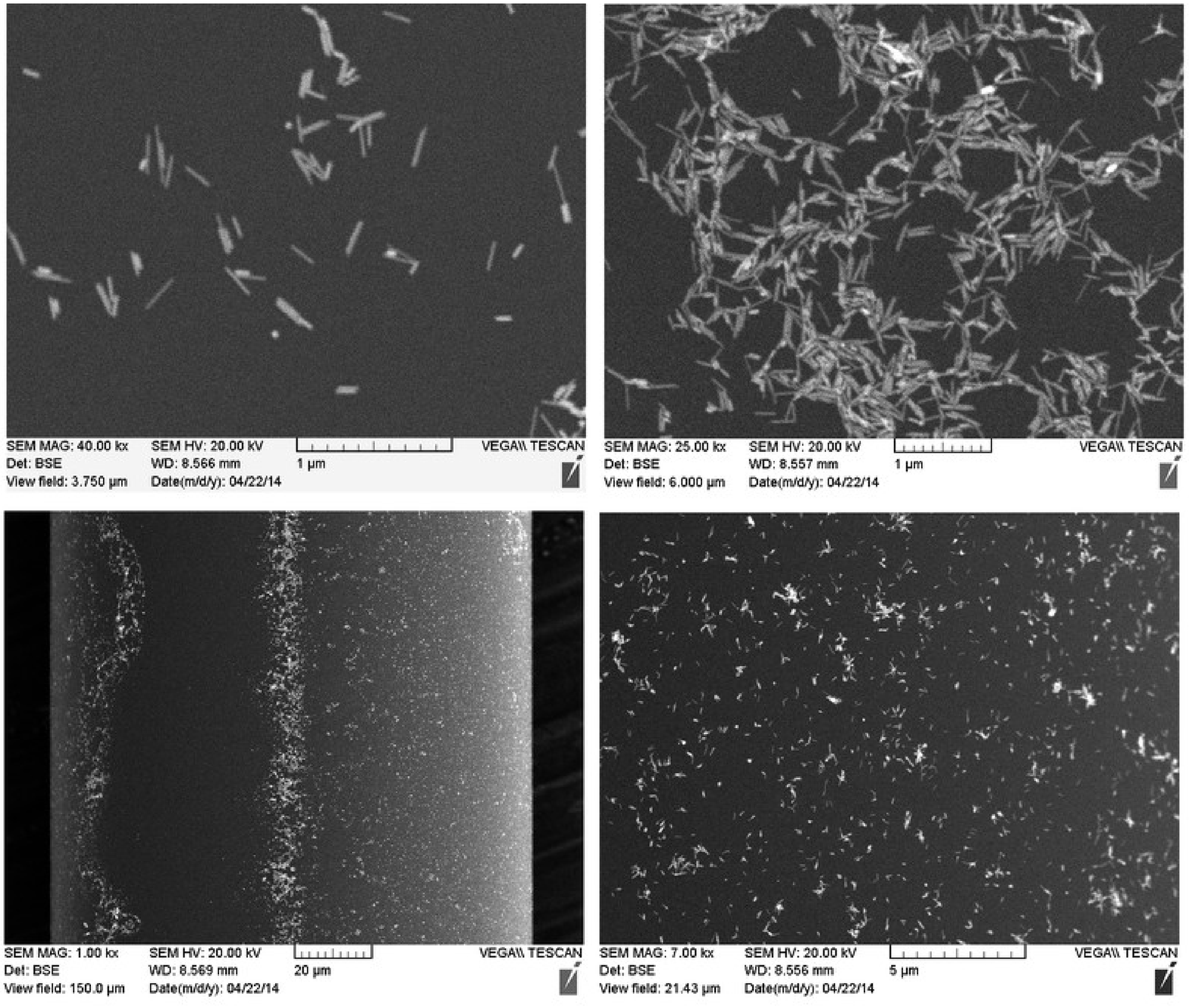}
{The SEM images of gold \C{nanorod} based coating}

\Fig{Depos_spheres_2}{0.9}{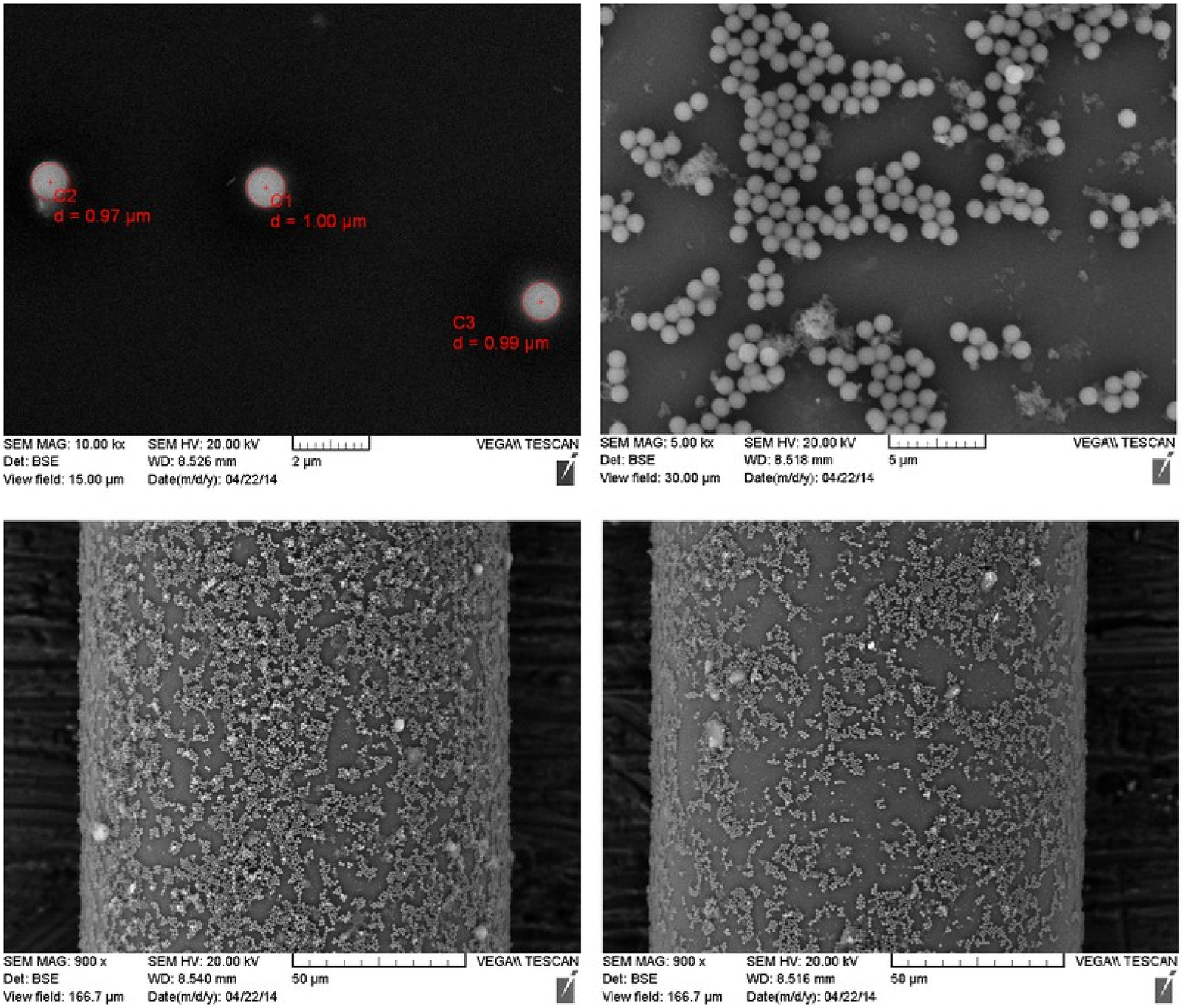}
{The SEM images of $Ti O_2$ spheres deposited on the fibre surface}

\Fig{Films_NC}{0.95}{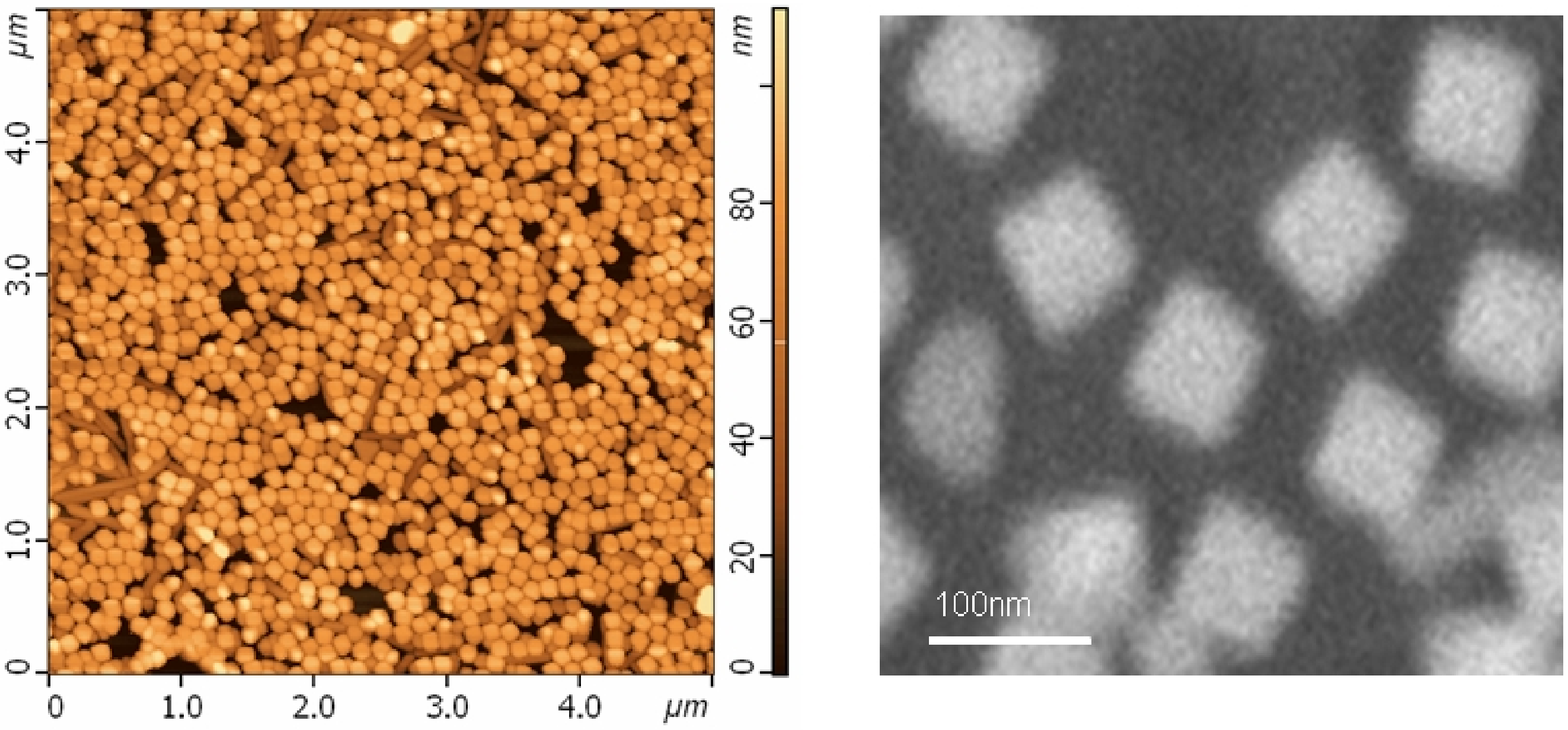}
{The AFM and SEM images of silver nanocubes (80nm)}

%% file: Chap_Chem_Deposition_Methods.tex
\section{Deposition and synthesis techniques}
In this section we briefly describe chemical methods of particles synthesis and deposition techniques that were used in the presented work.


\subsection{Synthesis of silver \C{nanorods} and nanowires}
Silver nanowires were chemically synthesized as described in~\cite{Sanders:2006}. 
The procedure results in highly crystalline nanowires with smooth surfaces. 
All the reagents used for synthesis were obtained from Sigma-Aldrich. 
All glassware was cleaned using aqua regia, rinsed in $18.2$~$M\Omega$ deionized water, and placed in an oven to dry prior to experimentation.

A $50$~mL round bottom flask containing $24.0~mL$ of anhydrous $99.8\%$ ethylene glycol (EG) and a clean stir bar were placed in an oil bath set to $150^o~C$ and allowed to sit for $1$ hour. 
Using a micropipette, $400~\mu L$ of $3~mM$ sodium sulfide dissolved in EG was added to the flask. 
Ten minutes later $6~mL$ of EG containing $0.12~g$ of dissolved poly(vinylpyrolidone) (PVP) with a molecular weight of $55000~amu$ was injected using a glass syringe immediately followed by the injection of $0.5~mL$ of $6~mM$ HCl. 
After an additional $5$ minutes, $2.0~mL$ of $282~mM$ $99\%$ + silver nitrate dissolved in EG was injected slowly using a glass syringe. 
Upon addition of the silver nitrate the solution immediately turned black and slowly became a transparent yellow, then changed to an ochre color as some plating in the flask occurred. 
The reaction was allowed to continue until the solution became white. 

The reaction was monitored by periodically taking small samples out of the reaction flask using a pasteur pipette and dispersing it in a quartz cuvette filled with $95\%$ ethanol for UV-visible spectroscopy. 
The reaction was quenched by placing the flask in an ice bath after the solution and had become fully white and turbid in appearance.
Purification of the nanowires was achieved by adding $20~mL$ of ethanol to the solution and centrifuging it at $13800$~g for $20$ minutes to remove the excess PVP, EG, and any reaction by-products. 
The supernatant was then discarded and the rods re-dispersed in ethanol by sonication. 
This process was then repeated several times at $400~g$ to separate out the heavier wires from the solution.

\subsection{Synthesis of gold nanoparticles}

The gold nanoparticles were synthesized by the reaction of citrate reduction of chloroauric Acid $(H[Au Cl_4])$ in water, as was proposed by Turkevich~\cite{Turkevich:1951, Kimling:2006, Daniel:2004}.
The reducing agent sodium citrate  $(Na_3 C_6 H_5 O_7)$ reduces gold ions to neutral gold atoms, when the solution becomes supersaturated gold nanoparticles start to form.
Citrate ions act as both a reducing agent and a capping agent. 
The repulsion of the negatively charged citrate ions prevents the nanoparticles from forming aggregates~\cite{Daniel:2004}. 
As the result of this process spherical gold nanoparticles in the range from $10$ to $20$ nm in diameter suspended in water were created.

\subsection{Deposition of nanoparticles}

At the first step, the fibre surface was cleaned and prepared for chemical deposition. 
The plastic jacket around the fibre was removed by soaking it into dichloromethane ($CH_2 Cl_2$) and further cleaned to remove organic residue through the following multistep approach. 
The uncoated area of the fibre was rinsed with methanol and subsequently treated with freshly prepared piranha solution (mixture of $H_2O: NH_3: H_2O_2 = 5:1:1$) at $70^o~C$ for $10$ minutes to remove the remaining organic residues, and rinsed in $18.2~M\Omega$ deionized water.

The cleaned fibre was then submersed in a $1\%$~(3-aminopropyl)trime-thoxysilane (APTMS, Aldrich, $97\%$, 281778) in methanol for $20$ minutes in order to form a uniform a positively charged monolayer on the fibre surface. 
The APTMS molecules assembled on the glass by covalent bonding were exposed to hydroxyl sites ($Si-OH$) on the glass, thus forming a cross-linked self-assembled monolayer (SAM).
The APTMS treatment replaces the hydroxyl groups (OH) adsorbed on the glass (fibre) substrate  $(Si O_2)$ with APTMS molecules forming a siloxane bond between the Si on one end of the APTMS molecules and an oxygen atom on the $Si O_2$ surface~\cite{Sato:1997}.
As a consequence, the amino group attached on the other end of the APTMS molecule is oriented away from the substrate.
These amino groups on the APTMS molecules immobilise gold particles onto the substrate because of the affinity of the amino group to the gold~\cite{Sato:1997}.

Finally, the APTMS modified fibre was rinsed with methanol and deionized water followed by drying with a flow of $N_2$ gas. 
After drying, the modified fibre was submerged into a freshly prepared colloidal solution of gold nanoparticles and left for $24$~hours. 
The results of nanoparticles deposition are shown in Figures~\ref{film_wires2},~\ref{Depos_nanorod_2} and~\ref{Films_NC},.

\subsection{Electroless metal coating}

The fact that a cylindrically symmetric fibre is used makes the deposition of an uniform nanometer-scale film quite challenging.
Previously in our group the time calibrated two-step sputtering process was used in order to deposit gold films~\cite{Shevchenko:2007,Kim:2008}, but we faced some reproducibility issues that could only come from coating errors. 
In an effort to improve this situation we had investigated electroless plating~\cite{Lo:2005} because of its potential for highly uniform metal coatings on non-planar surfaces, its low cost and simplicity of implementation (with a potential for batch production), and its relatively low deposition rate, which facilitates the control of the process duration.
The electroless gold and silver coatings can be deposited reliably and accurately for the fabrication of near infrared TFBG-SPR sensors, but most importantly the TFBG itself can be used to monitor the deposition process and to stop it at the optimum film thickness, for example for SPR excitation of metal coated fibres in water.
While coating conventional FBGs with copper and nickel had been described earlier for the purpose of making the FBG response athermal, the exact thickness and the uniformity of the films was not critical for those applications~\cite{Lo:2005}. Further studies of thermal stress between such metal coatings and optical fibres were performed in~\cite{Feng:2010}, where it was shown that the metal coating of FBG sensors can be used not only as a protective layer but also as a temperature sensitivity booster.

The method we were using to coat the TFBG sensor relies on the attachment of gold nanoparticles on a suitably prepared bare fibre surface, followed by the electroless plating process, based on the reduction of metallic ions from a solution to a solid surface without applying an electrical potential~\cite{Hrapovic:2003,Paunovic:2006}.

To initiate the reaction the fibre surface was precoated with gold nanoparticles, with the method described in the previous section.
The electroless plating is based on the process of negative charge production on the surface of gold nanoparticles through the oxidation reaction assisted by a reducing agent. 
The gold nanoparticles immobilized on the glass surface constitute excellent sites for the reduction of gold or silver ions by the hydroxyl amine solution.
Thus, the nanoparticles attached to the fibre surface play a catalytic role during the process of electroless plating.

The modified with gold nanoparticles fibre was next dipped into the plating bath.
For the silver coating the plating solution consisted of silver nitrate ($Ag NO_3$, $0.01~M$), ammonium nitrate  ($NH_4 NO_3$, $8.96$M), acetic acid  ($CH_3 COOH $,  $2.24$M), and hydrazine hydrate ($H_2 NH_2 .H_2 O$, $0.4~M$) at a volumetric ratio of $1:1:1:2$, respectively.
For the gold plating the solution consisted of $0.01\%$ chloroauric acid $(H Au Cl_4 .3H_2O)$ and the reducing agent hydroxylamine hydrochloride ($NH_2 OH HCl$, $0.4~M$ ) in a $1:1$ volumetric ratio.

The electroless plating occurs on the surfaces of already immobilized gold nanoparticles, eventually merging into a continuous thin metallic film, as shown in Figure~\ref{film_my_gold}.
The film thickness can be controlled in real-time with high precision~\cite{Bialiayeu:2011}. 
The film morphology can also be controlled by varying the deposition time, the solution composition, or the type and size of nanoparticles.

As a separate check we further calibrated the monitoring process by performing Atomic Force Measurements (AFM) on sample films to correlate the TFBG response with the physical thickness of the films.
The AFM measurements provided evidence that the electroless-deposited gold layer retained some significant granularity. 
During the deposition, we monitored the wavelengths and amplitudes of several resonances in the transmission spectrum.

%% file: Chap_Chem_Results.tex
\section{The optimal metal coating for Surface Plasmon Resonance (SPR) excitation}

The sensitivity of TFBG sensor can be enhanced by coating its surface with a thin metal film, of a proper thickness, thus allowing the energy coupling between the backward-propagating cladding modes and the collective electron oscillation on the metal surface, the so-called surface plasmon resonances (SPR)~\cite{Matsubara:1988}.
The sensitivity enhancement of the TFBG-SPR sensor was previously reported in~\cite{Shevchenko:2007, Caucheteur:11}. Here we focus on optimization of the coating process. 

In order to optimize the SPR sensitivity and resolution, the metal coating thickness and uniformity must be controlled very accurately around the fibre circumference, for thicknesses of the order of a few tens of nm. 
For coatings that are too thin, the SPR resonances broaden and weaken. 
On the other hand, when the thickness is too large, light cannot tunnel across the metal to excite the plasmon polariton on the outer surface. 

Here we present a novel method for the optimal metal coating of TFBG sensors by electroless plating,  thus creating a Surface Plasmon Resonance (SPR) finely tuned into the operational range of the TFBG sensor~\cite{Bialiayeu:2011}.

The electroless plating of gold or silver metals from solution is a simple and efficient way to deposit a conformal coating of the required thickness and uniformity for fibre SPR applications, and furthermore, the deposition process can be controlled in real-time by means of a TFBG sensor and stopped at the optimum film thickness for SPR excitation. Therefore, a precise control of the plating process (temperature, chemical concentrations) is not necessary and reproducible coatings with thicknesses on the order of 50 nm can be obtained regardless of the speed of the plating process. 

A quantitative measure of the quality of the SPR response of the TFBG in the plating solution is provided by tracking the polarization-dependent loss (PDL) in real time and used as a criterion to end the plating. 
When the metal coating has a thickness for which a plasmon wave can be excited by cladding modes, the PDL spectrum acquires a deep “notch” that reveals the SPR location~\cite{Caucheteur:11}. 
Using the polarization-based optical sensing method, presented in Chapter~\ref{Chap_polarization}, we can identify the precise moment of the plating process at which the SPR signature is observed. When the spectral notch is maximized, the thickness is optimal for SPR operation. 
This is best observed in the density plot of Figure~\ref{Exp_res3}, where the envelopes of the PDL spectra are represented as a colour scale in the horizontal direction at different moments of the plating process, the vertical direction. 

\Fig{Exp_res3}{1}{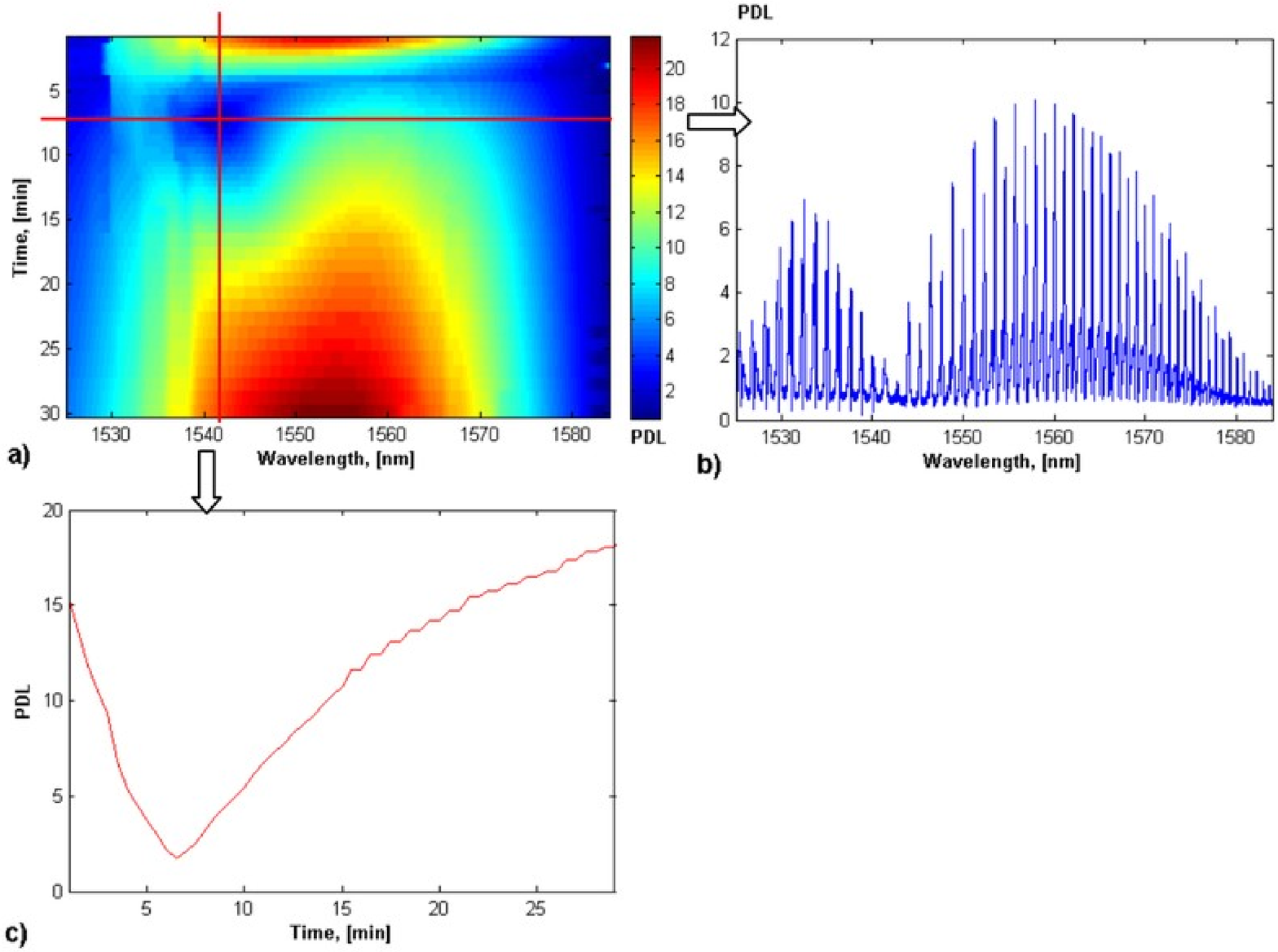}
{(a) The envelope of PDL spectra, taken continuously along the course of gold film deposition, and cross sections centered at the point of the deepest notch: wavelength $= 1542~nm$  (b) and time $= 7~min$ (c)}

Indeed, it can be clearly seen that after only approximately $7$~minutes of gold film deposition a deep notch in the PDL envelope occurs (darkest blue colour in Figure~\ref{Exp_res3}(a)). 
The individual PDL spectrum corresponding to this plating time is shown as Figure~\ref{Exp_res3}(b), while Figure~\ref{Exp_res3}(c) extracts the time evolution of the PDL value at the wavelength where the SPR is observed. 
The optimum point is clearly seen in the latter figure, hence the deposition process can be interrupted as soon as a local PDL envelope minimum is detected, regardless of the plating rate. 

We should note that the SPR signature appears suddenly, hence the monitoring process is necessary. 
Figure~\ref{Exp_res3}(c) further shows that the PDL envelope evolution slows down gradually for longer plating times. These effects are due to both the self-termination of the plating process and to the fact that as the metal layer becomes thicker it eventually shields the light from the cladding modes from the outside medium so that further thickness growth is not detected. 

We conclude that using TFBG's, we have presented a new method to monitor the growth of the plated film in situ and in real time.
Most importantly however, we showed that by monitoring a simple parameter in the polarization-dependent Loss of the TFBG during plating, the optimum film thickness for SPR operation can be found with great accuracy, regardless of the plating rate.
It is also likely that other metals could be plated and monitored in similar fashion.
The method is inherently limited in thickness since the evanescent field of the cladding modes must tunnel across to the outer surface of the metal film in order to detect changes in thickness.
When the thickness exceeds the penetration depth of the light, the thickness growth appears to saturate.
Obviously for the application of fibre SPR, the thickness needed must be smaller than the penetration depth, so this does not present an actual limitation in this case.


\section{Sensitivity enhancement with a nanoparticle based coating}

As we discussed in the previous section the sensitivity of a TFBG sensor can be increased by coating its surface with a metal film of the appropriate thickness, thus creating conditions for SPR excitation.
However, such improvement occur only over a narrow spectral range for which the phase matching condition between fibres cladding modes and the SPR wavevector is satisfied. 
In this section we present a new technique for increasing the sensor sensitivity. 
Instead of a resonance in continuous metallic film we proposing to use resonances of nanoparticles deposited on the sensor surface~\cite{Bialiayeu:2012}. 

We showed in~\cite{Bialiayeu:2012} that by sparsely coating the TFBG sensor with randomly oriented silver nanowires, a large number of fibre modes with different wavevectors and electric field distribution can resonate with nanowires, exciting localized surface plasmon resonances (LSPR). 
LSPR are collective oscillations of conducting electrons in the nanoparticles giving rise to strongly enhanced and highly localized electromagnetic fields, which can be utilized in various sensing devices~\cite{Shevchenko:2007}.

\subsection{Coating with silver nanowires}

\Fig{film_wires4}{0.65}{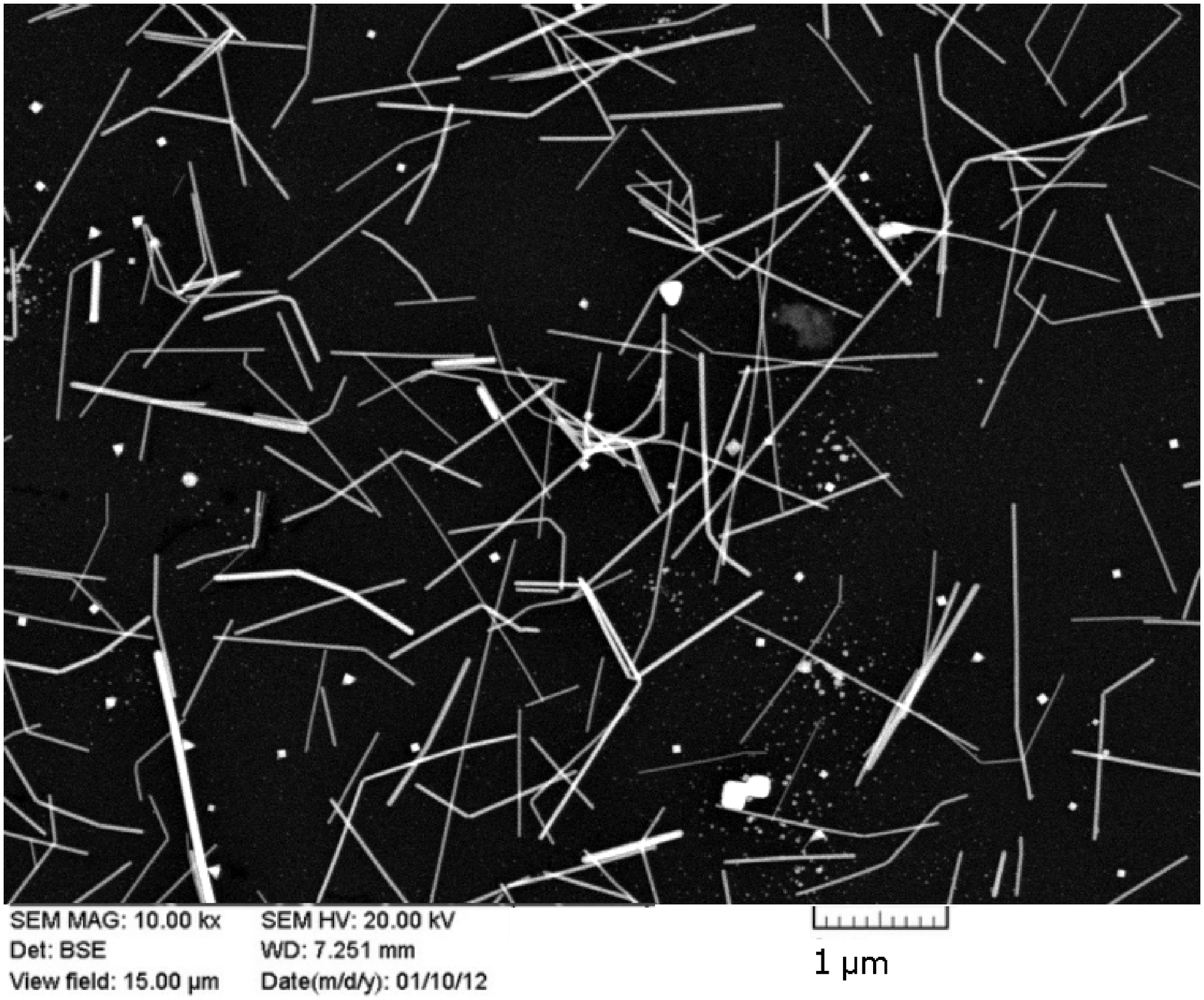}
{The SEM (b) images of the fibre surface coated with silver nanowires~\cite{Bialiayeu:2012}.}

The sensor was coated with chemically synthesized silver nanowires ${\sim 100~nm}$ in diameter and several micrometres in length, as shown in Figure~\ref{film_wires2}.
A UV-vis-NIR absorption spectrum was collected on a similar sample deposited on a flat glass slide. 
The spectrum shown in Figure~\ref{Depos_nanorod_3} contains two main features: a peak at $\sim 380-400~nm$  corresponding to the excitation of transverse plasmon resonances along the short axis and a broad peak in the NIR region corresponding to the excitation of longitudinal plasmon resonances along the long axis. 

\Fig{Depos_nanorod_3}{0.55}{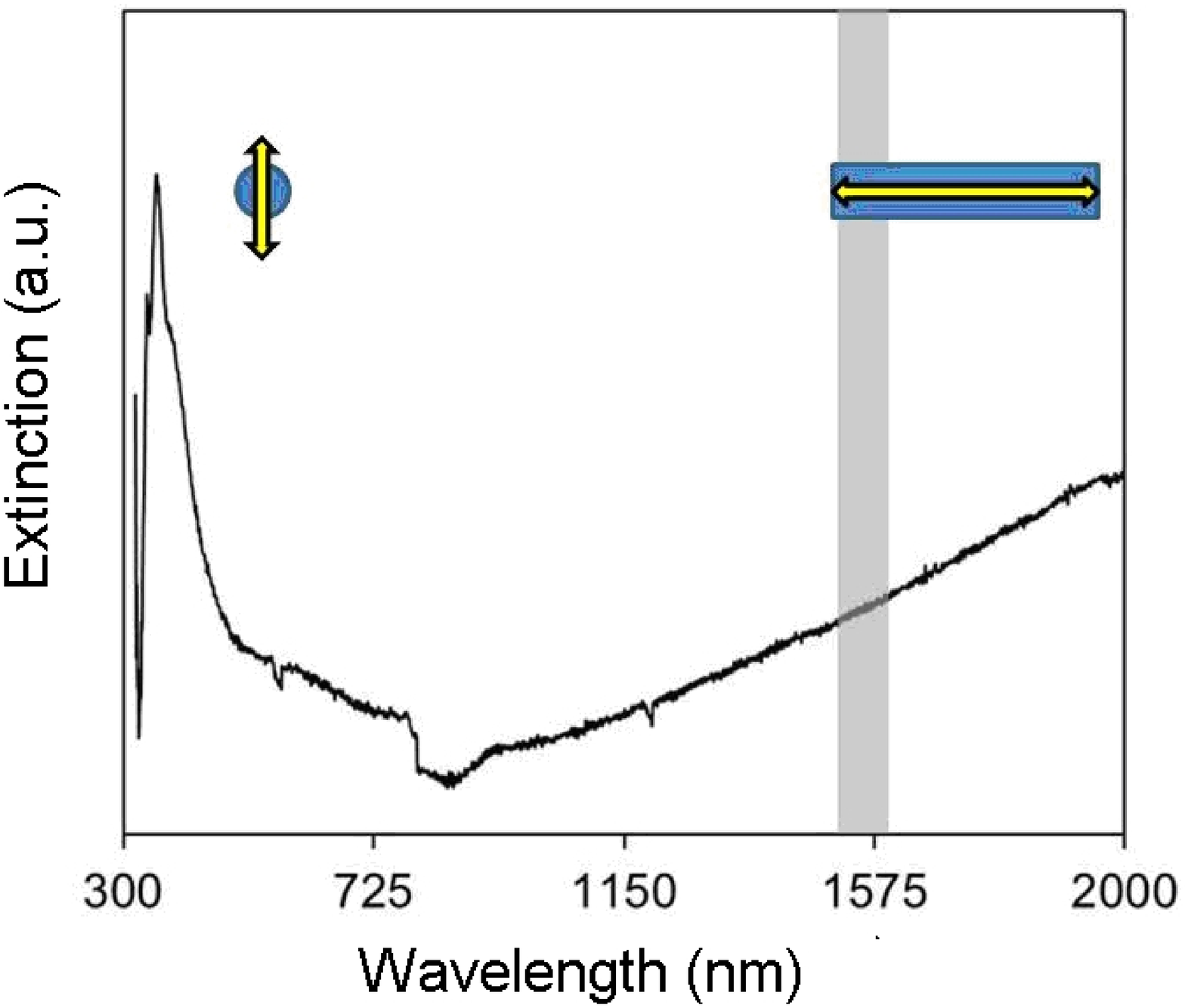}
{UV-vis-NIR absorption spectrum of synthesized nanowire coating (deposited on a flat glass substrate). The insets illustrate the relative polarization of the Plasmon oscillations that give rise to the absorption.}

The spectral region of interest for TFBG ($\lambda \in [1525,1590]~nm$) is highlighted in grey on the figure. 
It is seen that a strong extinction signal for silver nanowires exists at these frequencies. However, since the nanowires were deposited on the glass surfaces (fibre or flat substrate) using the self-assembly approach~\cite{Grabar:1995} (made possible by their partially negative charge~\cite{Badawy:32}), their orientation was not controlled (Figure~\ref{film_wires2}). Therefore the excitation of the longitudinal plasmons in the $1550-1600~nm$  band by polarized light sources can only occur for those wires that are parallel to the incident light polarization (i.e. about half on average). It appears impossible to excite the short (radial) plasmons of these nanowires at the same wavelengths.

At the first step of the characterization process, the TFBG sensor was tested to determine its refractometric operational range before and after the nanowire coating deposition. Mixtures of ethylene glycol (EG) and water were used with different volume ratios, varying from pure water to pure EG, allowing to cover a range of $\Delta n = 3.81 \times 10^{-2}$ for the refractive index of the surrounding medium. 
A custom-made Teflon cell was used to hold the fibre still and straight while being submerged in varying solutions of ethylene glycol (EG) and water while transmission measurements were carried out with an optical vector analyser (OVA).
Figure~\ref{Exp_res6} shows the average insertion loss of the coated and uncoated sensor for various intermediate values of the surrounding refractive index.

\Fig{Exp_res6}{1}{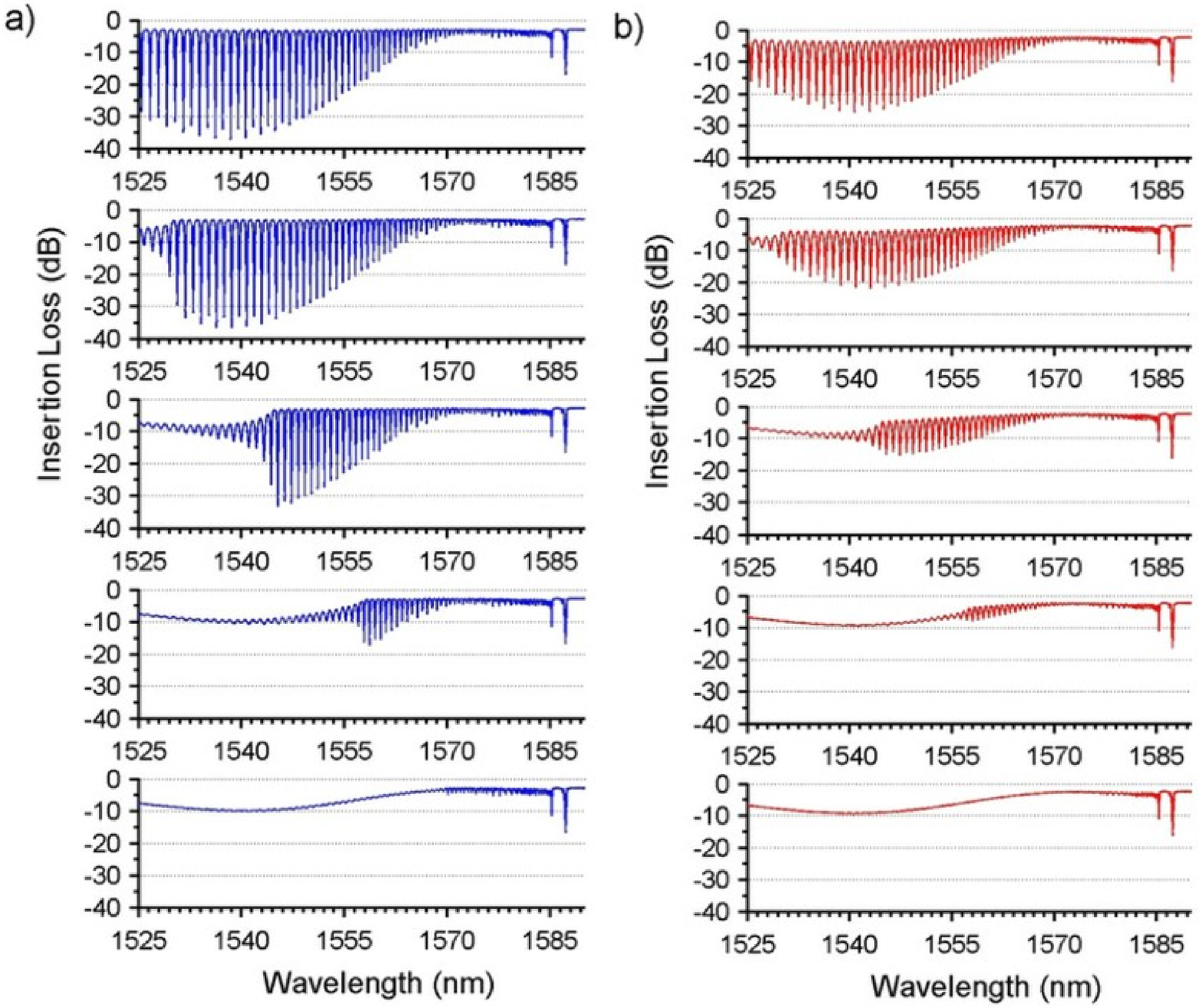}
{The TFBG spectrum evolution before (a) and after (b) deposition of nanowires, for several values of the refractive index of the solution. The concentration of the Ethylene Glycol in water goes from $0\%$, $25\%$, $50\%$, $75\%$, $100\%$ from the top to the bottom of the figure. 
The corresponding total refractive index change is $\Delta n = 3.81 \times 10^{-2}$. }

In those spectra, each downward peak corresponds to loss of light in the core, due to coupling to cladding modes. A discontinuity in the amplitudes of the resonances is observed on all spectra and its spectral position shifts towards longer wavelengths as the surrounding index increases. 
This discontinuity corresponds to the cut-off condition where the cladding modes become leaky because their effective index becomes larger than the surrounding index~\cite{Chan:2007}. 
The behavior of both groups of spectra is similar, apart from the fact that the amplitudes of the resonances are always smaller for the coated grating. This is consistent with the assumption that plasmons can be excited in the nanowires in the $1525-1590~nm$  range, thereby inducing additional loss in the cladding modes, which results in a decrease of the coupling strength (and of the resonance amplitude).
The observation of the shift in the cut-off wavelength of the average insertion loss provides a relatively rough (but absolute) estimate of the refractive index of the solution. 

For finer measurements (of small index change increments), observations on a single resonance provides better accuracy. 
The refractometric sensitivity measurements were performed on the TFBGs before and after the deposition of silver nanowires by monitoring the response following small changes of the relative EG concentration.
The initial measurements were taken using a well mixed $1:1$ mixture of EG and H2O prepared in a large $200~mL$ beaker and transferred to the $5~ml$  Teflon cell.
Subsequent measurements were taken with solutions obtained by adding $10~\mu l$  of water to the cell and continuous stirring. The mechanical stirring was stopped prior to each measurement to eliminate potentially detrimental fluid motions and acoustic disturbances around the fibre that could result in false readings.
After each dilution, the refractive index of the solution was determined from Lorentz-Lorenz mixing equation~\ref{eq:eqLL}~\cite{Tasic:1992}:
\Eq{}
{\frac{n^2 - 1}{n^2 +2} = p_1 \frac{n_1^2 - 1}{n_1^2 +2} + p_2 \frac{n_2^2 - 1}{n_2^2 +2}.}
Here, $n$ represents the refractive index of the mixture, $n_1$ and $n_2$ are refractive indices of pure components (water and EG, $1.318$ and $1.394$, respectively at wavelengths near $1550~nm$~\cite{Weber:2003}), and $p_1 = V1/V$, $p_2 = V2/V$ are the volume fractions of the components.

Figure~\ref{Exp_res2} (a and b) depicts the response of an individual resonance to small index variations, before and after the sensor surface was modified with silver nanowires, and for the two singular values of the transfer function. 

\Fig{Exp_res2}{0.9}{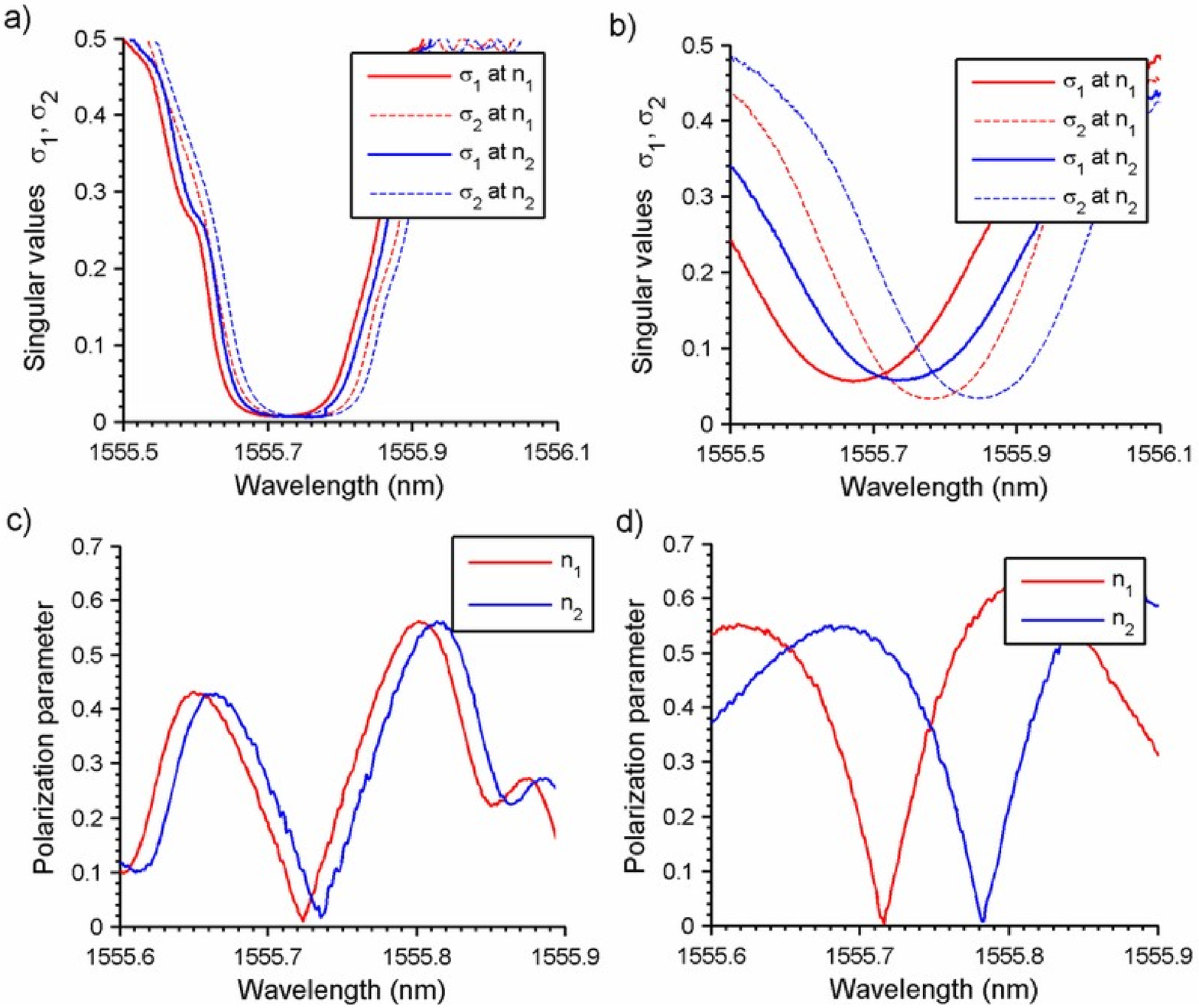}
{Singular values (a,b) and polarization-dependent loss parameter (c,d) (linear scale) changes due to a small refractive index change of $\Delta n = 3.77 \times 10^{-4}$ , before (a,c) and after (b,d) deposition, corresponding to a single resonance taken at ${\lambda = 1555.7~nm}$ .}

We note that the most obvious effect resulting from the nanowire coating is the increase in the splitting between the two singular values, from $\sim 30~pm$  to $\sim 100~pm$ (comparing the central wavelengths of $\lambda_1$ and $\lambda_2$ for $n_1$ or $n_2$). 
This increase should be associated with an increase in the polarization dependence of the boundary condition for the cladding modes, as expected for metal particles. A similar effect had been observed previously by our group for gratings covered by a sparse layer of spherical copper nanoparticles deposited by a pulsed chemical vapor depositio~(CVD) technique~\cite{Shao:2011}.

It was also observed that the singular value spectra shifted to shorter wavelengths with decreasing refractive index. 
However, this shift is somewhat difficult to quantify precisely because the resonances are rather broad and flat bottomed. 
Instead of isolating one of the singular value to quantify the shifts (and thereby throwing away half the data, corresponding to the other singular value) we will follow zeros in PDL spectra, corresponding to resonances of interest,  as we discussed in Chapter~\ref{Chap_polarization}.
We are basically measure differential spectrum between polarization states, which provides extremely sharp peaks (Figure~\ref{Exp_res2}c and~\ref{Exp_res2}d) that can be tracked down with much greater accuracy than the insertion loss or singular value resonance position.

Conducting subsequent measurements for fine variations in the refractive index obtained by the step-wise addition of $10~\mu l$ quantities of water to the solution we determine the refractometric sensitivity for several peaks. 
Each addition of water caused a small decrement of ${\Delta n \sim 7.5 \times 10^{-5}}$ in the refractive index of the surrounding medium. 
Those measurements can be carried out on any of the resonances. The results of these detailed measurements are shown in Figure~\ref{res_wires2}, where a set of $5$ different individual resonances were measured in order to determine if the spectral position of a resonance had an impact on its sensitivity, shown in Figure~\ref{res_wires2}b.

\Fig{res_wires2}{0.8}{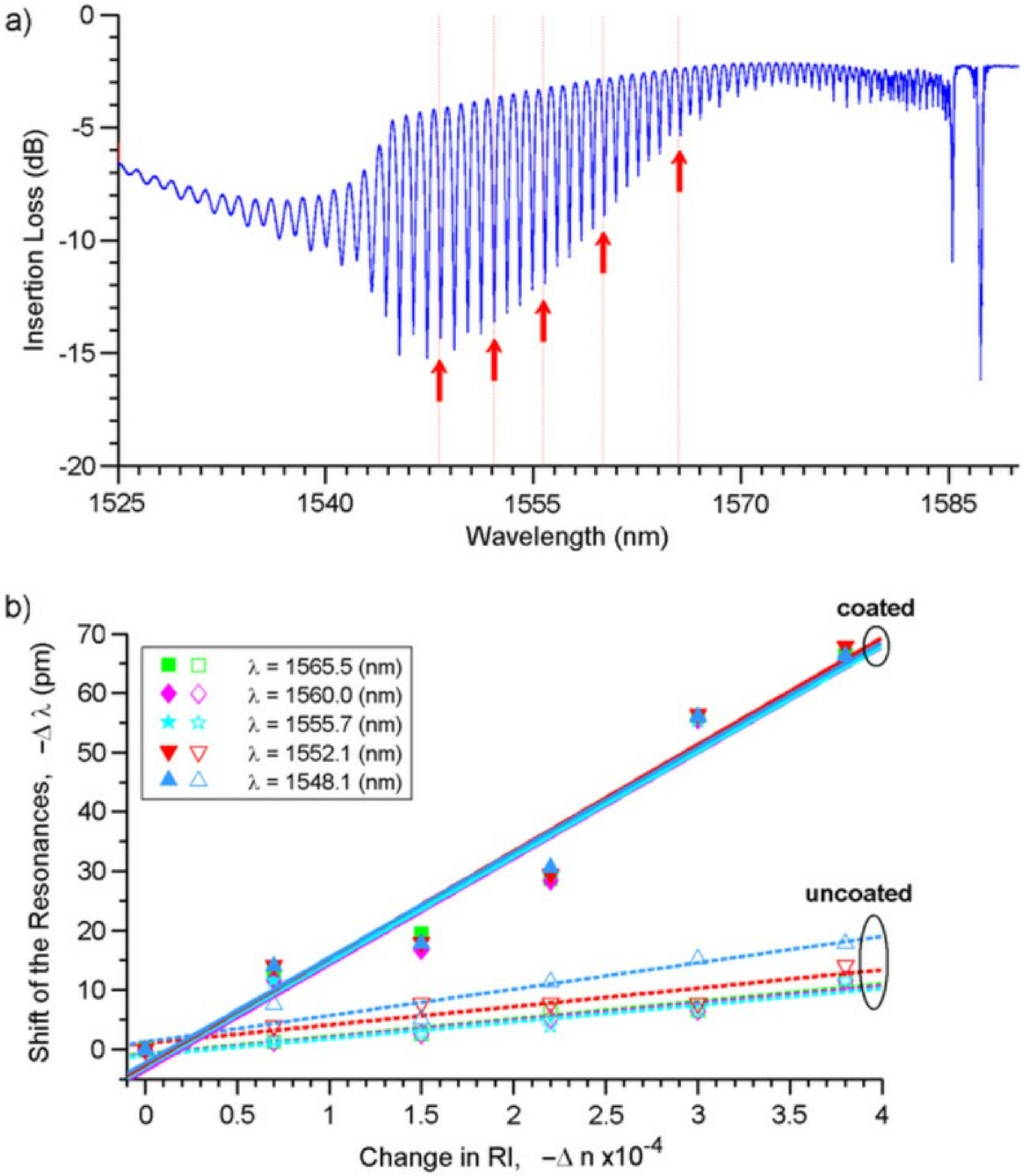}
{(a) Wavelength shift $\Delta \lambda$ of several individual resonances of the TFBG sensor due to the surrounding refractive index change $\Delta n$.
(b) The sensetivity before (open marks) and after (closed marks) deposition~\cite{Bialiayeu:2012}.}

Comparing the slopes in Figure~\ref{res_wires2}b for the resonances of the coated and uncoated TFBG, we obtain an increase in sensitivity from $53~nm$  $RIU^{-1}$ for the bare uncoated TFBG sensor to $185~nm$ $RIU^{-1}$ for the sensor coated with silver nanowires (RIU stands for \C{``Refractive index unit''}). 
\C{Although the reported} sensitivity for metal coated SPR TFBG sensors is reported to have a higher value, of about $555~nm$ $RIU^{-1}$~\cite{Shevch:2010}, the operational range of SPR sensors is limited to a narrow range of few nanometers where the resonant phenomena of SPR excitation is observed.
The proposed sensor with a coating of nanowires has an advantage over a SPR sensor as the increase in sensitivity is observed over the whole operational range of the sensor, as shown in Figure~\ref{res_wires2}a.

\clearpage
\subsection{Discussion}

The observed enhancement in sensitivity was investigated by taking a close look at the electromagnetic field interaction with the silver nanowires.
The device is capable of sensing the external medium by means of the evanescent field of cladding modes, leaking outside the fibre cladding, as shown in Figure~\ref{E_boundary_1nm}.
We note that the evanescent field decays slowly enough to extend into the thin layer of silver nanowires, thus conditions for coupling between the evanescent field incident from the fibre and the nanoparticles on its surface are created.
Perturbations of the evanescent field modify how the grating couples the light from the core to each cladding mode and hence the resonances observed in the transmission spectrum.
Knowledge of the fibre geometry and refractive indices, as well as of the grating period, tilt angle and modulation amplitude, allow us to uniquely assign specific cladding modes to the resonances observed experimentally.
The electric field distribution corresponding to two closely positioned TM--like and TE--like modes of the same azimuthal family at ${\lambda = 1548.1~nm}$ and ${\lambda = 1552.1~nm}$ are shown in Figure~\ref{NP_rod_assymetry2}.

\Fig{NP_rod_assymetry2}{0.9}{NP_rod_assymetry.eps}
{Vectorial $\V{E}$ field structure for two modes with almost identical propagation constants (hence resonance wavelengths), but different polarization states ((a)—TM--like mode, (b)—TE--like mode). Silver nanowires are shown schematically on top.}

As discussed in Section~\ref{compute_ellipsoidal}, the coatings with cylindrical particles give rise to very prominent and sharp longitudinal resonances, which shift toward the infrared band with an increase of particle elongation. 
The transversely polarized electric field excites a series of blue-shifted transverse resonances in the case of elongated nanoparticles. 
The particles should be chosen in such a way that the resonances are positioned far away from the operational range of the sensor, and hence the dielectric permittivity $\epsilon_\rho$ is a relatively constant function in the operational range. 
However, the longitudinal resonances of the particles, excited by the tangentially polarized electric field $\V{E}_\phi$, are positioned exactly in the range where the sensor operates, so that the energy coupling can occur between the TE--like modes of the sensor and local resonances of the particles, as shown in Figure~\ref{NP_rod_assymetry2}. 
The imaginary part of the $\epsilon_\phi$ dielectric permittivity function has a resonance in the operational range. The position of the resonance is defined, among other parameter, by the external refractive index of the surrounding medium. 
Therefore the differential sensitivity with the TE--like and TM--like modes is possible, due to the fact that the modes are affected differently by changes in the external refractive index.

The TFBG couples core-guided light to two different families of cladding modes according to the polarization of the input light relative to the orientation of the tilt plane, as shown in Figure~\ref{E_boundary_All}.
To be precise, when the input core-guided light is polarized linearly in the plane of the tilt (corresponding to P-polarized light), the cladding modes that are excited by the TFBG belong to the TM--like family of modes and have their electric field polarized predominantly in the radial direction at the cladding boundary.
On the other hand, the S-polarized light (again relative to the plane of tilt), and the grating couples the core light to TE--like cladding modes whose electric field at the cladding boundary is mostly azimuthal (i.e. tangential). The calculated wavelength spacing and refractometric properties of the two modes in each pair reveal that they can be associated with the observed singular values of the insertion loss resonances measured above.
Therefore, we can explain the observed splitting of the singular value pair by noting how the electric field of the corresponding mode probes the nanowires: since the TM--like modes have electric fields that are predominantly radial at the cladding boundary while TE--like modes are mostly tangential, TM--like modes cannot excite the \C{``long axis''} plasmons of the nanowires while the TE--like modes can.

We also note that these results have been obtained in a relatively stable temperature environment, even though no active temperature control was used. 
In was reported~\cite{Chan:2007} that the underlying TFBG platform can be made temperature-independent over several tens of degrees by referencing all wavelengths to the Bragg resonance, bounded the fibre's core and completely insensitive to changes outside the fibre.
With this referencing scheme the only impact of temperature on the positions of the resonances arises from the change in the refractive index of the nanoparticles and of the medium that surrounds them. 
The temperature effect was studied with application to TFBGs coated with uniform gold films and it was found to be insignificant (about $10~pm / ^{\circ}\mathrm{C}$~\cite{Erdogan:1997})  compared to the shifts observed here.

\subsection{Conclusion}

The sensitivity of a TFBG refractometer can be increased at least $3.5$-fold by the addition of a sparse coating of silver nanowires. 
The coating is created by simple liquid phase self-assembly process that does not require special deposition tools, unlike other methods used to fabricate plasmon assisted devices (e-beam lithography, chemical vapour deposition~\cite{Shevchenko:2007, Shao:11, Berini:2011} or real time monitoring of the deposition process~\cite{Bialiayeu:2011} to achieve very specific layer thicknesses). 

The high sensitivity of the sensor is explained by the fact that the resonances of a $1(cm)$ long grating have $Q$ values in excess of $15000$ at near infrared wavelengths, due to full widths of $100(pm)$. 
As a result, the positions of these resonances can be followed accurately, even for shifts of the order of several picometres (reproducibility better than $3~pm$ was recently demonstrated experimentally with a similar measuring system~\cite{Caucheteur:11}). 

Associated with our $185~nm$ $RIU^{-1}$ sensitivity, a $3(pm)$  resonance position accuracy yields a minimum detection level of $1.6 \times 10^{−5}$ RIU.

The second advantage is based on the fact that different resonances have similar sensitivity and any of them can be used for refractive index sensing, as
opposed to other kinds of SPR sensor, where only one or a few resonances satisfying the phase matching condition have high sensitivity~\cite{Caucheteur:11}.
The high Q value of these resonances is a key factor in taking advantage of plasmonic effects because of the inherent trade-off between strong plasmon excitation (and enhanced sensitivity) and increased loss, which tends to broaden resonances: the TFBG configuration allows the device designer a great amount of control to adjust the amount of overlap between the evanescent field and the metal particles and also in the relative orientation of the light polarization and particle geometry.

%% file: Chap_Summary.tex
\chapter{Conclusion}

The primary focus of this book has been set on theoretical analysis of the TFBG sensor and proposing methods of enhancing its sensitivity by surface modification.

The TFBG sensor is not a trivial object. The sensor has a complex polarization-dependent response dependent on more than a thousand interacting modes. 
To address the challenge of the theoretical analysis, we developed a highly efficient and fast numerical solver, capable of computing a TFBG spectra in approximately $3$~minutes for a given state of incident light polarization.

\subsubsection*{Theoretical analysis}
The solver we developed consists of two major parts: the mode solver and the numerical integrator, which solve a system of coupled differential equations.

The mode solver is the key element, as it is required to compute more than a thousand modes at various wavelengths. We developed a simple yet efficient and fast full-vectorial complex mode solver, capable of handling waveguides of an arbitrary complex refractive index profile.
First, the system of Maxwell's equations was reduced to only two coupled ordinary differential equations for the electric field. 
Next, the equations were transformed into a system of algebraic equations with the help of a finite difference method. 
Finally the problem was reduced to the problem of finding eigenvalues and eigenvectors of a five-diagonal sparse matrix. The eigenvalue problem was effectively solved with the standard iterative method. 

At the next step the modes obtained for the non-perturbed case were used as the basis functions to solve the problem of tilted grating inscribed along the optical axis. The problem of small perturbations along the optical axis was handled with the coupled mode theory. As a result, the problem was reduced to a set of coupled differential equations, representing the energy transfer between the modes. 
Due to the grating tilt, the energy transfer depends on the core mode polarization. 
The polarization-dependent effects were thoroughly analysed in this work. Interesting insights into the properties of the electric field at the TFBG sensor boundary were gained.

The results of the simulation were shown to be in a good accordance with the experimental measurements.

\subsubsection*{Polarization-based experimental measurements}
Along with the theoretical study, we conducted an extensive experimental study of the optical properties of the TFBG sensor.  
We based our study on the Jones matrix and Stokes vectors data made available by means of Optical Vector Analyser (OVA).

We proposed two alternative approaches towards the TFBG sensor data analysis.
The first method was based on tracking the grating transmission of two orthogonal states of linear polarized light that were extracted from the measured Jones matrix or Stokes vectors of the TFBG transmission spectra. 
The second method was based on the measurements taken along the system principle axes and polarization-dependent loss (PDL) parameter, also calculated from measured data. 
It was shown that the frequent crossing of the Jones matrix eigenvalues as a function of wavelength leads to a non-physical interchange of the calculated principal axes. A method to remove this unwanted mathematical artefact and to restore the order of the system eigenvalues and the corresponding principal axes was provided.
A comparison of the two approaches revealed that the PDL method provides a smaller standard deviation and therefore a lower limit of detection in refractometric sensing. 

Furthermore, the polarization analysis of the measured spectra allows for the identification of the principal states of polarization of the sensor system and consequently for the calculation of the transmission spectrum for any incident polarization state. 
The stability of the orientation of the system principal axes was also investigated as a function of wavelength. 
A small oscillation in the orientation of the principal axes as a function of wavelength was observed.

\subsubsection*{Sensitivity enhancement}
In the presented work we investigate the sensor response to various types of nano-scale film coatings.
We have proposed the method of resonant coupling between the high quality factor resonances of the TFBG structure and the local resonances of nanoparticles deposited on the sensor surface.
The problem was investigated theoretically for spherical and elliptical particles made of various materials. The optimal parameters for particles were determined. 
A $3.5$-fold increase in the TFBG sensor sensitivity was observed experimentally by coating the sensor surface with silver nanowires.

\vspace{10 mm}

This doctoral project resulted in the following publications:
\newcounter{itemcounter}
\begin{list}
{\textbf{\arabic{itemcounter}.}}
{\usecounter{itemcounter}\leftmargin=1.4em}
\item A. Bialiayeu, C. Caucheteur, N. Ahamad, A. Ianoul, and J. Albert, ''Self-optimized metal coatings for fiber plasmonics by electroless deposition,'' \emph{Optics express} 19, 18742–-18753 (2011).

In this paper we presented a novel method to prepare optimized metal coatings for infrared Surface Plasmon Resonance sensors by electroless plating. 
We show that Tilted Fiber Bragg grating sensors can be used to monitor in real-time the growth of gold nanofilms up to $70~nm$ in thickness and to stop the deposition of the gold at a thickness that maximizes the SPR (near $55~nm$ for sensors operating in the near infrared at wavelengths around $1550~nm$). The deposited films are highly uniform around the fiber circumference and in spite of some nanoscale roughness (RMS surface roughness of $5.17~nm$) the underlying gratings show high quality SPR responses in water. 

\item A. Bialiayeu, A. Bottomley, D. Prezgot, A. Ianoul, and J. Albert, ``Plasmon-enhanced refractometry using silver nanowire coatings on tilted fibre Bragg gratings,'' \emph{Nanotechnology} 23, 444012 (2012).

In this paper we presented a novel technique for increasing the sensitivity of tilted fibre Bragg grating (TFBG) based refractometers. 
The TFBG sensor was coated with chemically synthesized silver nanowires $\sim 100$ nm in diameter and several micrometres in length. 
A $3.5$-fold increase in sensor sensitivity was obtained relative to the uncoated TFBG sensor. 
This increase was associated with the excitation of surface plasmons by orthogonally polarized fibre cladding
modes at wavelengths near $1.5~\mu m$. Refractometric information was extracted from the sensor
via the strong polarization-dependence of the grating resonances using a Jones matrix analysis of the transmission spectrum of the fibre.

\item W. Zhou, D. J. Mandia, M. B. Griffiths, A. Bialiayeu, Y. Zhang, P. G. Gordon, S. T. Barry, and J. Albert, ``Polarization-dependent properties of the cladding modes of a single mode fiber covered with gold nanoparticles,'' \emph{Optics express} 21, 245–-255 (2013).

In this paper the properties of the high order cladding modes of a standard optical fiber were measured in real-time during the deposition of gold nanoparticle layers by chemical vapor deposition. 
A correlation between the transmission spectra of the $10^o$ TFBG and the optical properties of gold particle coatings was established. 

My contribution to this work allowed to explain the observed effects by the numerical FDTD modelling of the gold particle coating layer, as described in Section~\ref{NP_FDTD_sim}.

\item A. Bialiayeu, A. Ianoul, and J. Albert, ``Engineering a resonant nanocoating for an optical refractive index sensor,'' in \emph{``Electronic, photonic, plasmonic, phononic and magnetic properties
of nanomaterials,''} , vol. 1590 (AIP Publishing, 2014), vol. 1590, pp. 68–-70.

In this work we proposed an idea to boost the performance of refractive index sensor based on the tilted fiber Bragg grating structure by resonant coupling of small spherical nanoparticles to the TFBG resonances. The optimal choice of nanoparticle parameters was discussed.

\item A. Bialiayeu, J. Albert, A. Ianoul, A. Bottomley, and D. Prezgot, ``Silver nanowire coated tilted fibre Bragg gratings,'' in \emph{``Bragg Gratings, Photo-sensitivity, and Poling in Glass Waveguides,''} (Optical Society of America, 2012), pp. BW2E–1.

In this work we reported a $3.5$-fold increase in sensitivity of TFBG based refractometers by coating the sensor surface with silver nanowires. A strong polarization dependence of the grating resonances was observed and analyzed.

\end{list}

%% file: Apendix_Code_MatLab.tex
\label{chAp}

The following code had been developed to obtain the exact full vectorial solution to the problem of cylindrical and slab waveguides of an arbitrary refractive index profile, including lossy waveguides with a non-zero imaginary part of the refractive index.

\subsubsection*{The main routine.}
\begin{lstlisting}
% r in [-R, R]

clear all
clc
clf

%============== Input =====================================
lam = 1.6;          % wavelength um
ko = 2*pi/lam;      % free space wave vector

% layer vectors
%RL = [4.1, 62.5, 70];            
%nL = [1.4492, 1.444, 1.3556];
%nL = [1.4492+ 0.001, 1.444 + 0.001, 1.33];

R = 1.5;
% n1 = 1.4492; n2 = 1.3556;   % refractive index
n1 = 3; n2 = 1.3556;   % refractive index
RL = [R, R + 3];
nL = [n1, n2];

m=1;
Neig = 40;            % number of eigenvalues to compute
N = 10000;

%============== Potential =================================

h = RL(end)/(N-0.5);
r = linspace(0.5*h, RL(end), N);
r = [- fliplr(r), r].';
r = r - 0.1*h; 

ni = f_construct_ni2(RL, nL, N);
ni = [fliplr(ni), ni].';

e = ni.^2;
%============== Solve ===============================

%[Ls, U] = f_FD_Construct_Matrix_Cyl_Exact_Vect(r, e, ko, m);
[Ls, U] = f_FD_Construct_Matrix_Cyl_WGA_Vect(r, e, ko, m);
%[Ls, U] = f_FD_Construct_Matrix_Cyl_Exact_Vect(r, e, ko, m);
[V, Eig] = f_FD_eig(Ls, Neig, U);

%============== Results ===================================
Neff = sqrt(Eig./ko^2);
U = sqrt(U/ko^2);

f_Plot_vec(m, r, U, Neff, V, 3, 1, 3.2)
%f_Plot_Scal_Single(12, 3, -7, 7, 2.5, 3.1, m, r, sqrt(U)./ko, sqrt(Eig)./ko, V, h)
\end{lstlisting}

\subsubsection*{The matrix assembly for the weakly guided approximation of the scalar problem.}
\begin{lstlisting}
function [Ls, u] = f_FD_Construct_Matrix_Cyl_WGA_Scalar(r, e, ko, m)
% My working code
% r - grid vector
% ki = k_o n
% m - if zero we get symmetric function: d_r u(r) = 0

h = r(3) - r(2);             % the step size
N = length(r);               % number of elements r

%================== Construct d_x matrix ==================================
B = zeros(N,3);
B(:,1)= -1;
B(:,3)= 1;
D = 1/(2*h)*spdiags(B,[-1,0,1],N,N); 

%================== Construct d_x^2 matrix ================================
B = ones(N,3);
B(:,2)= -2;
D2 = 1/(h^2)*spdiags(B,[-1,0,1],N,N);

%========================= Construct matrix ===============================
% (d_r^2 + 1/r d_r + [k^2 - m^2/r^2])u(r) = b^2 u(r)

u = e*ko^2 - m^2./r.^2;
U = spdiags(u,0,N,N);

% sparse matrices
Ri = spdiags(1./r,0,N,N);

Ls = D2 + Ri*D + U;

\end{lstlisting}

\subsubsection*{The matrix assembly for the weakly guided approximation of the vectorial problem.}
\begin{lstlisting}
function [Ls, u] = f_FD_Construct_Matrix_Cyl_WGA_Vect(r, e, ko, m)

h = r(3) - r(2);             % the step size
N = length(r);               % number of elements r

%================== Construct d_x matrix ==================================
B = zeros(N,3);
B(:,1)= -1;
B(:,3)= 1;
D = 1/(2*h)*spdiags(B,[-1,0,1],N,N); 

%================== Construct d_x^2 matrix ================================
B = ones(N,3);
B(:,2)= -2;
D2 = 1/(h^2)*spdiags(B,[-1,0,1],N,N); 

%================== Construct diagonal matrices ===========================
% vectors
u = e*ko^2 - m^2./r.^2;

u1 = -(1+m^2)./r.^2 + e*ko^2;
u2 = (m*2)./r.^2;

% sparse matrices
Ri = spdiags(1./r,0,N,N);
U1 = spdiags(u1,0,N,N);
U2 = spdiags(u2,0,N,N);

L = D2 + Ri*D; 
Ls = [L+U1, U2; U2, L+U1];
\end{lstlisting}

\subsubsection{The matrix assembly for the exact solution of waveguides with cylindrical symmetry.}
\begin{lstlisting}
function [Ls, u] = f_FD_Construct_Matrix_Cyl_Exact_Vect(r, e, ko, m)
% puled r under E_r, Need it as the dispersion curves are more realistic !!!

h = r(3) - r(2);             % the step size
N = length(r);               % number of elements r

%================== Construct d_x matrix ==================================
B = zeros(N,3);
B(:,1)= -1;
B(:,3)= 1;
D = 1/(2*h)*spdiags(B,[-1,0,1],N,N); 

%================== Construct d_x^2 matrix ================================
B = ones(N,3);
B(:,2)= -2;
D2 = 1/(h^2)*spdiags(B,[-1,0,1],N,N); 

%====================== The operator =============================
ri = 1./r;
ri2 = ri.^2;

u = e*ko^2 - m^2*ri2;
U = spdiags(u,0,N,N);
Ri = spdiags(ri,0,N,N);

%============== TE-like modes
L2 = D2 - Ri*D + U;

%============== TM-like modes
ei = 1./e;
de = D*e;
dlne = de.*ei;
p = ri + dlne;
P = spdiags(p,0,N,N); 
L1 = D2 - P*D + U;

%================ off diagonal terms
a = 2*m*ri2;
u1 = e.*a;
u2 = ei.*(a + m*dlne.*ri);
U1 = spdiags(u1,0,N,N);
U2 = spdiags(u2,0,N,N);

%================ Final assembly
Ls = [L1, U1; U2, L2];
\end{lstlisting}

\subsubsection*{The matrix assembly for the exact solution of slab waveguides, TE and TM modes.}
\begin{lstlisting}
function [Ls, u] = f_FD_Construct_Matrix_SlabTE(r, e, ko)
% TE modes, BC missing for m = 0

% r - grid vector
% ki = k_o n
% m - if zero we get symmetric function: d_r u(r) = 0

N = length(r)               % number of elements r
h = (r(end) - r(1))/N;      % the step size

%================== Construct d_x matrix ==================================
B = zeros(N,3);
B(:,1)= -1;
B(:,3)= 1;
D = 1/(2*h)*spdiags(B,[-1,0,1],N,N); 

%================== Construct d_x^2 matrix ================================
B = ones(N,3);
B(:,2)= -2;
D2 = 1/(h^2)*spdiags(B,[-1,0,1],N,N); 

%================== Construct dioganal matrices ===========================
% vectors
u = e*ko^2;		         % Potential barier
% sparse matrices
U = spdiags(u,0,N,N);

%=================== The Operator Matrix ==================================
% (d_r^2 + U)u(r) = b^2 u(r)
% U = e*ko^2

Ls = D2 + U;
%Ls = D*D + U;		% Fail
\end{lstlisting}
\begin{lstlisting}
function [Ls, u] = f_FD_Construct_Matrix_SlabTM(r, k2)
%TM modes, BC missing for m = 0

% r - grid vector
% ki = k_o n
% m - if zero we get symmetric function: d_r u(r) = 0

N = length(r)               % number of elements r
h = (r(end) - r(1))/N;      % the step size

%================== Construct d_x matrix ==================================
B = zeros(N,3);
B(:,1)= -1;
B(:,3)= 1;
D = 1/(2*h)*spdiags(B,[-1,0,1],N,N); 

%================== Construct d_x^2 matrix ================================
B = ones(N,3);
B(:,2)= -2;
D2 = 1/(h^2)*spdiags(B,[-1,0,1],N,N); 

%=================== The Operator Matrix ==================================

u = k2;		         % Potential barier
e = k2.';

U = spdiags(e,0,N,N);

de = D*e;
d2e = D2*e;

dlne = de./e;
Dlne = spdiags(dlne,0,N,N);

d2lne = d2e./e - dlne.^2;
D2lne = spdiags(d2lne,0,N,N);

Ls = D2 + Dlne*D + D2lne + U;
\end{lstlisting}

\subsubsection*{The matrix diagonalization routine.}
\begin{lstlisting}
function [V, Eig] = f_FD_eig(Ls, Neig, U)
% find eigenvalues and sort them

% Ls - sparse operator matrix
% U - potential barrier, to cut unbounded eigenvalues
% Neig - number of eigenvalues to find.

Umax = max(U);    % EXTREAMLY IMPORTENT: find them close to the potential at the centre.
Umin = U(end);

% Find eigenvalues
tic; 
[V,Eig] = eigs(Ls, Neig, Umax);          % [L] u = lam u
toc;
Eig = diag(Eig);                          % creates vector from L's  diagonal elements

%Eig

% need only with positive real part 
b = real(Eig)>0;
Eig = Eig(b);
V = V(:,b);

% need only bounded eigenvalues
b = ((real(Eig)>Umin) & (real(Eig) < Umax))
Eig = Eig(b)
V = V(:,b);

% Sort Eiganvalues and eigenvectors
[t, ind] = sort(Eig, 'descend');
Eig = Eig(ind);                          % sorted eigenvalues
V = V(:,ind);                            % correspondent eigenvectors to sorted lambdas

% Remove Degenerecy
% v = real(Eig);
% d = v(1:end-1) - v(2:end);
% b = abs(d) > 0.0001;
% Eig = Eig(b);
\end{lstlisting}

\subsubsection*{Construct the waveguide profile.}
\begin{lstlisting}
function ni = f_construct_ni(RL, nL, N)
% Construct step index potential barrier
h = RL(end)/N;

ni = nL(1)*ones(1,N);
for k=1:length(nL)-1
    round(RL(k)/h)
    ni(round(RL(k)/h):end) = nL(k+1);
end
\end{lstlisting}

\begin{lstlisting}
function ni = f_construct_ni2(RL, nL, N)
% ramp instead of steps potential barrier

h = RL(end)/N;
d=7;            % number of steps for jump

ni = nL(1)*ones(1,N);
for k=1:length(nL)-1
    p = round(RL(k)/h);
    n1 = nL(k);
    n2 = nL(k+1);
    ni(p:end) = n2;     % next layer

    % slow ramp, nit a jump.    
    ni(p-d:p+d-1) = linspace(n1, n2, d*2);
end
\end{lstlisting}

\subsubsection*{Plot eigenfunctions(modes) and eigenvalues (effective refractive indexes) for scalar and vectorial cases.}
\begin{lstlisting}
function f_Plot_vec(m, r, U, Eig, Veig, scale, ymin, ymax)
% plot potential well, eigenvalues and eigenfunctions

N = length(r);
h = r(3)-r(2);

clf
%plot(r, ki2, 'k')
h1 = plot(r, U, 'g','LineWidth',1)

hline(real(Eig))
hold on

Neig = length(Eig)
for k=1:Neig
       
        y = abs(Veig(1:N,k)).^2;    % E_r component
        y = Eig(k) + y/sqrt(h)*scale;
        plot(r, real(y), 'r-','LineWidth', 0.5)
        hold on
       
        y = abs(Veig(N+1:end,k)).^2;    % E_f component
        y = Eig(k) + y/sqrt(h)*scale;
        plot(r, real(y), 'b','LineWidth', 0.5)
        hold on
       
end

%================== Labling ===============================================
hOx = ylabel('n, N_{eff}');             
hOy = xlabel('\rho,  \mu m');

s = sprintf('m = %u',m)
hTitle = title(s);
set( [hOx, hOy, hTitle], 'FontWeight','normal', 'FontSize', 12); 

dy = 0.04;
Vy = ymin:dy:ymax-dy;                % Y ticks

%axis([r(1),r(end),ymin,ymax]);
axis([-3.5,3.5,ymin,ymax]);

Vy = ymin:dy:ymax-dy;                % Y ticks

%==================== Axis =================================================
set(gca, ...
  'Box'         , 'off'     , ...
  'FontSize'    , 8         , ...
  'FontName'    , 'Arial'   , ...
  'TickDir'     , 'out'     , ...
  'TickLength'  , [.02 .02] , ...
  'XMinorTick'  , 'on'      , ...
  'YMinorTick'  , 'on'      , ...
  'YGrid'       , 'off'      , ...
  'XGrid'       , 'off'     , ...
  'YTick'       , Vy, ...
  'XScale'      , 'linear'   ,...        
  'LineWidth'   , 1         );

%=================== Print ================================================
set(gcf, 'PaperPositionMode', 'manual');
set(gcf, 'PaperUnits', 'inches');
set(gcf, 'PaperPosition', [0 0 4 3]);       % defines the aspekt ratio
print('-dpng', '-r300', 'test.png')         % saves file in cur. dir
set(gcf, 'Color', 'w');
\end{lstlisting}
\subsubsection*{Orthogonality verification.}
\begin{lstlisting}
function f_Plot_Orthogonality_Check(V,r)
% plot overlap matrix

dim = size(V)                       % number of points, and eigenfunctions

rsqr = sqrt(r)

for k=1:dim(2)
    v = V(:,k);
    v = v.*rsqr';                   % r - is weighted function
    v = v/sqrt(sum(v.*v));          % normed, already weighted functions
    V(:,k) = v;
end 

M = V'*V;
figure
clf, imagesc(M), colorbar           % orthogonality check
norm = sqrt(sum(sum(M)))
\end{lstlisting}

\subsubsection*{Dispersion curves.}
\begin{lstlisting}
clear all
clc

%for m = 2:6
m = 2

N = 15000;      % number of elements in the grid
Neig = 15;     % number of eigenvalues
Np = 30;      % number of points along x (40 is OK)

%n1 = 1.4492; n2 = 1.3556;   % Glass in Si in water, refractive index
n1=3; n2 = 1.3556;           % Si in water

Vmin = 0.1;
Vmax = 10;

lam = 1.6;          % wavelength um
ko = 2*pi/lam;      % free space wave vector

V = linspace(Vmin, Vmax, Np); 
x = V/sqrt(n1^2 - n2^2);

M = zeros(Neig+1, Np);
M(end,:) = V;                 % save position in the same matrix

for i=1:Np
    i
    
    R = x(i)/ko
    RL = [R, R + 10];
    nL = [n1, n2];
    %============== Potential =================================

    h = RL(end)/(N-0.5);
    r = linspace(0.5*h, RL(end), N);
    r = [- fliplr(r), r].';
    r = r - 0.1*h; 

    ni = f_construct_ni2(RL, nL, N);
    ni = [fliplr(ni), ni].';

    e = ni.^2;

    %============== Solve =================================
    %[Ls, U] = f_FD_Construct_Matrix_Cyl_WGA_Scalar(r, e, ko, m);
    %[Ls, U] = f_FD_Construct_Matrix_Cyl_WGA_Vect(r, e, ko, m);
    [Ls, U] = f_FD_Construct_Matrix_Cyl_Exact_Vect(r, e, ko, m);
    
    %[Ls, U] = f_FD_Construct_Matrix_SlabTM(r, e, ko);
    %[Ls, U] = f_FD_Construct_Matrix_SlabTE(r, e, ko);
    
    [V, Eig] = f_FD_eig(Ls, Neig, U);

    neff2 = Eig/ko^2
    
    b = (neff2 - n2^2)/(n1^2 - n2^2);
    M(1:length(neff2),i) = b;

end

save(['GL_E_', int2str(m)], 'M');
%save(['SiEx_', int2str(m)], 'M');
%save(['TE_', int2str(m)], 'M');
%save(['TM_', int2str(m)], 'M');
\end{lstlisting}
\begin{lstlisting}
clear all
clc
clf

s = 'GL_S_1'

C = 'r-'
N = 16;
n=1;
for k=1:N
    f_dispersion_Plot(s,C,n)
    n = n+1;
end    

%=================== Set Plot ===============================
axis([0,10,0,1])

hOy = ylabel('Normalized propagation constant b');             
hOx = xlabel('Normalized frequency V');
%hL = legend('Exact', '', 'WGA','location', 'NorthWest')
%legend('boxoff')

%s = sprintf('Dispersion curves')
hTitle = title('Dispersion curves');
set( [hOx, hOy, hTitle], 'FontWeight','normal', 'FontSize', 10); 

set(gca, ...
  'Box'         , 'off'     , ...
  'FontSize'    , 8         , ...
  'FontName'    , 'Arial'   , ...
  'TickDir'     , 'out'     , ...
  'TickLength'  , [.02 .02] , ...
  'XMinorTick'  , 'on'      , ...
  'YMinorTick'  , 'on'      , ...
  'YGrid'       , 'off'      , ...
  'XGrid'       , 'off'     , ...
  'YTick'       , 0:0.2:1, ...
  'LineWidth'   , 1         );

set(gcf, 'PaperPositionMode', 'manual');
set(gcf, 'PaperUnits', 'inches');
set(gcf, 'PaperPosition', [0 0 5 3]);       % defines the aspekt ratio
print('-dpng', '-r300', 'test.png')         % saves file in cur. dir
set(gcf, 'Color', 'w');

\end{lstlisting}
\begin{lstlisting}
function f_dispersion_Plot(path, col, N)
% path   the path to data
% col    color

load(path);

V = M(end,:);
y = real(M(N,:));

plot(V, y, col)
hold on 
\end{lstlisting}


%% file: Append_Code_Mie.tex
\label{Mathematica_code}


The following code was used to calculate absorption, scattering and extinction coefficients of a sphere particle submerged into solvent.

\begin{figure}[!htb]\centering 
	\includegraphics[width=1\textwidth]{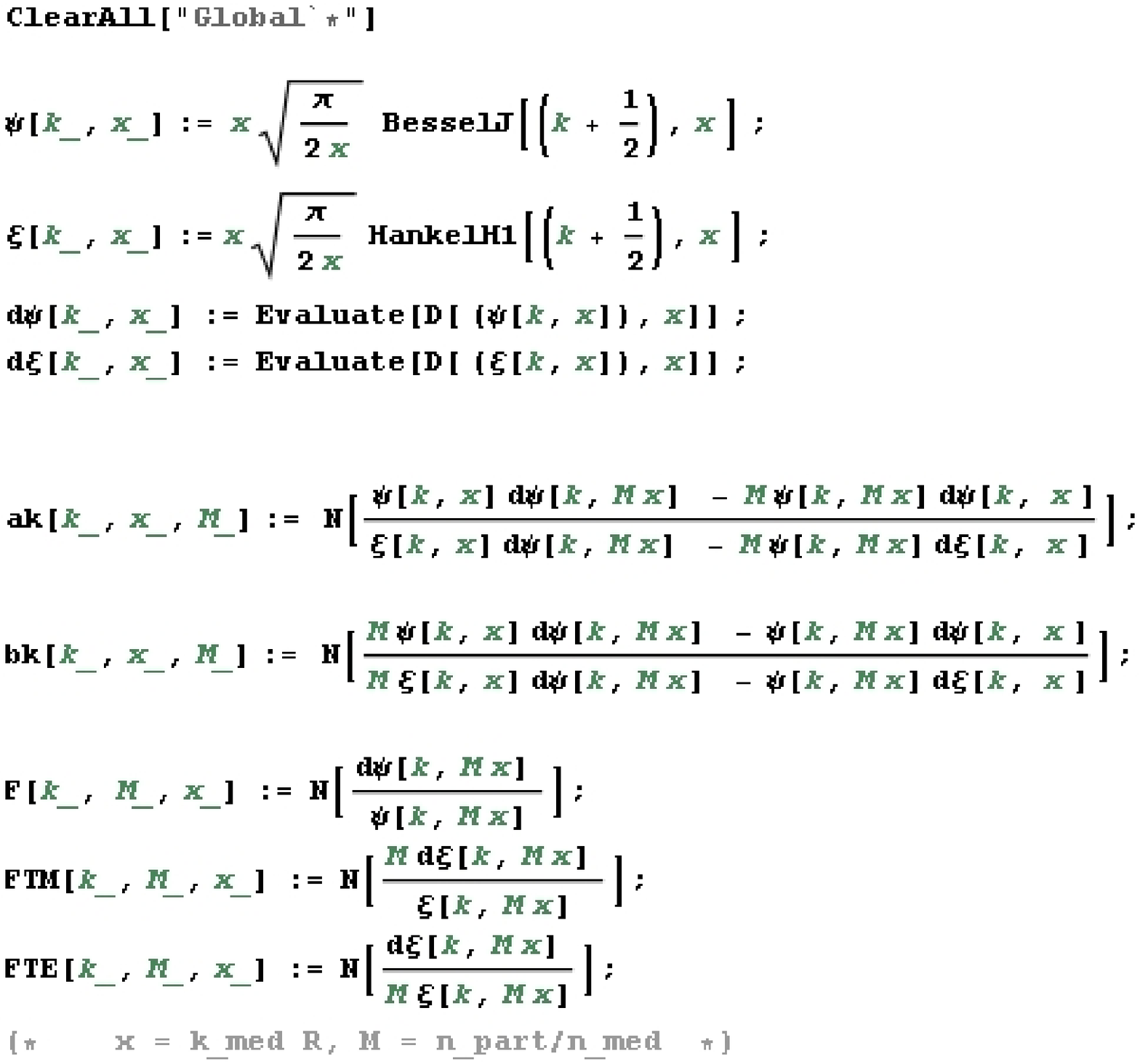}
		\label{Append_Mie} 
\end{figure}

\begin{figure}[!htb]\centering 
	\includegraphics[width=1\textwidth]{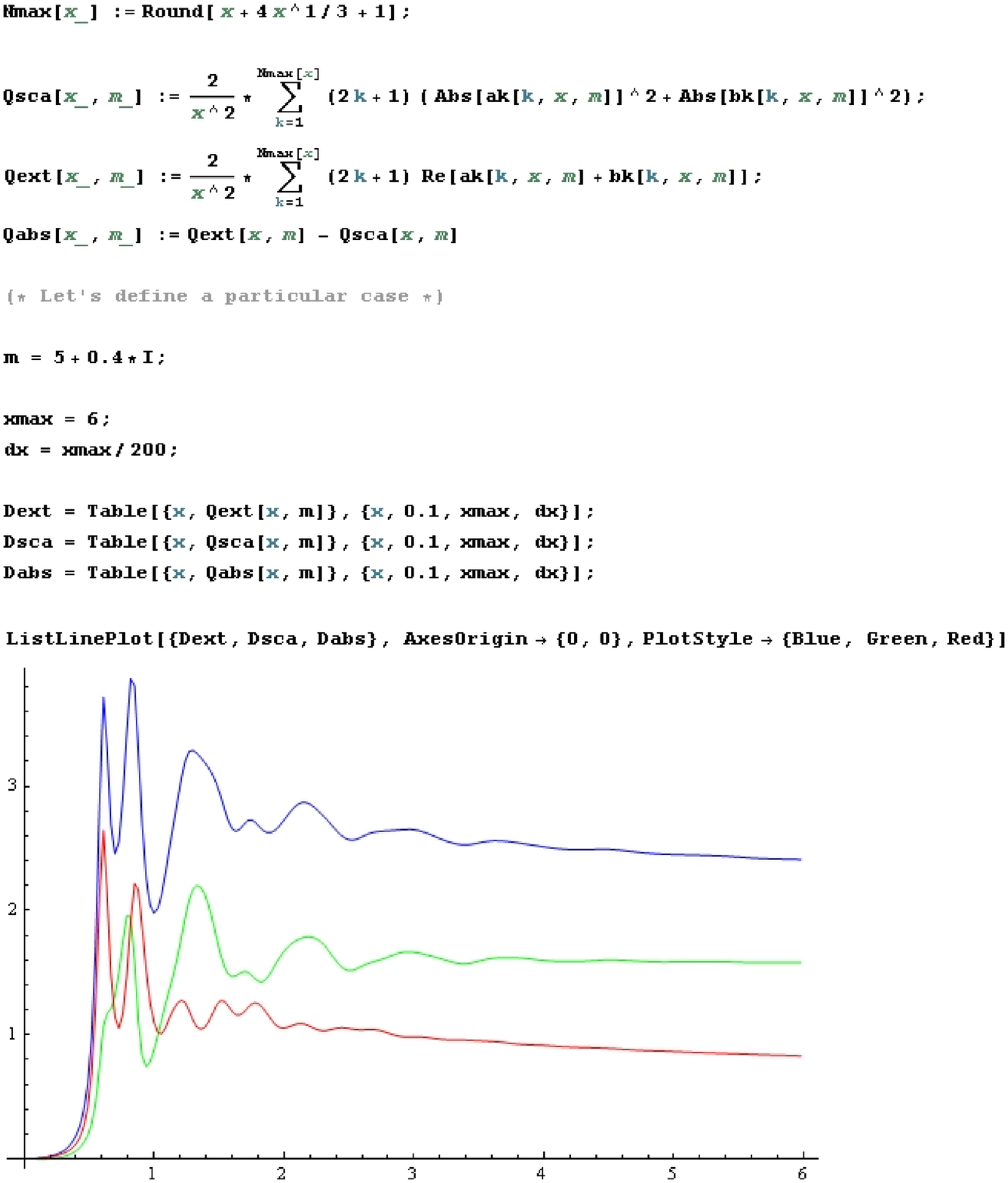}
\end{figure}